\documentclass[aps,prd]{revtex4-2}
\usepackage{graphicx}
\usepackage{amssymb}
\usepackage{amsmath}
\usepackage{float}

\begin{document}

\title{Gravitational Lensing in the Kerr Spacetime: An Analytic Approach for Light and High-Frequency Gravitational Waves}

\author{Torben C. Frost}

\affiliation{Kavli Institute for Astronomy and Astrophysics, Peking University, 100871 Beijing, China.\\
e-mail: torben.frost@pku.edu.cn}

\date{November 20th, 2024}

\begin{abstract}
The Kerr spacetime is one of the most widely known solutions to Einstein's vacuum field equations and is commonly used to describe a black hole with mass $m$ and spin $a$. Astrophysical observations in the electromagnetic spectrum as well as detected gravitational wave signals indicate that it can be used to describe the spacetime around candidates for rotating black holes. While the geodesic structure of the Kerr spacetime is already well known for decades, using exact analytic solutions to the equations of motion for applications to astrophysical problems has only attracted attention relatively recently. Here, these applications mainly focus on predicting observations for the shadow, the photon rings, and characteristic structures in the accretion disk. Using the exact analytic solutions to investigate exact gravitational lensing of light and gravitational waves emitted by sources outside the accretion disk has only received limited attention so far. Therefore, the focus of this paper will be to address this question.\\
For this purpose we assume that we have a standard observer in the domain of outer communication. We introduce a local orthonormal tetrad to relate the constants of motion of light rays and high-frequency gravitational waves detected by the observer to latitude-longitude coordinates on the observer's celestial sphere. In this parameterisation we derive the radius coordinates of the photon orbits and their latitudinal projections onto the observer's celestial sphere as functions of the celestial longitude. We use the latitude-longitude coordinates to classify the different types of motion, and solve the equations of motion analytically using elementary and Jacobi's elliptic functions as well as Legendre's elliptic integrals. We use the analytic solutions to write down an exact lens equation, and to calculate the redshift and the travel time. We discuss their sky projections for different stationary sources and the implications of the results in the context of conventional astronomical observations and the observation of high-frequency gravitational waves gravitationally lensed by supermassive black hole candidates. 
\end{abstract}

\maketitle
\section{Introduction}
In the last ten years observational studies of astrophysical black hole candidates made significant advances. The first direct detection of gravitational waves by the advanced Laser Interferometer Gravitational-Wave Observatory (LIGO) \cite{Abbott2016a} did not only open up the possibility to observe the Universe through a new window, it also allowed direct access to probing gravity in the strong field regime for the first time. The detected gravitational wave signal indicated that it was emitted during the final merger stage of a stellar mass binary black hole system and subsequent analyses showed that within the estimated uncertainties the components of the binary black hole system, the merger itself, and the ringdown signal from the merger product can be described by general relativity \cite{Abbott2016b}. After the first discovery so far only a handfull other observing runs have been conducted. However, already during the first three runs alone more than 90 gravitational wave signals from merging compact objects, predominantly black hole candidates but also a few neutron stars, have been detected \cite{Abbott2019a,Abbott2021a,Abbott2023}. Using the detected signals intensive tests of general relativity have been performed but so far no significant deviations from Einstein's theory of general relativity have been found (see, e.g., the works of B. P. Abbott \emph{et al.} and R. Abbott \emph{et al.} for tests of general relativity using the gravitational wave transient catalogs GWTC-1 \cite{Abbott2019b} and GWTC-2 \cite{Abbott2021b}).\\
While the detected gravitational wave signals were all emitted by systems containing compact objects in the stellar mass range, in 2019 and in 2022 the Event Horizon Telescope Collaboration published the first images of the shadows of the supermassive black hole candidates at the centres of the galaxy M87 \cite{EHTCollaboration2019a} and the Milky Way \cite{EHTCollaboration2022}, respectively. Observed using Very Large Baseline Interferometry (VLBI) with a nominal angular resolution of about 25$\mu$as at about 230~GHz \cite{EHTCollaboration2019b} both images show a dark area, interpreted as the shadow of a black hole, surrounded by a bright ring which is commonly interpreted as light emitted by an accretion disk. While the shadow itself is a strong hint that both supermassive compact objects are black holes, for both objects the boundary of the shadow is strongly blurred and thus we cannot determine its exact shape. Therefore, in both cases only the accretion disks and their structures allowed to determine, in comparison with results from numerical accretion disk models, that they agree with those expected for a Kerr black hole \cite{Kerr1963} from general relativity.\\
While currently the information we can infer from gravitational wave and Event Horizon Telescope data is rather limited, the next generations of gravitational wave detectors such as the Einstein Telescope \cite{Punturo2010} and the Cosmic Explorer \cite{Evans2021}, and of the Event Horizon Telescope \cite{Johnson2023} are already in their planning stage and will help to improve the sensitivity, the spectral range, and the sky localisation (for an estimation for gravitational waves detected by a third generation three-detector network see, e.g., Li \emph{et al.} \cite{Li2022}) or the angular resolution (for the next-generation Event Horizon Telescope) of the observations. However, while the information we can infer from gravitational wave signals is mainly determined by the detector specifications, their arm length, the strength of the detected gravitational wave signals, and only to a limited degree their distribution on Earth, the  resolution very large baseline interferometric observations can achieve is limited by the length of the longest baseline, which roughly corresponds to the diameter of Earth. Therefore, a significant increase in resolution can only be achieved by extending VLBI to space. For this purpose recently the Black Hole Explorer was proposed \cite{Johnson2024}. While the Black Hole Explorer has a broad range of scientific targets one of its main objectives is to observe what is commonly referred to as \emph{photon rings}. They are formed by light rays emitted in the accretion disk surrounding a black hole passing close to unstable photon orbits and then leaving the close vicinity of the black hole. That gravitational lensing of light emitted by accretion disks around black holes leads to the formation of ring-like images is already known since the 1970s, see, e.g., the work of Luminet \cite{Luminet1979}, and received increased attention after the first Event Horizon Telescope images of the shadows of the supermassive black hole candidates at the centres of the Milky Way and the galaxy M87 were released, see, e.g., Johnson \emph{et al.} \cite{Johnson2020}. The photon rings are expected to carry characteristic signatures of the spacetime describing the black hole, see, e.g., Johannsen and Psaltis \cite{Johannsen2010}, and thus will not only help us to characterise the nature of astrophysical black hole candidates but also to probe gravity in the strong field regime.\\
However, while the observation of the photons rings has the potential to be a characteristic probe of gravity in the strong field regime it is still limited by the sensitivity and the angular resolution of the single telescopes comprising the VLBI array. Therefore, the observation of electromagnetic radiation and gravitational waves (in the case of gravitational waves in the LIGO/Virgo/KAGRA band gravitationally lensed by a supermassive black hole we are in the high-frequency limit and thus the gravitational waves move along lightlike geodesics \cite{Isaacson1968}) emitted by isolated sources such as stars, neutron stars as well as stellar mass binary compact object systems orbiting a supermassive black hole will provide important complementary information to determine the correct spacetime and the correct theory of gravitation for describing the spacetime of the black hole. Therefore, the main aim of this paper will be to investigate how gravitationally lensed images (or in the case of gravitational waves signals) of isolated sources can be used to probe gravity in the strong field regime, and, in our particular case, which information they can provide to determine the spin of a black hole.\\
For this purpose we will assume that we have a rotating black hole described by the Kerr spacetime from general relativity \cite{Kerr1963}. The Kerr spacetime is axisymmetric and stationary and contains two different physical parameters, the mass parameter $m$ and the spin parameter $a$. While in general the spin parameter $a$ can be arbitrarily large in this paper we are only interested in black hole spacetimes and using the symmetries of the Kerr spacetime we limit its range to $0\leq a\leq m$.\\
Starting with the seminal works of Carter \cite{Carter1968} on the separability of the equations of motion and their basic properties, and of Bardeen \cite{Bardeen1973} on timelike and lightlike geodesic motion, including lightlike geodesics on photon orbits, timelike and lightlike geodesic motion in the Kerr spacetime has been extensively discussed in the literature. In the following we will provide a brief overview of the most relevant analytically exact approaches for investigating geodesic motion along lightlike geodesics. \\
While the fact that the Kerr spacetime possesses unstable photon orbits in the domain of outer communication and that these give rise to a deformed shadow is known since the seminal work of Bardeen \cite{Bardeen1973}, Teo \cite{Teo2003} has been one of the first to systematically investigate the motion on such orbits in the domain of outer communication. However, it has only been recently that Tavlayan and Tekin \cite{Tavlayan2020} derived an exact analytic expression for the conditional equation for the radius coordinates of the photon orbits in the form of a polynomial of sixth order containing the mass and the spin parameters as well as an effective inclination angle.\\
Also the investigation of lightlike geodesic motion in the Kerr spacetime, in particular solving the equations of motion, received a lot of attention. A summary of the work until the 1980s can be found in the book of Chandrasekhar \cite{Chandrasekhar1992}. However, even after these early works methods for solving the equations of motion of the Kerr spacetime analytically have been extensively discussed in the literature. While in some special cases the solutions to the equations of motion can be derived solely in terms of elementary functions in most cases they can only be obtained in terms of elementary and elliptic functions as well as elliptic integrals. Here, in general we have two main options. The first option is to use Jacobi's elliptic functions and Legendre's elliptic integrals, see, e.g., the work of Slez\'{a}kov\'{a} \cite{Slezakova2006} and Gralla and Lupsasca \cite{Gralla2020b}. The second option is to use Weierstra\ss' elliptic $\wp$ function and the associated $\zeta$ and $\sigma$ functions, see, e.g., Hackmann \emph{et al.} \cite{Hackmann2010a,Hackmann2010b} and Cie\'{s}lik, Hackmann, and Mach \cite{Cieslik2023}. While in principle we can use both methods to solve the equations of motion each has certain advantages and disadvantages. Jacobi's elliptic functions and Legendre's elliptic integrals on one hand have the advantage that the solutions to the equations of motion can be written such that they are explicitly real without any internally cancelling imaginary parts. However, they have the disadvantage that one has to derive the solutions for all types of motion separately. Using Weierstra\ss' elliptic $\wp$ function and the associated $\zeta$ and $\sigma$ functions on the other hand has the advantage that the solutions for the different types of motion can be written down in one unified form either with internally cancelling imaginary parts as in Hackmann \emph{et al.} \cite{Hackmann2010a,Hackmann2010b} or without them as in Cie\'{s}lik, Hackmann, and Mach \cite{Cieslik2023}. However, the approach of Hackmann \emph{et al.} \cite{Hackmann2010a,Hackmann2010b} has the disadvantage that for integrating the elliptic integrals associated with the time coordinate $t$ and the $\varphi$ coordinate we have to manually adjust the branches of the natural logarithm which occurs as part of the solutions while the approach of Cie\'{s}lik, Hackmann, and Mach \cite{Cieslik2023} has the disadvantage that many of the solutions become structurally very complex.\\
In this paper we want to apply the analytic solutions to gravitational lensing and thus we are only interested in the start and end points of the lightlike geodesics. Therefore, having to adjust the branch cuts along the whole geodesic is rather inconvenient and thus in addition to elementary functions we will use Jacobi's elliptic functions and Legendre's elliptic integrals to write down the analytic solutions to the equations of motion. However, while the authors of most previous studies used the constants of motion along the lightlike geodesics, namely the energy $E$, the angular momentum about the $z$-axis $L_{z}$, and the Carter constant $K$, in combination with the number and the positions of the real roots to classify the different types of motion, for our purpose it is more convenient to use a different approach. Since we are interested in investigating gravitational lensing and thus in how images of different sources are projected onto the celestial sphere of an observer in the domain of outer communication, instead of the constants of motion as far as possible we will use latitude-longitude coordinates on the observer's celestial sphere.\\
While in theory there is a whole range of different observers, in this paper we will use a standard observer (sometimes also referred to as Carter observer) as specified in Grenzebach, Perlick, and L\"{a}mmerzahl \cite{Grenzebach2015}. Compared to other observers it has two main advantages. On one hand it orbits the black hole with a constant angular velocity and thus can be placed in the whole domain of outer communication including the ergoregion. In addition, our Solar System, and with it Earth, orbits the centre of the Milky Way and thus it is more natural to assume a moving observer. Thus the work presented in this paper is also a first step towards a full analytic treatment of gravitational lensing of light rays and high-frequency gravitational waves detected by observers on arbitrary geodesic orbits around a Kerr black hole. On the other hand the latitude-longitude coordinates on the observer's celestial sphere are defined such that the celestial latitude is measured from the line connecting the observer and $r=0$. For light rays and gravitational waves observed tangentially to this line the Carter constant vanishes. In addition, for a fixed celestial longitude we can use the latitudinal projections of the photon orbits onto the observer's celestial sphere to distinguish the different types of $r$ motion. As we will see in the course of this paper both points together have as result that we obtain a straightforward and systematic classification scheme which will also be useful for ray tracing codes. Finally, when we place the observer at radius coordinates very far away from the black hole the orthonormal tetrad of the standard observer asymptotically approaches the orthonormal tetrad of an observer on a $t$-line. As a consequence it can be used to investigate lightlike geodesic motion and gravitational lensing in the weak and strong field regimes without the need to introduce different orthonormal tetrads. Therefore, discussing and applying this classification scheme and deriving the analytic solutions to the equations of motion for light rays and high-frequency gravitational waves travelling along lightlike geodesics will be the main objective of the first part of this paper. In the second part of this paper we will then use the derived solutions to discuss gravitational lensing of light and high-frequency gravitational waves in the Kerr spacetime. \\
Due to its astrophysical relevance for the Kerr spacetime gravitational lensing in the weak and strong field limits as well as exact has already been extensively discussed by several authors. We provide a brief overview of the most important works in the following.\\
Cunningham and Bardeen \cite{Cunningham1973} were among the first to investigate exact gravitational lensing in the Kerr spacetime. For stars on selected circular orbits in the equatorial plane of an extremal Kerr black hole they calculated the projection of the trajectories of the direct and of what they referred to as one-orbit images (the image for which the light ray crosses the equatorial plane exactly once) onto the celestial sphere of a distant observer and also different light curves. Rauch and Blandford \cite{Rauch1994} investigated the caustic surfaces for a distant observer in the equatorial plane of the Kerr spacetime. They found that it is a tube which has a rhombus-like cross section. In addition, they also calculated the light curves of a star crossing the caustic. V\'{a}zquez and Esteban \cite{Vazquez2004} used a combination of numerical, analytically exact, and approximative methods to investigate gravitational lensing in the Kerr spacetime. For a distant observer and distant light sources they derived the position of the lowest order relativistic images on the observer's celestial sphere as well as their magnification as a function of the spin. Vincent \emph{et al.} \cite{Vincent2011} implemented a multi-purpose numerical ray tracing code which is able to calculate gravitationally lensed images of a star orbiting and different accretion disks surrounding a Kerr black hole. While many of the early works were based on numerical approximation methods Yang and Wang \cite{Yang2013} were among the first who used elliptic integrals and functions in their ray tracing code.\\
Kraniotis \cite{Kraniotis2011} used Weierstra\ss' elliptic $\wp$ function and Appell-Lauricella's hypergeometric functions to write down the components of a lens equation and applied it to calculate the positions of the light sources and their projections onto the observer's sky for light rays solving the lens equation. In addition, Kraniotis also analytically calculated the magnification factor for generic photon orbits. Bohn \emph{et al.} \cite{Bohn2015} investigated numerically how the merger of a binary black hole looks like. For this purpose they first placed an observer in the domain of outer communication outside the ergoregion of a Kerr black hole. Then they defined a lens map for light sources located on a two-sphere surrounding the observer and the black hole. They divided the two-sphere into four differently coloured quadrants and calculated their projection onto the celestial sphere of the observer. As first step they calculated lens maps for isolated Schwarzschild and Kerr black holes before they calculated lens maps for different binary black hole systems. While the shadow is one of the most characteristic features of a black hole until recently for the Kerr black hole only very few specific cases were known for which one can derive the boundary curve analytically exact. However, Cunha and Herdeiro \cite{Cunha2018} showed that this is generally possible for an observer on a $t$-line at arbitrary inclinations at spatial infinity.\\ 
Most recently Gralla and Lupsasca \cite{Gralla2020a} used the analytic solutions to the equations of motion in terms of elementary and Jacobi's elliptic functions and Legendre's elliptic integrals they derived in \cite{Gralla2020b} to investigate how a distant observer would see light rays emitted by sources close to the event horizon or at spatial infinity. In addition, they calculated the front and back side images of light sources at constant radius coordinates in the equatorial plane, and approximative relations for the Lyapunov exponent, the lapse in the azimuthal angle, and the lapse in the travel time.\\
While usually the investigation of gravitational lensing in the Kerr spacetime, in particular in the exact geometric optics formalism, is limited to light rays, Zhang and Chen \cite{Zhang2023} used it to numerically investigate gravitational lensing of high-freqency gravitational waves emitted by a stellar mass binary black hole system orbiting a supermassive black hole. In particular they addressed the question if the detected signal of gravitationally lensed gravitational waves emitted by a low mass binary black hole system orbiting a supermassive black hole within a few gravitational radii can appear as if it was a signal from unlensed gravitational waves emitted by a higher mass stellar mass binary black hole system and thus explain corresponding LIGO/Virgo/KAGRA gravitational wave detections.\\
However, all previous investigations mentioned above have at least one of the following limitations. They calculate the trajectories of the light rays or gravitational waves using numerical approximation methods, they assume that the observer is at rest and located in the asymptotically flat region of the spacetime far away from the central black hole, or they combine analytically exact with approximated results, and in particular they only investigate lower order images (commonly only up to order two or three).\\ 
Therefore, in this paper we will not use these assumptions and approaches in our investigation of gravitational lensing in the Kerr spacetime and proceed as follows. First of all we will use the exact analytic solutions to the equations of motion of the Kerr spacetime in terms of elementary and Jacobi's elliptic functions and Legendre's elliptic integrals. We assume that the observer as well as the light or gravitational wave sources are located at finite radius coordinates with respect to the black hole. Since we assume that we have a standard observer this implies that the observer orbits the black hole at a constant angular velocity at a constant radius coordinate $r_{O}$ and a constant spacetime latitude $\vartheta_{O}$. In addition, we will assume that we have three different types of stationary sources, namely sources moving along $t$-lines (in the following we will refer to them as \emph{static} to distinguish them from the other stationary sources, however, we note that since their four-velocity is not orthogonal to the spacelike hypersurfaces they are actually stationary), zero angular momentum (or Bardeen) sources, and standard (or Carter) sources.\\
For this purpose in the first part of this paper we will write down the equations of motion for lightlike geodesics and relate the constants of motion of light rays and gravitational waves travelling along the geodesics to latitude-longitude coordinates on the observer's celestial sphere. For this purpose we will place the standard observer in the domain of outer communication and introduce an orthonormal tetrad following the approach of Grenzebach, Perlick, and L\"{a}mmerzahl \cite{Grenzebach2015}. We will then use the obtained relations to derive the radius coordinates of the photon orbits in the stationary regions of the spacetime as functions of the celestial longitude. Note that since we also include the extremal Kerr spacetime in some cases the radius coordinates of the photon orbits coincide with the radius coordinate of the horizon. In Boyer-Lindquist coordinates this is a result of the choice of the coordinate system. As pointed out by Bardeen, Press, and Teukolsky \cite{Bardeen1972} while in terms of the Boyer-Lindquist coordinates these photon orbits have the same radius coordinate as the horizon the embedding diagram shows that they are clearly distinct from the horizon and thus in the following when we refer to photon orbits in the stationary regions this statements will also include them. We will show that for special cases the radius coordinates of the photon orbits as well as the boundary of the shadow can always be calculated purely analytically. In particular we will see that this is the case for an observer in the equatorial plane at arbitrary radius coordinates in the domain of outer communication.\\
We will then use this parameterisation to classify the different types of motion. We will solve the equations of motion analytically using elementary and Jacobi's elliptic functions as well as Legendre's elliptic integrals following the approaches presented in Gralla and Lupsasca \cite{Gralla2020b} and Frost \cite{Frost2022}. Note that Gralla and Lupsasca \cite{Gralla2020b} (and also Slez\'{a}kov\'{a} \cite{Slezakova2006}) already derived solutions to the equations of motion in terms of Legendre's elliptic integrals and Jacobi's elliptic functions. However, although their work is very thorough in some sense it is incomplete. They only derived the solutions for the most general cases and classified them in terms of the constants of motion and the roots. However, as we will see during the course of this paper when we deal with gravitational lensing classifying the types of motion, in particular the different types of $r$ motion, as far as possible in terms of the latitude-longitude coordinates on the observer's celestial sphere allows to perform very high-resolution calculations and a much more straightforward implementation of ray tracing codes. In addition, Gralla and Lupsasca \cite{Gralla2020b} rescaled the angular momentum about the $z$-axis $L_{z}$ and the Carter constant $K$ by the energy $E$ which does not allow a straighforward charactersation when the energy along the lightlike geodesics $E$ is either zero or negative. Thus in this paper we will keep the constants of motion $E$, $L_{z}$, and $K$ without any further rescalings.\\
In this paper we will now complement the work of Slez\'{a}kov\'{a} \cite{Slezakova2006} and Gralla and Lupsasca \cite{Gralla2020b} in several ways. First of all we will parameterise the equations of motion using latitude-longitude coordinates on the celestial sphere of the standard observer. In this parameterisation we then classify the different types of motion. Here, we are mainly interested in investigating gravitational lensing of light rays and gravitational waves. Therefore, we can safely assume that the observer is located outside the ergoregion and thus in this paper we focus on lightlike geodesics with $E>0$. We systematically solve the equations of motion for all lightlike geodesics with $E>0$ using elementary and elliptic functions and Legendre's elliptic integrals. Here, we include the most general cases already discussed by Slez\'{a}kov\'{a} \cite{Slezakova2006} and Gralla and Lupsasca \cite{Gralla2020b}, the special cases which occur for a vanishing Carter constant, and the lightlike geodesics which are characterised by the same constants of motion as lightlike geodesics on stable and unstable photon orbits in the domain of outer communication and in the stationary region inside the Cauchy horizon. As a result this paper will provide a comprehensive overview of the solutions to the equations of motion for light rays and high-frequency gravitational waves with $E>0$.\\
In the second part of this paper we will then use the analytic solutions to investigate gravitational lensing of light rays and high-frequency gravitational waves in the Kerr spacetime. We will write down a lens equation, derive the redshift, and the travel time. We will discuss their interpretation for different stationary light or gravitational wave sources with particular focus on higher order images and gravitational wave signals for different values of the spin. With respect to the lens equation we will see that with increasing spin more and more higher order images become visible. We will discuss the physical reasons for their presence and if and how one potentially may be able to observe them using Event Horizon Telescope or Black Hole Explorer observations. In addition, we will also discuss implications for the observation of gravitationally lensed high-frequency gravitational waves emitted by merging stellar mass binary black holes, binary neutron stars, and black hole-neutron star binaries, as well as continuous gravitational waves emitted by neutron stars with asymmetries with respect to their axis of rotation orbiting a supermassive black hole.\\
The remainder of the paper is structured as follows. In Section~\ref{Sec:KST} we will briefly discuss the Kerr spacetime and its basic physical properties. In Section~\ref{Sec:EoM} we will first write down the equations of motion and relate the constants of motion to latitude-longitude coordinates on the celestial sphere of a standard observer. Then we will derive the radius coordinates of the photon orbits and their latitudinal projections onto the celestial sphere of the observer as functions of the celestial longitude. In Section~\ref{Sec:SolEoM} we will then use them to classify and discuss the different types of lightlike geodesic motion in the domain of outer communication, and solve the equations of motion. In Section~\ref{Sec:Lensing} we will investigate gravitational lensing in the Kerr spacetime. We will write down a lens equation, derive the redshift for a standard observer and static, zero angular momentum (or Bardeen), and standard (or Carter) sources, and calculate the travel time. We will plot them as functions on the observer's celestial sphere and discuss the visible features and their implications for astrophysical observations of the spin. In particular we will discuss the implications of the potential observability of very high order images for Kerr black holes with extremal spin and differences between gravitationally lensed electromagnetic radiation and gravitational waves. In Section~\ref{Sec:Sum} we will summarise our results and conclusions. Throughout the whole paper we will us the metric signature $(-,+,+,+)$ and geometric units such that $c=G=1$.\\
 
\section{The Kerr Spacetime}\label{Sec:KST}
\begin{figure}\label{fig:Horizons}
  \includegraphics[width=1.0\textwidth]{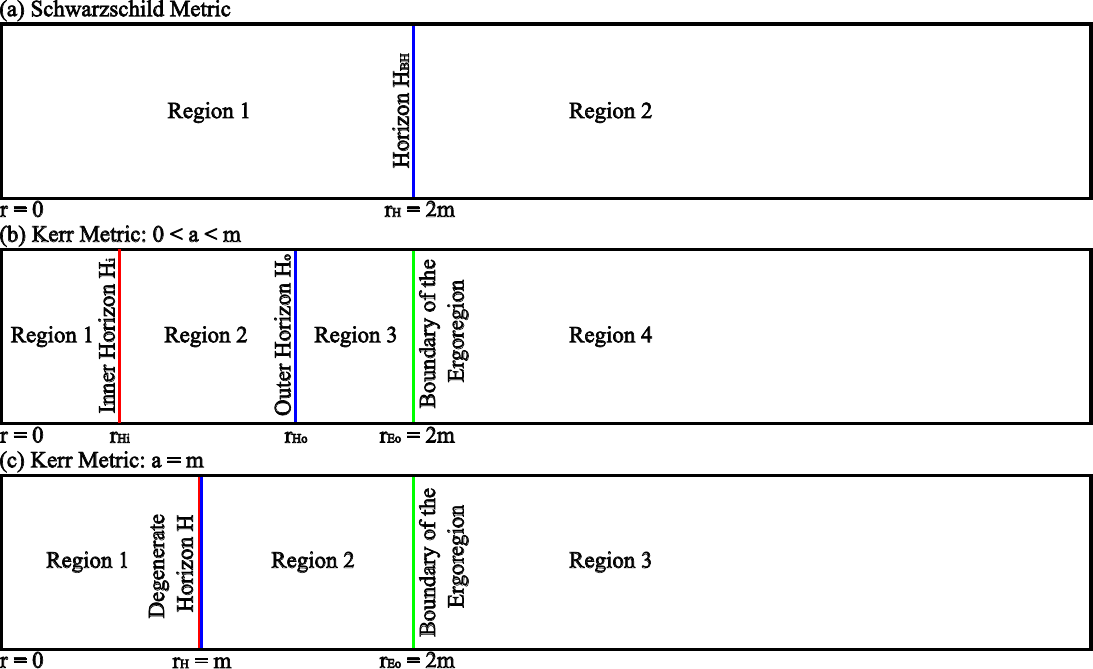}
\caption{Illustration of the spacetime structures of (a) the Schwarzschild spacetime and the Kerr spacetime in the equatorial plane for (b) $0<a<m$ and (c) $a=m$. In the equatorial plane of the Kerr spacetime the boundary of the inner ergoregion is located at $r_{\text{E}_{\text{i}}}=0$ and thus at the same radius coordinate as the curvature singularity.}
\end{figure}
The Kerr spacetime is axisymmetric and stationary and contains two physical parameters. The first parameter is the mass parameter $m$ and the second parameter is the spin parameter $a$. In Boyer-Lindquist coordinates \cite{Boyer1967} it reads
\begin{eqnarray}\label{eq:KerrMetric}
&g_{\mu\nu}\text{d}x^{\mu}\text{d}x^{\nu}=\frac{a^2\sin^2\vartheta-P(r)}{\rho(r,\vartheta)}\text{d}t^2+\frac{2a\sin^2\vartheta(P(r)-(r^2+a^2))}{\rho(r,\vartheta)}\text{d}t\text{d}\varphi\\
&+\frac{\sin^2\vartheta((r^2+a^2)^2-a^2\sin^2\vartheta P(r))}{\rho(r,\vartheta)}\text{d}\varphi^2+\frac{\rho(r,\vartheta)}{P(r)}\text{d}r^2+\rho(r,\vartheta)\text{d}\vartheta^2,\nonumber
\end{eqnarray}
where 
\begin{eqnarray}\label{eq:KerrCoeff}
P(r)=r^2-2mr+a^2~~~\text{and}~~~\rho(r,\vartheta)=r^2+a^2\cos^2\vartheta.
\end{eqnarray}
In the limit $a=0$ the line element of the Kerr spacetime (\ref{eq:KerrMetric}) reduces to the line element of the Schwarzschild spacetime.\\
Since we want the line element to represent black holes in space in this paper we assume that the mass parameter $m$ is always positive and thus we have $0<m$. In addition, from the line element (\ref{eq:KerrMetric}) we can easily see that when we replace $a \rightarrow -a$ and substitute $\varphi \rightarrow -\varphi$ the line element is invariant and thus we can choose the spin parameter $a$ such that we have $0\leq a$. In addition, the spacetime is invariant under time translations and rotations about the axis of rotation (usually referred to as $z$-axis). Here the invariance of the line element with respect to time translations is associated with the existence of a Killing vector field $\xi_{t}$ and the invariance of the line element with respect to rotations about the $z$-axis is associated with the existence of a Killing vector field $\xi_{\varphi}$. Both Killing vector fields are associated with the existence of a conserved quantity. In the case of the Killing vector field $\xi_{t}$ this quantity is the energy $E$ and in the case of the Killing vector field $\xi_{\varphi}$ this quantity is the angular momentum about the $z$-axis $L_{z}$.\\
In the following we will now briefly discuss and compare the spacetime structures of the Schwarzschild spacetime and the Kerr spacetime. As mentioned above in this paper we assume that the spacetime represents a black hole and therefore we limit our discussion to this case for the remainder of this paper.\\
The spacetime structures of the Schwarzschild spacetime and the equatorial plane of the Kerr spacetime are shown in Fig.~1. We start with discussing the structure of the Schwarzschild spacetime. It is shown in Fig.~1(a). The Schwarzschild spacetime is spherically symmetric and static and its line element has two singularities at the radius coordinates $r=0$ and $r_{\text{H}}=2m$. Here, the singularity at $r=0$ is a curvature singularity. On the other hand using an appropriate coordinate transformation the singularity at $r_{\text{H}}=2m$ can be transformed away and thus it is a coordinate singularity (we can also show this by calculating the Kretschmann scalar which at $r_{\text{H}}=2m$ remains finite). It marks the position of an event horizon which separates two different spacetime regions. Region 1 is located between the radius coordinates $r=0$ and $r_{\text{H}}$. In this region the vector field $\partial_{t}$ is spacelike while the vector field $\partial_{r}$ is timelike. Therefore, in this region the spacetime is nonstatic and the curvature singularity at $r=0$ is spacelike. Region 2 is located at radius coordinates $r_{\text{H}}<r$. In this region the vector field $\partial_{t}$ is timelike while the vector field $\partial_{r}$ is spacelike. Thus this region is static. This means that in region 2 massless (light rays and in our case also gravitational waves) and massive particles can move freely while when they cross the event horizon and enter region 1 they have to move inward towards the curvature singularity.\\
Figs.~1 (b) and (c) show the spacetime structures for the Kerr spacetime with $0<a<m$ and $a=m$ in the equatorial plane, respectively. For the chosen values of the spin parameter the spacetime can have up to two horizons. They are located at $r_{\text{H}_{\text{i}}}=m-\sqrt{m^2-a^2}$ and $r_{\text{H}_{\text{o}}}=m+\sqrt{m^2-a^2}$. For $0<m<a$ we have $r_{\text{H}_{\text{i}}}<r_{\text{H}_{\text{o}}}$. Here, the horizon at the radius coordinate $r_{\text{H}_{\text{o}}}$ is an event horizon and the horizon at the radius coordinate $r_{\text{H}_{\text{i}}}$ is a Cauchy horizon.
The horizons separate the region 1 located at the radius coordinates $r<r_{\text{H}_{\text{i}}}$ (note that in the equatorial plane we have $0<r<r_{\text{H}_{\text{i}}}$) and the regions 3 and 4 located at the radius coordinates $r_{\text{H}_{\text{o}}}<r$, in which the vector field $\partial_{r}$ is spacelike, from the region 2 located at the radius coordinates $r_{\text{H}_{\text{i}}}<r<r_{\text{H}_{\text{o}}}$, in which the vector field $\partial_{r}$ is timelike. For $a=m$ we have $r_{\text{H}}=r_{\text{H}_{\text{i}}}=r_{\text{H}_{\text{o}}}=m$ and thus we only have one horizon separating two regions in which $\partial_{r}$ is spacelike. When $m<a$ the spacetime does not possess horizons and thus we have a naked singularity. Since in this paper we want the spacetime to represent a black hole we now limit the choice of the spin parameter $a$ to $0\leq a\leq m$.\\
Because in the region outside the event horizon the motion of massless and massive particles is generally not limited it is commonly also referred to as \emph{domain of outer communication}. However, similarly to the Schwarzschild spacetime any particle which crosses the event horizon has to move inward through region 2 (for $0<a<m$) across the Cauchy horizon into region 1 (note that although it is an important question in this paper we do not concern ourselves with the stability of the Cauchy horizon).\\
While in the Schwarzschild spacetime the horizon is the only special feature we have to consider, this is different for the Kerr spacetime. In addition to the horizons the Kerr spacetime also possesses two ergoregions. The boundaries of the ergoregions are determined by the equation $a^2\sin^2\vartheta-P(r)=0$. When we solve this equation for $r$ we find that they are located at the radius coordinates
\begin{eqnarray}\label{eq:ergor}
r_{\text{E}_{\text{i}}}=m-\sqrt{m^2-a^2\cos^2\vartheta}~~~\text{and}~~~r_{\text{E}_{\text{o}}}=m+\sqrt{m^2-a^2\cos^2\vartheta}.
\end{eqnarray}
Thus, we have an inner ergoregion for $r_{\text{E}_{\text{i}}}<r<r_{\text{H}_{\text{i}}}$ [region 1 in Figs.~1(b) and 1(c); note that in the equatorial plane of the Kerr spacetime we have $r_{\text{E}_{\text{i}}}=0$] and an outer ergoregion for $r_{\text{H}_{\text{o}}}<r<r_{\text{E}_{\text{o}}}$ [region 3 in Fig.~1(b) and region 2 in Fig.~1(c)]. When we approach the axes at $\vartheta=0$ or $\vartheta=\pi$ the boundary of the inner ergoregion approaches the Cauchy horizon while the boundary of the outer ergoregion approaches the event horizon. On the axes the radius coordinates of the Cauchy horizon and the boundary of the inner ergoregion, and the radius coordinates of the event horizon and the boundary of the outer ergoregion coincide. Note that the latter two scenarios are not shown in Fig.~1. Here, for $r<r_{\text{E}_{\text{i}}}$ [note that in the equatorial plane the curvature singularity at $r=0$ separates the inner ergoregion from the region $r<r_{\text{E}_{\text{i}}}=0$] and $r_{\text{E}_{\text{o}}}<r$ [region 4 in Fig.~1(b) and region 3 in Fig.~1(c)] the vector field $\partial_{t}$ is timelike while for $r_{\text{E}_{\text{i}}}<r<r_{\text{E}_{\text{o}}}$ it is spacelike. Thus while in general in both ergoregions the direction of the $r$ motion is not limited any massless or massive particle entering one of the ergoregions cannot prevent to be dragged along by the black hole and thus in the ergoregions sources or observers moving along $t$-lines cannot exist.\\
This leads us directly to the question of the nature of the different spacetime patches. In the regions $r<r_{\text{E}_{\text{i}}}$ and $r_{\text{E}_{\text{o}}}<r$ the Killing vector field $\xi_{t}$ is timelike but not hypersurface orthogonal. Therefore, these patches of the spacetime are stationary. As already mentioned above in the patches $r_{\text{E}_{\text{i}}}<r<r_{\text{H}_{\text{i}}}$ and $r_{\text{H}_{\text{o}}}<r<r_{\text{E}_{\text{o}}}$ the Killing vector field $\xi_{t}$ is now spacelike, however, we can still construct a new vector field $\xi_{t\varphi}$ from the two Killing vector fields $\xi_{t}$ and $\xi_{\varphi}$ which is again a Killing vector field and timelike. Therefore, these patches of the spacetime are also stationary with respect to the new Killing vector field $\xi_{t\varphi}$. In the region $r_{\text{H}_{\text{i}}}<r<r_{\text{H}_{\text{o}}}$ on the other hand we cannot construct a timelike Killing vector field anymore and thus this region is nonstationary. For the remainder of this paper we will now agree on the following convention. Whenever we refer to the stationary parts of the spacetime we refer to both, the ergoregions and the parts of the spacetime in which $\xi_{t}$ is timelike. In the case that we refer to one of them specifically we will explicitly mention them.\\
In addition to the coordinate singularities the Kerr spacetime possesses a curvature singularity for $\rho(r,\vartheta)=0$. This condition is only fulfilled for $r=0$ and $\vartheta=\pi/2$. Therefore, while in the Schwarzschild spacetime the curvature singularity is spacelike and blocks all geodesics from passing from $0<r$ to $r<0$ in the Kerr spacetime this is different. In the Kerr spacetime the curvature singularity is timelike and has the shape of a ring and as long as massless or massive particles do not hit the curvature singularity they can smoothly pass through it.\\
Now the only thing left to discuss are the allowed coordinate ranges of the spacetime coordinates. In the domain of outer communication of the Kerr spacetime the allowed range for the time coordinate $t$ is not limited and thus we have $t\in \mathbb{R}$. Similarly as long as massless or massive particles do not hit the curvature singularity they can pass from $0<r$ through $r=0$ to $r<0$. However, in the equatorial plane the spacetime also possesses lightlike and timelike geodesics which end at the curvature singularity at $r=0$. Therefore, for the former we have $r\in \mathbb{R}$ and for the latter we have $r<0$ or $0<r$. Now the only coordinates which are not yet fixed are the angular coordinates $\vartheta$ and $\varphi$. The Kerr spacetime does not possess conical singularities and thus we can safely assume that they represent the coordinates on the two-sphere $S^{2}$ and thus we have $\vartheta\in[0,\pi]$ and $\varphi\in[0,2\pi)$. However, the $\varphi$ coordinate becomes pathological for $\sin\vartheta=0$ and thus geodesics passing through $\vartheta=0$ or $\vartheta=\pi$ require special consideration. Please also note that some authors simply assume that one can simply ignore the time coordinate and extend the geodesics across the horizons. However, this is not the case. When a massless or a massive particle approaches the event horizon in Boyer-Lindquist coordinates its time coordinate diverges and thus for these particles they do not properly cover this scenario. This scenario is only properly covered, e.g., in the maximal analytic extension.\\

\section{The Equations of Motion and the Observer's Celestial Sphere}\label{Sec:EoM}
In this section we will first briefly outline how to derive the equations of motion for light rays and high-frequency gravitational waves in the Kerr spacetime. Then we will place a standard observer in the domain of outer communication and introduce an orthonormal tetrad to relate the constants of motion of light rays and high-frequency gravitational waves detected by the observer to latitude-longitude coordinates on the observer's celestial sphere. Then we use this parameterisation to derive the radius coordinates of the photon orbits in the domain of outer communication and the stationary region inside the Cauchy horizon and their latitudinal projections onto the observer's celestial sphere as functions of the celestial longitude. In the next section we will then use them to classify the different types of motion in the Kerr spacetime.
\subsection{The Equations of Motion}\label{Subsec:EoM}
In this subsection we briefly outline how to derive the equations of motion for light rays and high-frequency gravitational waves in the Kerr spacetime. For this purpose let us start with the line element given by (\ref{eq:KerrMetric}). As first step we write down the associated Langrangian
\begin{eqnarray}
\mathcal{L}=\frac{1}{2}g_{\mu\nu}\dot{x}^{\mu}\dot{x}^{\nu},
\end{eqnarray}
where the dot indicates that we differentiate with respect to the affine parameter $s$. In the second step we use the Euler-Lagrange equations 
\begin{eqnarray}
\frac{\text{d}}{\text{d}s}\frac{\partial \mathcal{L}}{\partial \dot{x}^{\mu}}-\frac{\partial \mathcal{L}}{\partial x^{\mu}}=0,
\end{eqnarray}
to derive the equations of motion for the time coordinate $t$ and the $\varphi$ coordinate. In the third step we write down the Hamiltonian 
\begin{eqnarray}
\mathcal{H}=\frac{1}{2}g^{\mu\nu}p_{\mu}p_{\nu},
\end{eqnarray}
and introduce an action function such that it fulfills the separation Ansatz $S(t,r,\vartheta,\varphi)=S_{t}(t)+S_{r}(r)+S_{\vartheta}(\vartheta)+S_{\varphi}(\varphi)$. Here the components of the four-momentum are related to the action function and the components of the four-velocity of the light rays and gravitational waves along the lightlike geodesics $\dot{x}$ via $p_{\mu}=\partial S(x)/\partial x^{\mu}=g_{\mu\nu}\dot{x}^{\nu}$. The line element of the Kerr spacetime is independent of $t$ and $\varphi$ and thus we have $p_{t}=-E$ and $p_{\varphi}=L_{z}$. In addition, for lightlike geodesics we have $\mathcal{L}=\mathcal{H}=0$. Thus we can introduce the Carter constant $K$ to separate the equations of motion for $r$ and $\vartheta$. After some rearranging we bring them into their final form. Together with the equations of motion for $t$ and $\varphi$ they form a set of four differential equations. This set reads
\begin{eqnarray}\label{eq:EoMt}
\frac{\text{d}t}{\text{d}\lambda}=(r^2+a^2)\frac{(r^2+a^2)E-aL_{z}}{P(r)}+a\left(L_{z}-a\sin^2\vartheta E\right),
\end{eqnarray}
\begin{eqnarray}\label{eq:EoMr}
\left(\frac{\text{d}r}{\text{d}\lambda}\right)^2=((r^2+a^2)E-aL_{z})^2-P(r)K,
\end{eqnarray}
\begin{eqnarray}\label{eq:EoMtheta}
\left(\frac{\text{d}\vartheta}{\text{d}\lambda}\right)^2=K-\frac{(a\sin^2\vartheta E-L_{z})^2}{\sin^2\vartheta},
\end{eqnarray}
\begin{eqnarray}\label{eq:EoMphi}
\frac{\text{d}\varphi}{\text{d}\lambda}=a\frac{(r^2+a^2)E-aL_{z}}{P(r)}+\frac{L_{z}-a\sin^2\vartheta E}{\sin^2\vartheta},
\end{eqnarray}
where we already introduced the Mino-parameter $\lambda$ \cite{Mino2003} which is related to the affine parameter $s$ via
\begin{eqnarray}\label{eq:MiPa}
\frac{\text{d}\lambda}{\text{d}s}=\frac{1}{\rho(r,\vartheta)}.
\end{eqnarray}
As we can see in this parameterisation the equations of motion for $r$ and $\vartheta$ are completely separated. With respect to the Mino parameter for the remainder of this paper we now agree on the following convention: We choose the Mino parameter such that it increases along future-directed lightlike geodesics and decreases along past-directed lightlike geodesics.

\subsection{The Observer's Celestial Sphere}
\begin{figure}\label{fig:CSphere}
  \includegraphics[width=0.7\textwidth]{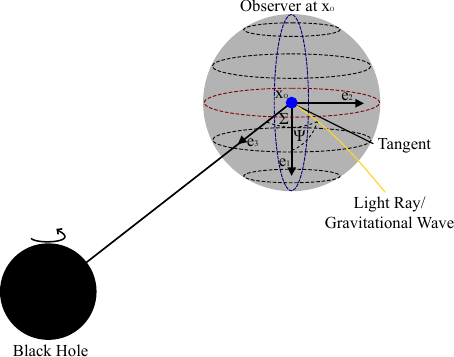}
\caption{Illustration of the celestial sphere of a standard observer in the domain of outer communication in the Kerr spacetime, and the orientation of the tetrad vectors $e_{1}$, $e_{2}$, and $e_{3}$. At an event marked by the spacetime coordinates $x_{O}=(t_{O},r_{O},\vartheta_{O},\varphi_{O})$ (marked by a blue dot) the observer detects a light ray or a gravitational wave (yellow line) on his celestial sphere. The direction on the observer's sky from which the light ray or gravitational wave is detected is marked by the celestial latitude $\Sigma$ and the celestial longitude $\Psi$. Here, we follow the conventions of Grenzebach, Perlick, and L\"{a}mmerzahl \cite{Grenzebach2015} and measure the celestial latitude $\Sigma\in[0,\pi]$ from the tetrad vector $e_{3}$ and the celestial longitude $\Psi\in[0,2\pi)$ from the tetrad vector $e_{1}$ in the direction of the tetrad vector $e_{2}$.}
\end{figure}
In Sec.~\ref{Subsec:EoM} we separated the equations of motion and saw that they depend on the energy $E$ along the lightlike geodesics, the angular momentum about the $z$-axis $L_{z}$, and the Carter constant $K$. While we can also solve the equations of motion in their current form in this paper we want to use the exact solutions to the equations of motion to investigate gravitational lensing in the Kerr spacetime. Here, we will see below that it will be much more convenient to characterise the different types of motion using the latitude-longitude coordinates from which an observer detects a gravitationally lensed light ray or a gravitationally lensed high-frequency gravitational wave on his celestial sphere. With this in mind we now first fix our observation geometry before we proceed to solving the equations of motion.\\
As first step we place an observer in the domain of outer communication. While in general we can choose between several different observers (note, however, that observers on $t$-lines can only exist outside the ergoregion), in this paper we will use a so-called standard observer (sometimes also referred to as Carter observer) following the approach of Grenzebach, Perlick, and L\"{a}mmerzahl \cite{Grenzebach2015}. We assume that the black hole is located at the centre of the spacetime and place the standard observer in the domain of outer communication between the outer boundary of the ergoregion (this ensures that all observed light rays and gravitational waves have $0<E$) and spatial infinity such that the observer detects a light ray or a gravitational wave at an event marked by the coordinates $x_{O}=(t_{O},r_{O},\vartheta_{O},\varphi_{O})$.  The local frame of the standard observer is now spanned by an orthonormal tetrad defined by the tetrad vectors $e_{0}$, $e_{1}$, $e_{2}$, and $e_{3}$. They read
\begin{eqnarray}\label{eq:ONT1}
e_{0}=\left.\frac{(r^2+a^2)\partial_{t}+a\partial_{\varphi}}{\sqrt{\rho(r,\vartheta)P(r)}}\right|_{x_{O}},~~~e_{1}=\left.\frac{\partial_{\vartheta}}{\sqrt{\rho(r,\vartheta)}}\right|_{x_{O}},
\end{eqnarray}
\begin{eqnarray}\label{eq:ONT2}
e_{2}=-\left.\frac{\partial_{\varphi}+a\sin^2\vartheta\partial_{t}}{\sqrt{\rho(r,\vartheta)}\sin\vartheta}\right|_{x_{O}},~~~e_{3}=-\left.\sqrt{\frac{P(r)}{\rho(r,\vartheta)}}\partial_{r}\right|_{x_{O}}.
\end{eqnarray}
Here, the tetrad vector $e_{0}$ is also the four-velocity of the observer and we can read from (\ref{eq:ONT1}) that the standard observer is stationary and moves along lines with constant $r$ and $\vartheta$. This implies two things. On one hand that in the Kerr spacetime the observer is orbiting the black hole with a constant angular velocity and thus its azimuthal coordinate $\varphi_{O}$ is not constant. Luckily, we can interpret the observations on the celestial sphere of this observer such as if they were evaluated at the time coordinate $t_{O}$ at the coordinates $r_{O}$, $\vartheta_{O}$, and $\varphi_{O}$. On the other hand this observer can also be placed inside the ergoregion and thus the whole domain of outer communication.\\
Now we have to fix the angles on the observer's celestial sphere. While in general there is no preferred choice, when we observe astrophysical sources it is a common convention to select one source and take it as reference direction. The position of all other sources on the celestial sphere are then measured in relation to this source. We now transfer this approach to our case at hand and choose the direction to the black hole marked by the tetrad vector $e_{3}$ as reference direction. As depicted in Fig.~2 we fix the celestial coordinates such that we measure the celestial latitude $\Sigma\in [0,\pi]$ from the tetrad vector $e_{3}$ and the celestial longitude $\Psi\in [0,2\pi)$ from the tetrad vector $e_{1}$ in the direction of the tetrad vector $e_{2}$.\\
Now we need to relate the constants of motion $E$, $L_{z}$, and $K$ to the angles $\Sigma$ and $\Psi$ on the observer's celestial sphere. For this purpose let us assume that the observer detects a light ray or a high-frequency gravitational wave travelling along a lightlike geodesic at an event marked by the spacetime coordinates $x_{\text{O}}=(t_{O},r_{O},\vartheta_{O},\varphi_{O})$. In Mino parameterisation the tangent vector to this lightlike geodesic reads
\begin{eqnarray}\label{eq:TanVec}
\left.\frac{\text{d}\eta}{\text{d}\lambda}\right|_{x_{O}}=\left.\frac{\text{d}t}{\text{d}\lambda}\partial_{t}\right|_{x_{O}}+\left.\frac{\text{d}r}{\text{d}\lambda}\partial_{r}\right|_{x_{O}}+\left.\frac{\text{d}\vartheta}{\text{d}\lambda}\partial_{\vartheta}\right|_{x_{O}}+\left.\frac{\text{d}\varphi}{\text{d}\lambda}\partial_{\varphi}\right|_{x_{O}}.
\end{eqnarray}
At the position of the observer we can also write the tangent vector in terms of the orthonormal tetrad, the latitude-longitude coordinates on the observer's celestial sphere, and a normalisation constant $\tilde{\sigma}$. It reads
\begin{eqnarray}\label{eq:TanVecObs}
\left.\frac{\text{d}\eta}{\text{d}\lambda}\right|_{x_{O}}=\tilde{\sigma}\left(-e_{0}+\sin\Sigma\cos\Psi e_{1}+\sin\Sigma\sin\Psi e_{2}+\cos\Sigma e_{3}\right),
\end{eqnarray}
where the normalisation constant $\tilde{\sigma}$ is defined by
\begin{eqnarray}\label{eq:NormCons}
\tilde{\sigma}=g\left(\left.\frac{\text{d}\eta}{\text{d}\lambda}\right|_{x_{O}},e_{0}\right).
\end{eqnarray}
Since the Mino parameter is defined up to an affine transformation we can without loss of generality choose $\tilde{\sigma}=-\rho(r_{O},\vartheta_{O})$. Now we insert the chosen value for the normalisation constant in (\ref{eq:TanVecObs}) and (\ref{eq:NormCons}). For obtaining the relations between the constants of motion and the latitude-longitude coordinates on the celestial sphere of the observer we now first evaluate (\ref{eq:NormCons}) to obtain a relation between $E$ and $L_{z}$. In the next step we compare coefficients between (\ref{eq:TanVec}) and (\ref{eq:TanVecObs}) and insert the results in the equations of motion. After a short calculation we obtain for the relations between the constants of motion $E$, $L_{z}$, and $K$ and the latitude-longitude coordinates $\Sigma$ and $\Psi$ on the observer's celestial sphere
\begin{eqnarray}\label{eq:ConsE}
E=\frac{\sqrt{P(r_{O})}+a\sin\vartheta_{O}\sin\Sigma\sin\Psi}{\sqrt{\rho(r_{O},\vartheta_{O})}},
\end{eqnarray}
\begin{eqnarray}\label{eq:ConsLz}
L_{z}=\frac{(r_{O}^2+a^2)\sin\vartheta_{O}\sin\Sigma\sin\Psi+a\sqrt{P(r_{O})}\sin^2\vartheta_{O}}{\sqrt{\rho(r_{O},\vartheta_{O})}},
\end{eqnarray}
\begin{eqnarray}\label{eq:ConsK}
K=\rho(r_{O},\vartheta_{O})\sin^2\Sigma.
\end{eqnarray}
As these expressions are relatively long we will now agree on the following convention. When we refer to the actual constants of motion we use the usual notation $E$, $L_{z}$, and $K$ while when we only use them for brevity and actually express them in terms of the celestial coordinates we will denote them as $E_{\text{C}}$, $L_{z\text{C}}$, and $K_{\text{C}}$.
\begin{figure}\label{fig:POStruc}
  \includegraphics[width=\textwidth]{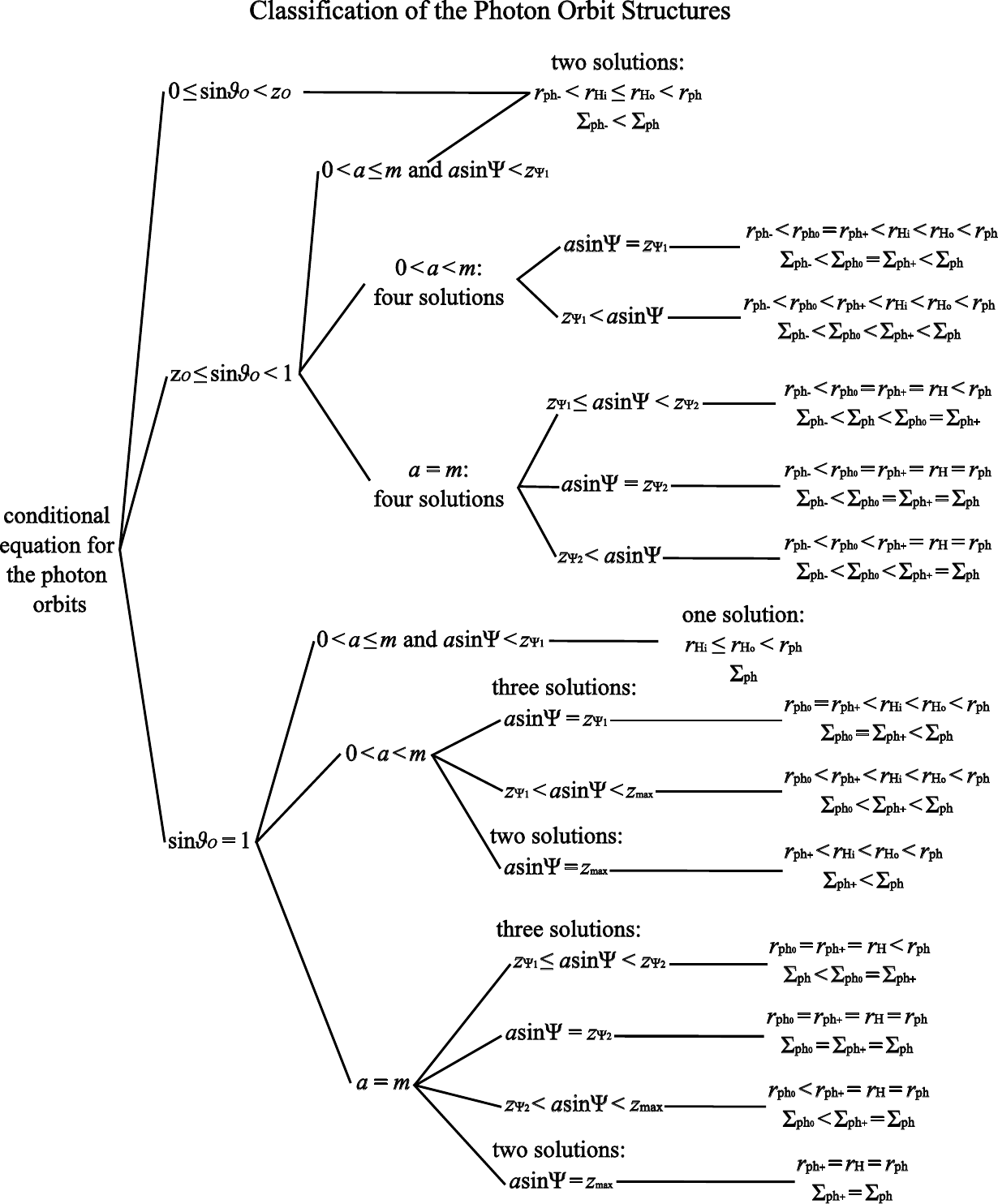}
\caption{Classification of the photon orbit structures of the Kerr spacetime in the parameterisation using the latitude-longitude coordinates on the celestial sphere of the standard observer.}
\end{figure}
\subsection{Photon Orbits and Their Celestial Projections}\label{Sec:POCP}
In general black hole spacetimes can contain light rays and gravitational waves on orbits with constant $r$. These orbits can either be unstable or stable. Here unstable means that when the light rays and gravitational waves on these orbits are infinitesimally perturbed in radial direction they move away from the photon orbits. When the unstable photon orbits are located in the domain of outer communication this means that the light rays and gravitational waves either fall into the black hole or escape to spatial infinity. Vice versa when light rays and gravitational waves on stable photon orbits are perturbed in radial direction they begin to oscillate between two turning points $r_{\text{min}}$ and $r_{\text{max}}$.\\  
In black hole spacetimes the photon orbits play two very special roles. On one hand the unstable photon orbits in the domain of outer communication give rise to the shadow of the black hole. On the other hand the photon orbits, unstable as well as stable, or better their projections onto the observer's celestial sphere, also provide a very convenient way to classify the different types of $r$ motion.\\ 
While it is a common practise to characterise the photon orbits in terms of their angular momentum about the $z$-axis $L_{z}$, for our purpose this is rather inconvenient since these orbits are associated with different longitudes on the observer's celestial sphere (vice versa in general photon orbits projected onto the observer's celestial sphere at the same celestial longitude $\Psi$ do not have the same angular momentum about the $z$-axis $L_{z}$). It will be much more convenient to characterise them in terms of a fixed celestial longitude $\Psi$. Therefore, in the following we will calculate the radius coordinates of the unstable and stable photon orbits in the stationary regions of the Kerr spacetime (here we recall that for $a=m$ this includes the photon orbits which have the same radius coordinate as the horizon) and their latitudinal projections onto the observer's celestial sphere in dependence on the position of the observer and the celestial longitude $\Psi$. \\
For this purpose we use the fact that light rays and gravitational waves, observed at the same celestial coordinates at which a photon orbit is projected onto the observer's celestial sphere, have the exact same constants of motion as light rays and gravitational waves on these photon orbits. Here, in the case that the photon orbits lie in the domain of outer communication or at the same radius coordinate as the horizon these light rays and gravitational waves are asymptotically coming from or going to these photon orbits and give rise to the shadow of the black hole (note though that there will be one exception as we will see below). For this purpose let us now first rewrite (\ref{eq:EoMr}) in terms of a potential $V_{\text{C}}(r)$
\begin{eqnarray}
\left(\frac{\text{d}r}{\text{d}\lambda}\right)^2=V_{\text{C}}(r),
\end{eqnarray}
where the potential $V_{\text{C}}(r)$ reads
\begin{eqnarray}
V_{\text{C}}(r)=((r^2+a^2)E_{\text{C}}-aL_{z\text{C}})^2-P(r)K_{\text{C}}.
\end{eqnarray}
Stable and unstable photon orbits are orbits with constant $r$. This requires that $\text{d}r/\text{d}\lambda=\text{d}^2r/\text{d}\lambda^2=0$. In terms of the potential this translates to $V_{\text{C}}(r)=\partial_{r}V_{\text{C}}(r)=0$ and thus we have
\begin{eqnarray}\label{eq:POE1}
((r^2+a^2)E_{\text{C}}-aL_{z\text{C}})^2-P(r)K_{\text{C}}=0
\end{eqnarray}
and 
\begin{eqnarray}\label{eq:POE2}
2E_{\text{C}}r((r^2+a^2)E_{\text{C}}-aL_{z\text{C}})+(m-r)K_{\text{C}}=0.
\end{eqnarray}
Now we combine both equations and obtain 
\begin{eqnarray}\label{eq:POE3}
\left(2aE_{\text{C}}\left(aE_{\text{C}}-L_{z\text{C}}\right)-K_{\text{C}}\right)r^2+3mK_{\text{C}}r+2a^2\left(aE_{\text{C}}-L_{z\text{C}}\right)^2-2a^2K_{\text{C}}=0.
\end{eqnarray}
Depending on the position of the obsever and the celestial longitude $\Psi$ we can now combine two of these equations to derive the radius coordinates of the photon orbits in dependence on the radius coordinate $r_{O}$ and the spacetime latitude $\vartheta_{O}$ of the observer, the mass and spin parameters $m$ and $a$, and the celestial longitude $\Psi$. While in the most general case we will see that they are determined by a polynomial of sixth order and thus the radius coordinates of the photon orbits can only be determined numerically, it is possible to identify several special cases for which the radius coordinates of the photon orbits can be calculated analytically. In particular we will see that it is possible to calculate the radius coordinates of the photon orbits and their latitudinal projections onto the celestial sphere of the observer analytically when have $\vartheta_{O}=0$, $\vartheta_{O}=\pi$, or $\vartheta_{O}=\pi/2$. Here, the first two cases correspond to an observer on one of the axes and the third case corresponds to an observer located in the equatorial plane. Note that in the case of the photon orbit at the largest radius coordinate, which we will label $r_{\text{ph}}$ in the following, its latitudinal projection onto the celestial sphere of the observer corresponds to the angular radius of the shadow at the specified celestial longitude $\Psi$. Here we emphasize again that since we derive the radius coordinates of the photon orbits and their projections as functions of $\Psi$ the corresponding lightlike geodesics have different angular momenta $L_{z}$ about the $z$-axis. In addition, we will assume that the reader is aware that the radius coordinates of the photon orbits and their projections onto the observer's celestial sphere depend on the celestial longitude $\Psi$ and thus for brevity we will not write the dependency in the following. We will also assume that the reader is aware that we calculate the latitudinal projections of the radius coordinates of the photon orbits onto the celestial sphere of the observer at a fixed celestial longitude $\Psi$ and thus often simply refer to them as \emph{the latitudinal projections of the photon orbits}. \\
While for the spherically symmetric and static Schwarzschild spacetime the photon orbit structure is very simple for the Kerr spacetime it is rather complex and varies with $a$, $\vartheta_{O}$, and $\Psi$. An overview of the different photon orbit structures can be found in Fig.~3. Here in Fig.~3 we introduced four different quantities $z_{O}$, $z_{\Psi_{1}}$, $z_{\Psi_{2}}$, and $z_{\text{max}}$ which have to be defined before we can proceed to discuss the photon orbits. For this purpose we already need some knowledge about the number of the photon orbits we can have for a fixed celestial longitude $\Psi$ and their stability, and we have to define their labelling. We will see that overall we can have up to four photon orbits at different radius coordinates. Let us for now assume that we have the maximal number. In this case we will label them such that we have $r_{\text{ph}_{-}}<r_{\text{ph}_{0}}\leq r_{\text{ph}_{+}}\leq r_{\text{H}_{\text{i}}}\leq r_{\text{H}_{\text{o}}}\leq r_{\text{ph}}$. In the case that we have four distinct photon orbits we have three unstable photon orbits at the radius coordinates $r_{\text{ph}_{-}}$, $r_{\text{ph}_{+}}$, and $r_{\text{ph}}$, and one stable photon orbit at the radius coordinate $r_{\text{ph}_{0}}$. Now the first quantity $z_{O}$ marks the lowest and the highest spacetime latitudes $\vartheta_{O}$ for which we can have a photon orbit at the radius coordinate $r_{\text{ph}_{0}}$ or a photon orbit at the radius coordinate $r_{\text{ph}_{+}}$ inside the Cauchy horizon or at the same radius coordinate as the horizon. The second quantity $z_{\Psi_{1}}=a\sin\Psi_{1}$ marks the longitudes $\Psi_{1}$ on the observer's celestial sphere for which $a\sin\Psi$ takes the smallest possible value for which the photon orbit at the radius coordinate $r_{\text{ph}_{0}}$ and the photon orbit at the radius coordinate $r_{\text{ph}_{+}}$ can coincide and we have $r_{\text{ph}_{0}}=r_{\text{ph}_{+}}\neq r_{\text{ph}}$. The quantity $z_{\Psi_{2}}=a\sin\Psi_{2}$ on the other hand marks the longitudes $\Psi_{2}$ for which we have $r_{\text{ph}_{0}}=r_{\text{ph}_{+}}=r_{\text{H}}=r_{\text{ph}}$. Finally, the quantity $z_{\text{max}}=a\sin\Psi_{\text{max}}$ marks corotating light rays and gravitational waves in the equatorial plane. Note that here we chose the classifier $a\sin\Psi$ such that although in this paper we only consider positive spins the whole classification scheme in Fig.~3 can also be applied to negative spins. Note that in general for a fixed celestial longitude $\Psi$ the existence of the two classifiers $z_{\Psi_{1}}$ and $z_{\Psi_{2}}$ and their exact value depend on the spin parameter $a$ and $\vartheta_{O}$. The classifier $z_{\text{max}}$ on the other hand only exists for corotating light rays in the equatorial plane.\\ 
Before we proceed to deriving the radius coordinates of the photon orbits and their latitudinal projections onto the observer's celestial sphere, and to discuss the photon orbit structures we would like to add one additional remark. Since this paper will mainly consider lightlike geodesic motion with $r_{\text{ph}}<r$ in the following we will only explicitly write down the latitudinal projection of $r_{\text{ph}}$ onto the observer's celestial sphere for this case. In the case that we have $r_{\text{H}_{\text{o}}}<r<r_{\text{ph}}$ we can easily obtain it by using the symmetry with respect to $\Sigma=\pi/2$.
\subsubsection{$\vartheta_{O}=0$ or $\vartheta_{O}=\pi$}
When we have $\vartheta_{O}=0$ or $\vartheta_{O}=\pi$ and thus $\sin\vartheta_{O}=0$ (\ref{eq:POE3}) reduces to 
\begin{eqnarray}\label{eq:POA}
(3mr-2a^2-r^2)(r_{O}^2+a^2)^2\sin^2\Sigma+2a^2(r^2+a^2)P\left(r_{O}\right)=0.
\end{eqnarray}
We solve for $\sin^2\Sigma$ and insert the obtained expression into (\ref{eq:POE2}). We sort all terms and see that the radius coordinates of the photon orbits are determined by a polynomial of third order. It reads
\begin{eqnarray}\label{eq:PODE1}
r^3-3mr^2+a^2r+ma^2=0.
\end{eqnarray}
Now we use Cardano's formula to calculate the roots and obtain three different solutions. However, only two of these solutions correspond to radius coordinates of photon orbits in the stationary regions of the spacetime. The first solution lies in the stationary region inside the Cauchy horizon while the second solution lies in the domain of outer communication. For both we have $\partial^2_{r}V_{\text{C}}(r)>0$ (note that here and in the following whenever we write "$V_{\text{C}}(r)$" this means that the potential is evaluated for the radius coordinates of several photon orbits separately) and thus they are unstable. We sort and label them such that we have $r_{\text{ph}_{-}}<r_{\text{H}_{\text{i}}}\leq r_{\text{H}_{\text{o}}}<r_{\text{ph}}$. Note that the radius coordinates of these photon orbits can never be equal.\\
We can now obtain their latitudinal projections onto the celestial sphere of the observer by solving (\ref{eq:POA}) for $\Sigma$ and inserting the determined radius coordinates $r_{\text{ph}}$ and $r_{\text{ph}_{-}}$. We will denote them as $\Sigma_{\text{ph}}$ and $\Sigma_{\text{ph}_{-}}$, respectively. For observers on the axes we always have $\Sigma_{\text{ph}_{-}}<\Sigma_{\text{ph}}$. In the case of $r_{\text{ph}}$ the derived equation determines the angular radius of the shadow and reads (to obtain the projection $\Sigma_{\text{ph}_{-}}$ of the radius coordinate $r_{\text{ph}_{-}}$ we simply have to replace $r_{\text{ph}}$ by $r_{\text{ph}_{-}}$)
\begin{eqnarray}
\Sigma_{\text{ph}}=\arcsin\left(\sqrt{\frac{2a^2(r_{\text{ph}}^2+a^2)P(r_{O})}{(r_{\text{ph}}^2+2a^2-3mr_{\text{ph}})(r_{O}^2+a^2)^2}}\right).
\end{eqnarray}
Note that in this case the radius coordinates $r_{\text{ph}}$ and $r_{\text{ph}_{-}}$ and thus the angular radius of the shadow $\Sigma_{\text{ph}}$, and $\Sigma_{\text{ph}_{-}}$ do not depend on the celestial longitude $\Psi$.

\subsubsection{$\vartheta_{O}=\pi/2$ and $\Psi=\pi/2$ or $\Psi=3\pi/2$}\label{Sec:POEP}
We have $\vartheta_{O}=\pi/2$ and either $\Psi=\pi/2$ or $\Psi=3\pi/2$ and thus $\sin\vartheta_{O}=1$ and $a\sin\Psi=z_{\text{max}}$ for corotating light rays and gravitational waves ($\Psi=\pi/2$), or $a\sin\Psi<0$ for counterrotating light rays and gravitational waves ($\Psi=3\pi/2$). We follow the same procedure as in the last section. This time the radius coordinates of the photon orbits are determined by
\begin{eqnarray}
r^3-6mr^2+9m^2r-4ma^2=0.
\end{eqnarray}
Again we use Cardano's method to calculate the roots. We again obtain three different solutions. Two of the solutions are the radius coordinates of the photon orbits for corotating light rays and gravitational waves while the third solution is the radius coordinate of the photon orbit for counterrotating light rays and gravitational waves. For all of them we have $\partial^2_{r}V_{\text{C}}(r)>0$ and thus all three are unstable. The photon orbit for counterrotating light rays and gravitational waves (note that in the following we will use the convention that when we use the singular we refer to a photon orbit for a fixed $\Psi$) lies in the domain of outer communication and we label it $r_{\text{ph}}$. For the discussion of the photon orbits for the corotating light rays and gravitational waves we first label and sort them such that we have $r_{\text{ph}_{+}}\leq r_{\text{H}_{\text{o}}}\leq r_{\text{H}_{\text{i}}}\leq r_{\text{ph}}$. When we have $0<a<m$ $r_{\text{ph}_{+}}$ lies inside the Cauchy horizon while $r_{\text{ph}}$ lies in the domain of outer communication. When we have $a=m$ we have $r_{\text{ph}_{+}}=r_{\text{H}}=r_{\text{ph}}$ and thus in Boyer-Lindquist coordinates we only have one photon orbit with the same radius coordinate as the horizon. In reality we cannot have unstable photon orbits at the horizon and it can be shown that this problem arises due to the choice of the coordinate system. As shown by Bardeen, Press, and Teukolsky \cite{Bardeen1972} in the embedding diagram the orbits are clearly distinct and do not coincide with the horizon. However, since in this paper it is our goal to investigate gravitational lensing of light rays and high-frequency gravitational waves in the domain of outer communication we will keep using the Boyer-Lindquist coordinates and do not dive into the realm of the full Kerr geometry. However, when we discuss gravitational lensing in the Kerr spacetime we will briefly comment on how the position of the photon orbits may affect the calculated lensing features for the extremal Kerr black hole.\\
Now the last thing we have to do is to calculate the latitudinal projections of the photon orbits onto the observer's celestial sphere. For this purpose we simply set $r=r_{\text{ph}}$ in the reduced form of (\ref{eq:POE1}) and solve for $\Sigma$. For counterrotating light rays and gravitational waves the projection is simply given by $\Sigma_{\text{ph}}$. For corotating light rays and gravitational waves on the other hand the projections are given by $\Sigma_{\text{ph}_{+}}\leq\Sigma_{\text{ph}}$. In both cases $\Sigma_{\text{ph}}$ is the angular radius of the shadow of the black hole at the celestial longitude $\Psi$ (in our case either $\Psi=\pi/2$ or $\Psi=3\pi/2$). It reads (to obtain the projection $\Sigma_{\text{ph}_{+}}$ of the radius coordinate $r_{\text{ph}_{+}}$ we simply have to replace $r_{\text{ph}}$ by $r_{\text{ph}_{+}}$)
\begin{eqnarray}\label{eq:POEPSigmaph}
\Sigma_{\text{ph}}=\arcsin\left(\frac{2ar_{\text{ph}}^2\sqrt{P(r_{O})}\sin\Psi}{(3mr_{\text{ph}}-2a^2-r_{\text{ph}}^2)r_{O}^2+2a^2(r_{O}^2-r_{\text{ph}}^2)\sin^2\Psi}\right).
\end{eqnarray}
\subsubsection{$\vartheta_{O}=\pi/2$ and $\Psi=0$ or $\Psi=\pi$}
In this case we have $\vartheta_{O}=\pi/2$ and $\Psi=0$ or $\Psi=\pi$ and thus $\sin\vartheta_{O}=1$ and $a\sin\Psi=0$. We can easily see that we have $aE_{\text{C}}-L_{z\text{C}}=0$. Since we will see in Sec.~\ref{Sec:EoMr} that for $\Sigma=0$ or $\Sigma=\pi$ we cannot have turning points in the stationary regions of the spacetime for light rays and gravitational waves on these photon orbits we can safely assume that we have $0<\Sigma_{\text{ph}}$. Thus we divide (\ref{eq:POE3}) by $K_{\text{C}}$. In this case the radius coordinates of the photon orbits are determined by a polynomial of second order. It reads
\begin{eqnarray}
r^2-3mr+2a^2=0.
\end{eqnarray}
The polynomial has two solutions. In our case we only have one solution that represents a photon orbit and this photon orbit lies in the domain of outer communication. The solution reads
\begin{eqnarray}
r_{\text{ph}}=\frac{3m+\sqrt{9m^2-8a^2}}{2}.
\end{eqnarray}
Again we have $\partial^2_{r}V_{\text{C}}(r_{\text{ph}})>0$ and thus the photon orbit at this radius coordinate is unstable. In this case the angular projection of the photon orbit onto the observer's celestial sphere and thus the angular radius of the shadow at the celestial longitude $\Psi$ is easy to calculate. We set $r=r_{\text{ph}}$ in the reduced form of (\ref{eq:POE1}) and solve for $\Sigma$. The result reads
\begin{eqnarray}
\Sigma_{\text{ph}}=\arcsin\left(\frac{r_{\text{ph}}^2}{r_{O}^2}\sqrt{\frac{P(r_{O})}{P(r_{\text{ph}})}}\right).
\end{eqnarray}
Here it is interesing to note that in this case the radius coordinate of the photon orbit as well as the angular radius of the shadow have the same dependence on the spin parameter $a$ as the radius coordinate of the unstable photon orbit and the angular radius of the shadow on the electric charge $e$ in the Reissner-Nordstr\"{o}m spacetime.

\subsubsection{$\vartheta_{O}=\pi/2$: Orbits with Other Values for $\Psi$}\label{Sec:POEAG}
In this case we have $\vartheta_{O}=\pi/2$ and $0<\Psi<\pi/2$, $\pi/2<\Psi<\pi$, $\pi<\Psi<3\pi/2$, or $3\pi/2<\Psi<2\pi$. Again we follow the steps outlined above. In this case the radius coordinates of the photon orbits are determined by a polynomial of fourth order. It reads
\begin{eqnarray}\label{eq:OEPAoE}
r^4-6mr^3+(9m^2+4a^2\cos^2\Psi)r^2-4ma^2(1+2\cos^2\Psi)r+4a^4\cos^2\Psi=0.
\end{eqnarray}
This time we use Ferrari's method to calculate the roots. Note that because the equation is degenerate with respect to $\cos\Psi$ for a fixed celestial longitude $\Psi$ not all real solutions to (\ref{eq:OEPAoE}) correspond to photon orbits for the specified longitude $\Psi$. Thus in the following when we list the number of the real solutions for the photon orbits we only refer to the photon orbits for the case at hand, not all photon orbits. In addition, in all cases that the roots represent photon orbits they are either located in one of the stationary regions of the spacetime or, when the spin takes the extremal value $a=m$, at the radius coordinate of the horizon. As for photon orbits in the equatorial plane we can calculate their projections onto the celestial sphere of the observer by inserting the calculated radius coordinates for the photon orbits into (\ref{eq:POEPSigmaph}).\\
In general the number and the nature of the photon orbits depend on the spin parameter $a$, $\vartheta_{O}$, and $\Psi$. Here, we have to distinduish between six different photon orbit structures. For their classification we will now use the quantities $z_{O}$, $z_{\Psi_{1}}$, $z_{\Psi_{2}}$, and $z_{\Psi_{\text{max}}}$ which we defined above. When we have $0<a\leq m$ and $a\sin\Psi<z_{\Psi_{1}}$ (\ref{eq:OEPAoE}) has one real solution which corresponds to the radius coordinate of a photon orbit in the domain of outer communication. We label it $r_{\text{ph}}$. We have $\partial^2_{r}V_{\text{C}}(r_{\text{ph}})>0$ and thus the photon orbit at this radius coordinate is unstable. In this case the latitudinal projection $\Sigma_{\text{ph}}$ of the photon orbit onto the celestial sphere of the observer is also the angular radius of the shadow for the respective celestial longitude $\Psi$.  When we have $0<a<m$ and $a\sin\Psi=z_{\Psi_{1}}$ (\ref{eq:OEPAoE}) has three real solutions which correspond to photon orbits. We sort and label them such that we have $r_{\text{ph}_{0}}=r_{\text{ph}_{+}}<r_{\text{H}_{\text{i}}}<r_{\text{H}_{\text{o}}}<r_{\text{ph}}$. Two of these solutions are located at the same radius coordinate and thus we only have two distinct photon orbits. For the photon orbit at the radius coordinate $r_{\text{ph}}$ we have $\partial^2_{r}V_{\text{C}}(r_{\text{ph}})>0$ and thus this photon orbit is unstable. For the photon orbit at the radius coordinate $r_{\text{ph}_{0}}=r_{\text{ph}_{+}}$ we have $\partial^2_{r}V_{\text{C}}(r_{\text{ph}_{+}})=0$ and $\partial^3_{r}V_{\text{C}}(r_{\text{ph}_{+}})\neq 0$ and thus we have a saddle. The projections of these photon orbits onto the celestial sphere of the observer are given by $\Sigma_{\text{ph}_{0}}=\Sigma_{\text{ph}_{+}}<\Sigma_{\text{ph}}$ and the angular radius of the shadow at the celestial longitude $\Psi$ is given by $\Sigma_{\text{ph}}$. When we have $0<a<m$ and $z_{\Psi_{1}}<a\sin\Psi<z_{\text{max}}$ (\ref{eq:OEPAoE}) has three distinct real solutions which correspond to photon orbits. We label and sort them such that we have $r_{\text{ph}_{0}}<r_{\text{ph}_{+}}<r_{\text{H}_{\text{i}}}<r_{\text{H}_{\text{o}}}<r_{\text{ph}}$. For the photon orbits at the radius coordinates $r_{\text{ph}_{+}}$ and $r_{\text{ph}}$ we have $\partial^2_{r}V_{\text{C}}(r)> 0$ and thus both photon orbits are unstable. For the photon orbit at the radius coordinate $r_{\text{ph}_{0}}$ we have $\partial^2_{r}V_{\text{C}}(r_{\text{ph}_{0}})< 0$ and thus this photon orbit is stable. In this case the projections of these photon orbits onto the celestial sphere of the observer are given by $\Sigma_{\text{ph}_{0}}<\Sigma_{\text{ph}_{+}}<\Sigma_{\text{ph}}$ and the angular radius of the shadow at the celestial longitude $\Psi$ is given by $\Sigma_{\text{ph}}$.\\
When we have $a=m$ and $z_{\Psi_{1}}\leq a\sin\Psi<z_{\Psi_{2}}$ (\ref{eq:OEPAoE}) has three real solutions which correspond to photon orbits. We label and sort them such that we have $r_{\text{ph}_{0}}=r_{\text{ph}_{+}}=r_{\text{H}}<r_{\text{ph}}$. As we can see two of the solutions are located at the same radius coordinate and at the same time in Boyer-Lindquist coordinates at the same radius coordinate as the horizon. In the case of the photon orbit at the radius coordinate $r_{\text{ph}}$ we have $\partial^2_{r}V_{\text{C}}(r_{\text{ph}})> 0$ and thus it is unstable. For the photon orbit at the radius coordinate $r_{\text{ph}_{0}}=r_{\text{ph}_{+}}=r_{\text{H}}$ on the other hand we have $\partial^2_{r}V_{\text{C}}(r_{\text{ph}_{+}})<0$ and thus with respect to the Boyer-Lindquist coordinates this photon orbit is stable. However, as in the case when we have $r_{\text{ph}_{+}}=r_{\text{H}}=r_{\text{ph}}$ this is very likely just a projection effect due to our coordinate choice. In this case the projections of the photon orbits onto the observer's celestial sphere are given by $\Sigma_{\text{ph}}<\Sigma_{\text{ph}_{0}}=\Sigma_{\text{ph}_{+}}$. Note though that although we have $\Sigma_{\text{ph}}<\Sigma_{\text{ph}_{0}}=\Sigma_{\text{ph}_{+}}$ also in this case the angular radius of the shadow at the celestial longitude $\Psi$ is given by $\Sigma_{\text{ph}}$.\\
When we have $a=m$ and $a\sin\Psi=z_{\Psi_{2}}$ (\ref{eq:OEPAoE}) has again three real solutions which correspond to photon orbits. We label and sort them such that we have $r_{\text{ph}_{0}}=r_{\text{ph}_{+}}=r_{\text{H}}=r_{\text{ph}}$. As we can see in Boyer-Lindquist coordinates all three solutions have the same value $r_{\text{ph}}$ and again they have the same radius coordinate as the horizon. Formally we have $\partial^2_{r}V_{\text{C}}(r_{\text{ph}})=0$ and $\partial^3_{r}V_{\text{C}}(r_{\text{ph}})\neq 0$ and thus again we have a saddle. In this case the projection of the photon orbit onto the observer's celestial sphere and thus also the angular radius of the shadow at the celestial longitude $\Psi$ is given by $\Sigma_{\text{ph}}$.\\
In the last case we have $a=m$ and $z_{\Psi_{2}}<a\sin\Psi<z_{\text{max}}$. Again (\ref{eq:OEPAoE}) has three real solutions which correspond to photon orbits. We label and sort them such that we have $r_{\text{ph}_{0}}<r_{\text{ph}_{+}}=r_{\text{H}}=r_{\text{ph}}$. Also in this case two of the solutions, this time $r_{\text{ph}_{+}}$ and $r_{\text{ph}}$, are located at the radius coordinate of the horizon $r_{\text{H}}$. In this case we have $\partial^2_{r}V_{\text{C}}(r_{\text{ph}})>0$ and thus these photon orbits are unstable. For the photon orbit at the radius coordinate $r_{\text{ph}_{0}}$ we have $\partial^2_{r}V_{\text{C}}(r_{\text{ph}_{0}})<0$ and thus this photon orbit is stable. The projections of these photon orbits onto the celestial sphere of the observer are given by $\Sigma_{\text{ph}_{0}}<\Sigma_{\text{ph}_{+}}=\Sigma_{\text{ph}}$. Also in this case the angular radius of the shadow at the celestial longitude $\Psi$ is given by $\Sigma_{\text{ph}}$. 

\subsubsection{$0<\vartheta_{O}<\pi/2$ or $\pi/2<\vartheta_{O}<\pi$ and $\Psi=0$ or $\Psi=\pi$}
In this case we have $0<\vartheta_{O}<\pi/2$ or $\pi/2<\vartheta_{O}<\pi$ and $\Psi=0$ or $\Psi=\pi$. Again we follow the steps outlined above. In this case the radius coordinates of the photon orbits are determined by a polynomial of third order. It reads
\begin{eqnarray}
r^3-3mr^2+a^2(1+\sin^2\vartheta_{O})r+ma^2\cos^2\vartheta_{O}=0.
\end{eqnarray}
We again use Cardano's formula to calculate the roots. This time only two of these roots lie either in the stationary region inside the Cauchy horizon or in the domain of outer communication. We label and sort them such that we have $r_{\text{ph}_{-}}<r_{\text{H}_{\text{i}}}\leq r_{\text{H}_{\text{o}}}<r_{\text{ph}}$. Since in both cases we have $\partial^2_{r}V_{\text{C}}(r)>0$ both photon orbits are unstable. As in the cases for $\vartheta_{O}=0$ or $\vartheta_{O}=\pi$ their projections onto the celestial sphere of the observer are denoted $\Sigma_{\text{ph}}$ and $\Sigma_{\text{ph}_{-}}$. For $r_{\text{ph}}$ the derived expression determines the angular radius of the shadow and reads (as before to obtain the latitudinal projection $\Sigma_{\text{ph}_{-}}$ of the radius coordinate $r_{\text{ph}_{-}}$ we have to replace $r_{\text{ph}}$ by $r_{\text{ph}_{-}}$)
\begin{eqnarray}\label{eq:OEP0}
\Sigma_{\text{ph}}=\arcsin\left(\sqrt{\frac{2a^2 \cos^2\vartheta_{O}(r_{\text{ph}}^2+a^2\cos^2\vartheta_{O})P(r_{O})}{(r_{\text{ph}}^2+2a^2-3mr_{\text{ph}})\rho(r_{O},\vartheta_{O})^2}}\right).
\end{eqnarray}
As for observers on the axes we always have $\Sigma_{\text{ph}_{-}}<\Sigma_{\text{ph}}$. 

\subsubsection{Orbits with Other $\vartheta_{O}$ and $\Psi$}
For all other geodesics the unstable and stable photon orbits are determined by the roots of a polynomial of sixth order. We obtain it by first solving (\ref{eq:POE3}) for $\sin\Sigma$. Then we insert the obtained result for $\sin\Sigma$ into (\ref{eq:POE2}) and simplify all terms. The result reads
\begin{eqnarray}
&r^6-6mr^5+\left(9m^2+2a^2\left(\cos^2\vartheta_{O}+2\sin^2\vartheta_{O}\cos^2\Psi\right)\right)r^4-4ma^2\left(1+2\sin^2\vartheta_{O}\cos^2\Psi\right)r^3\\&+a^2\left(a^2\left(\cos^4\vartheta_{O}+4\sin^2\vartheta_{O}\cos^2\Psi\right)-6m^2\cos^2\vartheta_{O}\right)r^2+2ma^4\cos^2\vartheta_{O}\left(1+\sin^2\vartheta_{O}\right)r+m^2a^4\cos^4\vartheta_{O}=0.\nonumber
\end{eqnarray}
Since we do not have analytical solution methods for polynomials of sixth order in this case the roots and thus the radius coordinates of the photon orbits have to be calculated numerically. The basic structure is nearly identical to the one we found in Section~\ref{Sec:POEAG} with only one distinct difference. In this case we have another unstable photon orbit at the radius coordinate $r_{\text{ph}_{-}}$ (with $\partial_{r}^2V_{\text{C}}(r_{\text{ph}_{-}})>0$) in the stationary region inside the Cauchy horizon. It is always the smallest radius coordinate and does never coincide with $r_{\text{ph}_{0}}$, $r_{\text{ph}_{+}}$, or $r_{\text{ph}}$. As a consequence it is always projected to smaller latitudes $\Sigma_{\text{ph}_{-}}$ on the observer's celestial sphere than $r_{\text{ph}_{0}}$, $r_{\text{ph}_{+}}$, and $r_{\text{ph}}$.\\
In this case the expression for the angular radius of the shadow is determined up to a sign. \emph{A priori} this sign is not fixed and thus for each photon orbit we have to fix it such that the latitudinal projection is in the allowed range for the celestial latitude $\Sigma\in [0,\pi]$ and the conditions $V_{\text{C}}(r)=0$ and $\partial_{r}V_{\text{C}}(r)=0$ are fulfilled. In its general form the expression for the angular radius of the shadow reads (when we want to get the latitudinal projections $\Sigma_{\text{ph}_{+}}$, $\Sigma_{\text{ph}_{0}}$, and $\Sigma_{\text{ph}_{-}}$ of the other three radius coordinates onto the celestial sphere of the observer we have to replace $r_{\text{ph}}$ by $r_{\text{ph}_{+}}$, $r_{\text{ph}_{0}}$, and $r_{\text{ph}_{-}}$, respectively) 
\begin{eqnarray}
&\hspace{-0.7cm}\Sigma_{\text{ph}}=\arcsin\left(\frac{a\sqrt{P(r_{O})}\left(\sin\vartheta_{O}\sin\Psi(r_{O}^2\rho(r_{\text{ph}},\vartheta_{O})+a^2\cos^2\vartheta_{O}(r_{O}^2-r_{\text{ph}}^2))\pm\rho(r_{O},\vartheta_{O})\sqrt{\sin^2\vartheta_{O}\sin^2\Psi r_{\text{ph}}^4-2\cos^2\vartheta_{O}\rho(r_{\text{ph}},\vartheta_{O})\sigma(r_{\text{ph}})}\right)}{\sigma(r_{\text{ph}})\rho(r_{O},\vartheta_{O})^2+2a^2r_{O}^2(r_{O}^2-r_{\text{ph}}^2)\sin^2\vartheta_{O}\sin^2\Psi}\right),
\end{eqnarray}
where we defined 
\begin{eqnarray}
\sigma(r_{\text{ph}})=3mr_{\text{ph}}-2a^2-r_{\text{ph}}^2.
\end{eqnarray}
\section{Solving the Equations of Motion}\label{Sec:SolEoM}
In the last section we already derived the equations of motion for lightlike geodesics and related the constants of motion of light rays and gravitational waves travelling along these geodesics to latitude-longitude coordinates on the celestial sphere of a standard observer in the domain of outer communication. In this section we will now use this parameterisation to classify, discuss, and solve the equations of motion for light rays and gravitational waves with $0<E$ travelling along lightlike geodesics in the domain of outer communication. While in this paper we parameterise the light rays and gravitational waves travelling along these lightlike geodesics using the latitude-longitude coordinates on the celestial sphere of a standard observer who detects the light rays and gravitational waves at an event marked by the coordinates $x_{O}=(t_{O},r_{O},\vartheta_{O},\varphi_{O})$ we will derive the solutions to the equations of motion in their most general form so that readers who are interested in using them for other purposes can easily transfer them. For this purpose we will assume that we have arbitrary initial conditions $x(\lambda_{i})=x_{i}=(t_{i},r_{i},\vartheta_{i},\varphi_{i})$. While throughout the paper we assume that we have a positive spin parameter $0<a\leq m$ we will derive all solutions such that they are also valid for negative spin parameters $-m\leq a<0$.\\ 
Before we proceed it is also worth noting that some of the solutions have already been derived for the redefined angular momentum $l=L_{z}/E$ and the redefined Carter constant $\eta=Q/E^2$, e.g., in the work of Slez\'{a}kov\'{a} \cite{Slezakova2006} and Gralla and Lupsasca \cite{Gralla2020b}, however, we can easily see from the derived relation between the energy $E$ along the lightlike geodesics and the latitude-longitude coordinates on the observer's celestial sphere that the redefined constants of motion become singular at the boundary of the ergoregion and therefore when one wants to complement our solutions for $0<E$ with the solutions for $E\leq 0$ this is easier using the parameterisation we use in this paper. In addition, using the parameterisation in this paper has the advantage that the results are easier to compare with the results for massive particles travelling along timelike geodesics. Furthermore, we will see that one of the solutions of Gralla and Lupsasca \cite{Gralla2020b} is rather impractical and that the solutions are in general still incomplete and therefore in this part of the paper we aim at providing a full set of exact analytic solutions to the equations of motion for light rays and gravitational waves with $0<E$ travelling along lightlike geodesics in the domain of outer communication for the Kerr spacetime valid for $-m\leq a< 0$ and $0<a\leq m$.\\

\subsection{The $r$ Motion}\label{Sec:EoMr}
\subsubsection{Types of Motion} 
Using the radius coordinates of the unstable and stable photon orbits derived in Section~\ref{Sec:POCP} and their latitudinal projections onto the observer's celestial sphere we can now distinguish the different types of $r$ motion for light rays and gravitational waves with $0<E$ travelling along lightlike geodesics in the domain of outer communication of the Kerr spacetime. Here, depending on the spacetime latitude $\vartheta_{O}$ of the observer and the celestial longitude $\Psi$ from which the observer detects a light ray or a gravitational wave we can have 14 different structures. These depend on the number and types of the unstable and stable photon orbits which can occur for a fixed celestial longitude $\Psi$. As we saw in Section~\ref{Sec:POCP} for a fixed spacetime latitude $\vartheta_{O}$ and a fixed celestial longitude $\Psi$ the maximal number of photon orbits is four (note that here in the counting we include distinct and degenerate photon orbits) and thus in the following we will classify the different types of motion based on the connected photon orbit structures. In the most general case that we have four unstable and stable photon orbits we have $r_{\text{ph}_{-}}<r_{\text{ph}_{0}}\leq r_{\text{ph}_{+}}\leq r_{\text{H}_{\text{i}}}\leq r_{\text{H}_{\text{o}}}\leq r_{\text{ph}}$ and $\Sigma_{\text{ph}_{-}}<\Sigma_{\text{ph}_{0}}\leq \Sigma_{\text{ph}_{+}}\leq \Sigma_{\text{ph}}$ and the special case $r_{\text{ph}_{-}}<r_{\text{ph}_{0}}= r_{\text{ph}_{+}}=r_{\text{H}}< r_{\text{ph}}$ and $\Sigma_{\text{ph}_{-}}<\Sigma_{\text{ph}}<\Sigma_{\text{ph}_{0}}=\Sigma_{\text{ph}_{+}}$ (for a certain longitude range in the case $a=m$). We will now first classify all types of motion based on these two latitude schemes. Note that in this case the observer is not located in the equatorial plane and while in general all types of motion which occur for light rays and gravitational waves detected by an observer in the equatorial plane also occur for the latitude schemes specified above, the type of motion (and as a consequence also the solution to the equation of motion for $r$) along lightlike geodesics with vanishing Carter constant is different. Thus we will add it directly at the beginning as case 1a. In addition, for an observer in the equatorial plane the types of motion classified as case 2 and case 3 do not occur. After classifying the different types of motion we will briefly describe how this scheme changes when we have degenerate or less than four photon orbits. Note that here the boundaries depend on the celestial longitude $\Psi$, however, we omit it in the following for brevity. \\
Now when we take into account the different photon orbit structures and the latitudinal projections of the radius coordinates of the photon orbits onto the celestial sphere of the observer for a fixed celestial longitude $\Psi$, the different types of motion are:
\begin{itemize}
\item Case 1a: In this case we have lightlike geodesics characterised by $\vartheta_{i}=\pi/2$ and $\Sigma=0$ or $\Sigma=\pi$. This translates to $K_{\text{C}}=0$ and thus we can easily see from (\ref{eq:EoMtheta}) that for these geodesics we have $\text{d}\vartheta/\text{d}\lambda=0$. Thus these lightlike geodesics lie in the equatorial plane. We can easily read from (\ref{eq:EoMtheta}) that we have $L_{z\text{C}}=aE_{\text{C}}$ and thus the right-hand side of (\ref{eq:EoMr}) has four real and equal roots at $r_{1}=r_{2}=r_{3}=r_{4}=0$. Thus in the domain of outer communication light rays and gravitational waves travelling along these geodesics cannot pass through a turning point.  
\item Case 1b: In this case we have lightlike geodesics outside the equatorial plane characterised by $\vartheta_{i}\neq\pi/2$ and $\Sigma=0$ or $\Sigma=\pi$. As for case 1a we have $K_{\text{C}}=0$ and thus these lightlike geodesics lie on cones. We can easily read from (\ref{eq:EoMtheta}) that we have $L_{z\text{C}}=a\sin^2\vartheta_{i} E_{\text{C}}$ and thus the right-hand side of (\ref{eq:EoMr}) has a pair of complex conjugate purely imaginary double roots at $r_{1}=r_{3}=\bar{r}_{2}=\bar{r}_{4}=ia\cos\vartheta_{i}$. Again in the domain of outer communication light rays and gravitational waves travelling along these geodesics cannot pass through a turning point.
\item Case 2: In this case we have lightlike geodesics characterised by $0<\Sigma<\Sigma_{\text{ph}_{-}}$ or $\pi-\Sigma_{\text{ph}_{-}}<\Sigma<\pi$. They only exist for $\vartheta_{i}\neq\pi/2$. In this case the right-hand side of (\ref{eq:EoMr}) has two distinct pairs of complex conjugate roots. In this paper we label and sort them such that $r_{1}=\bar{r}_{2}=R_{1}+iR_{2}$ and $r_{3}=\bar{r}_{4}=R_{3}+iR_{4}$, where we choose $R_{1}<R_{3}$, and $0<R_{2}$ and $0<R_{4}$. Light rays and gravitational waves travelling along these geodesics cannot pass through a turning point.
\item Case 3: In this case we have lightlike geodesics characterised by $\Sigma=\Sigma_{\text{ph}_{-}}$ or $\Sigma=\pi-\Sigma_{\text{ph}_{-}}$ and thus light rays and gravitational waves travelling along these geodesics have the same constants of motion as light rays and gravitational waves on the unstable photon orbit at the radius coordinate $r_{\text{ph}_{-}}$. In this case the right-hand side of (\ref{eq:EoMr}) has a real double root and a pair of complex conjugate roots. We label and sort them such that $r_{\text{ph}_{-}}=r_{2}=r_{1}<r_{\text{H}_{\text{i}}}$ and $r_{3}=\bar{r}_{4}=R_{3}+iR_{4}$, where we choose $0<R_{4}$. In the domain of outer communication light rays and gravitational waves travelling along these geodesics cannot pass through a turning point.
\item Case 4: In this case we have lightlike geodesics characterised by $\Sigma_{\text{ph}_{-}}<\Sigma<\Sigma_{\text{ph}_{0}}$ or $\pi-\Sigma_{\text{ph}_{0}}<\Sigma<\pi-\Sigma_{\text{ph}_{-}}$. In this case the right-hand side of (\ref{eq:EoMr}) has two distinct real roots and a pair of complex conjugate roots. We label and sort them such that $r_{2}<r_{1}<r_{\text{H}_{\text{i}}}$ and $r_{3}=\bar{r}_{4}=R_{3}+iR_{4}$, where we choose $0<R_{4}$. In the domain of outer communication light rays and gravitational waves travelling along these geodesics cannot pass through a turning point.
\item Case 5: In this case we have lightlike geodesics characterised by $\Sigma=\Sigma_{\text{ph}_{0}}$ or $\Sigma=\pi-\Sigma_{\text{ph}_{0}}$ and thus light rays and gravitational waves travelling along these geodesics have the same constants of motion as light rays and gravitational waves on the stable photon orbit at the radius coordinate $r_{\text{ph}_{0}}$. In this case the right-hand side of (\ref{eq:EoMr}) has four real roots. One of these roots is a double root. We label and sort them such that $r_{4}<r_{\text{ph}_{0}}=r_{3}=r_{2}<r_{1}<r_{\text{H}_{\text{i}}}$. In the domain of outer communication light rays and gravitational waves travelling along these geodesics cannot pass through a turning point.
\item Case 6: In this case we have lightlike geodesics characterised by $\Sigma_{\text{ph}_{0}}<\Sigma<\Sigma_{\text{ph}_{+}}$ or $\pi-\Sigma_{\text{ph}_{+}}<\Sigma<\pi-\Sigma_{\text{ph}_{0}}$. In this case the right-hand side of (\ref{eq:EoMr}) has four distinct real roots inside the Cauchy horizon. We label and sort them such that $r_{4}<r_{3}<r_{2}<r_{1}<r_{\text{H}_{\text{i}}}$. Since these roots lie inside the Cauchy horizon in the domain of outer communication light rays and gravitational waves travelling along these geodesics cannot pass through a turning point.
\item Case 7a: In this case we have lightlike geodesics characterised by $\Sigma=\Sigma_{\text{ph}_{+}}$ or $\Sigma=\pi-\Sigma_{\text{ph}_{+}}$ and thus light rays and gravitational waves travelling along these geodesics have the same constants of motion as light rays and gravitational waves on the unstable photon orbit at the radius coordinate $r_{\text{ph}_{+}}$. In this case the right-hand side of (\ref{eq:EoMr}) has four real roots and all of them lie inside the Cauchy horizon. In addition, two of them are equal. We label and sort the roots such that $r_{4}<r_{3}<r_{\text{ph}_{+}}=r_{2}=r_{1}<r_{\text{H}_{\text{i}}}$. Since all roots lie inside the Cauchy horizon in the domain of outer communication light rays and gravitational waves travelling along these geodesics cannot pass through a turning point. 
\item Case 7b: In this case we have $0<a<m$ and lightlike geodesics characterised by $\Sigma=\Sigma_{\text{ph}_{0}}=\Sigma_{\text{ph}_{+}}$ or $\Sigma=\pi-\Sigma_{\text{ph}_{+}}=\pi-\Sigma_{\text{ph}_{0}}$. Light rays and gravitational waves travelling along these geodesics have the same constants of motion as light rays and gravitational waves on the photon orbit at the radius coordinate $r_{\text{ph}_{0}}=r_{\text{ph}_{+}}$. In this case the right-hand side of (\ref{eq:EoMr}) has four real roots and all of them lie inside the Cauchy horizon. Three of these roots are equal. Now we label and sort the roots such that $r_{4}<r_{\text{ph}_{0}}=r_{\text{ph}_{+}}=r_{3}=r_{2}=r_{1}<r_{\text{H}_{\text{i}}}$. Since again all roots lie inside the Cauchy horizon in the domain of outer communication light rays and gravitational waves travelling along these geodesics cannot pass through a turning point.
\item Case 7c: In this case we have $a=m$ and lightlike geodesics characterised by $\Sigma=\Sigma_{\text{ph}_{0}}=\Sigma_{\text{ph}_{+}}$ or $\Sigma=\pi-\Sigma_{\text{ph}_{+}}=\pi-\Sigma_{\text{ph}_{0}}$. In this case we have photon orbits at the radius coordinates $r_{\text{ph}_{0}}=r_{\text{ph}_{+}}=r_{\text{H}}<r_{\text{ph}}$. However, for the latitudinal projections of the photon orbits we have $\Sigma_{\text{ph}}<\Sigma_{\text{ph}_{0}}=\Sigma_{\text{ph}_{+}}$. Light rays and gravitational waves travelling along these geodesics have the same constants of motion as light rays and gravitational waves on the photon orbit at the radius coordinate $r_{\text{ph}_{0}}=r_{\text{ph}_{+}}$. In this case the right-hand side of (\ref{eq:EoMr}) has four real roots. Two of these roots are equal. Now we label and sort them such that $r_{4}<r_{\text{ph}_{0}}=r_{\text{ph}_{+}}=r_{\text{H}}=r_{3}=r_{2}<r_{1}$. The root $r_{1}=r_{\text{min}}$ lies outside the horizon and thus in the domain of outer communication light rays and gravitational waves travelling along these geodesics can pass through a turning point.
\item Case 8: In this case we have lightlike geodesics characterised by $\Sigma_{\text{ph}_{+}}<\Sigma<\Sigma_{\text{ph}}$ or $\pi-\Sigma_{\text{ph}}<\Sigma<\pi-\Sigma_{\text{ph}_{+}}$. As for case 4 in this case the right-hand side of (\ref{eq:EoMr}) has two distinct real roots and a pair of complex conjugate roots. Again we label and sort them such that $r_{2}<r_{1}<r_{\text{H}_{\text{i}}}$ and $r_{3}=\bar{r}_{4}=R_{3}+iR_{4}$, where we choose $0<R_{4}$. In the domain of outer communication light rays and gravitational waves travelling along these geodesics cannot pass through a turning point.
\item Case 9a: In this case we have lightlike geodesics characterised by $\Sigma=\Sigma_{\text{ph}}$ or $\Sigma=\pi-\Sigma_{\text{ph}}$. Light rays and gravitational waves travelling along these geodesics are either travelling on the unstable photon orbit at the radius coordinate $r_{\text{ph}}$ or they are either asymptotically coming from or going to the unstable photon orbit at the radius coordinate $r_{\text{ph}}$. In the latter case they have the same constants of motion as light rays and gravitational waves on the unstable photon orbit. In this case the right-hand side of (\ref{eq:EoMr}) has four real roots and the two largest roots are equal. We label and sort them such that $r_{4}<r_{3}<r_{\text{H}_{\text{i}}}\leq r_{\text{H}_{\text{o}}}<r_{\text{ph}}=r_{2}=r_{1}$. Since the roots $r_{1}$ and $r_{2}$ coincide with the radius coordinate of the unstable photon orbit $r_{\text{ph}}$, in the domain of outer communication light rays and gravitational waves travelling along these geodesics cannot pass through a turning point.
\item Case 9b: In this case we have $a=m$ and lightlike geodesics characterised by $\Sigma=\Sigma_{\text{ph}_{+}}=\Sigma_{\text{ph}}$ or $\Sigma=\pi-\Sigma_{\text{ph}}=\pi-\Sigma_{\text{ph}_{+}}$. Again light rays and gravitational waves travelling along these geodesics are either asymptotically coming from or going to the photon orbit at the radius coordinate $r_{\text{ph}_{+}}=r_{\text{H}}=r_{\text{ph}}$ (note though that in this case the radius coordinate of the photon orbit has the same value as the radius coordinate of the horizon and thus as mentioned in Sec.~\ref{Sec:POCP} in Boyer-Lindquist coordinates this case is not well-represented). In this case the right-hand side of (\ref{eq:EoMr}) has four real roots and two of these roots are equal. We label and sort them such that $r_{4}<r_{3}<r_{\text{ph}_{+}}=r_{\text{H}}=r_{\text{ph}}=r_{2}=r_{1}$. In the domain of outer communication light rays and gravitational waves travelling along these geodesics cannot pass through a turning point.
\item Case 9c: In this case we have $a=m$ and lightlike geodesics characterised by $\Sigma=\Sigma_{\text{ph}_{0}}=\Sigma_{\text{ph}_{+}}=\Sigma_{\text{ph}}$ or $\Sigma=\pi-\Sigma_{\text{ph}}=\pi-\Sigma_{\text{ph}_{+}}=\pi-\Sigma_{\text{ph}_{0}}$. Again light rays and gravitational waves travelling along these geodesics are either asymptotically coming from or going to the photon orbit at the radius coordinate $r_{\text{ph}_{0}}=r_{\text{ph}_{+}}=r_{\text{H}}=r_{\text{ph}}$ (as in case 9b in this case the radius coordinate of the photon orbit has the same value as the radius coordinate of the horizon and thus as mentioned in Sec.~\ref{Sec:POCP} in Boyer-Lindquist coordinates this case is not well-represented). In this case the right-hand side of (\ref{eq:EoMr}) has four real roots and three of these roots are equal. We label and sort them such that $r_{4}<r_{\text{ph}_{0}}=r_{\text{ph}_{+}}=r_{\text{H}}=r_{\text{ph}}=r_{3}=r_{2}=r_{1}$. In the domain of outer communication light rays and gravitational waves travelling along these geodesics cannot pass through a turning point.
\item Case 10: In this case for $r_{\text{ph}}<r$ we have lightlike geodesics characterised by $\Sigma_{\text{ph}}<\Sigma<\pi-\Sigma_{\text{ph}}$ and for $r_{\text{H}_{\text{o}}}<r<r_{\text{ph}}$ we have lightlike geodesics characterised by $\pi-\Sigma_{\text{ph}}<\Sigma<\Sigma_{\text{ph}}$. In this case the right-hand side of (\ref{eq:EoMr}) has four distinct real roots. We label and sort them such that $r_{4}<r_{3}<r_{2}<r_{1}$. Note that at least one of these roots lies in the domain of outer communication. For light rays and gravitational waves travelling along lightlike geodesics at radius coordinates $r_{\text{ph}}<r$ this is necessarily a minimum at $r_{\text{min}}=r_{1}$. For light rays and gravitational waves travelling along lightlike geodesics between the event horizon and the photon orbit at the radius coordinate $r_{\text{ph}}$ this is necessarily a maximum at $r_{\text{max}}=r_{2}$. Thus in the domain of outer communication light rays and gravitational waves travelling along these geodesics can pass through a turning point.
\end{itemize}
After classifying the different types of motion we now have to briefly discuss how this scheme changes when we have degenerate or less than four photon orbits. We start with the case that we have degenerate photon orbits. With the exception of $r_{\text{ph}_{0}}=r_{\text{ph}_{+}}$ for $a=m$ in this case the type of $r$ motion which would occur at and between the boundaries simply does not occur, e.g., when we have $0<a<m$ and $\Sigma_{\text{ph}_{0}}=\Sigma_{\text{ph}_{+}}$ the types of $r$ motion classified as cases 5, 6, and 7a do not occur and we only have the type of $r$ motion classified as case 7b instead. Similarly when we have a triple-degenerate photon orbit at $r_{\text{ph}_{0}}=r_{\text{ph}_{+}}=r_{\text{ph}}$ we have $\Sigma_{\text{ph}_{0}}=\Sigma_{\text{ph}_{+}}=\Sigma_{\text{ph}}$ and the types of $r$ motion classified as cases 5, 6, 7a, 8, and 9a do not occur and we only have the type of $r$ motion classified as case 9c instead. However, this is slightly different when we have $r_{\text{ph}_{0}}=r_{\text{ph}_{+}}<r_{\text{ph}}$ for $a=m$. In this case the types of $r$ motion classified as cases 5, 6, 7a, and 8 do not occur and we have to replace for case 4 $\Sigma_{\text{ph}_{0}}$ by $\Sigma_{\text{ph}}$. In addition, the type of motion along lightlike geodesics classified as case 10 is now characterised by $\Sigma_{\text{ph}}<\Sigma<\Sigma_{\text{ph}_{0}}=\Sigma_{\text{ph}_{+}}$, $\Sigma_{\text{ph}_{0}}=\Sigma_{\text{ph}_{+}}<\Sigma<\pi-\Sigma_{\text{ph}_{+}}=\pi-\Sigma_{\text{ph}_{0}}$, and $\pi-\Sigma_{\text{ph}_{+}}=\pi-\Sigma_{\text{ph}_{0}}<\Sigma<\pi-\Sigma_{\text{ph}}$ while the type of motion for lightlike geodesics characterised by $\Sigma=\Sigma_{\text{ph}_{0}}=\Sigma_{\text{ph}_{+}}$ is classified as case 7c.\\
The last question we need to address is now how this classification scheme changes when we have less than four photon orbits. Here, we have to distinguish between observers outside the equatorial plane and observers inside the equatorial plane. We start with the former. For observers outside the equatorial plane we only have one more photon orbit constellation with $r_{\text{ph}_{-}}<r_{\text{H}_{\text{i}}}\leq r_{\text{H}_{\text{o}}}<r_{\text{ph}}$ and $\Sigma_{\text{ph}_{-}}<\Sigma_{\text{ph}}$. In this case we only have the types of motion classified as cases 1b, 2, 3, 4 (with $\Sigma_{\text{ph}_{0}}$ and $\pi-\Sigma_{\text{ph}_{0}}$ replaced by $\Sigma_{\text{ph}}$ and $\pi-\Sigma_{\text{ph}}$, respectively), 9a, and 10.\\
For observers inside the equatorial plane we have three different photon orbit constellations. First we have $0<a\leq m$ and counterrotating or corotating light rays and gravitational waves travelling along lightlike geodesics outside the equatorial plane characterised by $a\sin\Psi<z_{\Psi_{1}}$. For all geodesics which fulfill this criterion we only have the projection $\Sigma_{\text{ph}}$ of the unstable photon orbit at the radius coordinate $r_{\text{ph}}$ onto the observer's celestial sphere. In this case we only have the types of $r$ motion classified as cases 1a, 8 (with $\Sigma_{\text{ph}_{+}}$ and $\pi-\Sigma_{\text{ph}_{+}}$ replaced by $0$ and $\pi$, respectively), 9a, and 10. In the second case we have corotating light rays and gravitational waves travelling along lightlike geodesics in the equatorial plane characterised by $a\sin\Psi=z_{\text{max}}$. For all geodesics which fulfill this criterion we only have the projections $\Sigma_{\text{ph}_{+}}$ and $\Sigma_{\text{ph}}$ of the unstable photon orbits at the radius coordinates $r_{\text{ph}_{+}}$ and $r_{\text{ph}}$ onto the observer's celestial sphere. Here, for $0<a<m$ we have only the types of $r$ motion classified as cases 1a, 6 (with $\Sigma_{\text{ph}_{0}}$ and $\pi-\Sigma_{\text{ph}_{0}}$ replaced by $0$ and $\pi$, respectively), 7a, 8, 9a, and 10. For $a=m$ on the other hand we have only the types of $r$ motion classified as cases 1a, 6 (with $\Sigma_{\text{ph}_{0}}$ and $\pi-\Sigma_{\text{ph}_{0}}$ replaced by $0$ and $\pi$, respectively), 9b, and 10. In the third case we have $\vartheta_{i}=\pi/2$ and light rays and gravitational waves travelling along lightlike geodesics outside the equatorial plane. This time the geodesics are characterised by $z_{\Psi_{1}}\leq a\sin\Psi<z_{\text{max}}$ and thus we have up to three distinct photon orbits. For these geodesics the overall classification for the different types of motion is the same as for four photon orbits, however, case 1b is replaced by case 1a, and the cases 2 and 3 do not occur. In addition, for case 4 we have to replace $\Sigma_{\text{ph}_{-}}$ and $\pi-\Sigma_{\text{ph}_{-}}$ by $0$ and $\pi$, respectively. 

\subsubsection{Solving the Equation of Motion}
Case 1a: In this case we have lightlike geodesics in the equatorial plane characterised by $\Sigma=0$ or $\Sigma=\pi$. This implies that $K_{\text{C}}=0$. Since we have $\text{d}\vartheta/\text{d}\lambda=0$ the right-hand side of (\ref{eq:EoMtheta}) has to vanish. Thus we can easily see that we have $L_{z\text{C}}=aE_{\text{C}}$ and thus (\ref{eq:EoMr}) reduces to
\begin{eqnarray}
\left(\frac{\text{d}r}{\text{d}\lambda}\right)^2=E_{\text{C}}^2r^4.
\end{eqnarray}
Here, in our convention for the Mino parameter we can easily read from (\ref{eq:EoMt}) that when we have $K_{\text{C}}=0$ the energy $E$ and thus also $E_{\text{C}}$ is always positive. Now we separate variables and integrate from $r(\lambda_{i})=r_{i}$ to $r(\lambda)=r$. We get
\begin{eqnarray}
\lambda-\lambda_{i}=\frac{i_{r_{i}}}{E_{\text{C}}}\int_{r_{i}}^{r}\frac{\text{d}r'}{r'^2}=\frac{i_{r_{i}}}{E_{\text{C}}}\left(\frac{1}{r_{i}}-\frac{1}{r}\right),
\end{eqnarray}
where $i_{r_{i}}=\text{sgn}\left(\left.\text{d}r/\text{d}\lambda\right|_{r=r_{i}}\right)$. We solve for $r$ and obtain as solution for $r(\lambda)$
\begin{eqnarray}\label{eq:rSolEq}
r(\lambda)=\frac{r_{i}}{1-i_{r_{i}}E_{\text{C}}r_{i}\left(\lambda-\lambda_{i}\right)}. 
\end{eqnarray}
Case 1b: In this case we have lightlike geodesics outside the equatorial plane characterised by $\Sigma=0$ or $\Sigma=\pi$. Again this implies that we have $K_{\text{C}}=0$ and thus these geodesics lie on cones. From the right-hand side of (\ref{eq:EoMtheta}) we can easily see that in this case we have $\text{d}\vartheta/\text{d}\lambda=0$ and thus $\vartheta(\lambda)=\vartheta_{i}$. As a consequence we have $L_{z\text{C}}=a\sin^2\vartheta_{i}E_{\text{C}}$. Thus in this case (\ref{eq:EoMr}) reduces to
\begin{eqnarray}
\left(\frac{\text{d}r}{\text{d}\lambda}\right)^2=\left(r^2+a^2\cos^2\vartheta_{i}\right)^2 E_{\text{C}}^2
\end{eqnarray}
Now we separate variables and integrate from $r(\lambda_{i})=r_{i}$ to $r(\lambda)=r$ and get
\begin{eqnarray}
\lambda-\lambda_{i}=\frac{i_{r_{i}}}{E_{\text{C}}}\int_{r_{i}}^{r}\frac{\text{d}r'}{r'^2+a^2\cos^2\vartheta_{i}}.
\end{eqnarray}
The right-hand side is an elementary integral. Now we evaluate it and solve for $r$. We obtain as as solution for $r(\lambda)$
\begin{eqnarray}
r(\lambda)=a\cos\vartheta_{i}\tan\left(\arctan\left(\frac{r_{i}}{a\cos\vartheta_{i}}\right)+i_{r_{i}}a\cos\vartheta_{i}E_{\text{C}}\left(\lambda-\lambda_{i}\right)\right).
\end{eqnarray}
Case 2: In this case we have lightlike geodesics characterised by $0<\Sigma<\Sigma_{\text{ph}_{-}}$ or $\pi-\Sigma_{\text{ph}_{-}}<\Sigma<\pi$. As already mentioned above they only exist for $\vartheta_{i}\neq \pi/2$. We recall that in this case the right-hand side of (\ref{eq:EoMr}) has two distinct pairs of complex conjugate roots and that we labelled and sorted them such that $r_{1}=\bar{r}_{2}=R_{1}+iR_{2}$ and $r_{3}=\bar{r}_{4}=R_{3}+iR_{4}$, where we chose $R_{1}<R_{3}$, and $0<R_{2}$ and $0<R_{4}$. Now we first rewrite the right-hand side of (\ref{eq:EoMr}) in terms of the roots. Then we separate variables and integrate from $r(\lambda_{i})=r_{i}$ to $r(\lambda)=r$. We get
\begin{eqnarray}\label{eq:EoMrEI1}
\lambda-\lambda_{i}=\frac{i_{r_{i}}}{E_{\text{C}}}\int_{r_{i}}^{r}\frac{\text{d}r'}{\sqrt{\left(\left(R_{1}-r'\right)^2+R_{2}^2\right)\left(\left(R_{3}-r'\right)^2+R_{4}^2\right)}}.
\end{eqnarray}
In the next step we use the real ($R_{1}$ and $R_{3}$) and imaginary ($R_{2}$ and $R_{4}$) parts of the roots to define two new constants of motion. They read
\begin{eqnarray}\label{eq:CR1}
R=\sqrt{\left(R_{2}-R_{4}\right)^2+\left(R_{1}-R_{3}\right)^2}~~~\text{and}~~~\bar{R}=\sqrt{\left(R_{2}+R_{4}\right)^2+\left(R_{1}-R_{3}\right)^2}.
\end{eqnarray}
Then we use the newly defined constants of motion to define a third constant of motion $g_{0}$. It reads
\begin{eqnarray}\label{eq:g0}
g_{0}=\sqrt{\frac{4R_{2}^2-(R-\bar{R})^2}{(R+\bar{R})^2-4R_{2}^2}}.
\end{eqnarray}
In the next step we substitute \cite{Gralla2020b,Byrd1954}
\begin{eqnarray}\label{eq:elsub1}
r=R_{1}-R_{2}\frac{g_{0}-\tan\chi}{1+g_{0}\tan\chi}
\end{eqnarray}
to put (\ref{eq:EoMrEI1}) into the Legendre form given by (\ref{eq:LFI}) in Appendix~\ref{Sec:EFD}. Now we follow the steps outlined in Appendix~\ref{Sec:EFD} and obtain the solution for $r(\lambda)$ in terms of Jacobi's elliptic sc function. It reads
\begin{eqnarray}
r(\lambda)=R_{1}-R_{2}\frac{g_{0}-\text{sc}\left(i_{r_{i}}\frac{R+\bar{R}}{2}E_{\text{C}}\left(\lambda-\lambda_{i}\right)+\lambda_{r_{i},k_{1}},k_{1}\right)}{1+g_{0}\text{sc}\left(i_{r_{i}}\frac{R+\bar{R}}{2}E_{\text{C}}\left(\lambda-\lambda_{i}\right)+\lambda_{r_{i},k_{1}},k_{1}\right)},
\end{eqnarray}
where the quantity $\lambda_{r_{i},k_{1}}$, the square of the elliptic modulus $k_{1}$, and $\chi_{i}$ are given by
\begin{eqnarray}
\lambda_{r_{i},k_{1}}=F_{L}(\chi_{i},k_{1}),
\end{eqnarray}
\begin{eqnarray}\label{eq:EM1}
k_{1}=\frac{4R\bar{R}}{\left(R+\bar{R}\right)^2},
\end{eqnarray}
and
\begin{eqnarray}\label{eq:elchi1}
\chi_{i}=\arctan\left(\frac{r_{i}-R_{1}}{R_{2}}\right)+\arctan\left(g_{0}\right).
\end{eqnarray}
Case 3: In this case we have lightlike geodesics characterised by $\Sigma=\Sigma_{\text{ph}_{-}}$ or $\Sigma=\pi-\Sigma_{\text{ph}_{-}}$. Like the geodesics in case 2 they only exist for $\vartheta_{i}\neq\pi/2$. Light rays and gravitational waves travelling along these geodesics have the same constants of motion as light rays and gravitational waves on the unstable photon orbit at the radius coordinate $r_{\text{ph}_{-}}$. We recall that in this case the right-hand side of (\ref{eq:EoMr}) has a real double root at $r_{\text{ph}_{-}}=r_{2}=r_{1}<r_{\text{H}_{\text{i}}}$ and a pair of complex conjugate roots given by $r_{3}=\bar{r}_{4}=R_{3}+iR_{4}$, where we chose $0<R_{4}$. We first rewrite the right-hand side of (\ref{eq:EoMr}) in terms of the roots. Then we separate variables and integrate from $r(\lambda_{i})=r_{i}$ to $r(\lambda)=r$.
We get
\begin{eqnarray}
\lambda-\lambda_{i}=\frac{i_{r_{i}}}{E_{\text{C}}}\int_{r_{i}}^{r}\frac{\text{d}r'}{\left(r'-r_{\text{ph}_{-}}\right)\sqrt{\left(\left(R_{3}-r'\right)^2+R_{4}^2\right)}}.
\end{eqnarray}
We see that we can rewrite the right-hand side in terms of the elementary integral $I_{3}$ given by (\ref{eq:I3}) in Appendix~\ref{Sec:ELI1}. Now we evaluate the integral and insert the result. We solve for $r$ and get as solution for $r(\lambda)$
\begin{eqnarray}
r(\lambda)=r_{\text{ph}_{-}}+\frac{\left(R_{3}-r_{\text{ph}_{-}}\right)^2+R_{4}^2}{R_{3}-r_{\text{ph}_{-}}+R_{4}\sinh\left(a_{r_{\text{ph}_{-}}}-i_{r_{i}}E_{\text{C}}\sqrt{(R_{3}-r_{\text{ph}_{-}})^2+R_{4}^2}\left(\lambda-\lambda_{i}\right)\right)},
\end{eqnarray}
where the coefficient $a_{r_{\text{ph}_{-}}}$ is given by
\begin{eqnarray}
a_{r_{\text{ph}_{-}}}=\text{arsinh}\left(\frac{(r_{\text{ph}_{-}}-R_{3})(r_{i}-r_{\text{ph}_{-}})+(R_{3}-r_{\text{ph}_{-}})^2+R_{4}^2}{R_{4}\left(r_{i}-r_{\text{ph}_{-}}\right)}\right).
\end{eqnarray}
Case 4: In this case we have lightlike geodesics characterised by $\Sigma_{\text{ph}_{-}}<\Sigma<\Sigma_{\text{ph}_{0}}$ or $\pi-\Sigma_{\text{ph}_{0}}<\Sigma<\pi-\Sigma_{\text{ph}_{-}}$. We recall that in this case the right-hand side of (\ref{eq:EoMr}) has two distinct real roots $r_{2}<r_{1}<r_{\text{H}_{\text{i}}}$ and a pair of complex conjugate roots given by $r_{3}=\bar{r}_{4}=R_{3}+iR_{4}$, where we chose $0<R_{4}$. Again we rewrite the right-hand side of (\ref{eq:EoMr}) in terms of the roots. In the next step we integrate from $r(\lambda_{i})=r_{i}$ to $r(\lambda)=r$ and get
\begin{eqnarray}\label{eq:EoMrEI2}
\lambda-\lambda_{i}=\frac{i_{r_{i}}}{E_{\text{C}}}\int_{r_{i}}^{r}\frac{\text{d}r'}{\sqrt{\left(r'-r_{1}\right)\left(r'-r_{2}\right)\left(\left(R_{3}-r'\right)^2+R_{4}^2\right)}}.
\end{eqnarray}
Then we define two new constants of motion $\tilde{R}$ and $\hat{R}$. They read
\begin{eqnarray}\label{eq:CR2}
\tilde{R}=\sqrt{(R_{3}-r_{1})^2+R_{4}^2}~~~\text{and}~~~\hat{R}=\sqrt{(R_{3}-r_{2})^2+R_{4}^2}.
\end{eqnarray}
In the next step we substitute \cite{Hancock1917,Gralla2020b}
\begin{eqnarray}\label{eq:elsub2}
r=\frac{r_{1}\hat{R}-r_{2}\tilde{R}+(r_{1}\hat{R}+r_{2}\tilde{R})\cos\chi}{\hat{R}-\tilde{R}+(\hat{R}+\tilde{R})\cos\chi}
\end{eqnarray}
to put (\ref{eq:EoMrEI2}) into the Legendre form given by (\ref{eq:LFI}) in Appendix~\ref{Sec:EFD}. We follow the steps outlined in Appendix~\ref{Sec:EFD} and obtain the solution for $r(\lambda)$ in terms of Jacobi's elliptic cn function. It reads
\begin{eqnarray}\label{eq:solcase4}
r(\lambda)=\frac{r_{1}\hat{R}-r_{2}\tilde{R}+(r_{1}\hat{R}+r_{2}\tilde{R})\text{cn}\left(i_{r_{i}}E_{\text{C}}\sqrt{\tilde{R}\hat{R}}\left(\lambda-\lambda_{i}\right)+\lambda_{r_{i},k_{2}},k_{2}\right)}{\hat{R}-\tilde{R}+(\hat{R}+\tilde{R})\text{cn}\left(i_{r_{i}}E_{\text{C}}\sqrt{\tilde{R}\hat{R}}\left(\lambda-\lambda_{i}\right)+\lambda_{r_{i},k_{2}},k_{2}\right)},
\end{eqnarray}
where $\lambda_{r_{i},k_{2}}$, the square of the elliptic modulus $k_{2}$, and $\chi_{i}$ are given by 
\begin{eqnarray}\label{eq:lamcase4}
\lambda_{r_{i},k_{2}}=F_{L}(\chi_{i},k_{2}),
\end{eqnarray}
\begin{eqnarray}\label{eq:EM2}
k_{2}=\frac{(\tilde{R}+\hat{R})^2-(r_{1}-r_{2})^2}{4\tilde{R}\hat{R}},
\end{eqnarray}
and
\begin{eqnarray}\label{eq:elchi2}
\chi_{i}=\arccos\left(\frac{(r_{i}-r_{2})\tilde{R}-(r_{i}-r_{1})\hat{R}}{(r_{i}-r_{2})\tilde{R}+(r_{i}-r_{1})\hat{R}}\right).
\end{eqnarray}
Case 5: In this case we have lightlike geodesics characterised by $\Sigma=\Sigma_{\text{ph}_{0}}$ or $\Sigma=\pi-\Sigma_{\text{ph}_{0}}$. Light rays and gravitational waves travelling along these geodesics have the same constants of motion as light rays and gravitational waves on the stable photon orbit at the radius coordinate $r_{\text{ph}_{0}}$. We recall that in this case the right-hand side of (\ref{eq:EoMr}) has four real roots and that we labelled and sorted them such that $r_{4}<r_{\text{ph}_{0}}=r_{3}=r_{2}<r_{1}<r_{\text{H}_{\text{i}}}$. Again we first rewrite the right-hand side of (\ref{eq:EoMr}) in terms of the roots. We separate variables and integrate from $r(\lambda_{i})=r_{i}$ to $r(\lambda)=r$. We get
\begin{eqnarray}
\lambda-\lambda_{i}=\frac{i_{r_{i}}}{E_{\text{C}}}\int_{r_{i}}^{r}\frac{\text{d}r'}{\left(r'-r_{\text{ph}_{0}}\right)\sqrt{\left(r'-r_{1}\right)\left(r'-r_{4}\right)}}.
\end{eqnarray}
The integral on the right-hand side has the same form as the elementary integral $I_{7}$ given by (\ref{eq:I7}) in Appendix~\ref{Sec:ELI2}. We evaluate the integral and obtain $I_{7_{1}}$ given by (\ref{eq:I71S}) with $r_{a}=r_{\text{ph}_{0}}$. Now we insert the obtained expression and solve for $r$. We get as solution for $r(\lambda)$
\begin{eqnarray}\label{eq:solcase5}
r(\lambda)=r_{\text{ph}_{0}}-\frac{2(r_{1}-r_{\text{ph}_{0}})(r_{\text{ph}_{0}}-r_{4})}{r_{1}+r_{4}-2r_{\text{ph}_{0}}+(r_{1}-r_{4})\sin\left(a_{r_{\text{ph}_{0}}}+i_{r_{i}}E_{\text{C}}\sqrt{(r_{1}-r_{\text{ph}_{0}})(r_{\text{ph}_{0}}-r_{4})}\left(\lambda-\lambda_{i}\right)\right)},
\end{eqnarray}
where the coefficient $a_{r_{\text{ph}_{0}}}$ is given by
\begin{eqnarray}\label{eq:coeffcase5}
a_{r_{\text{ph}_{0}}}=\arcsin\left(\frac{(2r_{\text{ph}_{0}}-r_{1}-r_{4})(r_{i}-r_{\text{ph}_{0}})-2(r_{1}-r_{\text{ph}_{0}})(r_{\text{ph}_{0}}-r_{4})}{(r_{1}-r_{4})(r_{i}-r_{\text{ph}_{0}})}\right).
\end{eqnarray}
Case 6: In this case we have lightlike geodesics characterised by $\Sigma_{\text{ph}_{0}}<\Sigma<\Sigma_{\text{ph}_{+}}$ or $\pi-\Sigma_{\text{ph}_{+}}<\Sigma<\pi-\Sigma_{\text{ph}_{0}}$. We recall that in this case the right-hand side of (\ref{eq:EoMr}) has four distinct real roots, that we labelled and sorted them such that $r_{4}<r_{3}<r_{2}<r_{1}<r_{\text{H}_{\text{i}}}$, and that all of these roots lie inside the Cauchy horizon. Again we first rewrite the right-hand side of (\ref{eq:EoMr}) in terms of the roots. Then we separate variables and integrate from $r(\lambda_{i})=r_{i}$ to $r(\lambda)=r$ and get
\begin{eqnarray}\label{eq:MinoC6}
\lambda-\lambda_{i}=\frac{i_{r_{i}}}{E_{\text{C}}}\int_{r_{i}}^{r}\frac{\text{d}r'}{\sqrt{\left(r'-r_{1}\right)\left(r'-r_{2}\right)\left(r'-r_{3}\right)\left(r'-r_{4}\right)}}.
\end{eqnarray}
Then we substitute \cite{Gralla2020b,Hancock1917}
\begin{eqnarray}\label{eq:elsub3}
r=r_{2}+\frac{(r_{1}-r_{2})(r_{2}-r_{4})}{r_{2}-r_{4}-(r_{1}-r_{4})\sin^2\chi}
\end{eqnarray}
to put (\ref{eq:MinoC6}) into the Legendre form given by (\ref{eq:LFI}) in Appendix~\ref{Sec:EFD}. Now we follow the steps outlined in Appendix~\ref{Sec:EFD} and obtain the solution for $r(\lambda)$. This time the solution for $r(\lambda)$ is given in terms of Jacobi's elliptic sn function and reads 
\begin{eqnarray}\label{eq:solcase6}
r(\lambda)=r_{2}+\frac{(r_{1}-r_{2})(r_{2}-r_{4})}{r_{2}-r_{4}-(r_{1}-r_{4})\text{sn}^2\left(\frac{i_{r_{i}}E_{\text{C}}}{2}\sqrt{(r_{1}-r_{3})(r_{2}-r_{4})}\left(\lambda-\lambda_{i}\right)+\lambda_{r_{i},k_{3}},k_{3}\right)},
\end{eqnarray}
where $\lambda_{r_{i},k_{3}}$, the square of the elliptic modulus $k_{3}$, and $\chi_{i}$ are given by 
\begin{eqnarray}\label{eq:lamcase6}
\lambda_{r_{i},k_{3}}=F_{L}(\chi_{i},k_{3}),
\end{eqnarray}
\begin{eqnarray}\label{eq:EM3}
k_{3}=\frac{(r_{2}-r_{3})(r_{1}-r_{4})}{(r_{1}-r_{3})(r_{2}-r_{4})},
\end{eqnarray}
and
\begin{eqnarray}\label{eq:elchi3}
\chi_{i}=\arcsin\left(\sqrt{\frac{(r_{i}-r_{1})(r_{2}-r_{4})}{(r_{i}-r_{2})(r_{1}-r_{4})}}\right).
\end{eqnarray}
Case 7a: In this case we have lightlike geodesics characterised by $\Sigma=\Sigma_{\text{ph}_{+}}$ or $\Sigma=\pi-\Sigma_{\text{ph}_{+}}$. Light rays and gravitational waves travelling along these geodesics have the same constants of motion as light rays and gravitational waves on the unstable photon orbit at the radius coordinate $r_{\text{ph}_{+}}$. We recall that in this case the right-hand side of (\ref{eq:EoMr}) has four real roots and that we labelled and sorted them such that $r_{4}<r_{3}<r_{\text{ph}_{+}}=r_{2}=r_{1}<r_{\text{H}_{\text{i}}}$. Again all roots are located inside the Cauchy horizon. As before we first rewrite the right-hand side of (\ref{eq:EoMr}) in terms of the roots and integrate from $r(\lambda_{i})=r_{i}$ to $r(\lambda)=r$. This time we get
\begin{eqnarray}
\lambda-\lambda_{i}=\frac{i_{r_{i}}}{E_{\text{C}}}\int_{r_{i}}^{r}\frac{\text{d}r'}{\left(r'-r_{\text{ph}_{+}}\right)\sqrt{\left(r'-r_{3}\right)\left(r'-r_{4}\right)}}.
\end{eqnarray}
In the second step we substitute 
\begin{eqnarray}\label{eq:elm1}
r=r_{3}+\frac{a_{3r}}{4\left(y-\frac{a_{2r}}{12}\right)},
\end{eqnarray}
where the coefficients $a_{2r}$ and $a_{3r}$ read
\begin{eqnarray}\label{eq:coeffa2r}
a_{2r}=6E_{\text{C}}^2r_{3}^2+2aE_{\text{C}}\left(aE_{\text{C}}-L_{z\text{C}}\right)-K_{\text{C}}
\end{eqnarray}
and
\begin{eqnarray}\label{eq:coeffa3r}
a_{3r}=2\left(2E_{\text{C}}^2r_{3}^3+\left(2aE_{\text{C}}\left(aE_{\text{C}}-L_{z\text{C}}\right)-K_{\text{C}}\right)r_{3}+mK_{\text{C}}\right),
\end{eqnarray}
and get
\begin{eqnarray}
\lambda-\lambda_{i}=-\frac{i_{r_{i}}}{2}\int_{y_{i}}^{y}\frac{\text{d}y'}{\sqrt{(y'-y_{\text{ph}_{+}})^2(y'-y_{1})}},
\end{eqnarray}
where $y_{i}$, $y$, $y_{\text{ph}_{+}}$, and $y_{1}$ are related to $r_{i}$, $r$, $r_{\text{ph}_{+}}$, and $r_{4}$ by (\ref{eq:elm1}), respectively.
We can now rewrite the right-hand side in terms of the elementary integral $I_{14}$ given by (\ref{eq:I14}) in Appendix~\ref{Sec:ELI4}. We evaluate $I_{14}$ and obtain $I_{14_{3}}$ given by (\ref{eq:I143}). We insert the result and solve for $r$. The obtained solution for $r(\lambda)$ reads
\begin{eqnarray}
&r(\lambda)=r_{3}-\frac{(r_{\text{ph}_{+}}-r_{3})(r_{3}-r_{4})}{r_{\text{ph}_{+}}-r_{3}-(r_{\text{ph}_{+}}-r_{4})\tanh^2\left(\text{artanh}\left(\sqrt{\frac{(r_{\text{ph}_{+}}-r_{3})(r_{i}-r_{4})}{(r_{\text{ph}_{+}}-r_{4})(r_{i}-r_{3})}}\right)-i_{r_{i}}\sqrt{\frac{a_{3r}(r_{\text{ph}_{+}}-r_{4})}{4(r_{\text{ph}_{+}}-r_{3})(r_{3}-r_{4})}}\left(\lambda-\lambda_{i}\right)\right)}.
\end{eqnarray}
Case 7b: In this case we have $0<a<m$ and lightlike geodesics characterised by $\Sigma=\Sigma_{\text{ph}_{0}}=\Sigma_{\text{ph}_{+}}$ or $\Sigma=\pi-\Sigma_{\text{ph}_{+}}=\pi-\Sigma_{\text{ph}_{0}}$. Light rays and gravitational waves travelling along these geodesics have the same constants of motion as light rays and gravitational waves on the photon orbit at the radius coordinate $r_{\text{ph}_{0}}=r_{\text{ph}_{+}}$. We recall that in this case the right-hand side of (\ref{eq:EoMr}) has four real roots and that we labelled and sorted them such that $r_{4}<r_{\text{ph}_{0}}=r_{\text{ph}_{+}}=r_{3}=r_{2}=r_{1}<r_{\text{H}_{\text{i}}}$ and that all of these roots lie inside the Cauchy horizon. Again we first separate variables. Then we integrate from $r(\lambda_{i})=r_{i}$ to $r(\lambda)=r$. We get
\begin{eqnarray}
\lambda-\lambda_{i}=\frac{i_{r_{i}}}{E_{\text{C}}}\int_{r_{i}}^{r}\frac{\text{d}r'}{\sqrt{\left(r'-r_{\text{ph}_{+}}\right)^3\left(r'-r_{4}\right)}}.
\end{eqnarray}
We see that the integral on the right-hand side can be rewritten in terms of the elementary integral $I_{12}$ given by (\ref{eq:I12}) in Appendix~\ref{Sec:ELI3}. We use (\ref{eq:I12S}) to evaluate $I_{12}$. We insert the result and solve for $r$. We obtain as result for $r(\lambda)$
\begin{eqnarray}\label{eq:solcase7b}
r(\lambda)=\frac{r_{\text{ph}_{+}}\left(\sqrt{r_{i}-r_{4}}-\frac{i_{r_{i}}E_{\text{C}}}{2}\sqrt{(r_{i}-r_{\text{ph}_{+}})}\left(r_{\text{ph}_{+}}-r_{4}\right)\left(\lambda-\lambda_{i}\right)\right)^2-(r_{i}-r_{\text{ph}_{+}})r_{4}}{\left(\sqrt{r_{i}-r_{4}}-\frac{i_{r_{i}}E_{\text{C}}}{2}\sqrt{(r_{i}-r_{\text{ph}_{+}})}\left(r_{\text{ph}_{+}}-r_{4}\right)\left(\lambda-\lambda_{i}\right)\right)^2+r_{\text{ph}_{+}}-r_{i}}.
\end{eqnarray}
Case 7c: In this case we have $a=m$ and lightlike geodesics characterised by $\Sigma_{\text{ph}}<\Sigma=\Sigma_{\text{ph}_{0}}=\Sigma_{\text{ph}_{+}}$ or $\Sigma=\pi-\Sigma_{\text{ph}_{+}}=\pi-\Sigma_{\text{ph}_{0}}<\pi-\Sigma_{\text{ph}}$. Light rays and gravitational waves travelling along these geodesics have the same constants of motion as light rays and gravitational waves on the photon orbit at the radius coordinate $r_{\text{ph}_{0}}=r_{\text{ph}_{+}}=r_{\text{H}}$. We recall that in this case the right-hand side of (\ref{eq:EoMr}) has four real roots, that we labelled and sorted them such that $r_{4}<r_{\text{ph}_{0}}=r_{\text{ph}_{+}}=r_{\text{H}}=r_{3}=r_{2}<r_{1}$, and that the root $r_{1}=r_{\text{min}}$ lies in the domain of outer communication and is a turning point. We see that we have the same root structure as for case 5 and thus also in this case the solution for $r(\lambda)$ is given by (\ref{eq:solcase5}) [note that in Boyer-Lindquist coordinates we can choose both $r_{\text{ph}_{0}}$ and $r_{\text{ph}_{+}}$ to write down this solution but this may be different for other coordinate systems which allow a more appropriate projection of the region at the horizons].\\
Case 8: In this case we have lightlike geodesics characterised by $\Sigma_{\text{ph}_{+}}<\Sigma<\Sigma_{\text{ph}}$ or $\pi-\Sigma_{\text{ph}}<\Sigma<\pi-\Sigma_{\text{ph}_{+}}$. We recall that in this case the right-hand side of (\ref{eq:EoMr}) has two distinct real roots $r_{2}<r_{1}<r_{\text{H}_{\text{i}}}$ and a pair of complex conjugate roots given by $r_{3}=\bar{r}_{4}=R_{3}+iR_{4}$, where we chose $0<R_{4}$. We easily see that we have the same root structure as for case 4. Therefore, also in this case the solution for $r(\lambda)$ is given by (\ref{eq:solcase4}), where the coefficients $\tilde{R}$ and $\hat{R}$ are given by (\ref{eq:CR2}) and $\lambda_{r_{i},k_{2}}$, the square of the elliptic modulus $k_{2}$, and $\chi_{i}$ are given by (\ref{eq:lamcase4}), (\ref{eq:EM2}), and (\ref{eq:elchi2}), respectively.\\
Case 9a: In this case we have lightlike geodesics characterised by $\Sigma=\Sigma_{\text{ph}}$ or $\Sigma=\pi-\Sigma_{\text{ph}}$. Light rays and gravitational waves travelling along these geodesics are either travelling on the photon orbit at the radius coordinate $r_{\text{ph}}$, or they are either asymptotically coming from or going to the photon orbit. We recall that in this case the right-hand side of (\ref{eq:EoMr}) has four real roots and that we labelled and sorted them such that $r_{4}<r_{3}<r_{\text{H}_{\text{i}}}\leq r_{\text{H}_{\text{o}}}<r_{\text{ph}}=r_{2}=r_{1}$. In this case the roots $r_{1}=r_{2}=r_{\text{ph}}$ are located outside the event horizon. In this case we have to distinguish three different subcases. In the first subcase we have $r_{i}=r_{\text{ph}}$ and thus light rays and gravitational waves on the photon orbit. In this case we have $\text{d}r/\text{d}\lambda=0$ and thus the solution to the equation of motion simply reads $r(\lambda)=r_{i}$. For the other two subcases we again follow the steps already outlined for case 7a. We first rewrite the right-hand side of (\ref{eq:EoMr}) in terms of the roots and integrate from $r(\lambda_{i})=r_{i}$ to $r(\lambda)=r$. Then we substitute using (\ref{eq:elm1}) and obtain
\begin{eqnarray}
\lambda-\lambda_{i}=-\frac{i_{r_{i}}}{2}\int_{y_{i}}^{y}\frac{\text{d}y'}{\sqrt{(y'-y_{\text{ph}})^2(y'-y_{1})}},
\end{eqnarray}
where $y_{i}$, $y$, $y_{\text{ph}}$, and $y_{1}$ are related to $r_{i}$, $r$, $r_{\text{ph}}$, and $r_{4}$ by (\ref{eq:elm1}), respectively. As in case 7a we see that we can rewrite the right-hand side in terms of the elementary integral $I_{14}$ given by (\ref{eq:I14}) in Appendix~\ref{Sec:ELI4}. Now we have to distinguish between the remaining two different subcases. In the first subcase we have $r_{\text{ph}}<r$. In this subcase $I_{14}$ is again evaluated using $I_{14_{3}}$ given by (\ref{eq:I143}). We insert the obtained result and solve for $r$. We obtain as solution for $r(\lambda)$
\begin{eqnarray}\label{eq:solcase9a}
&r(\lambda)=r_{3}-\frac{(r_{\text{ph}}-r_{3})(r_{3}-r_{4})}{r_{\text{ph}}-r_{3}-(r_{\text{ph}}-r_{4})\tanh^2\left(\text{artanh}\left(\sqrt{\frac{(r_{\text{ph}}-r_{3})(r_{i}-r_{4})}{(r_{\text{ph}}-r_{4})(r_{i}-r_{3})}}\right)-i_{r_{i}}\sqrt{\frac{a_{3r}(r_{\text{ph}}-r_{4})}{4(r_{\text{ph}}-r_{3})(r_{3}-r_{4})}}\left(\lambda-\lambda_{i}\right)\right)},
\end{eqnarray}
where $a_{3r}$ is given by (\ref{eq:coeffa3r}).\\
In the second subcase we have $r_{\text{H}_{\text{o}}}<r<r_{\text{ph}}$. It only exists when $r_{\text{H}_{\text{o}}}<r_{\text{ph}}$ and thus it is not relevant for the investigation of gravitational lensing in the next section, however, we include it in our discussion for the sake of completeness. In this subcase $I_{14}$ has to be evaluated using $I_{14_{2}}$ given by (\ref{eq:I142}). We again insert the obtained result and solve for $r$. This time we obtain as solution for $r(\lambda)$
\begin{eqnarray}
&r(\lambda)=r_{3}-\frac{(r_{\text{ph}}-r_{3})(r_{3}-r_{4})}{r_{\text{ph}}-r_{3}-(r_{\text{ph}}-r_{4})\coth^2\left(\text{arcoth}\left(\sqrt{\frac{(r_{\text{ph}}-r_{3})(r_{i}-r_{4})}{(r_{\text{ph}}-r_{4})(r_{i}-r_{3})}}\right)+i_{r_{i}}\sqrt{\frac{a_{3r}(r_{\text{ph}}-r_{4})}{4(r_{\text{ph}}-r_{3})(r_{3}-r_{4})}}\left(\lambda-\lambda_{i}\right)\right)},
\end{eqnarray}
where, again $a_{3r}$ is given (\ref{eq:coeffa3r}).\\
Case 9b: In this case we have $a=m$ and lightlike geodesics characterised by $\Sigma=\Sigma_{\text{ph}_{+}}=\Sigma_{\text{ph}}$ or $\Sigma=\pi-\Sigma_{\text{ph}}=\pi-\Sigma_{\text{ph}_{+}}$. Again light rays and gravitational waves travelling along these geodesics are either asymptotically coming from or going to the photon orbit at the radius coordinate $r_{\text{ph}_{+}}=r_{\text{H}}=r_{\text{ph}}$. We recall that in this case the right-hand side of (\ref{eq:EoMr}) has four real roots and that we labelled and sorted them such that $r_{4}<r_{3}<r_{\text{ph}_{+}}=r_{\text{H}}=r_{\text{ph}}=r_{2}=r_{1}$. We can easily see that this is the same root structure as for case 9a. Since we have $r_{\text{ph}_{+}}=r_{\text{H}}=r_{\text{ph}}$ in this case the solution $r(\lambda)$ is again given by (\ref{eq:solcase9a}).\\
Case 9c: In this case we have $a=m$ and lightlike geodesics characterised by $\Sigma=\Sigma_{\text{ph}_{0}}=\Sigma_{\text{ph}_{+}}=\Sigma_{\text{ph}}$ or $\Sigma=\pi-\Sigma_{\text{ph}}=\pi-\Sigma_{\text{ph}_{+}}=\pi-\Sigma_{\text{ph}_{0}}$. Again light rays and gravitational waves travelling along these geodesics are either asymptotically coming from or going to the photon orbit at the radius coordinate $r_{\text{ph}_{0}}=r_{\text{ph}_{+}}=r_{\text{H}}=r_{\text{ph}}$. We recall that in this case the right-hand side of (\ref{eq:EoMr}) has four real roots and that we labelled and sorted them such that $r_{4}<r_{\text{ph}_{0}}=r_{\text{ph}_{+}}=r_{\text{H}}=r_{\text{ph}}=r_{3}=r_{2}=r_{1}$. We can easily see that this is the same root structure as for case 7b. Thus also in this case the solution $r(\lambda)$ is given by (\ref{eq:solcase7b}).\\
Case 10: In this case for $r_{\text{ph}}<r$ we have lightlike geodesics characterised by $\Sigma_{\text{ph}}<\Sigma<\pi-\Sigma_{\text{ph}}$ and for $r_{\text{H}_{\text{o}}}<r<r_{\text{ph}}$ we have lightlike geodesics characterised by $\pi-\Sigma_{\text{ph}}<\Sigma<\Sigma_{\text{ph}}$. We recall that in this case the right-hand side of (\ref{eq:EoMr}) has four distinct real roots, that we labelled and sorted them such that $r_{4}<r_{3}<r_{2}<r_{1}$, and that at least one of these roots lies in the domain of outer communication. When we have light rays and gravitational waves travelling along lightlike geodesics at radius coordinates $r_{\text{ph}}<r$ they can pass through a turning point at the radius coordinate $r_{\text{min}}=r_{1}$. We can easily see that although at least one of the roots lies in the domain of outer communication in this case the root structure is the same as for case 6. Therefore, also in this case we can write down the solution for $r(\lambda)$ in terms of Jacobi's elliptic sn function and as for case 6 it is given by (\ref{eq:solcase6}), where $\lambda_{r_{i},k_{3}}$, the square of the elliptic modulus $k_{3}$, and $\chi_{i}$ are given by (\ref{eq:lamcase6}), (\ref{eq:EM3}), and (\ref{eq:elchi3}), respectively. Here we have to note that although the original integral was just valid up to the first turning point due to the periodicity of Jacobi's elliptic sn function the solution for $r(\lambda)$ is also valid beyond the first turning point.\\
While for our investigation of gravitational lensing in the Kerr spacetime we only need the solution to the equation of motion for $r$ discussed above there is another class of geodesics which is structurally very similar and we include it in our discussion for the sake of completeness. It describes the motion along lightlike geodesics between the event horizon and the photon orbit at the radius coordinate $r_{\text{ph}}$. These geodesics only exist when $r_{\text{H}_{\text{o}}}<r_{\text{ph}}$. In this case light rays and gravitational waves travelling along these geodesics can pass through a turning point at the radius coordinate $r_{\text{max}}=r_{2}$ and this turning point is always a maximum of the $r$ motion. Now for deriving the solution for $r(\lambda)$ we first rewrite the right-hand side of (\ref{eq:EoMr}) in terms of the roots and get (\ref{eq:MinoC6}). Then we substitute \cite{Hancock1917}
\begin{eqnarray}\label{eq:elsub4}
r=r_{1}-\frac{(r_{1}-r_{2})(r_{1}-r_{3})}{r_{1}-r_{3}-(r_{2}-r_{3})\sin^2\chi}
\end{eqnarray}
to put (\ref{eq:MinoC6}) into the Legendre form given by (\ref{eq:LFI}) in Appendix~\ref{Sec:EFD}. Now as for case 6 we follow the steps outlined in Appendix~\ref{Sec:EFD} and obtain the solution for $r(\lambda)$. Again the solution for $r(\lambda)$ is given in terms of Jacobi's elliptic sn function and reads 
\begin{eqnarray}
r(\lambda)=r_{1}-\frac{(r_{1}-r_{2})(r_{1}-r_{3})}{r_{1}-r_{3}-(r_{2}-r_{3})\text{sn}^2\left(\frac{i_{r_{i}}E_{\text{C}}}{2}\sqrt{(r_{1}-r_{3})(r_{2}-r_{4})}\left(\lambda_{i}-\lambda\right)+\lambda_{r_{i},k_{3}},k_{3}\right)},
\end{eqnarray}
where $\lambda_{r_{i},k_{3}}$ and $k_{3}$ are given by (\ref{eq:lamcase6}) and (\ref{eq:EM3}), respectively. $\chi_{i}$ is related to the radius coordinate $r_{i}$ by
\begin{eqnarray}\label{eq:elchi4}
\chi_{i}=\arcsin\left(\sqrt{\frac{(r_{2}-r_{i})(r_{1}-r_{3})}{(r_{1}-r_{i})(r_{2}-r_{3})}}\right).
\end{eqnarray}
Here we have to note again that although the original integral was just valid up to the first turning point due to the periodicity of Jacobi's elliptic sn function the solution for $r(\lambda)$ is also valid beyond the first turning point. In addition, we have to note that on the first view the substitution (\ref{eq:elsub4}) looks similar to the one used by Gralla and Lupsasca \cite{Gralla2020b}, however, they labelled the roots opposite to our convention and thus they are actually not the same. In addition, the substitution of Gralla and Lupsasca has the disadvantage that in the case at hand the $r$ dependent components of the $\varphi$ coordinate and the time coordinate $t$ diverge when we integrate over the horizons. The substitution given by (\ref{eq:elsub4}) completely avoids this problem. 
\subsection{The $\vartheta$ Motion}\label{Sec:EoMtheta}
Now we turn to the $\vartheta$ motion. Before we will solve the equation of motion we will first derive the conditions for light rays and gravitational waves travelling at constant spacetime latitudes. In analogy to the photon orbits we will call the surfaces associated with these spacetime latitudes \emph{photon plane} and \emph{photon cones}. Afterwards we will then proceed to solving the equation of motion.
\subsubsection{Photon Plane and Photon Cones}
For light rays and gravitational waves propagating at constant spacetime latitudes the equation of motion for $\vartheta$ has to fulfill the conditions $\text{d}\vartheta/\text{d}\lambda=\text{d}^2\vartheta/\text{d}\lambda^2=0$. When we apply the condition $\text{d}^2\vartheta/\text{d}\lambda^2=0$ to (\ref{eq:EoMtheta}) we get three different solutions. Then we insert the obtained solutions into the condition $\text{d}\vartheta/\text{d}\lambda=0$ and derive the conditions for the constants of motion and the angles on the celestial sphere of the observer. The first solution is given by
\begin{eqnarray}
\vartheta_{\text{ph}}=\frac{\pi}{2}~~~\text{for}~~~K_{\text{C}}-(aE_{\text{C}}-L_{z\text{C}})^2=0.
\end{eqnarray}
In terms of the angles on the observer's celestial sphere this translates to an observer in the equatorial plane detecting light rays or gravitational waves from directions marked by the celestial longitudes $\Psi=\pi/2$ or $\Psi=3\pi/2$. Therefore, the equatorial plane of the Kerr spacetime is what one can refer to as a \emph{photon plane}. All light rays and gravitational waves starting or ending parallel to the equatorial plane remain in the equatorial plane.\\
The other two solutions are given by
\begin{eqnarray}
\vartheta_{\text{ph}_{-}}=\arccos\left(\sqrt{\frac{2aE_{\text{C}}\left(aE_{\text{C}}-L_{z\text{C}}\right)-K_{\text{C}}}{2a^2E_{\text{C}}^2}}\right)~~~\text{and}~~~\vartheta_{\text{ph}_{+}}=\arccos\left(-\sqrt{\frac{2aE_{\text{C}}\left(aE_{\text{C}}-L_{z\text{C}}\right)-K_{\text{C}}}{2a^2E_{\text{C}}^2}}\right),
\end{eqnarray}
where the constants of motion have to fulfill the condition
\begin{eqnarray}
K_{\text{C}}+4aE_{\text{C}}L_{z\text{C}}=0.
\end{eqnarray}
In this case relating the lightlike geodesics to the latitude-longitude coordinates on the observer's celestial sphere is slightly more difficult. However, we can easily see when we insert the relations between the constants of motion and the latitude-longitude coordinates on the observer's celestial sphere in (\ref{eq:EoMtheta}) that for $\Psi=\pi/2$ and $\Psi=3\pi/2$ we always have a turning point at $\vartheta_{O}$. Therefore, when we have $K_{\text{C}}+4aE_{\text{C}}L_{z\text{C}}=0$ in terms of the angles on the observer's celestial sphere this translates to an observer at either $\vartheta_{\text{ph}_{-}}$ or $\vartheta_{\text{ph}_{+}}$ detecting light rays or gravitational waves from the directions marked by the celestial longitudes $\Psi=\pi/2$ or $\Psi=3\pi/2$. Note that due to the dependence on the constants of motion these cones exist only for certain individual light rays and gravitational waves. Therefore, these cones are \emph{individual photon cones}.
\subsubsection{Solving the Equation of Motion}
For the $\vartheta$ motion we have to distinguish five different types of motion. In the first case we have $K_{\text{C}}=0$ and thus $\Sigma=0$ or $\Sigma=\pi$. We can easily see that in this case the right-hand side of (\ref{eq:EoMtheta}) has to be zero. Light rays and gravitational waves travelling along these geodesics either travel in the equatorial plane or on cones at the spacetime latitude $\vartheta_{i}$. Among these light rays and gravitational waves the ones travelling along the axes at $\vartheta_{i}=0$ or $\vartheta_{i}=\pi$ mark another special type of geodesics. As these geodesics are confined to the axis of rotation we have $L_{z\text{C}}=0$ and thus these geodesics are purely radial. The second type of geodesics are geodesics in the equatorial plane or on an individual photon cone. For lightlike geodesics in the equatorial plane we have $\vartheta_{i}=\pi/2$ and the constants of motion fulfill the relation $K_{\text{C}}-(aE_{\text{C}}-L_{z\text{C}})^2=0$ (or for an observer in the equatorial plane the observer detects light rays or gravitational waves coming from the directions marked by the longitudes $\Psi=\pi/2$ or $\Psi=3\pi/2$). For lightlike geodesics on one of the individual photon cones we have $\vartheta_{i}=\vartheta_{\text{ph}_{-}}$ or $\vartheta_{i}=\vartheta_{\text{ph}_{+}}$ and the constants of motion fulfill the relation $K_{\text{C}}+4aE_{\text{C}}L_{z\text{C}}=0$ (or for an observer located on one of the cones the observer detects light rays or gravitational waves coming from the directions marked by the longitudes $\Psi=\pi/2$ or $\Psi=3\pi/2$).\\ 
Thus for both types of motion we have $\text{d}\vartheta/\text{d}\lambda=0$ and thus the solution to the equation of motion reads
\begin{eqnarray}
\vartheta(\lambda)=\vartheta_{i}.
\end{eqnarray}
Before we discuss the other types of motion let us first rewrite the right-hand side of (\ref{eq:EoMtheta}) in terms of $x=\cos\vartheta$. For this purpose we first multiply (\ref{eq:EoMtheta}) with $\sin^2\vartheta$. Then we use $\text{d}\cos\vartheta=-\sin\vartheta\text{d}\vartheta$ and get
\begin{eqnarray}\label{eq:EoMx}
\left(\frac{\text{d}x}{\text{d}\lambda}\right)^2=-a^2E_{\text{C}}^2x^4+\left(2aE_{\text{C}}\left(aE_{\text{C}}-L_{z\text{C}}\right)-K_{\text{C}}\right)x^2+K_{\text{C}}-\left(aE_{\text{C}}-L_{z\text{C}}\right)^2.
\end{eqnarray}
From (\ref{eq:EoMx}) we can easily see that we can only have $\vartheta$ motion when the right-hand side has either two or four real roots. Here, we exclude the cases of light rays and gravitational waves in the equatorial plane or on individual photon cones. In this case we have three different types of motion which we will now discuss. Note that here we could also rewrite the conditions for which the three different types of motion occur in terms of the celestial angles, however, since this would require several case distinctions it is more straightforward to write the conditions in terms of the constants of motion rewritten in terms of the celestial angles $E_{\text{C}}$, $L_{z\text{C}}$, and $K_{\text{C}}$.\\
The third type of motion is given by lightlike geodesics characterised by $K_{\text{C}}-(aE_{\text{C}}-L_{z\text{C}})^2>0$. In this case the right-hand side of (\ref{eq:EoMx}) has two real roots $x_{2}=-x_{1}<0<x_{1}$ and a pair of purely imaginary complex conjugate roots given by $x_{3}=\bar{x}_{4}=iX_{4}$, where we choose $0<X_{4}$. In general light rays and gravitational waves travelling along these geodesics oscillate between the turning points $x_{1}=\cos\vartheta_{\text{min}}$ and $x_{2}=\cos\vartheta_{\text{max}}$. Now we rewrite the right-hand side of (\ref{eq:EoMx}) in terms of the roots, separate variables, and integrate from $x(\lambda_{i})=\cos\vartheta_{i}$ to $x(\lambda)=\cos\vartheta(\lambda)=\cos\vartheta$. We get
\begin{eqnarray}\label{eq:EoMxR}
\lambda-\lambda_{i}=-i_{\vartheta_{i}}\int_{\cos\vartheta_{i}}^{\cos\vartheta}\frac{\text{d}x'}{\sqrt{-a^2E_{\text{C}}^2(x'^2-x_{1}^2)(x'^2+X_{4}^2)}},
\end{eqnarray}
where we defined $i_{\vartheta_{i}}=\text{sgn}\left(\left.\text{d}\vartheta/\text{d}\lambda\right|_{\vartheta=\vartheta_{i}}\right)$. In the next step we substitute 
\begin{eqnarray}\label{eq:EoMxsub1}
x=\frac{x_{1}X_{4}\sin\chi}{\sqrt{X_{4}^2+x_{1}^2\cos^2\chi}}
\end{eqnarray}
to put (\ref{eq:EoMxR}) into the Legendre form given by (\ref{eq:LFI}) in Appendix~\ref{Sec:EFD}. Now we follow the steps outlined in Appendix~\ref{Sec:EFD} to derive the solution for $\vartheta(\lambda)$. It is given in terms of Jacobi's elliptic sd function and reads  
\begin{eqnarray}\label{eq:soleltheta1}
\vartheta(\lambda)=\arccos\left(\frac{x_{1}X_{4}}{\sqrt{x_{1}^2+X_{4}^2}}\text{sd}\left(i_{\vartheta_{i}}\sqrt{a^2(x_{1}^2+X_{4}^2)E_{\text{C}}^2}\left(\lambda_{i}-\lambda\right)+\lambda_{\vartheta_{i},k_{4}},k_{4}\right)\right),
\end{eqnarray}
where $\lambda_{\vartheta_{i},k_{4}}$, the square of the elliptic modulus $k_{4}$, and $\chi_{i}$ are given by
\begin{eqnarray}
\lambda_{\vartheta_{i},k_{4}}=F_{L}\left(\chi_{i},k_{4}\right),
\end{eqnarray}
\begin{eqnarray}\label{eq:EM4}
k_{4}=\frac{x_{1}^2}{x_{1}^2+X_{4}^2},
\end{eqnarray}
and
\begin{eqnarray}\label{eq:elchi5}
\chi_{i}=\arcsin\left(\frac{\sqrt{x_{1}^2+X_{4}^2}\cos\vartheta_{i}}{x_{1}\sqrt{\cos^2\vartheta_{i}+X_{4}^2}}\right).
\end{eqnarray}
Here we can easily see that (\ref{eq:elchi5}) projects the coordinate range $x_{2}\leq \cos\vartheta\leq x_{1}$ to $-\pi/2\leq\chi\leq \pi/2$ and thus the solution covers the whole coordinate range between both turning points. In addition, although the solution (\ref{eq:soleltheta1}) was only derived using an integral which was valid up to the first turning point due to the periodicity of Jacobi's elliptic sd function it is also valid beyond the first turning point. Note that here and also for the other two types of motion we keep $E_{\text{C}}^2$ inside the root since these types of motion can also occur for $E_{\text{C}}<0$. \\
The fourth type of motion is given by lightlike geodesics characterised by $2aE_{\text{C}}(aE_{\text{C}}-L_{z\text{C}})-K_{\text{C}}>0$ and $K_{\text{C}}-(aE_{\text{C}}-L_{z\text{C}})^2<0$. In this case the right-hand side of (\ref{eq:EoMx}) has four real roots and we label and sort them such that $x_{4}=-x_{1}<x_{3}=-x_{2}<0<x_{2}<x_{1}$ and the motion is either limited to spacetime latitudes above or spacetime latitudes below the equatorial plane. In the first case light rays and gravitational waves travelling along these geodesics generally oscillate between the turning points at $x_{1}=\cos\vartheta_{\text{min}}$ and $x_{2}=\cos\vartheta_{\text{max}}$. In the second case light rays and gravitational waves travelling along these geodesics generally oscillate between the turning points at $x_{3}=\cos\vartheta_{\text{min}}$ and $x_{4}=\cos\vartheta_{\text{max}}$. This type of motion is usually referred to as \emph{vortical} motion. \\
For solving (\ref{eq:EoMx}) we again rewrite the right-hand side in terms of the roots, separate variables, and integrate from $x(\lambda_{i})=\cos\vartheta_{i}$ to $x(\lambda)=\cos\vartheta(\lambda)=\cos\vartheta$. This time we get
\begin{eqnarray}\label{eq:EoMxRR}
\lambda-\lambda_{i}=-i_{\vartheta_{i}}\int_{\cos\vartheta_{i}}^{\cos\vartheta}\frac{\text{d}x'}{\sqrt{-a^2E_{\text{C}}^2(x'^2-x_{1}^2)(x'^2-x_{2}^2)}}.
\end{eqnarray}
In the next step we substitute 
\begin{eqnarray}\label{eq:EoMxsub2}
x=\sqrt{x_{1}^2-(x_{1}^2-x_{2}^2)\sin^2\chi},
\end{eqnarray}
for light rays and gravitational waves oscillating between the turning points $x_{1}=\cos\vartheta_{\text{min}}$ and $x_{2}=\cos\vartheta_{\text{max}}$, and 
\begin{eqnarray}\label{eq:EoMxsub3}
x=-\sqrt{x_{1}^2-(x_{1}^2-x_{2}^2)\sin^2\chi},
\end{eqnarray}
for light rays and gravitational waves oscillating between the turning points $x_{3}=\cos\vartheta_{\text{min}}$ and $x_{4}=\cos\vartheta_{\text{max}}$
to put (\ref{eq:EoMxRR}) into the Legendre form given by (\ref{eq:LFI}) in Appendix~\ref{Sec:EFD}. Now we again follow the steps described in Appendix~\ref{Sec:EFD} to derive the solutions for $\vartheta(\lambda)$. This time they are given in terms of Jacobi's elliptic dn function. For light rays and gravitational waves oscillating between the turning points $x_{1}=\cos\vartheta_{\text{min}}$ and $x_{2}=\cos\vartheta_{\text{max}}$ it reads
\begin{eqnarray}\label{eq:soleltheta2}
\vartheta(\lambda)=\arccos\left(x_{1}\text{dn}\left(i_{\vartheta_{i}}x_{1}\sqrt{a^2E_{\text{C}}^2}\left(\lambda-\lambda_{i}\right)+\lambda_{\vartheta_{i},k_{5}},k_{5}\right)\right).
\end{eqnarray}
On the other hand for light rays and gravitational waves oscillating between the turning points $x_{3}=\cos\vartheta_{\text{min}}$ and $x_{4}=\cos\vartheta_{\text{max}}$ the solution reads
\begin{eqnarray}\label{eq:soleltheta3}
\vartheta(\lambda)=\arccos\left(x_{4}\text{dn}\left(i_{\vartheta_{i}}x_{4}\sqrt{a^2E_{\text{C}}^2}\left(\lambda-\lambda_{i}\right)+\lambda_{\vartheta_{i},k_{5}},k_{5}\right)\right),
\end{eqnarray}
where in both cases $\lambda_{\vartheta_{i},k_{5}}$, the square of the elliptic modulus $k_{5}$, and $\chi_{i}$ are given by
\begin{eqnarray}
\lambda_{\vartheta_{i},k_{5}}=F_{L}\left(\chi_{i},k_{5}\right),
\end{eqnarray}
\begin{eqnarray}\label{eq:EM5}
k_{5}=\frac{x_{1}^2-x_{2}^2}{x_{1}^2},
\end{eqnarray}
and
\begin{eqnarray}\label{eq:elchi6}
\chi_{i}=\arcsin\left(\sqrt{\frac{x_{1}^2-\cos^2\vartheta_{i}}{x_{1}^2-x_{2}^2}}\right).
\end{eqnarray}
Note that as for (\ref{eq:soleltheta1}), although the solutions (\ref{eq:soleltheta2}) and (\ref{eq:soleltheta3}) were only derived using an integral which was valid up to the first turning point due to the periodicity of Jacobi's elliptic dn function they are also valid beyond the first turning point.\\
The fifth and last type of motion we have to discuss is a special type of vortical motion. In the case that we have $2aE_{\text{C}}(aE_{\text{C}}-L_{z\text{C}})-K_{\text{C}}>0$ and $K_{\text{C}}-(aE_{\text{C}}-L_{z\text{C}})^2=0$ we have $x_{2}=x_{3}=0$. The solutions derived above are still valid, however, in this case we can also derive the solutions to (\ref{eq:EoMtheta}) in terms of elementary functions. For this purpose we first insert $x_{2}=0$ in (\ref{eq:EoMxRR}) and get
\begin{eqnarray}\label{eq:EoMxR0}
\lambda-\lambda_{i}=\mp i_{\vartheta_{i}}\int_{\cos\vartheta_{i}}^{\cos\vartheta}\frac{\text{d}x'}{x'\sqrt{-a^2E_{\text{C}}^2(x'^2-x_{1}^2)}},
\end{eqnarray}
where we have to choose the $-$ sign for lightlike geodesics above the equatorial plane and the $+$ sign for lightlike geodesics below the equatorial plane. Now the integral on the right-hand side is an elementary integral which can be evaluated using
\begin{eqnarray}\label{eq:intVG}
\int\frac{\text{d}x'}{x'\sqrt{b^2-x'^2}}=-\frac{1}{b}\ln\left(\frac{b+\sqrt{b^2-x'^2}}{x'}\right)=-\frac{1}{b}\text{arsech}\left(\frac{x'}{b}\right).
\end{eqnarray}
Here for lightlike geodesics above the equatorial plane we can use this formula as is, while for lightlike geodesics below the equatorial plane we first substitute $x'=-z'$ in (\ref{eq:EoMxR0}) and only then use (\ref{eq:intVG}) to evaluate the integral. Then we solve for $\vartheta$ and get the solutions for $\vartheta(\lambda)$. For lightlike geodesics above the equatorial plane it reads
\begin{eqnarray}\label{eq:soleltheta4}
\vartheta(\lambda)=\arccos\left(x_{1}\text{sech}\left(\text{arsech}\left(\frac{\cos\vartheta_{i}}{x_{1}}\right)+i_{\vartheta_{i}}x_{1}\sqrt{a^2E_{\text{C}}^2}\left(\lambda-\lambda_{i}\right)\right)\right).
\end{eqnarray}
On the other hand for lightlike geodesics below the equatorial plane it reads
\begin{eqnarray}\label{eq:soleltheta5}
\vartheta(\lambda)=\arccos\left(x_{4}\text{sech}\left(\text{arsech}\left(\frac{\cos\vartheta_{i}}{x_{4}}\right)+i_{\vartheta_{i}}x_{4}\sqrt{a^2E_{\text{C}}^2}\left(\lambda-\lambda_{i}\right)\right)\right).
\end{eqnarray}
Note that in both cases for $\lambda \rightarrow \infty$ light rays and gravitational waves travelling along these geodesics are asymptotically approaching the equatorial plane.

\subsection{The $\varphi$ Motion}\label{Sec:EoMphi}
Next we turn to the $\varphi$ motion. We can see from (\ref{eq:EoMphi}) that the right-hand side of the equation of motion consists of two terms. The first term only depends on the radius coordinate $r$ and the second term only depends on the spacetime latitude $\vartheta$. For deriving the solutions describing the $\varphi$ motion we now explicitly write the dependency of $r$ and $\vartheta$ on the Mino parameter $\lambda$. Then we separate variables and integrate from $\varphi(\lambda_{i})=\varphi_{i}$ to $\varphi(\lambda)=\varphi$. The result reads
\begin{eqnarray}
\varphi\left(\lambda\right)=\varphi_{i}+a\varphi_{r}(\lambda)+\varphi_{\vartheta}(\lambda),
\end{eqnarray}
where we defined two functions $\varphi_{r}(\lambda)$ and $\varphi_{\vartheta}(\lambda)$ which describe the $r$ dependent part and the $\vartheta$ dependent part as functions of the Mino parameter $\lambda$, respectively. They read
\begin{eqnarray}\label{eq:phirlam}
\varphi_{r}(\lambda)=\int_{\lambda_{i}}^{\lambda}\frac{(r(\lambda')^2+a^2)E_{\text{C}}-aL_{z\text{C}}}{P(r(\lambda'))}\text{d}\lambda'
\end{eqnarray}
and
\begin{eqnarray}\label{eq:EoMphiinttheta}
\varphi_{\vartheta}(\lambda)=\int_{\lambda_{i}}^{\lambda}\frac{L_{z\text{C}}-a\sin^2\vartheta(\lambda')E_{\text{C}}}{\sin^2\vartheta\left(\lambda'\right)}\text{d}\lambda'.
\end{eqnarray}
In the following we will now evaluate $\varphi_{r}(\lambda)$ and $\varphi_{\vartheta}(\lambda)$ separately. Here, due to the large number of different cases and the length of the solutions for $\varphi_{r}(\lambda)$ we will only explicitly present the solutions for light rays and gravitational waves travelling along lightlike geodesics characterised by $\Sigma=0$ or $\Sigma=\pi$ (corresponding to $K_{\text{C}}=0$), and the solutions for light rays and gravitational waves travelling along lightlike geodesics on the unstable photon orbit at the radius coordinate $r_{\text{ph}}$ in the domain of outer communication. For all other other cases we will only outline the steps for their derivation.
\subsubsection{Calculating $\varphi_{\vartheta}(\lambda)$}
For the $\vartheta$ motion we had to distinguish five different types of motion. For $\varphi_{\vartheta}(\lambda)$ we generally have to distinguish the same different types of motion, however, due to the pathology of the $\varphi$ coordinate at $\vartheta=0$ and $\vartheta=\pi$ radial lightlike geodesics on the axes and lightlike geodesics crossing the axes will be excluded from our discussion.\\
As first type of motion we have light rays and gravitational waves travelling along lightlike geodesics characterised by $\Sigma=0$ or $\Sigma=\pi$ (corresponding to $K_{\text{C}}=0$). Here we have to distinguish between radial lightlike geodesics on the axes with $\vartheta_{i}=0$ or $\vartheta_{i}=\pi$ and lightlike geodesics on cones with $0<\vartheta_{i}<\pi$. As already mentioned above for radial lightlike geodesics on the axes the $\varphi$ coordinate becomes pathological. Therefore, in this case we cannot derive $\varphi_{\vartheta}(\lambda)$ and thus a solution for $\varphi(\lambda)$.\\
For all other geodesics with $0<\vartheta_{i}<\pi$ we saw when we derived the solutions to the equation of motion for $r$ that we have $L_{z\text{C}}=a\sin^2\vartheta_{i} E_{\text{C}}$. When we insert this in (\ref{eq:EoMphiinttheta}) we see that the nominator vanishes and thus in this case we get 
\begin{eqnarray}
\varphi_{\vartheta}(\lambda)=0.
\end{eqnarray}
As second type of motion we have light rays and gravitational waves travelling along lightlike geodesics in the equatorial plane or on individual photon cones. For these geodesics we have $\vartheta(\lambda)=\vartheta_{i}$ and thus the integrand of (\ref{eq:EoMphiinttheta}) is constant. Therefore, we integrate directly over $\lambda$. Here, lightlike geodesics in the equatorial plane are characterised by $\vartheta_{i}=\vartheta_{\text{ph}}=\pi/2$ and $\Psi=\pi/2$ or $\Psi=3\pi/2$ (corresponding to $K_{\text{C}}-(aE_{\text{C}}-L_{z\text{C}})^2=0$ in terms of the constants of motion). Thus we get as result for light rays and gravitational waves travelling along lightlike geodesics in the equatorial plane 
\begin{eqnarray}
\varphi_{\vartheta}(\lambda)=\left(L_{z\text{C}}-aE_{\text{C}}\right)\left(\lambda-\lambda_{i}\right).
\end{eqnarray}
Analogously when we have $\vartheta(\lambda)=\vartheta_{i}=\vartheta_{\text{ph}_{\mp}}$ and $\Psi=\pi/2$ or $\Psi=3\pi/2$ (corresponding to $K_{\text{C}}+4aE_{\text{C}}L_{z\text{C}}=0$ in terms of the constants of motion) the solution for $\varphi_{\vartheta}(\lambda)$ reads
\begin{eqnarray}
\varphi_{\vartheta}(\lambda)=\frac{L_{z\text{C}}-a\sin^2\vartheta_{\text{ph}_{\mp}}E_{\text{C}}}{\sin^2\vartheta_{\text{ph}_{\mp}}}\left(\lambda-\lambda_{i}\right).
\end{eqnarray}
For the remaining three types of motion we have to explicitly evaluate the right-hand side of (\ref{eq:EoMphiinttheta}) in terms of elementary functions and elliptic integrals. Here we note again that for light rays and gravitational waves starting on or crossing the axes at $\vartheta=0$ or $\vartheta=\pi$ the $\varphi$ coordinate becomes pathological and thus we will not cover them in the following discussion. As first step we now rewrite the integrand in the form of a constant term and a term containing $1/(1-\cos^2\vartheta(\lambda'))$. Then we split the integral into two separate integrals and evaluate them. We first evaluate the integral over the constant term. In the case of the second integral we rewrite it such that we get the product of a constant and the integral
\begin{eqnarray}\label{eq:I1t}
I_{\vartheta_{1}}(\lambda)=\int_{\lambda_{i}}^{\lambda}\frac{\text{d}\lambda'}{1-\cos^2\vartheta\left(\lambda'\right)}.
\end{eqnarray}
In this notation $\varphi_{\vartheta}(\lambda)$ now reads
\begin{eqnarray}\label{eq:phitheta1}
\varphi_{\vartheta}(\lambda)=L_{z\text{C}}I_{\vartheta_{1}}(\lambda)-aE_{\text{C}}\left(\lambda-\lambda_{i}\right).
\end{eqnarray}
In the next step we will now evaluate the integral $I_{\vartheta_{1}}(\lambda)$. Here, the exact evaluation procedure we have to use depends on which type of motion we have. \\
We start with the third type of motion. We recall from Section~\ref{Sec:EoMtheta} that in this case we have light rays and gravitational waves travelling along lightlike geodesics characterised by $K_{\text{C}}-(aE_{\text{C}}-L_{z\text{C}})^2>0$. In general the light rays and gravitational waves travelling along these geodesics oscillate between the turning points $x_{1}=\cos\vartheta_{\text{min}}$ and $x_{2}=\cos\vartheta_{\text{max}}$. We recall that in this case the right-hand side of (\ref{eq:EoMx}) has two real roots $x_{2}=-x_{1}=\cos\vartheta_{\text{max}}<0<x_{1}=\cos\vartheta_{\text{min}}$ and a pair of purely imaginary complex conjugate roots given by $x_{3}=\bar{x}_{4}=iX_{4}$, where we chose $0<X_{4}$. In this case we first split the integral $I_{\vartheta_{1}}(\lambda)$ at the turning points of the $\vartheta$ motion into separate integrals. For each integral we now substitute using (\ref{eq:EoMxsub1}). Then we rewrite the integrand in the form of a constant term and a term containing the expression $1/(1-n_{1}\sin^2\chi')$. Then we split the integral again into two integrals and evaluate each separately. In the case of the integral with the constant integrand we integrate directly over $\lambda$. In the case of the other integral we use (\ref{eq:LFD}), adapted to the type of motion at hand, to rewrite it as an integral over $\chi'$. We now immediately see that we can rewrite this integral in terms of two elliptic integrals of the third kind. We insert the obtained result for $I_{\vartheta_{1}}(\lambda)$ in (\ref{eq:phitheta1}) and obtain for $\varphi_{\vartheta}(\lambda)$ 
\begin{eqnarray}\label{eq:phithetaEL1}
\varphi_{\vartheta}(\lambda)=\frac{L_{z\text{C}}-aE_{\text{C}}\left(1+X_{4}^2\right)}{1+X_{4}^2}\left(\lambda-\lambda_{i}\right)+\sum_{n=1}^{N}\frac{i_{\vartheta_{i}n}L_{z\text{C}}X_{4}^2\left(\Pi_{L}\left(\chi_{n-1},k_{4},n_{1}\right)-\Pi_{L}\left(\chi_{n},k_{4},n_{1}\right)\right)}{\left(1+X_{4}^2\right)\sqrt{a^2E_{\text{C}}^2\left(x_{1}^2+X_{4}^2\right)}},
\end{eqnarray}
where $N$ is the number of terms into which we had to split $I_{\vartheta_{1}}(\lambda)$. In addition, we have $i_{\vartheta_{i}n}=-i_{\vartheta_{i}n-1}$ with $i_{\vartheta_{i}1}=i_{\vartheta_{i}}$, $\chi_{0}=\chi_{i}$, $\chi_{N}=\chi(\lambda)$, and for $1<n<N$ $\chi_{n}$ and $\chi_{n-1}$ are given by $\chi_{\text{min}}=\pi/2$ or $\chi_{\text{max}}=-\pi/2$. Here, $\chi_{i}$, $\chi_{\text{min}}$, $\chi_{\text{max}}$, and $\chi(\lambda)$ are related to $\cos\vartheta_{i}$, the turning points $x_{1}=\cos\vartheta_{\text{min}}$ and $x_{2}=\cos\vartheta_{\text{max}}$, and $\cos\vartheta(\lambda)$ by (\ref{eq:elchi5}), respectively. The square of the elliptic modulus $k_{4}$ is given by (\ref{eq:EM4}), and the parameter $n_{1}$ reads 
\begin{eqnarray}\label{eq:parn1}
n_{1}=\frac{x_{1}^2\left(1+x_{4}^2\right)}{x_{1}^2+X_{4}^2}.
\end{eqnarray}
Now we turn to the fourth type of motion. It describes light rays and gravitational waves travelling along vortical lightlike geodesics characterised by $2aE_{\text{C}}\left(aE_{\text{C}}-L_{z\text{C}}\right)-K_{\text{C}}>0$ and $K_{\text{C}}-(aE_{\text{C}}-L_{z\text{C}})^2<0$. The light rays and gravitational waves travelling along these geodesics generally oscillate between two turning points either above or below the equatorial plane. We recall that in this case we always have four real roots and that we labelled and sorted them such that $x_{4}=-x_{1}<x_{3}=-x_{2}<0<x_{2}<x_{1}$. Again we first split the integral $I_{\vartheta_{1}}(\lambda)$ at the turning points of the $\vartheta$ motion into separate integrals. Then we evaluate each integral separately. For light rays and gravitational waves oscillating between the turning points $x_{1}=\cos\vartheta_{\text{min}}$ and $x_{2}=\cos\vartheta_{\text{max}}$ we substitute using (\ref{eq:EoMxsub2}). Then we rewrite the integrand such that it contains the expression $1/(1-n_{2}\sin^2\chi')$. Now we use (\ref{eq:LFD}), adapted to the type of motion at hand, to rewrite $I_{\vartheta_{1}}(\lambda)$ as an integral over $\chi'$. We immediately see that the obtained result can be rewritten in terms of Legendre's elliptic integral of the third kind and insert it in (\ref{eq:phitheta1}). Now the result for $\varphi_{\vartheta}(\lambda)$ reads
\begin{eqnarray}\label{eq:phithetaEL2}
\hspace{-0.5cm}\varphi_{\vartheta}\left(\lambda\right)=\sum_{n=1}^{N}\frac{i_{\vartheta_{i}n}L_{z\text{C}}\left(\Pi_{L}\left(\chi_{n},k_{5},n_{2}\right)-\Pi_{L}\left(\chi_{n-1},k_{5},n_{2}\right)\right)}{x_{1}(1-x_{1}^2)\sqrt{a^2E_{\text{C}}^2}}-aE_{\text{C}}\left(\lambda-\lambda_{i}\right),
\end{eqnarray}
where $N$ is the number of terms into which we had to split $I_{\vartheta_{1}}(\lambda)$. In addition, we have $i_{\vartheta_{i}n}=-i_{\vartheta_{i}n-1}$ with $i_{\vartheta_{i}1}=i_{\vartheta_{i}}$, $\chi_{0}=\chi_{i}$, $\chi_{N}=\chi(\lambda)$, and for $1<n<N$ $\chi_{n}$ and $\chi_{n-1}$ are given by $\chi_{\text{min}}=0$ or $\chi_{\text{max}}=\pi/2$. Here, $\chi_{i}$, $\chi_{\text{min}}$, $\chi_{\text{max}}$, and $\chi(\lambda)$ are related to $\cos\vartheta_{i}$, the turning points $x_{1}=\cos\vartheta_{\text{min}}$ and $x_{2}=\cos\vartheta_{\text{max}}$, and $\cos\vartheta(\lambda)$ by (\ref{eq:elchi6}), respectively, the square of the elliptic modulus $k_{5}$ is given by (\ref{eq:EM5}), and the parameter $n_{2}$ reads
\begin{eqnarray}\label{eq:parn2}
n_{2}=\frac{x_{1}^2-x_{2}^2}{x_{1}^2-1}.
\end{eqnarray}
For light rays and gravitational waves oscillating between the turning points $x_{3}=\cos\vartheta_{\text{min}}$ and $x_{4}=\cos\vartheta_{\text{max}}$ on the other hand we substitute using (\ref{eq:EoMxsub3}). Again we rewrite the integrand such that it contains the expression $1/(1-n_{2}\sin^2\chi')$. Again we use (\ref{eq:LFD}), adapted to the type of motion at hand, to rewrite $I_{\vartheta_{1}}(\lambda)$ as an integral over $\chi'$. Again we rewrite the obtained result in terms of Legendre's elliptic integral of the third kind. Finally, we insert the obtained result for $I_{\vartheta_{1}}(\lambda)$ in (\ref{eq:phitheta1}). This time we get as result for $\varphi_{\vartheta}(\lambda)$ 
\begin{eqnarray}\label{eq:phithetaEL3}
\hspace{-0.5cm}\varphi_{\vartheta}\left(\lambda\right)=\sum_{n=1}^{N}\frac{i_{\vartheta_{i}n}L_{z\text{C}}\left(\Pi_{L}\left(\chi_{n-1},k_{5},n_{2}\right)-\Pi_{L}\left(\chi_{n},k_{5},n_{2}\right)\right)}{x_{1}(1-x_{1}^2)\sqrt{a^2E_{\text{C}}^2}}-aE_{\text{C}}\left(\lambda-\lambda_{i}\right),
\end{eqnarray}
where as above $N$ is the number of terms into which we had to split $I_{\vartheta_{1}}(\lambda)$. In addition, we have $i_{\vartheta_{i}n}=-i_{\vartheta_{i}n-1}$ with $i_{\vartheta_{i}1}=i_{\vartheta_{i}}$, $\chi_{0}=\chi_{i}$, $\chi_{N}=\chi(\lambda)$, and for $1<n<N$ $\chi_{n}$ and $\chi_{n-1}$ are given by $\chi_{\text{min}}=\pi/2$ or $\chi_{\text{max}}=0$. Again $\chi_{i}$, $\chi_{\text{min}}$, $\chi_{\text{max}}$, and $\chi(\lambda)$ are related to $\cos\vartheta_{i}$, the turning points $x_{3}=\cos\vartheta_{\text{min}}$ and $x_{4}=\cos\vartheta_{\text{max}}$, and $\cos\vartheta(\lambda)$ by (\ref{eq:elchi6}), respectively. The square of the elliptic modulus $k_{5}$ is given by (\ref{eq:EM5}) and the parameter $n_{2}$ is given by (\ref{eq:parn2}).\\
The fifth and last type of motion describes light rays and gravitational waves travelling along the special vortical geodesics characterised by $2aE_{\text{C}}\left(aE_{\text{C}}-L_{z\text{C}}\right)-K_{\text{C}}>0$ and $K_{\text{C}}-(aE_{\text{C}}-L_{z\text{C}})^2=0$. In this case we have $x_{2}=x_{3}=0$. For lightlike geodesics above the equatorial plane we now insert (\ref{eq:soleltheta4}) in (\ref{eq:I1t}). For lightlike geodesics below the equatorial plane on the other hand we insert (\ref{eq:soleltheta5}). We rewrite the resulting integrals in terms of the cosine hyperbolicus and simplify the integrands such that we again have a constant term, which we can easily integrate, and a term which contains the square of the cosine hyperbolicus in the denominator. In the next step for light rays and gravitational waves travelling along lightlike geodesics above the equatorial plane we substitute $\tilde{\lambda}'=\text{arsech}\left(\frac{\cos\vartheta_{i}}{x_{1}}\right)+i_{\vartheta_{i}}x_{1}\sqrt{a^2E_{\text{C}}^2}\left(\lambda'-\lambda_{i}\right)$. For light rays and gravitational waves travelling along lightlike geodesics below the equatorial plane on the other hand we substitute $\tilde{\lambda}'=\text{arsech}\left(\frac{\cos\vartheta_{i}}{x_{4}}\right)+i_{\vartheta_{i}}x_{4}\sqrt{a^2E_{\text{C}}^2}\left(\lambda'-\lambda_{i}\right)$. In both cases we get an integral of the form (note that we do not explicitly write the integration constant)
\begin{eqnarray}
\int\frac{\text{d}\tilde{\lambda}'}{a_{\vartheta_{1}}+\text{cosh}^2\tilde{\lambda}'}=-\frac{1}{\sqrt{-a_{\vartheta_{1}}\left(1+a_{\vartheta_{1}}\right)}}\arctan\left(\sqrt{-\frac{1+a_{\vartheta_{1}}}{a_{\vartheta_{1}}}}\text{coth}\tilde{\lambda}'\right),
\end{eqnarray}
where we have $a_{\vartheta_{1}}=-x_{1}^2$ for light rays and gravitational waves travelling along lightlike geodesics above the equatorial plane, and $a_{\vartheta_{1}}=-x_{4}^2$ for light rays and gravitational waves travelling along lightlike geodesics below the equatorial plane. We evaluate the integral and obtain as solution for light rays and gravitational waves travelling along lightlike geodesics above the equatorial plane 
\begin{eqnarray}
&\hspace{-0.5cm}\varphi_{\vartheta}(\lambda)=\left(L_{z\text{C}}-aE_{\text{C}}\right)\left(\lambda-\lambda_{i}\right)+\frac{i_{\vartheta_{i}}L_{z\text{C}}}{\sqrt{a^2E_{\text{C}}^2(1-x_{1}^2)}}\left(\arctan\left(\frac{\sqrt{1-x_{1}^2}}{x_{1}}\text{coth}\left(\text{arsech}\left(\frac{\cos\vartheta_{i}}{x_{1}}\right)\right)\right)\right.\\
&\left.-\arctan\left(\frac{\sqrt{1-x_{1}^2}}{x_{1}}\text{coth}\left(\text{arsech}\left(\frac{\cos\vartheta_{i}}{x_{1}}\right)+i_{\vartheta_{i}}x_{1}\sqrt{a^2E_{\text{C}}^2}\left(\lambda-\lambda_{i}\right)\right)\right)\right).\nonumber
\end{eqnarray}
On the other hand for light rays and gravitational waves travelling along lightlike geodesics below the equatorial plane we obtain
\begin{eqnarray}
&\hspace{-0.5cm}\varphi_{\vartheta}(\lambda)=\left(L_{z\text{C}}-aE_{\text{C}}\right)\left(\lambda-\lambda_{i}\right)+\frac{i_{\vartheta_{i}}L_{z\text{C}}}{\sqrt{a^2E_{\text{C}}^2(1-x_{4}^2)}}\left(\arctan\left(\frac{\sqrt{1-x_{4}^2}}{x_{4}}\text{coth}\left(\text{arsech}\left(\frac{\cos\vartheta_{i}}{x_{4}}\right)\right)\right)\right.\\
&\left.-\arctan\left(\frac{\sqrt{1-x_{4}^2}}{x_{4}}\text{coth}\left(\text{arsech}\left(\frac{\cos\vartheta_{i}}{x_{4}}\right)+i_{\vartheta_{i}}x_{4}\sqrt{a^2E_{\text{C}}^2}\left(\lambda-\lambda_{i}\right)\right)\right)\right).\nonumber
\end{eqnarray}
\subsubsection{Calculating $\varphi_{r}(\lambda)$}
Now we turn to the $r$ dependent part $\varphi_{r}(\lambda)$ given by (\ref{eq:phirlam}). We first rewrite it as an integral over $r$. For this purpose we first separate variables in (\ref{eq:EoMr}) and then insert the result in (\ref{eq:phirlam}). Now $\varphi_{r}(\lambda)$ reads
\begin{eqnarray}\label{eq:phirint}
\varphi_{r}(\lambda)=\int_{r_{i}...}^{...r(\lambda)}\frac{\left((r'^2+a^2)E_{\text{C}}-aL_{z\text{C}}\right)\text{d}r'}{P(r')\sqrt{\left(\left(r'^2+a^2\right)E_{\text{C}}-aL_{z\text{C}}\right)^2-P(r')K_{\text{C}}}},
\end{eqnarray}
where the dots in the limits indicate that we have to split the integral at turning points and the sign of the root in the denominator has to be chosen according to the direction of the $r$ motion. When the light rays or gravitational waves do not pass through a turning point we directly integrate from $r_{i}$ to $r(\lambda)$. However, when the light rays or gravitational waves pass through a turning point we have to split (\ref{eq:phirint}) into an integral from $r_{i}$ to the turning point and an integral from the turning point to $r(\lambda)$. For the evaluation of the integral we now have to distinguish the same different types of motion as for $r$. Here, we note again that due to the large number of different cases we will only explicitly present a few selected solutions, namely the solutions for light rays and gravitational waves travelling along lightlike geodesics characterised by $\Sigma=0$ or $\Sigma=\pi$ (corresponding to $K_{\text{C}}=0$), and the solutions for light rays and gravitational waves on the photon orbit at the radius coordinate $r_{\text{ph}}$. In all other cases we will only briefly describe the steps how the solution for $\varphi_{r}(\lambda)$ can be derived.\\
Case 1a and case 1b: In these cases we have light rays and gravitational waves characterised by $\Sigma=0$ or $\Sigma=\pi$. In these cases we have $K_{\text{C}}=0$ and thus $L_{z\text{C}}=a\sin^2\vartheta_{i} E_{\text{C}}$ (we again note that in this case we can see from (\ref{eq:EoMt}) that $E$, and thus also $E_{\text{C}}$, is always positive). We recall that in the case that we have lightlike geodesics on the axis of rotation with $\vartheta(\lambda)=0$ or $\vartheta(\lambda)=\pi$ the $\varphi$ coordinate becomes pathological and thus cannot be derived. Interestingly in all other cases (\ref{eq:phirint}) takes the same form and thus for all remaining lightlike geodesics classified as cases 1a and 1b we get the same results. When we have $0<a<m$ we first use a partial fraction decomposition of $P(r')^{-1}$ with respect to $r'$ to rewrite the integrand as two terms with $r'-r_{a}$ in the denominator. Here, $r_{a}$ can be $r_{\text{H}_{\text{o}}}$ or $r_{\text{H}_{\text{i}}}$. We can easily see that we can now rewrite (\ref{eq:phirint}) in terms of elementary integrals. Now we integrate and get for $\varphi_{r}(\lambda)$ the well-known form
\begin{eqnarray}
\varphi_{r}\left(\lambda\right)=\frac{i_{r_{i}}}{r_{\text{H}_{\text{o}}}-r_{\text{H}_{\text{i}}}}\left(\ln\left(\frac{r\left(\lambda\right)-r_{\text{H}_{\text{o}}}}{r_{i}-r_{\text{H}_{\text{o}}}}\right)+\ln\left(\frac{r_{i}-r_{\text{H}_{\text{i}}}}{r\left(\lambda\right)-r_{\text{H}_{\text{i}}}}\right)\right).
\end{eqnarray}
When we have $a=m$ on the other hand we can evaluate (\ref{eq:phirint}) directly. Again it takes the form of an elementary integral which is easy to calculate. In this case the result for $\varphi_{r}(\lambda)$ reads
\begin{eqnarray}
\varphi_{r}\left(\lambda\right)=i_{r_{i}}\left(\frac{1}{r_{i}-r_{\text{H}}}-\frac{1}{r\left(\lambda\right)-r_{\text{H}}}\right).
\end{eqnarray}
Case 2: In this case we have lightlike geodesics characterised by $0<\Sigma<\Sigma_{\text{ph}_{-}}$ or $\pi-\Sigma_{\text{ph}_{-}}<\Sigma<\pi$. We recall that in this case the right-hand side of (\ref{eq:EoMr}) has two distinct pairs of complex conjugate roots and that we labelled and sorted them such that we have $r_{1}=\bar{r}_{2}=R_{1}+iR_{2}$ and $r_{3}=\bar{r}_{4}=R_{3}+iR_{4}$, where we chose $R_{1}<R_{3}$, and $0<R_{2}$ and $0<R_{4}$. Now we first rewrite the polynomial inside the root in terms of its roots. Again when we have $0<a<m$ we first perform a partial fraction decomposition of $P(r')^{-1}$ with respect to $r'$. Then we rewrite the resulting terms outside the root such that we have a constant term and two terms with $r'-r_{a}$ in the denominator. In this case $r_{a}$ can be $r_{\text{H}_{\text{o}}}$ or $r_{\text{H}_{\text{i}}}$. Then we substitute using (\ref{eq:elsub1}) and rewrite (\ref{eq:phirint}) in terms of Legendre's elliptic integral of the first kind and the nonstandard elliptic integral $I_{L_{2}}(\chi_{i},\chi(\lambda),k_{1},n)$ given by (\ref{eq:EI2}) in Appendix~\ref{Sec:ELIL}. Now we use the first term in (\ref{eq:EI2S}) to completely rewrite $\varphi_{r}(\lambda)$ in terms of elementary functions and Legendre's elliptic integrals of the first and third kind.\\
When we have $a=m$ we proceed analogously, however, in this case we do not need to perform a partial fraction decomposition. We rewrite the term outside the root in terms of a constant term and two terms with powers of $r'-r_{\text{H}}$ in the denominator. Then we again substitute using (\ref{eq:elsub1}) and rewrite (\ref{eq:phirint}) in terms of Legendre's elliptic integral of the first kind and the two nonstandard elliptic integrals $I_{L_{2}}(\chi_{i},\chi(\lambda),k_{1},n)$ and $I_{L_{3}}(\chi_{i},\chi(\lambda),k_{1},n)$ given by (\ref{eq:EI2}) and (\ref{eq:EI3}) in Appendix~\ref{Sec:ELIL}, respectively. This time we use both terms in (\ref{eq:EI2S}) to completely rewrite $\varphi_{r}(\lambda)$ in terms of elementary functions and Legendre's elliptic integrals of the first, second, and third kind.\\
Case 3: In this case we have lightlike geodesics characterised by $\Sigma=\Sigma_{\text{ph}_{-}}$ or $\Sigma=\pi-\Sigma_{\text{ph}_{-}}$. We recall that in this case the right-hand side of (\ref{eq:EoMr}) has a real double root at $r_{\text{ph}_{-}}=r_{2}=r_{1}<r_{\text{H}_{\text{i}}}$ and a pair of complex conjugate roots given by $r_{3}=\bar{r}_{4}=R_{3}+iR_{4}$, where we chose $0<R_{4}$. Again we first rewrite the polynomial inside the root in terms of its roots. Then we pull the factor $r'-r_{\text{ph}_{-}}$ out of the root. For $0<a<m$ we first perform a partial fraction decomposition of $P(r')^{-1}(r'-r_{\text{ph}_{-}})^{-1}$ with respect to $r'$. Then we rewrite the resulting terms outside the root such that three terms with $r'-r_{a}$ in the denominator remain. Here $r_{a}$ can be $r_{\text{ph}_{-}}$, $r_{\text{H}_{\text{o}}}$, or $r_{\text{H}_{\text{i}}}$. We can easily see that we can rewrite (\ref{eq:phirint}) in terms of three different elementary integrals and that they are simply three different versions of the elementary integral $I_{3}$ given by (\ref{eq:I3}) in Appendix~\ref{Sec:ELI1}. We evaluate each integral using the right-hand side of (\ref{eq:I3}) and obtain $\varphi_{r}(\lambda)$ in terms of elementary functions.\\
When we have $a=m$ we proceed analogously. In this case we obtain two different integrals which have the form of the elementary integral $I_{3}$ and one integral which has the form of the elementary integral $I_{4}$ given by (\ref{eq:I3}) and (\ref{eq:I4}) in Appendix~\ref{Sec:ELI1}, respectively. This time we evaluate the integrals using the right-hand sides of (\ref{eq:I3}) and (\ref{eq:I4}) and obtain $\varphi_{r}(\lambda)$ in terms of elementary functions.\\
Case 4: In this case we have lightlike geodesics characterised by $\Sigma_{\text{ph}_{-}}<\Sigma<\Sigma_{\text{ph}_{0}}$ or $\pi-\Sigma_{\text{ph}_{0}}<\Sigma<\pi-\Sigma_{\text{ph}_{-}}$. We recall that in this case the right-hand side of (\ref{eq:EoMr}) has two distinct real roots given by $r_{2}<r_{1}<r_{\text{H}_{\text{i}}}$ and a pair of complex conjugate roots given by $r_{3}=\bar{r}_{4}=R_{3}+iR_{4}$, where we chose $0<R_{4}$. Again we first rewrite the polynomial inside the root in terms of its roots. For $0<a<m$ we perform a partial fraction decomposition of $P(r')^{-1}$ with respect to $r'$. Again we rewrite the resulting terms outside the root such that we have a constant term and two terms with $r'-r_{a}$ in the denominator. As for case 2 $r_{a}$ can be $r_{\text{H}_{\text{o}}}$ or $r_{\text{H}_{\text{i}}}$. In the next step we substitute using (\ref{eq:elsub2}) and rewrite (\ref{eq:phirint}) in terms of Legendre's elliptic integral of the first kind and the nonstandard elliptic integral $I_{L_{4}}(\chi_{i},\chi(\lambda),k_{2},n)$ given by (\ref{eq:EI4}) in Appendix~\ref{Sec:ELIL}. In the next step we use the first term in (\ref{eq:EI3S}) to completely rewrite $\varphi_{r}(\lambda)$ in terms of elementary functions and Legendre's elliptic integrals of the first and third kind. \\
For $a=m$ we proceed analogously, however, in this case we do not need to perform a partial fraction decomposition. We rewrite the term outside the root in terms of a constant term and two terms with powers of $r'-r_{\text{H}}$ in the denominator. In the next step we again substitute using (\ref{eq:elsub2}). Then we rewrite (\ref{eq:phirint}) in terms of Legendre's elliptic integral of the first kind and the two nonstandard elliptic integrals $I_{L_{4}}(\chi_{i},\chi(\lambda),k_{2},n)$ and $I_{L_{5}}(\chi_{i},\chi(\lambda),k_{2},n)$ given by (\ref{eq:EI4}) and (\ref{eq:EI5}) in Appendix~\ref{Sec:ELIL}, respectively. As last step we then use both terms in (\ref{eq:EI3S}) to completely rewrite $\varphi_{r}(\lambda)$ in terms of elementary functions and Legendre's elliptic integrals of the first, second, and third kind.\\
Case 5: In this case we have lightlike geodesics characterised by $\Sigma=\Sigma_{\text{ph}_{0}}$ or $\Sigma=\pi-\Sigma_{\text{ph}_{0}}$. We recall that in this case the right-hand side of (\ref{eq:EoMr}) has four real roots and that we labelled and sorted them such that $r_{4}<r_{\text{ph}_{0}}=r_{3}=r_{2}<r_{1}<r_{\text{H}_{\text{i}}}$. Again we first rewrite the polynomial inside the root in terms of its roots. Then we pull the term $r'-r_{\text{ph}_{0}}$ out of the root. For $0<a<m$ we perform a partial fraction decomposition of the term $P(r')^{-1}(r'-r_{\text{ph}_{0}})^{-1}$ with respect to $r'$. In the next step we again rewrite the resulting terms outside the root such that we have three terms with $r'-r_{a}$ in the denominator. This time $r_{a}$ can be $r_{\text{ph}_{0}}$, $r_{\text{H}_{\text{o}}}$, or $r_{\text{H}_{\text{i}}}$. Now we can easily see that we can rewrite (\ref{eq:phirint}) in terms of three elementary integrals. All of them have the form of the elementary integral $I_{7}$ given by (\ref{eq:I7}) in Appendix~\ref{Sec:ELI2}. We evaluate the integrals using $I_{7_{1}}$ and $I_{7_{2}}$ given by (\ref{eq:I71S}) and (\ref{eq:I72S}), respectively, and obtain $\varphi_{r}(\lambda)$ in terms of elementary functions.\\ 
For $a=m$ we proceed analogously. We perform a partial fraction decomposition of the term $P(r')^{-1}(r'-r_{\text{ph}_{0}})^{-1}$ with respect to $r'$. Then we rewrite the resulting terms outside the root such that we again have three terms with powers of $r'-r_{a}$ in the denominator. In this case $r_{a}$ can be $r_{\text{ph}_{0}}$ or $r_{\text{H}}$. Now we rewrite (\ref{eq:phirint}) in terms of the elementary integrals $I_{7}$ and $I_{8}$ given by (\ref{eq:I7}) and (\ref{eq:I8}) in Appendix~\ref{Sec:ELI2}, respectively. Then we evaluate the integrals using $I_{7_{1}}$, $I_{7_{2}}$, and $I_{8_{2}}$, given by (\ref{eq:I71S}), (\ref{eq:I72S}), and (\ref{eq:I82S}), respectively, and obtain $\varphi_{r}(\lambda)$ in terms of elementary functions.\\ 
Case 6: In this case we have lightlike geodesics characterised by $\Sigma_{\text{ph}_{0}}<\Sigma<\Sigma_{\text{ph}_{+}}$ or $\pi-\Sigma_{\text{ph}_{+}}<\Sigma<\pi-\Sigma_{\text{ph}_{0}}$. We recall that in this case the right-hand side of (\ref{eq:EoMr}) has four real roots and that we labelled and sorted them such that $r_{4}<r_{3}<r_{2}<r_{1}<r_{\text{H}_{\text{i}}}$. Again we first rewrite the polynomial inside the root in terms of its roots. For $0<a<m$ we perform a partial fraction decomposition of $P(r')^{-1}$ with respect to $r'$. Then we rewrite the resulting terms outside the root such that we have a constant term and two terms with $r'-r_{a}$ in the denominator. In this case $r_{a}$ can be $r_{\text{H}_{\text{o}}}$ or $r_{\text{H}_{\text{i}}}$. Then we substitute using (\ref{eq:elsub3}) and rewrite $\varphi_{r}(\lambda)$ in terms of Legendre's elliptic integrals of the first and third kind.\\
For $a=m$ on the other hand we first rewrite the term outside the root such that we have a constant term and two terms with powers of $r'-r_{\text{H}}$ in the denominator. Then we again substitute using (\ref{eq:elsub3}). We rewrite (\ref{eq:phirint}) in terms of Legendre's elliptic integrals of the first and third kind and the nonstandard elliptic integral $I_{L_{6}}(\chi_{i},\chi(\lambda),k_{3},n)$ given by (\ref{eq:EI6}) in Appendix~\ref{Sec:ELIL}. In the next step we use (\ref{eq:EI4S}) to completely rewrite $\varphi_{r}(\lambda)$ in terms of elementary functions and Legendre's elliptic integrals of the first, second, and third kind.\\
Note that in this case when we rewrite $\varphi_{r}(\lambda)$ in terms of elementary functions and Legendre's elliptic integrals of the first, second, and third kind we cannot avoid to integrate over the coordinate singularities at the radius coordinates of the horizons and thus for the associated terms we use (\ref{eq:PiN}) to rewrite Legendre's elliptic integral of the third kind.\\
Case 7a: In this case we have lightlike geodesics characterised by $\Sigma=\Sigma_{\text{ph}_{+}}$ or $\Sigma=\pi-\Sigma_{\text{ph}_{+}}$. We recall that in this case the right-hand side of (\ref{eq:EoMr}) has four real roots and that we labelled and sorted them such that $r_{4}<r_{3}<r_{\text{ph}_{+}}=r_{2}=r_{1}<r_{\text{H}_{\text{i}}}$. Again we first rewrite the polynomial inside the root in terms of its roots. Then we perform a partial fraction decomposition of $P(r')^{-1}$ with respect to $r'$. We rewrite the resulting terms outside the root such that a constant term and two terms with $r'-r_{a}$ in the denominator remain. This time $r_{a}$ can be $r_{\text{H}_{\text{o}}}$ or $r_{\text{H}_{\text{i}}}$. In the next step we substitute using (\ref{eq:elm1}) and pull the term $y'-y_{\text{ph}_{+}}$ under consideration of its sign out of the root.  We perform a second partial fraction decomposition with respect to $y'$. We simplify the result and rewrite (\ref{eq:phirint}) in terms of three different elementary integrals which all take the form of $I_{14}$ given by (\ref{eq:I14}) in Appendix~\ref{Sec:ELI4}. We follow the steps outlined in Appendix~\ref{Sec:ELI4} to evaluate the integrals. We obtain three different versions of $I_{14_{3}}$ given by (\ref{eq:I143}) as results. Using the obtained results we now rewrite $\varphi_{r}(\lambda)$ in terms of elementary functions.\\
Case 7b: In this case we have $0<a<m$ and lightlike geodesics characterised by $\Sigma=\Sigma_{\text{ph}_{0}}=\Sigma_{\text{ph}_{+}}$ or $\Sigma=\pi-\Sigma_{\text{ph}_{+}}=\pi-\Sigma_{\text{ph}_{0}}$. We recall that in this case the right-hand side of (\ref{eq:EoMr}) has four real roots and that we labelled and sorted them such that $r_{4}<r_{\text{ph}_{0}}=r_{\text{ph}_{+}}=r_{3}=r_{2}=r_{1}<r_{\text{H}_{\text{i}}}$. Again we first rewrite the polynomial inside the root in terms of its roots. We pull a factor $r'-r_{\text{ph}_{+}}$ out of the root and perform a partial fraction decomposition of $P(r')^{-1}(r'-r_{\text{ph}_{+}})^{-1}$ with respect to $r'$. Then we rewrite the terms outside the root such that three terms with $r'-r_{a}$ in the denominator remain. This time $r_{a}$ can be $r_{\text{ph}_{+}}$, $r_{\text{H}_{\text{o}}}$, or $r_{\text{H}_{\text{i}}}$. Then we rewrite (\ref{eq:phirint}) in terms of the elementary integrals $I_{12}$ and $I_{13}$ given by (\ref{eq:I12}) and (\ref{eq:I13}) in Appendix~\ref{Sec:ELI3}, respectively. Then we use $I_{12}$ and $I_{13}$ given by (\ref{eq:I12S}) and (\ref{eq:I13S}), respectively, to rewrite $\varphi_{r}(\lambda)$ in terms of elementary functions.\\
Case 7c: In this case we have $a=m$ and lightlike geodesics characterised by $\Sigma=\Sigma_{\text{ph}_{0}}=\Sigma_{\text{ph}_{+}}$ or $\Sigma=\pi-\Sigma_{\text{ph}_{+}}=\pi-\Sigma_{\text{ph}_{0}}$, where $\Sigma_{\text{ph}}<\Sigma_{\text{ph}_{0}}=\Sigma_{\text{ph}_{+}}$. We recall that in this case the right-hand side of (\ref{eq:EoMr}) has four real roots and that we labelled and sorted them such that $r_{4}<r_{\text{ph}_{0}}=r_{\text{ph}_{+}}=r_{\text{H}}=r_{3}=r_{2}<r_{1}$ and that the root $r_{1}=r_{\text{min}}$ lies in the domain of outer communication and is a turning point. Again we first rewrite the polynomial inside the root in terms of its roots. Then we pull the term $r'-r_{\text{H}}$ out of the root. We rewrite the term outside the root such that only three terms with powers of $r'-r_{\text{H}}$ in the denominator remain. Then we rewrite (\ref{eq:phirint}) in terms of the three elementary integrals $I_{7}$, $I_{8}$, and $I_{9}$ given by (\ref{eq:I7}), (\ref{eq:I8}), and (\ref{eq:I9}) in Appendix~\ref{Sec:ELI2}, respectively. Finally, we use $I_{7_{1}}$, $I_{8_{1}}$, and $I_{9}$, given by (\ref{eq:I71S}), (\ref{eq:I81S}), and (\ref{eq:I9S}), respectively, to rewrite $\varphi_{r}(\lambda)$ in terms of elementary functions.\\
Case 8: In this case we have lightlike geodesics characterised by $\Sigma_{\text{ph}_{+}}<\Sigma<\Sigma_{\text{ph}}$ or $\pi-\Sigma_{\text{ph}}<\Sigma<\pi-\Sigma_{\text{ph}_{+}}$. In this case the root structure and thus the evaluation procedure is the same as for case 4.\\
Case 9a: In this case we have lightlike geodesics characterised by $\Sigma=\Sigma_{\text{ph}}$ or $\Sigma=\pi-\Sigma_{\text{ph}}$. We recall that in this case the right-hand side of (\ref{eq:EoMr}) has four real roots and that we labelled and sorted them such that $r_{4}<r_{3}<r_{\text{H}_{\text{i}}}\leq r_{\text{H}_{\text{o}}}<r_{\text{ph}}=r_{2}=r_{1}$. Here we have three different subcases. In the first subcase we have $r(\lambda)=r_{i}=r_{\text{ph}}$. In this subcase the light rays and gravitational waves travel along the unstable photon orbit and thus their radius coordinate is constant. Thus we evaluate (\ref{eq:phirlam}) directly. We integrate over $\lambda$ and get
\begin{eqnarray}
\varphi_{r}(\lambda)=\frac{(r_{\text{ph}}^2+a^2)E_{\text{C}}-aL_{z\text{C}}}{P(r_{\text{ph}})}\left(\lambda-\lambda_{i}\right).
\end{eqnarray}
In the other two subcases we have motion at $r_{\text{ph}}<r$ or $r_{\text{H}_{\text{o}}}<r<r_{\text{ph}}$. Again we first rewrite the polynomial inside the root in terms of its roots. For $0<a<m$ we perform a partial fraction decomposition of $P(r')^{-1}$ with respect to $r'$. We rewrite the terms outside the root in the form of a constant term and two terms with $r'-r_{a}$ in the denominator. This time $r_{a}$ can be $r_{\text{H}_{\text{o}}}$ or $r_{\text{H}_{\text{i}}}$. Then we substitute using (\ref{eq:elm1}). Now we have to distinguish between the two subcases. In both subcases we first pull the term $y'-y_{\text{ph}}$ under consideration of its sign out of the root. Then we perform a second partial fraction decomposition with respect to $y'$. We simplify the terms and see that (\ref{eq:phirint}) can be rewritten in terms of three different versions of the elementary integral $I_{14}$ given by (\ref{eq:I14}) in Appendix~\ref{Sec:ELI4}. Now for $r_{\text{ph}}<r$ we use $I_{14_{3}}$, given by (\ref{eq:I143}) in Appendix~\ref{Sec:ELI4}, and for $r_{\text{H}_{\text{o}}}<r<r_{\text{ph}}$ we use $I_{14_{2}}$ and $I_{14_{3}}$, given by (\ref{eq:I142}) and (\ref{eq:I143}) in Appendix~\ref{Sec:ELI4}, respectively, to rewrite $\varphi_{r}(\lambda)$ in terms of elementary functions.\\
For $a=m$ we first rewrite the term outside the root in the form of a constant term and two terms with powers of $r'-r_{\text{H}}$ in the denominator. Again we substitute using (\ref{eq:elm1}) and pull the term $y'-y_{\text{ph}}$ under consideration of its sign out of the root. We perform a partial fraction decomposition with respect to $y'$ and rewrite (\ref{eq:phirint}) in terms of the elementary integrals $I_{14}$ and $I_{15}$ given by (\ref{eq:I14}) and (\ref{eq:I15}) in Appendix~\ref{Sec:ELI4}, respectively. Now for $r_{\text{ph}}<r$ we use $I_{14_{3}}$ and $I_{15_{2}}$, given by (\ref{eq:I143}) and (\ref{eq:I152}) in Appendix~\ref{Sec:ELI4}, respectively, and for $r_{\text{H}_{\text{o}}}<r<r_{\text{ph}}$ we use $I_{14_{2}}$, $I_{14_{3}}$, and $I_{15_{2}}$, given by (\ref{eq:I142}), (\ref{eq:I143}), and (\ref{eq:I152}) in Appendix~\ref{Sec:ELI4}, respectively, to rewrite $\varphi_{r}(\lambda)$ in terms of elementary functions.\\
Case 9b: In this case we have $a=m$ and lightlike geodesics characterised by $\Sigma=\Sigma_{\text{ph}_{+}}=\Sigma_{\text{ph}}$ or $\Sigma=\pi-\Sigma_{\text{ph}}=\pi-\Sigma_{\text{ph}_{+}}$. We recall that in this case the right-hand side of (\ref{eq:EoMr}) has four real roots and that we labelled and sorted them such that $r_{4}<r_{3}<r_{\text{ph}_{+}}=r_{\text{H}}=r_{\text{ph}}=r_{2}=r_{1}$. Again we first rewrite the polynomial inside the root in terms of its roots. Then we rewrite the term outside the root in terms of a constant term and two terms with powers of $r'-r_{\text{H}}$ in the denominator. Then we substitute using (\ref{eq:elm1}), pull the term $y'-y_{\text{H}}$ under consideration of its sign out of the root, and rewrite (\ref{eq:phirint}) in terms of the elementary integrals $I_{14}$, $I_{15}$, and $I_{16}$ given by (\ref{eq:I14}), (\ref{eq:I15}), and (\ref{eq:I16}) in Appendix~\ref{Sec:ELI4}, respectively. Then we use $I_{14_{3}}$, $I_{15_{2}}$, and $I_{16}$ given by (\ref{eq:I143}), (\ref{eq:I152}), and (\ref{eq:I16S}) in Appendix~\ref{Sec:ELI4}, respectively, to rewrite $\varphi_{r}(\lambda)$ in terms of elementary functions.\\
Case 9c: In this case we have $a=m$ and lightlike geodesics characterised by $\Sigma=\Sigma_{\text{ph}_{0}}=\Sigma_{\text{ph}_{+}}=\Sigma_{\text{ph}}$ or $\Sigma=\pi-\Sigma_{\text{ph}}=\pi-\Sigma_{\text{ph}_{+}}=\pi-\Sigma_{\text{ph}_{0}}$. We recall that in this case the right-hand side of (\ref{eq:EoMr}) has four real roots and that we labelled and sorted them such that $r_{4}<r_{\text{ph}_{0}}=r_{\text{ph}_{+}}=r_{\text{H}}=r_{\text{ph}}=r_{3}=r_{2}=r_{1}$. Again we first rewrite the polynomial inside the root in terms of its roots. We simplify the integrand such that we can rewrite (\ref{eq:phirint}) in terms of the three different elementary integrals $I_{19}$, $I_{20}$, and $I_{21}$ given by (\ref{eq:I19}), (\ref{eq:I20}), and (\ref{eq:I21}) in Appendix~\ref{Sec:ELI5}. Then we use $I_{19}$, $I_{20}$, and $I_{21}$ given by (\ref{eq:I19S}), (\ref{eq:I20S}), and (\ref{eq:I21S}) in Appendix~\ref{Sec:ELI5}, respectively, to rewrite $\varphi_{r}(\lambda)$ in terms of elementary functions.\\
Case 10: In this case for $r_{\text{ph}}<r$ we have lightlike geodesics characterised by $\Sigma_{\text{ph}}<\Sigma<\pi-\Sigma_{\text{ph}}$ and for $r_{\text{H}_{\text{o}}}<r<r_{\text{ph}}$ we have lightlike geodesics characterised by $\pi-\Sigma_{\text{ph}}<\Sigma<\Sigma_{\text{ph}}$. We recall that in this case the right-hand side of (\ref{eq:EoMr}) has four distinct real roots and that we labelled and sorted them such that $r_{4}<r_{3}<r_{2}<r_{1}$. At least one of these roots lies in the domain of outer communication. Again we first rewrite the polynomial inside the root in terms of its roots. For $0<a<m$ we perform a partial fraction decomposition of the term $P(r')^{-1}$ with respect to $r'$. For $a=m$ we directly proceed to the next step. In the next step we rewrite the terms outside the root such that only a constant term and two terms with powers of $r'-r_{a}$ in the denominator remain. This time for $0<a<m$ $r_{a}$ can be $r_{\text{H}_{\text{o}}}$ or $r_{\text{H}_{\text{i}}}$, while for $a=m$ $r_{a}$ can only be $r_{\text{H}}$. Now we have to distinguish the two different subcases. In the first subcase we have $r_{\text{ph}}<r$. Light rays and gravitational waves travelling along these geodesics can pass through a turning point at the radius coordinate $r_{\text{min}}=r_{1}$. We substitute using (\ref{eq:elsub3}) and rewrite (\ref{eq:phirint}) in terms of Legendre's elliptic integrals of the first and third kind, and, for $a=m$, the nonstandard elliptic integral $I_{L_{6}}(\chi_{i},\chi(\lambda),k_{3},n)$ given by (\ref{eq:EI6}) in Appendix~\ref{Sec:ELIL}. For $0<a<m$ we directly obtain $\varphi_{r}(\lambda)$ in terms of Legendre's elliptic integrals of the first and third kind. For $a=m$ on the other hand we use (\ref{eq:EI4S}) to completely rewrite $\varphi_{r}(\lambda)$ in terms of elementary functions and Legendre's elliptic integrals of the first, second, and third kind.\\
In the second subcase we have $r_{\text{H}_{\text{o}}}<r<r_{\text{ph}}$. Light rays and gravitational waves travelling along these geodesics can pass through a turning point at the radius coordinate $r_{\text{max}}=r_{2}$. This time we substitute using (\ref{eq:elsub4}). Again we rewrite (\ref{eq:phirint}) in terms of Legendre's elliptic integrals of the first and third kind, and, for $a=m$, the nonstandard elliptic integral $I_{L_{6}}(\chi_{i},\chi(\lambda),k_{3},n)$ given by (\ref{eq:EI6}) in Appendix~\ref{Sec:ELIL}. Again for $0<a<m$ we directly obtain $\varphi_{r}(\lambda)$ in terms of Legendre's elliptic integrals of the first and third kind. For $a=m$ on the other hand we again use (\ref{eq:EI4S}) to completely rewrite $\varphi_{r}(\lambda)$ in terms of elementary functions and Legendre's elliptic integrals of the first, second, and third kind.\\
Note that when we have light rays or gravitational waves passing through a turning point we have to split (\ref{eq:phirint}) at the turning point into two integrals. Then we evaluate both integrals separately. Here, for the first integral the sign of the root in the denominator is given by $i_{r_{i}}$. For the second integral on the other hand the sign of the root is given by $-i_{r_{i}}$.

\subsection{The Time Coordinate $t$}\label{sec:timec}
The last equation of motion we have to integrate describes the evolution of the time coordinate $t$. Again we can see that the right-hand side of the corresponding equation of motion (\ref{eq:EoMt}) consists of two terms. As for the $\varphi$ motion the first term only depends on the radius coordinate $r$, while the second term only depends on the spacetime latitude $\vartheta$. For deriving the solution describing the evolution of the time coordinate $t$ we now explicitly write the dependency of $r$ and $\vartheta$ on the Mino parameter $\lambda$. Then we separate variables and integrate from $t(\lambda_{i})=t_{i}$ to $t(\lambda)=t$. The result reads
\begin{eqnarray}\label{eq:tint}
t\left(\lambda\right)=t_{i}+t_{r}\left(\lambda\right)+a t_{\vartheta}\left(\lambda\right),
\end{eqnarray}
where we again defined two functions $t_{r}(\lambda)$ and $t_{\vartheta}(\lambda)$  which describe the $r$ dependent part and the $\vartheta$ dependent part as functions of the Mino parameter $\lambda$, respectively. They read
\begin{eqnarray}\label{eq:trlam}
t_{r}(\lambda)=\int_{\lambda_{i}}^{\lambda}\left(r(\lambda')^2+a^2\right)\frac{(r(\lambda')^2+a^2)E_{\text{C}}-aL_{z\text{C}}}{P(r(\lambda'))}\text{d}\lambda'
\end{eqnarray}
and
\begin{eqnarray}\label{eq:EoMtinttheta}
t_{\vartheta}(\lambda)=\int_{\lambda_{i}}^{\lambda}\left(L_{z\text{C}}-a\sin^2\vartheta(\lambda')E_{\text{C}}\right)\text{d}\lambda'.
\end{eqnarray}
In the following we will now evaluate $t_{r}(\lambda)$ and $t_{\vartheta}(\lambda)$ separately. As for $\varphi_{r}(\lambda)$ due to the large number of different cases and the length of the solutions for $t_{r}(\lambda)$ we will only explicitly present the solutions for light rays and gravitational waves travelling along lightlike geodesics characterised by $\Sigma=0$ or $\Sigma=\pi$ (corresponding to $K_{\text{C}}=0$), and the solutions for light rays and gravitational waves travelling along lightlike geodesics on the unstable photon orbit at the radius coordinate $r_{\text{ph}}$ in the domain of outer communication. Again in all other cases we will only outline the steps which are necessary for their derivation.

\subsubsection{Calculating $t_{\vartheta}(\lambda)$}\label{Sec:ttheta}
For $t_{\vartheta}(\lambda)$ we have to distinguish the same five different types of motion as for $\vartheta$. However, contrary to the $\vartheta$ dependent part of the $\varphi$ coordinate $\varphi_{\vartheta}(\lambda)$ for $t_{\vartheta}(\lambda)$ we can derive solutions for radial lightlike geodesics on the axes and lightlike geodesics crossing the axes.\\
We again start with the first type of motion describing light rays and gravitational waves travelling along lightlike geodesics characterised by $\Sigma=0$ or $\Sigma=\pi$ (corresponding to $K_{\text{C}}=0$). We saw that for these geodesics we have $\text{d}\vartheta/\text{d}\lambda=0$ and thus $\vartheta(\lambda)=\vartheta_{i}$. As a consequence we have  $L_{z\text{C}}=a\sin^2\vartheta_{i} E_{\text{C}}$. We can easily see from (\ref{eq:EoMtinttheta}) that in this case we get
\begin{eqnarray}
t_{\vartheta}(\lambda)=0.
\end{eqnarray}
The second type of motion describes light rays and gravitational waves travelling along lightlike geodesics in the equatorial plane or on individual photon cones. Again we have $\text{d}\vartheta/\text{d}\lambda=0$ and thus $\vartheta(\lambda)=\vartheta_{i}$ but this time the integrand is nonzero. We directly integrate the right-hand side of (\ref{eq:EoMtinttheta}) over $\lambda$. In the first case we have light rays and gravitational waves travelling along lightlike geodesics in the equatorial plane. Thus, we have $\vartheta_{i}=\vartheta_{\text{ph}}=\pi/2$ and $\Psi=\pi/2$ or $\Psi=3\pi/2$ (corresponding to $K_{\text{C}}-(aE_{\text{C}}-L_{z\text{C}})^2=0$ in terms of the constants of motion) and get as result for $t_{\vartheta}(\lambda)$
\begin{eqnarray}
t_{\vartheta}(\lambda)=\left(L_{z\text{C}}-aE_{\text{C}}\right)\left(\lambda-\lambda_{i}\right).
\end{eqnarray}
In the second case we have light rays and gravitational waves travelling along lightlike geodesics on the individual photon cones. In this case we have $\vartheta_{i}=\vartheta_{\text{ph}_{\mp}}$  and $\Psi=\pi/2$ or $\Psi=3\pi/2$ (corresponding to $K_{\text{C}}+4aE_{\text{C}}L_{z\text{C}}=0$ in terms of the constants of motion) and thus we get
\begin{eqnarray}
t_{\vartheta}(\lambda)=\left(L_{z\text{C}}-a\sin^2\vartheta_{\text{ph}_{\mp}}E_{\text{C}}\right)\left(\lambda-\lambda_{i}\right).
\end{eqnarray}
For the remaining three types of motion we have to explicitly evaluate the right-hand side of (\ref{eq:EoMtinttheta}) in terms of elementary functions and elliptic integrals. For this purpose we first rewrite the integrand in the form of a constant term and a term proportional to $\cos^2\vartheta(\lambda')$. Again we split the integral into two integrals and evaluate them separately. We first evaluate the integral over the constant term. In the case of the second term we rewrite it such that it is the product of a constant and the integral 
\begin{eqnarray}\label{eq:I2t}
I_{\vartheta_{2}}(\lambda)=\int_{\lambda_{i}}^{\lambda}\cos^2\vartheta\left(\lambda'\right)\text{d}\lambda'.
\end{eqnarray}
In this notation $t_{\vartheta}(\lambda)$ now reads
\begin{eqnarray}\label{eq:ttheta1}
t_{\vartheta}(\lambda)=\left(L_{z\text{C}}-aE_{\text{C}}\right)\left(\lambda-\lambda_{i}\right)+aE_{\text{C}}I_{\vartheta_{2}}(\lambda).
\end{eqnarray}
In the next step we have to evaluate the integral $I_{\vartheta_{2}}(\lambda)$. Here, depending on the type of motion for the evaluation of the integral we again have to use different evaluation procedures.\\ 
We will again start with the third type of motion. We recall from Section~\ref{Sec:EoMtheta} that in this case we have light rays and gravitational waves travelling along lightlike geodesics characterised by $K_{\text{C}}-(aE_{\text{C}}-L_{z\text{C}})^2>0$. We also recall that in this case the right-hand side of (\ref{eq:EoMx}) has two real roots at $x_{2}=-x_{1}=\cos\vartheta_{\text{max}}<0<x_{1}=\cos\vartheta_{\text{min}}$ and a pair of purely imaginary complex conjugate roots given by $x_{3}=\bar{x}_{4}=iX_{4}$, where we chose $0<X_{4}$. In general the light rays and gravitational waves travelling along these geodesics oscillate between the turning points $x_{1}=\cos\vartheta_{\text{min}}$ and $x_{2}=\cos\vartheta_{\text{max}}$. As for the evaluation of $I_{\vartheta_{1}}(\lambda)$ we first split the integral $I_{\vartheta_{2}}(\lambda)$ at the turning points of the $\vartheta$ motion into separate integrals. For each integral we now substitute using (\ref{eq:EoMxsub1}). Then we rewrite the integrand in the form of a constant term and a term containing the expression $1/(1-k_{4}\sin^2\chi')$. Then we again split the integral into two integrals. In the case of the integral with the constant integrand we integrate directly over $\lambda$. In the case of the other integral we use (\ref{eq:LFD}), adapted to the type of motion at hand, to rewrite it as an integral over $\chi'$. In this case we get the nonstandard elliptic integral $I_{L_{1}}(\chi_{i},\chi(\lambda),k_{4})$ given by (\ref{eq:EI1}) in Appendix~\ref{Sec:ELIL}. We now use (\ref{eq:EI1S}) to rewrite it in terms of elementary functions and Legendre's elliptic integral of the second kind. We insert the obtained result for $I_{\vartheta_{2}}(\lambda)$ in (\ref{eq:ttheta1}) and obtain for $t_{\vartheta}(\lambda)$ 
\begin{eqnarray}
&t_{\vartheta}(\lambda)=\left(L_{z\text{C}}-aE_{\text{C}}(1+X_{4}^2)\right)\left(\lambda-\lambda_{i}\right)+\sum_{n=1}^{N}\frac{i_{\vartheta_{i}n}aE_{\text{C}}X_{4}^2}{\sqrt{a^2 E_{\text{C}}^2\left(x_{1}^2+X_{4}^2\right)}(1-k_{4})}\Bigg(E_{L}\left(\chi_{n-1},k_{4}\right)-E_{L}\left(\chi_{n},k_{4}\right)\\
&+k_{4}\left(\frac{\sin\chi_{n}\cos\chi_{n}}{\sqrt{1-k_{4}\sin^2\chi_{n}}}-\frac{\sin\chi_{n-1}\cos\chi_{n-1}}{\sqrt{1-k_{4}\sin^2\chi_{n-1}}}\right)\Bigg),\nonumber
\end{eqnarray}
where $N$ is the number of terms into which we had to split $I_{\vartheta_{2}}(\lambda)$. In addition, we have $i_{\vartheta_{i}n}=-i_{\vartheta_{i}n-1}$ with $i_{\vartheta_{i}1}=i_{\vartheta_{i}}$, $\chi_{0}=\chi_{i}$, $\chi_{N}=\chi(\lambda)$, and for $1<n<N$ $\chi_{n}$ and $\chi_{n-1}$ are given by $\chi_{\text{min}}=\pi/2$ or $\chi_{\text{max}}=-\pi/2$. Here, $\chi_{i}$, $\chi_{\text{min}}$, $\chi_{\text{max}}$, and $\chi(\lambda)$ are related to $\cos\vartheta_{i}$, the turning points $x_{1}=\cos\vartheta_{\text{min}}$ and $x_{2}=\cos\vartheta_{\text{max}}$, and $\cos\vartheta(\lambda)$ by (\ref{eq:elchi5}), respectively, and the square of the elliptic modulus $k_{4}$ is given by (\ref{eq:EM4}).\\
Now we turn to the fourth type of motion. It describes light rays and gravitational waves travelling along vortical lightlike geodesics characterised by $2aE_{\text{C}}\left(aE_{\text{C}}-L_{z\text{C}}\right)-K_{\text{C}}>0$ and $K_{\text{C}}-(aE_{\text{C}}-L_{z\text{C}})^2<0$. We recall that in this case we have four real roots and that we labelled and sorted them such that $x_{4}=-x_{1}<x_{3}=-x_{2}<0<x_{2}<x_{1}$. In general light rays and gravitational waves travelling along these geodesics above the equatorial plane oscillate between the turning points $x_{1}=\cos\vartheta_{\text{min}}$ and $x_{2}=\cos\vartheta_{\text{max}}$. On the other hand light rays and gravitational waves travelling along these geodesics below the equatorial plane oscillate between the turning points $x_{3}=\cos\vartheta_{\text{min}}$ and $x_{4}=\cos\vartheta_{\text{max}}$.\\ 
As in the last case we first split the integral $I_{\vartheta_{2}}(\lambda)$ at the turning points of the $\vartheta$ motion into separate integrals. Then we evaluate each integral separately. For light rays and gravitational waves oscillating between the turning points $x_{1}=\cos\vartheta_{\text{min}}$ and $x_{2}=\cos\vartheta_{\text{max}}$ we now substitute using (\ref{eq:EoMxsub2}), while for light rays and gravitational waves oscillating between the turning points $x_{3}=\cos\vartheta_{\text{min}}$ and $x_{4}=\cos\vartheta_{\text{max}}$ we substitute using (\ref{eq:EoMxsub3}). In both cases we rewrite the integrand in terms of a constant term and a term containing $\sin^2\chi'$. Then we split the integral again into two integrals. Again we evaluate each integral separately. In the case of the integral with the constant integrand we integrate directly over $\lambda$. In the case of the other integral we use (\ref{eq:LFD}), adapted to the type of motion at hand, to rewrite it as an integral over $\chi'$. We then rearrange it such that we can rewrite it in terms of Legendre's elliptic integrals of the first and second kind. We insert the obtained result for $I_{\vartheta_{2}}(\lambda)$ into (\ref{eq:ttheta1}) and simplify the terms. In the case of light rays and gravitational waves  oscillating between the turning points $x_{1}=\cos\vartheta_{\text{min}}$ and $x_{2}=\cos\vartheta_{\text{max}}$ above the equatorial plane $t_{\vartheta}(\lambda)$ now reads
\begin{eqnarray}
&t_{\vartheta}\left(\lambda\right)=(L_{z\text{C}}-aE_{\text{C}}(1-x_{1}^2))\left(\lambda-\lambda_{i}\right)+\sum_{n=1}^{N}\frac{i_{\vartheta_{i}n}x_{1}aE_{\text{C}}}{\sqrt{a^2E_{\text{C}}^2}}\left(F_{L}(\chi_{n-1},k_{5})-F_{L}(\chi_{n},k_{5})\right.\\
&\left.+E_{L}\left(\chi_{n},k_{5}\right)-E_{L}\left(\chi_{n-1},k_{5}\right)\right),\nonumber
\end{eqnarray}
where $N$ is the number of terms into which we had to split $I_{\vartheta_{2}}(\lambda)$. In addition, we have $i_{\vartheta_{i}n}=-i_{\vartheta_{i}n-1}$ with $i_{\vartheta_{i}1}=i_{\vartheta_{i}}$, $\chi_{0}=\chi_{i}$, $\chi_{N}=\chi(\lambda)$, and for $1<n<N$ $\chi_{n}$ and $\chi_{n-1}$ are given by $\chi_{\text{min}}=0$ or $\chi_{\text{max}}=\pi/2$. Here, $\chi_{i}$, $\chi_{\text{min}}$, $\chi_{\text{max}}$, and $\chi(\lambda)$ are related to $\cos\vartheta_{i}$, the turning points $x_{1}=\cos\vartheta_{\text{min}}$ and $x_{2}=\cos\vartheta_{\text{max}}$, and $\cos\vartheta(\lambda)$ by (\ref{eq:elchi6}), respectively, and the square of the elliptic modulus $k_{5}$ is given by (\ref{eq:EM5}). For light rays and gravitational waves oscillating between the turning points $x_{3}=\cos\vartheta_{\text{min}}$ and $x_{4}=\cos\vartheta_{\text{max}}$ below the equatorial plane on the other hand $t_{\vartheta}(\lambda)$ reads
\begin{eqnarray}
&t_{\vartheta}\left(\lambda\right)=(L_{z\text{C}}-aE_{\text{C}}(1-x_{1}^2))\left(\lambda-\lambda_{i}\right)+\sum_{n=1}^{N}\frac{i_{\vartheta_{i}n}x_{1}aE_{\text{C}}}{\sqrt{a^2E_{\text{C}}^2}}\left(F_{L}(\chi_{n},k_{5})-F_{L}(\chi_{n-1},k_{5})\right.\\
&\left.+E_{L}\left(\chi_{n-1},k_{5}\right)-E_{L}\left(\chi_{n},k_{5}\right)\right),\nonumber
\end{eqnarray}
where as above $N$ is the number of terms into which we had to split $I_{\vartheta_{2}}(\lambda)$. In addition, we have $i_{\vartheta_{i}n}=-i_{\vartheta_{i}n-1}$ with $i_{\vartheta_{i}1}=i_{\vartheta_{i}}$, $\chi_{0}=\chi_{i}$, $\chi_{N}=\chi(\lambda)$, and for $1<n<N$ $\chi_{n}$ and $\chi_{n-1}$ are given by $\chi_{\text{min}}=\pi/2$ or $\chi_{\text{max}}=0$. Again $\chi_{i}$, $\chi_{\text{min}}$, $\chi_{\text{max}}$, and $\chi(\lambda)$ are related to $\cos\vartheta_{i}$, the turning points $x_{3}=\cos\vartheta_{\text{min}}$ and $x_{4}=\cos\vartheta_{\text{max}}$, and $\cos\vartheta(\lambda)$ by (\ref{eq:elchi6}), respectively, and the square of the elliptic modulus $k_{5}$ is given by (\ref{eq:EM5}).\\
As for $\varphi_{\vartheta}(\lambda)$ the fifth and last type of motion is given by the special vortical lightlike geodesics characterised by $2aE_{\text{C}}\left(aE_{\text{C}}-L_{z\text{C}}\right)-K_{\text{C}}>0$ and $K_{\text{C}}-(aE_{\text{C}}-L_{z\text{C}})^2=0$. In this case we have $x_{2}=x_{3}=0$. For lightlike geodesics above the equatorial plane we insert (\ref{eq:soleltheta4}) in (\ref{eq:I2t}) while for lightlike geodesics below the equatorial plane we insert (\ref{eq:soleltheta5}). Again we rewrite the resulting integral in terms of the cosinus hyperbolicus. We obtain an elementary integral which is easy to evaluate. For lightlike geodesics above the equatorial plane we now substitute $\tilde{\lambda}'=\text{arsech}\left(\frac{\cos\vartheta_{i}}{x_{1}}\right)+i_{\vartheta_{i}}x_{1}\sqrt{a^2E_{\text{C}}^2}\left(\lambda'-\lambda_{i}\right)$. For lightlike geodesics below the equatorial plane on the other hand we substitute $\tilde{\lambda}'=\text{arsech}\left(\frac{\cos\vartheta_{i}}{x_{4}}\right)+i_{\vartheta_{i}}x_{4}\sqrt{a^2E_{\text{C}}^2}\left(\lambda'-\lambda_{i}\right)$. Then we evaluate the integrals and insert the obtained results in (\ref{eq:I2t}). In the next step we insert the obtained results for $I_{\vartheta_{2}}(\lambda)$ in (\ref{eq:ttheta1}). For lightlike geodesics above the equatorial plane the obtained result for $t_{\vartheta}(\lambda)$ reads
\begin{eqnarray}
&\hspace{-0.5cm}t_{\vartheta}(\lambda)=\left(L_{z\text{C}}-aE_{\text{C}}\right)\left(\lambda-\lambda_{i}\right)+\frac{i_{\vartheta_{i}}x_{1}aE_{\text{C}}}{\sqrt{a^2E_{\text{C}}^2}}\left(\text{tanh}\left(\text{arsech}\left(\frac{\cos\vartheta_{i}}{x_{1}}\right)+i_{\vartheta_{i}}x_{1}\sqrt{a^2E_{\text{C}}^2}\left(\lambda-\lambda_{i}\right)\right)\right.\\
&\left.-\text{tanh}\left(\text{arsech}\left(\frac{\cos\vartheta_{i}}{x_{1}}\right)\right)\right).\nonumber
\end{eqnarray}
For lightlike geodesics below the equatorial plane on the other hand the obtained result for $t_{\vartheta}(\lambda)$ reads
\begin{eqnarray}
&\hspace{-0.5cm}t_{\vartheta}(\lambda)=\left(L_{z\text{C}}-aE_{\text{C}}\right)\left(\lambda-\lambda_{i}\right)+\frac{i_{\vartheta_{i}}x_{4}aE_{\text{C}}}{\sqrt{a^2E_{\text{C}}^2}}\left(\text{tanh}\left(\text{arsech}\left(\frac{\cos\vartheta_{i}}{x_{4}}\right)+i_{\vartheta_{i}}x_{4}\sqrt{a^2E_{\text{C}}^2}\left(\lambda-\lambda_{i}\right)\right)\right.\\
&\left.-\text{tanh}\left(\text{arsech}\left(\frac{\cos\vartheta_{i}}{x_{4}}\right)\right)\right).\nonumber
\end{eqnarray}

\subsubsection{Calculating $t_{r}(\lambda)$}
Next we turn to the $r$ dependent part $t_{r}(\lambda)$ given by (\ref{eq:trlam}). As for $\varphi_{r}(\lambda)$ we first rewrite it as an integral over $r$. For this purpose we again separate variables in (\ref{eq:EoMr}) and then insert the result in (\ref{eq:trlam}). Now $t_{r}(\lambda)$ reads
\begin{eqnarray}\label{eq:trint}
t_{r}(\lambda)=\int_{r_{i}...}^{...r(\lambda)}\frac{\left(r'^2+a^2\right)\left((r'^2+a^2)E_{\text{C}}-aL_{z\text{C}}\right)\text{d}r'}{P(r')\sqrt{\left(\left(r'^2+a^2\right)E_{\text{C}}-aL_{z\text{C}}\right)^2-P(r')K_{\text{C}}}},
\end{eqnarray}
where again the dots in the limits indicate that we have to split the integral at turning points and the sign of the root in the denominator has to be chosen according to the direction of the $r$ motion. Again when the light rays or gravitational waves do not pass through a turning point we directly integrate from $r_{i}$ to $r(\lambda)$. However, when the light rays or gravitational waves pass through a turning point we have to split (\ref{eq:trint}) into an integral from $r_{i}$ to the turning point and an integral from the turning point to $r(\lambda)$. For the evaluation of the integral we have to distinguish the same different types of motion as for $r$. As for $\varphi_{r}(\lambda)$ we have a large number of different cases and thus again we will only explicitly present the solutions for light rays and gravitational waves travelling along lightlike geodesics characterised by $\Sigma=0$ or $\Sigma=\pi$ (corresponding to $K_{\text{C}}=0$), and the solutions for light rays and gravitational waves on the photon orbit at the radius coordinate $r_{\text{ph}}$. Again in all other cases we will only briefly outline the steps for their derivation.\\
Case 1a and case 1b: In these cases we have light rays and gravitational waves travelling along lightlike geodesics characterised by $\Sigma=0$ or $\Sigma=\pi$. In these cases we have $K_{\text{C}}=0$ and thus $L_{z\text{C}}=a\sin^2\vartheta_{i} E_{\text{C}}$. Note that for $\vartheta=0$ and $\vartheta=\pi$ the time coordinate is completely well-defined and thus the presented solutions also include these cases. As for $\varphi_{r}(\lambda)$ (\ref{eq:trint}) takes the same form for all geodesics characterised by $\Sigma=0$ or $\Sigma=\pi$ and thus for all lightlike geodesics classified as cases 1a and 1b we obtain the same results.\\
When we have $0<a<m$ we first use a partial fraction decomposition of $P(r')^{-1}$ with respect to $r'$ to rewrite the integrand (we recall that in this case $E$, and thus also $E_{C}$, is always positive) in terms of a constant term and two terms containing $r'-r_{a}$ in the denominator. Here, $r_{a}$ can be $r_{\text{H}_{\text{o}}}$ or $r_{\text{H}_{\text{i}}}$. We now have several elementary integrals. We integrate and get for $t_{r}(\lambda)$ the well-known result
\begin{eqnarray}
t_{r}\left(\lambda\right)=i_{r_{i}}\left(r(\lambda)-r_{i}+\frac{r_{\text{H}_{\text{o}}}^2+a^2}{r_{\text{H}_{\text{o}}}-r_{\text{H}_{\text{i}}}}\ln\left(\frac{r\left(\lambda\right)-r_{\text{H}_{\text{o}}}}{r_{i}-r_{\text{H}_{\text{o}}}}\right)+\frac{r_{\text{H}_{\text{i}}}^2+a^2}{r_{\text{H}_{\text{o}}}-r_{\text{H}_{\text{i}}}}\ln\left(\frac{r_{i}-r_{\text{H}_{\text{i}}}}{r\left(\lambda\right)-r_{\text{H}_{\text{i}}}}\right)\right).
\end{eqnarray}
When we have $a=m$ we do not need to perform a partial fraction decomposition and only simplify all terms such that we have three simple elementary integrals. We evaluate the integrals and obtain as result for $t_{r}(\lambda)$
\begin{eqnarray}
t_{r}\left(\lambda\right)=i_{r_{i}}\left(r(\lambda)-r_{i}+2r_{\text{H}}\ln\left(\frac{r\left(\lambda\right)-r_{\text{H}}}{r_{i}-r_{\text{H}}}\right)+\left(r_{\text{H}}^2+a^2\right)\left(\frac{1}{r_{i}-r_{\text{H}}}-\frac{1}{r(\lambda)-r_{\text{H}}}\right)\right).
\end{eqnarray}
Case 2: In this case we have lightlike geodesics characterised by $0<\Sigma<\Sigma_{\text{ph}_{-}}$ or $\pi-\Sigma_{\text{ph}_{-}}<\Sigma<\pi$. We recall that in this case the right-hand side of (\ref{eq:EoMr}) has two distinct pairs of complex conjugate roots and that we labelled and sorted them such that we have $r_{1}=\bar{r}_{2}=R_{1}+iR_{2}$ and $r_{3}=\bar{r}_{4}=R_{3}+iR_{4}$, where we chose $R_{1}<R_{3}$, and $0<R_{2}$ and $0<R_{4}$. Now we first rewrite the polynomial inside the root in terms of its roots. When we have $0<a<m$ we perform a partial fraction decomposition of $P(r')^{-1}$ with respect to $r'$. In the case $a=m$ we do not need to perform a partial fraction decomposition. Then we rewrite the resulting terms outside the root such that we have two terms with powers of $r'$, a constant term, and two terms with powers of $r'-r_{a}$ in the denominator. In this case for $0<a<m$ $r_{a}$ can be $r_{\text{H}_{\text{o}}}$ or $r_{\text{H}_{\text{i}}}$ while for $a=m$ $r_{a}$ can only be $r_{\text{H}}$. Then we substitute using (\ref{eq:elsub1}) and rewrite (\ref{eq:trint}) in terms of Legendre's elliptic integral of the first kind and the nonstandard elliptic integrals $I_{L_{2}}(\chi_{i},\chi(\lambda),k_{1},n)$ and $I_{L_{3}}(\chi_{i},\chi(\lambda),k_{1},n)$ given by (\ref{eq:EI2}) and (\ref{eq:EI3}) in Appendix~\ref{Sec:ELIL}, respectively. Now we use both terms in (\ref{eq:EI2S}) to completely rewrite $t_{r}(\lambda)$ in terms of elementary functions and Legendre's elliptic integrals of the first, second, and third kind.\\
Case 3: In this case we have lightlike geodesics characterised by $\Sigma=\Sigma_{\text{ph}_{-}}$ or $\Sigma=\pi-\Sigma_{\text{ph}_{-}}$. We recall that in this case the right-hand side of (\ref{eq:EoMr}) has a real double root at $r_{\text{ph}_{-}}=r_{2}=r_{1}<r_{\text{H}_{\text{i}}}$ and a pair of complex conjugate roots given by $r_{3}=\bar{r}_{4}=R_{3}+iR_{4}$, where we chose $0<R_{4}$. Again we first rewrite the polynomial inside the root in terms of its roots. Then we pull the term $r'-r_{\text{ph}_{-}}$ out of the root. We perform a partial fraction decomposition of $P(r')^{-1}(r'-r_{\text{ph}_{-}})^{-1}$ with respect to $r'$. For $0<a<m$ we then rewrite the resulting terms outside the root such that we get one term with $r'$, a constant term, and three terms with $r'-r_{a}$ in the denominator. In this case $r_{a}$ can be $r_{\text{ph}_{-}}$, $r_{\text{H}_{\text{o}}}$, or $r_{\text{H}_{\text{i}}}$. Now we can easily see that we can rewrite (\ref{eq:trint}) in terms of the elementary integrals $I_{1}$, $I_{2}$, and $I_{3}$ given by (\ref{eq:I1}), (\ref{eq:I2}), and (\ref{eq:I3}) in Appendix~\ref{Sec:ELI1}, respectively. We evaluate the integrals and get the right-hand sides of (\ref{eq:I1}), (\ref{eq:I2}), and (\ref{eq:I3}). We insert the results and obtain $t_{r}(\lambda)$ in terms of elementary functions.\\
When we have $a=m$ we proceed analogously. In this case we can rewrite (\ref{eq:trint}) in terms of the elementary integrals $I_{1}$, $I_{2}$, $I_{3}$, and $I_{4}$ given by (\ref{eq:I1}), (\ref{eq:I2}), (\ref{eq:I3}), and (\ref{eq:I4}) in Appendix~\ref{Sec:ELI1}, respectively. Again we evaluate the integrals and get the right-hand sides of (\ref{eq:I1}), (\ref{eq:I2}), (\ref{eq:I3}), and (\ref{eq:I4}). We insert the results and obtain $t_{r}(\lambda)$ in terms of elementary functions.\\
Case 4: In this case we have lightlike geodesics characterised by $\Sigma_{\text{ph}_{-}}<\Sigma<\Sigma_{\text{ph}_{0}}$ or $\pi-\Sigma_{\text{ph}_{0}}<\Sigma<\pi-\Sigma_{\text{ph}_{-}}$. We recall that in this case the right-hand side of (\ref{eq:EoMr}) has two distinct real roots at $r_{2}<r_{1}<r_{\text{H}_{\text{i}}}$ and a pair of complex conjugate roots given by $r_{3}=\bar{r}_{4}=R_{3}+iR_{4}$, where we chose $0<R_{4}$. Again we first rewrite the polynomial inside the root in terms of its roots. For $0<a<m$ we perform a partial fraction decomposition of $P(r')^{-1}$ with respect to $r'$. For $a=m$ on the other hand we directly proceed to the next step. We rewrite the resulting terms outside the root such that we have two terms with powers of $r'$, a constant term, and two terms with powers of $r'-r_{a}$ in the denominator. In this case for $0<a<m$ $r_{a}$ can be $r_{\text{H}_{\text{o}}}$ or $r_{\text{H}_{\text{i}}}$, while for $a=m$ $r_{a}$ can only be $r_{\text{H}}$. In the next step we substitute using (\ref{eq:elsub2}) and rewrite (\ref{eq:trint}) in terms of Legendre's elliptic integral of the first kind and the nonstandard elliptic integrals $I_{L_{4}}(\chi_{i},\chi(\lambda),k_{2},n)$ and $I_{L_{5}}(\chi_{i},\chi(\lambda),k_{2},n)$ given by (\ref{eq:EI4}) and (\ref{eq:EI5}) in Appendix~\ref{Sec:ELIL}, respectively. In the next step we use both terms in (\ref{eq:EI3S}) to completely rewrite $t_{r}(\lambda)$ in terms of elementary functions and Legendre's elliptic integrals of the first, second, and third kind. \\
Case 5: In this case we have lightlike geodesics characterised by $\Sigma=\Sigma_{\text{ph}_{0}}$ or $\Sigma=\pi-\Sigma_{\text{ph}_{0}}$. We recall that in this case the right-hand side of (\ref{eq:EoMr}) has four real roots and that we labelled and sorted them such that $r_{4}<r_{\text{ph}_{0}}=r_{3}=r_{2}<r_{1}$. Again we first rewrite the polynomial inside the root in terms of its roots. Then we pull the term $r'-r_{\text{ph}_{0}}$ out of the root. For $0<a<m$ we perform a partial fraction decomposition of the term $P(r')^{-1}(r'-r_{\text{ph}_{0}})^{-1}$ with respect to $r'$. In the next step we again rewrite the resulting terms outside the root such that we have a term containing $r'$, a constant term, and three terms with $r'-r_{a}$ in the denominator. In this case $r_{a}$ can be $r_{\text{ph}_{0}}$, $r_{\text{H}_{\text{o}}}$, or $r_{\text{H}_{\text{i}}}$. Then we rewrite (\ref{eq:trint}) in terms of the elementary integrals $I_{5}$, $I_{6}$, and $I_{7}$ given by (\ref{eq:I5}), (\ref{eq:I6}), and (\ref{eq:I7}) in Appendix~\ref{Sec:ELI2}, respectively. Finally, we use $I_{5}$, $I_{6}$, $I_{7_{1}}$, and $I_{7_{2}}$ given by (\ref{eq:I5S}), (\ref{eq:I6S}), (\ref{eq:I71S}), and (\ref{eq:I72S}), respectively, to rewrite $t_{r}(\lambda)$ in terms of elementary functions.\\ 
For $a=m$ we proceed analogously. We first perform a partial fraction decomposition of the term $P(r')^{-1}(r'-r_{\text{ph}_{0}})^{-1}$ with respect to $r'$ and then we rewrite the resulting terms outside the root such that we have a term containing $r'$, a constant term, and three terms with powers of $r'-r_{a}$ in the denominator. In this case $r_{a}$ can be $r_{\text{ph}_{0}}$ or $r_{\text{H}}$. Then we rewrite (\ref{eq:trint}) in terms of the elementary integrals $I_{5}$, $I_{6}$, $I_{7}$, and $I_{8}$ given by (\ref{eq:I5}), (\ref{eq:I6}), (\ref{eq:I7}), and (\ref{eq:I8}) in Appendix~\ref{Sec:ELI2}, respectively. Finally, we use $I_{5}$, $I_{6}$, $I_{7_{1}}$, $I_{7_{2}}$, and $I_{8_{2}}$ given by (\ref{eq:I5S}), (\ref{eq:I6S}), (\ref{eq:I71S}), (\ref{eq:I72S}), and (\ref{eq:I82S}), respectively, to rewrite $t_{r}(\lambda)$ in terms of elementary functions.\\ 
Case 6: In this case we have lightlike geodesics characterised by $\Sigma_{\text{ph}_{0}}<\Sigma<\Sigma_{\text{ph}_{+}}$ or $\pi-\Sigma_{\text{ph}_{+}}<\Sigma<\pi-\Sigma_{\text{ph}_{0}}$. We recall that in this case the right-hand side of (\ref{eq:EoMr}) has four real roots and that we labelled and sorted them such that $r_{4}<r_{3}<r_{2}<r_{1}<r_{\text{H}_{\text{i}}}$. Again we first rewrite the polynomial inside the root in terms of its roots. For $0<a<m$ we perform a partial fraction decomposition of $P(r')^{-1}$ with respect to $r'$ while for $a=m$ we directly proceed to the next step. In the next step we then rewrite the resulting terms outside the root such that we have two terms with powers of $r'$, a constant term, and two terms with powers of $r'-r_{a}$ in the denominator. In this case for $0<a<m$ $r_{a}$ can be $r_{\text{H}_{\text{o}}}$ or $r_{\text{H}_{\text{i}}}$, while for $a=m$ $r_{a}$ can only be $r_{\text{H}}$. Then we substitute using (\ref{eq:elsub3}) and rewrite (\ref{eq:trint}) in terms of Legendre's elliptic integrals of the first and third kind, and the nonstandard elliptic integral $I_{L_{6}}(\chi_{i},\chi(\lambda),k_{3},n)$ given by (\ref{eq:EI6}) in Appendix~\ref{Sec:ELIL}. In the next step we use (\ref{eq:EI4S}) to completely rewrite $t_{r}(\lambda)$ in terms of elementary functions and Legendre's elliptic integrals of the first, second, and third kind.\\
Note that in this case also for the time coordinate $t$ we have the same problem as for the $\varphi$ coordinate. When we rewrite $t_{r}(\lambda)$ in terms of elementary functions and Legendre's elliptic integrals of the first, second, and third kind we cannot avoid to integrate over the coordinate singularities at the radius coordinates of the horizons and thus for the associated terms we use (\ref{eq:PiN}) to rewrite Legendre's elliptic integral of the third kind.\\
Case 7a: In this case we have lightlike geodesics characterised by $\Sigma=\Sigma_{\text{ph}_{+}}$ or $\Sigma=\pi-\Sigma_{\text{ph}_{+}}$. We recall that in this case the right-hand side of (\ref{eq:EoMr}) has four real roots and that we labelled and sorted them such that $r_{4}<r_{3}<r_{\text{ph}_{+}}=r_{2}=r_{1}<r_{\text{H}_{\text{i}}}$. Again we first rewrite the polynomial inside the root in terms of its roots. Then we pull the term $r'-r_{\text{ph}_{+}}$ out of the root. We perform a partial fraction decomposition of $P(r')^{-1}(r'-r_{\text{ph}_{+}})^{-1}$ with respect to $r'$ and rewrite the resulting terms outside the root such that one term with $r'$, a constant term, and three terms with $r'-r_{a}$ in the denominator remain. In this case $r_{a}$ can be $r_{\text{ph}_{+}}$, $r_{\text{H}_{\text{o}}}$, or $r_{\text{H}_{\text{i}}}$. In the next step we substitute using (\ref{eq:elm1}). We rewrite (\ref{eq:trint}) in terms of the elementary integrals $I_{14}$ and $I_{15}$ given by (\ref{eq:I14}) and (\ref{eq:I15}) in Appendix~\ref{Sec:ELI4}, respectively. Then we follow the steps outlined in Appendix~\ref{Sec:ELI4} to evaluate the integrals and obtain $I_{14_{1}}$, $I_{14_{3}}$, and $I_{15_{1}}$, given by (\ref{eq:I141}), (\ref{eq:I143}), and (\ref{eq:I151}), respectively. Finally, we use the obtained results to rewrite $t_{r}(\lambda)$ in terms of elementary functions.\\
Case 7b: In this case we have $0<a<m$ and lightlike geodesics characterised by $\Sigma=\Sigma_{\text{ph}_{0}}=\Sigma_{\text{ph}_{+}}$ or $\Sigma=\pi-\Sigma_{\text{ph}_{+}}=\pi-\Sigma_{\text{ph}_{0}}$. We recall that in this case the right-hand side of (\ref{eq:EoMr}) has four real roots and that we labelled and sorted them such that $r_{4}<r_{\text{ph}_{0}}=r_{\text{ph}_{+}}=r_{3}=r_{2}=r_{1}<r_{\text{H}_{\text{i}}}$. Again we first rewrite the polynomial inside the root in terms of its roots. Then we pull a factor $r'-r_{\text{ph}_{+}}$ out of the root. As for case 7a we perform a partial fraction decomposition of $P(r')^{-1}(r'-r_{\text{ph}_{+}})^{-1}$ with respect to $r'$. Then we again rewrite the terms outside the root such that one term with $r'$, a constant term, and three terms with $r'-r_{a}$ in the denominator remain. In this case $r_{a}$ can be $r_{\text{ph}_{+}}$, $r_{\text{H}_{\text{o}}}$, or $r_{\text{H}_{\text{i}}}$. We can now easily see that we can rewrite (\ref{eq:trint}) in terms of the elementary integrals $I_{10}$, $I_{11}$, $I_{12}$, and $I_{13}$ given by (\ref{eq:I10}), (\ref{eq:I11}), (\ref{eq:I12}), and (\ref{eq:I13}) in Appendix~\ref{Sec:ELI3}, respectively. Then we use $I_{10}$, $I_{11}$, $I_{12}$, and $I_{13}$ given by (\ref{eq:I10S}), (\ref{eq:I11S}), (\ref{eq:I12S}), and (\ref{eq:I13S}), respectively, to rewrite $t_{r}(\lambda)$ in terms of elementary functions.\\
Case 7c: In this case we have $a=m$ and lightlike geodesics characterised by $\Sigma=\Sigma_{\text{ph}_{0}}=\Sigma_{\text{ph}_{+}}$ or $\Sigma=\pi-\Sigma_{\text{ph}_{+}}=\pi-\Sigma_{\text{ph}_{0}}$, where $\Sigma_{\text{ph}}<\Sigma_{\text{ph}_{0}}=\Sigma_{\text{ph}_{+}}$. We recall that in this case the right-hand side of (\ref{eq:EoMr}) has four real roots, that we labelled and sorted them such that $r_{4}<r_{\text{ph}_{0}}=r_{\text{ph}_{+}}=r_{\text{H}}=r_{3}=r_{2}<r_{1}$, and that the root $r_{1}=r_{\text{min}}$ lies in the domain of outer communication and is a turning point. Again we first rewrite the polynomial inside the root in terms of its roots. Then we pull the term $r'-r_{\text{H}}$ out of the root. We rewrite the term outside the root such that only one term with $r'$, a constant term, and three terms with powers of $r'-r_{\text{H}}$ in the denominator remain. We can now easily see that we can rewrite (\ref{eq:trint}) in terms of the elementary integrals $I_{5}$, $I_{6}$, $I_{7}$, $I_{8}$, and $I_{9}$ given by (\ref{eq:I5}), (\ref{eq:I6}), (\ref{eq:I7}), (\ref{eq:I8}), and (\ref{eq:I9}) in Appendix~\ref{Sec:ELI2}, respectively. Now we use $I_{5}$, $I_{6}$, $I_{7_{1}}$, $I_{8_{1}}$, and $I_{9}$ given by (\ref{eq:I5S}), (\ref{eq:I6S}), (\ref{eq:I71S}), (\ref{eq:I81S}), and (\ref{eq:I9S}) to rewrite $t_{r}(\lambda)$ in terms of elementary functions.\\
Case 8: In this case we have lightlike geodesics characterised by $\Sigma_{\text{ph}_{+}}<\Sigma<\Sigma_{\text{ph}}$ or $\pi-\Sigma_{\text{ph}}<\Sigma<\pi-\Sigma_{\text{ph}_{+}}$. In this case the root structure and thus the evaluation procedure is the same as for case 4.\\
Case 9a: In this case we have lightlike geodesics characterised by $\Sigma=\Sigma_{\text{ph}}$ or $\Sigma=\pi-\Sigma_{\text{ph}}$. We recall that in this case the right-hand side of (\ref{eq:EoMr}) has four real roots and that we labelled and sorted them such that $r_{4}<r_{3}<r_{\text{H}_{\text{i}}}\leq r_{\text{H}_{\text{o}}}<r_{\text{ph}}=r_{2}=r_{1}$. Here we have three different subcases. In the first subcase we have $r(\lambda)=r_{i}=r_{\text{ph}}$. In this subcase the light rays and gravitational waves travel along the unstable photon orbit and thus their radius coordinate is constant. In this subcase we evaluate (\ref{eq:trlam}) directly. We integrate over $\lambda$ and get
\begin{eqnarray}
t_{r}(\lambda)=(r_{\text{ph}}^2+a^2)\frac{(r_{\text{ph}}^2+a^2)E_{\text{C}}-aL_{z\text{C}}}{P(r_{\text{ph}})}\left(\lambda-\lambda_{i}\right).
\end{eqnarray}
In the other two subcases we have motion with $r_{\text{ph}}<r$ or $r_{\text{H}_{\text{o}}}<r<r_{\text{ph}}$. Again we first rewrite the polynomial inside the root in terms of its roots. For $0<a<m$ we first perform a partial fraction decomposition of $P(r')^{-1}$ with respect to $r'$. We rewrite the resulting terms outside the root in the form of two terms with powers of $r'$, a constant term, and two terms with $r'-r_{a}$ in the denominator. In both subcases $r_{a}$ can be $r_{\text{H}_{\text{o}}}$ or $r_{\text{H}_{\text{i}}}$. Then we substitute using (\ref{eq:elm1}). Now we have to distinguish between both subcases. In both subcases we first pull the term $y'-y_{\text{ph}}$ under consideration of its sign out of the root. Then we perform a second partial fraction decomposition with respect to $y'$ and rewrite (\ref{eq:trint}) in terms of the elementary integrals $I_{14}$ and $I_{15}$ given by (\ref{eq:I14}) and (\ref{eq:I15}) in Appendix~\ref{Sec:ELI4}, respectively. For $r_{\text{ph}}<r$ we now use $I_{14_{1}}$, $I_{14_{3}}$, and $I_{15_{1}}$, given by (\ref{eq:I141}), (\ref{eq:I143}), and  (\ref{eq:I151}) in Appendix~\ref{Sec:ELI4}, respectively, to rewrite $t_{r}(\lambda)$ in terms of elementary functions. For $r_{\text{H}_{\text{o}}}<r<r_{\text{ph}}$ on the other hand we use $I_{14_{1}}$, $I_{14_{2}}$, $I_{14_{3}}$, and $I_{15_{1}}$, given by (\ref{eq:I141}), (\ref{eq:I142}), (\ref{eq:I143}), and (\ref{eq:I151}) in Appendix~\ref{Sec:ELI4}, respectively, to rewrite $t_{r}(\lambda)$ in terms of elementary functions.\\
For $a=m$ we rewrite the term outside the root in the form of two terms with powers of $r'$, a constant term, and two terms with powers of $r'-r_{\text{H}}$ in the denominator. Then we again substitute using (\ref{eq:elm1}) and pull the term $y'-y_{\text{ph}}$ under consideration of its sign out of the root.  We perform a partial fraction decomposition with respect to $y'$ and rewrite (\ref{eq:trint}) in terms of the elementary integrals $I_{14}$ and $I_{15}$ given by (\ref{eq:I14}) and (\ref{eq:I15}) in Appendix~\ref{Sec:ELI4}, respectively. Now for $r_{\text{ph}}<r$ we use $I_{14_{1}}$, $I_{14_{3}}$, $I_{15_{1}}$, and $I_{15_{2}}$, given by (\ref{eq:I141}), (\ref{eq:I143}), (\ref{eq:I151}), and (\ref{eq:I152}) in Appendix~\ref{Sec:ELI4}, respectively, and for $r_{\text{H}_{\text{o}}}<r<r_{\text{ph}}$ we use $I_{14_{1}}$, $I_{14_{2}}$, $I_{14_{3}}$, $I_{15_{1}}$, and $I_{15_{2}}$, given by (\ref{eq:I141}), (\ref{eq:I142}), (\ref{eq:I143}), (\ref{eq:I151}), and (\ref{eq:I152}) in Appendix~\ref{Sec:ELI4}, respectively, to rewrite $t_{r}(\lambda)$ in terms of elementary functions.\\
Case 9b: In this case we have $a=m$ and lightlike geodesics characterised by $\Sigma=\Sigma_{\text{ph}_{+}}=\Sigma_{\text{ph}}$ or $\Sigma=\pi-\Sigma_{\text{ph}}=\pi-\Sigma_{\text{ph}_{+}}$. We recall that in this case the right-hand side of (\ref{eq:EoMr}) has four real roots and that we labelled and sorted them such that $r_{4}<r_{3}<r_{\text{ph}_{+}}=r_{\text{H}}=r_{\text{ph}}=r_{2}=r_{1}$. Again we first rewrite the polynomial inside the root in terms of its roots. Then we rewrite the term outside the root in terms of two terms with powers of $r'$, a constant term, and two terms with powers of $r'-r_{\text{H}}$ in the denominator. Then we substitute using (\ref{eq:elm1}) and pull the term $y'-y_{\text{H}}$ under consideration of its sign out of the root. We perform two partial fraction decompositions with respect to $y'$ and rewrite (\ref{eq:trint}) in terms of the elementary integrals $I_{14}$, $I_{15}$, and $I_{16}$ given by (\ref{eq:I14}), (\ref{eq:I15}), and (\ref{eq:I16}) in Appendix~\ref{Sec:ELI4}, respectively. Then we use $I_{14_{1}}$, $I_{14_{3}}$, $I_{15_{1}}$, $I_{15_{2}}$, and $I_{16}$ given by (\ref{eq:I141}), (\ref{eq:I143}), (\ref{eq:I151}), (\ref{eq:I152}), and (\ref{eq:I16S}) in Appendix~\ref{Sec:ELI4}, respectively, to rewrite $t_{r}(\lambda)$ in terms of elementary functions.\\
Case 9c: In this case we have $a=m$ and lightlike geodesics characterised by $\Sigma=\Sigma_{\text{ph}_{0}}=\Sigma_{\text{ph}_{+}}=\Sigma_{\text{ph}}$ or $\Sigma=\pi-\Sigma_{\text{ph}}=\pi-\Sigma_{\text{ph}_{+}}=\pi-\Sigma_{\text{ph}_{0}}$. We recall that in this case the right-hand side of (\ref{eq:EoMr}) has four real roots and that we labelled and sorted them such that $r_{4}<r_{\text{ph}_{0}}=r_{\text{ph}_{+}}=r_{\text{H}}=r_{\text{ph}}=r_{3}=r_{2}=r_{1}$. Again we first rewrite the polynomial inside the root in terms of its roots. Then we rewrite the term outside the root in terms of two terms containing powers of $r'$, a constant term, and two terms with powers of $r'-r_{\text{H}}$ in the denominator. Then we rewrite (\ref{eq:trint}) in terms of the elementary integrals $I_{17}$, $I_{18}$, $I_{19}$, $I_{20}$, and $I_{21}$ given by (\ref{eq:I17}), (\ref{eq:I18}), (\ref{eq:I19}), (\ref{eq:I20}), and (\ref{eq:I21}) in Appendix~\ref{Sec:ELI5}, respectively. Finally, we use $I_{17}$, $I_{18}$, $I_{19}$, $I_{20}$, and $I_{21}$ given by (\ref{eq:I17S}), (\ref{eq:I18S}), (\ref{eq:I19S}), (\ref{eq:I20S}), and (\ref{eq:I21S}) in Appendix~\ref{Sec:ELI5}, respectively, to rewrite $t_{r}(\lambda)$ in terms of elementary functions.\\
Case 10: In this case for $r_{\text{ph}}<r$ we have lightlike geodesics characterised by $\Sigma_{\text{ph}}<\Sigma<\pi-\Sigma_{\text{ph}}$ and for $r_{\text{H}_{\text{o}}}<r<r_{\text{ph}}$ we have lightlike geodesics characterised by $\pi-\Sigma_{\text{ph}}<\Sigma<\Sigma_{\text{ph}}$. We recall that in this case the right-hand side of (\ref{eq:EoMr}) has four distinct real roots and that we labelled and sorted them such that $r_{4}<r_{3}<r_{2}<r_{1}$. In this case at least one of these roots lies in the domain of outer communication. Again we first rewrite the polynomial inside the root in terms of its roots. For $0<a<m$ we perform a partial fraction decomposition of the term $P(r')^{-1}$ with respect to $r'$. For $a=m$ we directly proceed to the next step. In the next step we rewrite the terms outside the root such that two terms with powers of $r'$, a constant term, and two terms with powers of $r'-r_{a}$ in the denominator remain. In this case for $0<a<m$ $r_{a}$ can be $r_{\text{H}_{\text{o}}}$ or $r_{\text{H}_{\text{i}}}$, while for $a=m$ $r_{a}$ can only be $r_{\text{H}}$. Now we have to distinguish the two different subcases. In the first subcase we have $r_{\text{ph}}<r$. In this subcase light rays and gravitational waves travelling along the lightlike geodesics can pass through a turning point at the radius coordinate $r_{\text{min}}=r_{1}$. We substitute using (\ref{eq:elsub3}). Then we rewrite (\ref{eq:trint}) in terms of Legendre's elliptic integrals of the first and third kind, and the nonstandard elliptic integral $I_{L_{6}}(\chi_{i},\chi(\lambda),k_{3},n)$ given by (\ref{eq:EI6}) in Appendix~\ref{Sec:ELIL}. Now we use (\ref{eq:EI4S}) to completely rewrite $t_{r}(\lambda)$ in terms of elementary functions and Legendre's elliptic integrals of the first, second, and third kind.\\
In the second subcase we have $r_{\text{H}_{\text{o}}}<r<r_{\text{ph}}$. In this subcase light rays and gravitational waves travelling along the lightlike geodesics can pass through a turning point at the radius coordinate $r_{\text{max}}=r_{2}$. This time we substitute using (\ref{eq:elsub4}). Again we rewrite (\ref{eq:trint}) in terms of Legendre's elliptic integrals of the first and third kind, and the nonstandard elliptic integral $I_{L_{6}}(\chi_{i},\chi(\lambda),k_{3},n)$ given by (\ref{eq:EI6}) in Appendix~\ref{Sec:ELIL}. Again we use (\ref{eq:EI4S}) to completely rewrite $t_{r}(\lambda)$ in terms of elementary functions and Legendre's elliptic integrals of the first, second, and third kind.\\
Note that when we have light rays and gravitational waves passing through a turning point we have to split (\ref{eq:trint}) at the turning point into two integrals. Then we evaluate both integrals separately. Here, for the first integral the sign of the root in the denominator is given by $i_{r_{i}}$. For the second integral on the other hand the sign of the root is given by $-i_{r_{i}}$.

\section{Gravitational Lensing}\label{Sec:Lensing}
In this section we will investigate gravitational lensing in the Kerr spacetime. Here, we will discuss three different quantities. The first is the lens equation. It relates the positions of the images of one or several sources in terms of latitude-longitude coordinates on the celestial sphere of an observer to their spacetime coordinates on a source surface. The second quantity is the redshift. It relates the energy of a light ray or a gravitational wave measured at its source to the energy measured by an observer. The third quantity is the travel time. It measures in terms of the time coordinate $t$ the time a light ray or a gravitational wave needs to travel from its source to an observer. However, before we proceed to discuss these quantities we first need to agree on the naming conventions we will use to discuss their projections onto the celestial sphere of the observer. We will define them in the next subsection. In the following subsections we will then discuss the lens equation, the redshift, and the travel time. We will conclude this section with a brief discussion of the astrophysical implications of the results.

\subsection{Dividing the Observer's Celestial Sphere}
In Section~\ref{Sec:EoM} we already defined the coordinate system on the celestial sphere of the standard observer. However, when we discuss the projections of the lens equation, the redshift, and the travel time onto the observer's celestial sphere always referring to specific latitude-longitude coordinates is rather inconvenient. Therefore, we will now agree on the following naming conventions. Commonly the direction towards the black hole shows the most interesting features and therefore, with the exception of three examples for the redshift, we will limit our discussion to the celestial latitudes $0\leq\Sigma\leq\pi/2$. On the associated maps we will now refer to the celestial longitudes $\pi/2<\Psi<3\pi/2$ as \emph{northern hemisphere} and to the celestial longitudes $0\leq\Psi<\pi/2$ and $3\pi/2<\Psi<2\pi$ as \emph{southern hemisphere}. Analogously, we will refer to the celestial longitudes $0<\Psi<\pi$ as western hemisphere and to the celestial longitudes $\pi<\Psi<2\pi$ as eastern hemisphere. In agreement with this terminology we will refer to the line marked by $\Psi=\pi/2$ and $\Psi=3\pi/2$ as \emph{celestial equator}. Similarly we will refer to the celestial longitude $\Psi=0$ as \emph{meridian} and to the celestial longitude $\Psi=\pi$ as \emph{antimeridian}.  In addition, in the cases in which we are discussing features on the whole celestial sphere we will refer to the celestial latitudes $0\leq\Sigma<\pi/2$, technically the real northern hemisphere, as \emph{direction towards the black hole} and to the celestial latitudes $\pi/2<\Sigma\leq\pi$, technically the real southern hemisphere, as \emph{direction away from the black hole}.

\subsection{The Lens Equation}\label{Subsec:LensEqua}
\begin{figure}\label{fig:SoLS}
  \includegraphics[width=0.7\textwidth]{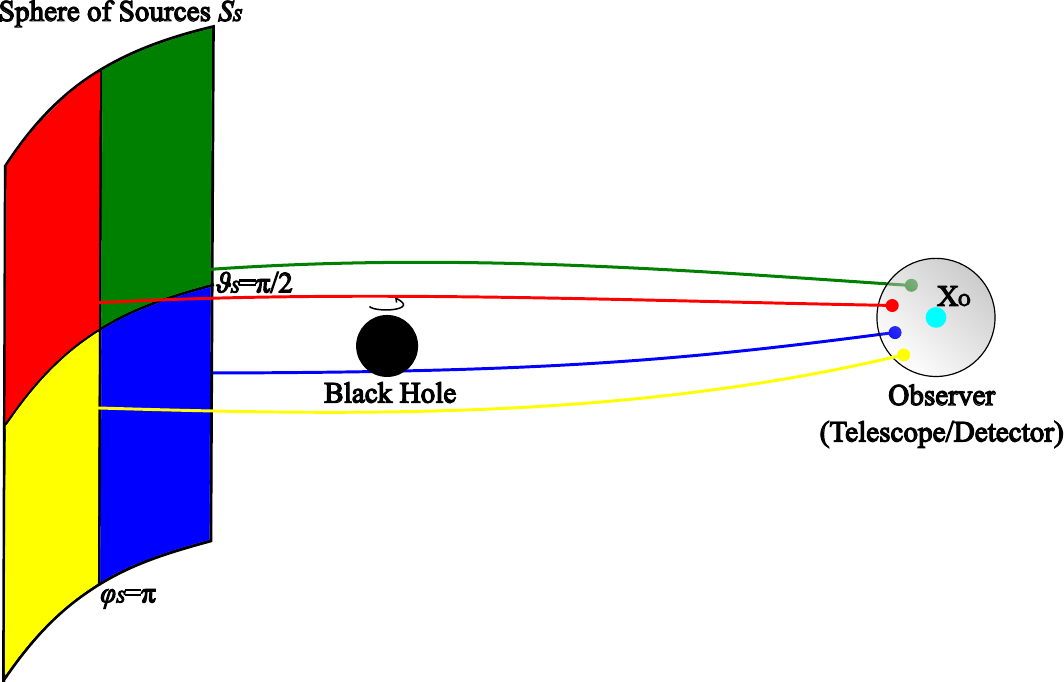}
\caption{Schematic illustration of the lens map. We assume that the observer is located in the domain of outer communication outside the ergoregion and the photon region at the coordinates $r_{O}$, $\vartheta_{O}$, and $\varphi_{O}$. Then we distribute sources on a two-sphere $S_{S}^2$ at the radius coordinate $r_{O}<r_{S}$. We divide the two-sphere into four different quadrants and assign to each quadrant a colour following the colour convention of Bohn \emph{et al.} \cite{Bohn2015}: $0\leq \vartheta_{S}\leq\pi/2$ and $0\leq\varphi_{S}<\pi$, green; $\pi/2< \vartheta_{S}\leq\pi$ and $0\leq\varphi_{S}<\pi$, blue; $0\leq \vartheta_{S}\leq\pi/2$ and $\pi\leq\varphi_{S}<2\pi$, red; $\pi/2< \vartheta_{S}\leq\pi$ and $\pi\leq\varphi_{S}<2\pi$, yellow. The sources emit light rays and gravitational waves, in the figure marked by the coloured lines, which are gravitationally lensed by the black hole (depicted by a black dot) and then detected by the observer from different directions on his celestial sphere (marked by the grey circle) at an event marked by the spacetime coordinates $x_{O}=(t_{O},r_{O},\vartheta_{O},\varphi_{O})$.}
\end{figure}
For a general relativistic spacetime an exact lens map or lens equation was first formulated by Frittelli and Newman \cite{Frittelli1999}. It was later adapted to spherically symmetric and static spacetimes by Perlick \cite{Perlick2004} and also to stationary and axisymmetric spacetimes, see, e.g., the work of Frost \cite{Frost2022}. We refer to something as a lens equation when it maps points, lines, or two-dimensional structures on the celestial sphere of an observer to their spacetime coordinates on a source surface. In this paper we construct it as follows. We place a standard observer in the domain of outer communication outside the ergoregion and the photon region. Then we distribute light and gravitational wave sources on a two-sphere $S_{S}^2$ at the radius coordinate $r_{S}$. Here, we choose the radius coordinate of the two-sphere such that it encloses the black hole and the obsever. Thus we have $r_{O}<r_{S}$. From the position of the observer we now follow lightlike geodesics in all directions into the past. Among these lightlike geodesics we now have to distinguish three different types. The first type of geodesics approaches the event horizon. The second type of geodesics asymptotically approaches the photon orbits and marks the boundary of the shadow. The third type of geodesics will at some point intersect with the two-sphere of sources $S_{S}^2$. The latter now constitute a map from the latitude-longitude coordinates $\Sigma$ and $\Psi$ on the celestial sphere of the observer to the spacetime latitudes $\vartheta_{S}$ and $\varphi_{S}$ on the two-sphere of sources $S_{S}^2$. This is our lens equation. In short it reads
\begin{eqnarray}\label{eq:LensMap}
(\Sigma,\Psi)\rightarrow (\vartheta_{S}(\Sigma,\Psi),\varphi_{S}(\Sigma,\Psi)).
\end{eqnarray}
In this paper we want to calculate $\vartheta_{S}(\Sigma,\Psi)$ and $\varphi_{S}(\Sigma,\Psi)$ from the exact analytic solutions to the equations of motion derived in Sec.~\ref{Sec:SolEoM}. For this purpose we first have to fix the initial conditions. In our setup the light rays and gravitational waves are detected by the observer at an event marked by the coordinates $x(\lambda_{O})=x_{O}$ and thus we set $\lambda_{i}=\lambda_{O}$ and $x_{i}=x_{O}$. Since the line element (\ref{eq:KerrMetric}) is invariant with respect to rotations about the $z$-axis we choose $\varphi_{O}=0$. Now we only need to fix the two quantities $\lambda_{O}$ and $\lambda_{S}$. The Mino parameter is defined up to an affine transformation and thus without loss of generality we can choose $\lambda_{O}=0$. The quantity $\lambda_{S}<\lambda_{O}=0$ on the other hand can be calculated from the radius coordinates of the observer and the sphere of sources $S_{S}^2$ $r_{O}$ and $r_{S}$, respectively. For this purpose we first separate variables in (\ref{eq:EoMr}), rewrite the constants of motion in terms of the latitude-longitude coordinates on the observer's celestial sphere by inserting (\ref{eq:ConsE})-(\ref{eq:ConsK}), and integrate. We get for $\lambda_{S}$
\begin{eqnarray}\label{eq:lambdaS}
\lambda_{S}=\int_{r_{O}...}^{...r_{S}}\frac{\sqrt{\rho(r_{O},\vartheta_{O})}\text{d}r'}{\sqrt{\left(\left(r'^2+a^2\cos^2\vartheta_{O}\right)\sqrt{P(r_{O})}+a(r'^2-r_{O}^2)\sin\vartheta_{O}\sin\Sigma\sin\Psi\right)^2-P(r')\rho(r_{O},\vartheta_{O})^2\sin^2\Sigma}},
\end{eqnarray}
where the dots in the limits indicate that when the light rays and gravitational waves pass through a turning point we have to split the integral at the turning point into one integral from the radius coordinate of the observer $r_{O}$ to the turning point and one integral from the turning point to the radius coordinate of the source $r_{S}$. In addition, the sign of the root in the denominator has to be chosen such that along each part of the geodesic it agrees with the direction of the $r$ motion. For the explicit calculation of the lens equation we now use the following procedure. We first rewrite (\ref{eq:lambdaS}) in terms of elementary functions or Legendre's elliptic integral of the first kind. Then we use $\lambda_{S}$ to derive $\vartheta_{S}(\Sigma,\Psi)$ and $\varphi_{S}(\Sigma,\Psi)$ from the solutions to the equations of motion for $\vartheta(\lambda)$ and $\varphi(\lambda)$ derived in Secs.~\ref{Sec:EoMtheta} and \ref{Sec:EoMphi}. Note that here for practical purposes for the calculation of $\varphi_{S}(\Sigma,\Psi)$ we first completely rewrite (\ref{eq:phithetaEL1}), (\ref{eq:phithetaEL2}), and (\ref{eq:phithetaEL3}) in terms of elliptic integrals (note that we will use the same approach for the calculation of the travel time $T(\Sigma,\Psi)$ in Sec.~\ref{Subsec:TravTime}). Then we determine the number of the turning points of the $\vartheta$ motion along the geodesic which connects the observer and the source, and finally calculate $\varphi_{S}(\Sigma,\Psi)$. The computational evaluation of the lens equation (as well as of the redshifts and the travel time, which will be discussed in the next two subsections) was carried out in the programming language Julia \cite{Bezanson2017}.\\ 
For the visualisation of the lens equation we now divide the sphere of sources into four different quadrants as depicted in Fig.~4. We assign to each quadrant a colour following the colour convention of Bohn \emph{et al.} \cite{Bohn2015}. We colour the first quadrant defined by $0\leq \vartheta_{S}\leq\pi/2$ and $0\leq\varphi_{S}<\pi$ green, the second quadrant defined by $\pi/2<\vartheta_{S}\leq\pi$ and $0\leq\varphi_{S}<\pi$ blue, the third quadrant defined by $0\leq \vartheta_{S}\leq\pi/2$ and $\pi\leq\varphi_{S}<2\pi$ red, and the fourth quadrant defined by $\pi/2< \vartheta_{S}\leq\pi$ and $\pi\leq\varphi_{S}<2\pi$ yellow. However, for our plots of the lens equation we slightly modify this convention. While we keep the general colour scheme we adapt it as follows. When we plot the lens equation we will see that we have images of different orders. Here we will define the orders of these images as follows. Images generated by light rays and gravitational waves for which the covered angle $\Delta\varphi_{S}$ takes values in the interval $0<\left|\Delta\varphi_{S}\right|<\pi$ are images of first order. Analogously images generated by light rays and gravitational waves for which the covered angle $\Delta\varphi_{S}$ takes values in the interval $\pi<\left|\Delta\varphi_{S}\right|<2\pi$ are images of second order and so on. We now modify the colour scheme of Bohn \emph{et al.} such that images of odd order are plotted in stronger colours than images of even order. Note that in the following we will refer to the lensing features in the calculated lens maps as \emph{images} independent of the fact whether they were generated by light rays, and thus can be visually observed, or gravitational waves, and thus are signals instead of images. Only when we explicitly discuss gravitational waves we will refer to them as signals.\\
Fig.~5 shows lens maps for the Schwarzschild spacetime (top left panel), and the Kerr spacetime with $a=m/4$ (top right panel), $a=m/2$ (middle left panel), $a=3m/4$ (middle right panel), and $a=m$ (bottom left panel) for an observer in the equatorial plane ($\vartheta_{O}=\pi/2$) at the radius coordinate $r_{O}=10m$ and sources distributed on the two-sphere $S_{S}^2$ at $r_{S}=20m$. The bottom right panel shows a lens map for the Kerr spacetime with $a=m$ for an observer at $r_{O}=10m$ and $\vartheta_{O}=\pi/4$. The two-sphere of sources $S_{S}^2$ is again located at $r_{S}=20m$. For all lens maps the observer looks in the direction of the black hole.\\
We start our discussion with the lens map for the Schwarzschild spacetime. At the centre of the map we see a black circle. This is the shadow of the black hole. The images of different orders form concentric rings around the shadow. In these rings the images of the sources on the different quadrants of the two-sphere of sources $S_{S}^2$ are clearly separated. At the outer boundary of the map we have images of first order, on the eastern hemisphere in green and blue and on the western hemisphere in red and yellow. Further in at lower latitudes we have images of second order, on the eastern hemisphere in bright yellow and bright red and on the western hemisphere in bright blue and bright green. At even lower latitudes we can also see images of third and, when we zoom in, very close to the boundary of the shadow barely visible images of fourth order. The width of the rings decreases with increasing order of the images and the boundaries between the rings with the images of the different orders mark the positions of the critical curves. Because of the spherical symmetry for the Schwarzschild spacetime they are circles.\\
When we turn on the spin we can already see quite significant changes for small spin values. We start with the lens map for $a=m/4$ in the top right panel of Fig.~5. As in the lens map for the Schwarzschild spacetime we see images up to fourth order on both hemispheres. The first and most obvious change is that the shadow takes a slightly asymmetric form and in general the symmetry with respect to the meridian and the antimeridian is broken. In addition, the formerly disconnected regions with images of first and second order of sources from the same quadrants on the two-sphere of sources $S_{S}^2$ on the eastern and western hemispheres, on the northern hemisphere first in green and then in blue, and on the southern hemisphere first in blue and then in green, connect. The same happens for the images of second and third order, and the images of third and fourth order. The sharp boundaries between the areas with images of first and second order, the areas with images of second and third order, and so on from the same quadrant on the two-sphere of sources $S_{S}^2$ mark light rays and gravitational waves crossing the axes at least once. To the east of these lines we have counterrotating light rays and gravitational waves and to the west we have corotating light rays and gravitational waves. In addition, at the boundaries between the areas with images of first and second order of sources on different quadrants on the two-sphere of sources $S_{S}^2$ we also see two more effects. On the eastern hemisphere we see images of sources from the other side of the equatorial plane (colour change from green to blue on the northern hemisphere and blue to green on the southern hemisphere) before we observe images of second order (colour change from blue to bright yellow on the northern hemisphere and green to bright red on the southern hemisphere). On the western hemisphere on the other hand we observe images of second order (colour change from red to bright green on the northern hemisphere and yellow to bright blue on the southern hemisphere) before we observe images of sources on the other side of the equatorial plane (colour change from bright green to bright blue on the northern hemisphere and bright blue to bright green on the southern hemisphere). The reason for this effect is simply that corotating light rays and gravitational waves cover the same angle $\left|\Delta \varphi\right|$ faster than counterrotating light rays and gravitational waves. In addition, on the western hemisphere the width of the half-circles with images of second and third order (and so on) is slightly larger than for the half-circles on the eastern hemisphere. In particular, we can see that images of second and third order (and so on) generated by corotating light rays and gravitational waves passing close to the axes $\vartheta=0$ and $\vartheta=\pi$ can already be observed at higher celestial latitudes than images of the same order generated by counterrotating light rays and gravitational waves.\\
When we increase the spin to $a=m/2$ the observed features become more pronounced. While on the eastern hemisphere, when we zoom in, we can see images up to fourth order on the western hemisphere close to the boundary of the shadow also a very thin band of images of fifth order becomes visible. When we increase the spin even further to $a=3m/4$ the observed effects again become more pronounced. In addition, on the western hemisphere close to the boundary of the shadow there are indicators that images of sixth order become visible, however, it also cannot be completely excluded that these features are plotting artefacts.\\
When we now increase the spin parameter to $a=m$ (in this case the Kerr spacetime becomes extremal) the observed effects get again more pronounced. On the western hemisphere we can clearly see areas with images of first, second, third, and fourth order generated by corotating light rays and gravitational waves. In addition, in close proximity to the shadow we can see a broad region which consists of very narrow semicircular-shaped bands formed by images of fifth order and higher. Because these bands are very narrow we show an enlarged view of this region in Fig.~6. The figure clearly shows that for $a=m$ in the case of corotating light rays and gravitational waves we can see areas with images up to and beyond 15th order. Beyond the 15th order it becomes very hard to count the order of the images but we can likely see images up to about 30th order or slightly higher. In addition, Fig.~6 shows another phenomenon. Close to the celestial equator for images of fifth to about 15th order the observed light rays and gravitational waves come from sources on the same spacetime hemisphere (from the northern spacetime hemisphere for images on the northern celestial hemisphere and from the southern spacetime hemisphere for images on the southern celestial hemisphere). This means that light rays and gravitational waves generating images of seventh and higher orders orbit the black hole several times on the same hemisphere before they cross the equatorial plane. In addition, the closer to the celestial equator (which for an observer in the equatorial plane also corresponds to the equatorial plane of the spacetime) the light rays and gravitational waves orbit the black hole the broader becomes the latitudinal range across which this effect occurs. In addition, we can see that this effect also occurs for images beyond 15th order. When we now have a look at the eastern hemisphere, we see that here the band with images of third order has a smaller latitudinal width than for $a=3m/4$ and also that images of fourth order are barely visible.\\
In addition, when we zoom in close to the axis crossing on the northern hemisphere we see east of it a small branch of images which can either be of second or fourth order. Here, an analysis of the output data revealed that these images are images of second order. These images are generated by counterrotating light rays and gravitational waves which were emitted by light sources on the southern hemisphere of the two-sphere of sources $S_{S}^{2}$. For these light rays and gravitational waves the spin of the black hole effectively reduces the covered angle $\left|\Delta\varphi\right|$ compared to light rays and gravitational waves emitted by the surrounding light sources. A second branch of the same type of images can be found on the southern hemisphere but this time the associated sources are located on the northern hemisphere of the two-sphere of sources $S_{S}^{2}$. \\
Last but not least we now turn to the bottom right panel of Fig.~5. It shows the lens map for an observer at $r_{O}=10m$ and $\vartheta_{O}=\pi/4$ and sources distributed on the two-sphere $S_{S}^2$ at the radius coordinate $r_{S}=20m$ in the Kerr spacetime. The spin is $a=m$. When we compare the lens map with the lens map for the observer in the equatorial plane in the bottom left panel we can easily see that the asymmetry of the shadow is less pronounced. In addition, the lens map now also shows an asymmetry with respect to the celestial equator. However, otherwise we see relatively similar structures. In particular, for corotating light rays and gravitational waves close to the shadow we see an area with bands with images of higher orders, however, compared to the lens map for the observer in the equatorial plane the maximal order of the visible image bands is lower.\\
Note that in general the discussed lensing patterns occur for static sources (we recall that for simplicity and to distinguish them from the other stationary sources we refer to sources on $t$-lines as static) and sources on orbits at the radius coordinate $r_{S}=20m$. However, it is intuitively clear that light rays and gravitational waves emitted by a source at the same time coordinate $t_{S}$ and generating images of different orders on the celestial sphere of the observer need different times to arrive at the observer. Therefore, only for static sources light rays and gravitational waves emitted at the same spacetime coordinates $\vartheta_{S}$ and $\varphi_{S}$ on the two-sphere of sources $S_{S}^2$ and arriving at the same time coordinate $t_{O}$ at the observer can generate images of different orders of the same source. When we have sources orbiting the black hole at a fixed radius coordinate $r_{S}$ even if they are emitted at the same spacetime coordinates $\vartheta_{S}$ and $\varphi_{S}$ the light rays and gravitational waves generating images of different orders on the observer's celestial sphere at the time coordinate $t_{O}$ are generally emitted by different sources. When we want to calculate the time coordinates $t_{O}$ at which the images of different orders generated by light rays and gravitational waves emitted by the same moving source at the time coordinate $t_{S}$ at the spacetime coordinates $\vartheta_{S}$ and $\varphi_{S}$ on the two-sphere of sources $S_{S}^2$ can be detected on the observer's celestial sphere, and the associated celestial coordinates at which these images can be observed, we have to combine the lens maps with the associated travel time maps and also consider the motion of the observer (note that we can always generate lens and travel time maps at different $t_{O}$ and $\varphi_{O}$ by simply adjusting the initial conditions). In general the images of a single source on the observer's celestial sphere at different time coordinates $t_{O}$ can be obtained analogously to the procedure described above, however, finding them is a rather tedious task. Thus it is beyond the scope of this paper and may be part of future work.\\
In addition, it is also interesting to compare the features in the lens maps discussed above with the lens maps from earlier studies, in particular the works from Bohn \emph{et al.} \cite{Bohn2015} and Cunha \emph{et al.} \cite{Cunha2015}. Note that the approaches the authors of both works used to derive the lens maps deviate from our approach in several ways. Here, the most important difference is certainly that in both works the lens maps were derived numerically while in this paper we derived them analytically. The remaining differences vary for each work. First, Bohn \emph{et al.} derived the lens map for the Kerr spacetime for a dimensionless spin of $a=0.95$. Second, they measure the distance to the black hole in terms of the Kerr-Schild coordinates and they use the orthonormal tetrad for an observer moving along a $t$-line. Third, they placed their sphere of light sources at spatial infinity. Cunha \emph{et al.} on the other hand use a zero angular momentum observer (note that this is only mentioned in their supplementary material and their follow up paper Cunha \emph{et al.} \cite{Cunha2016}) located at the radius coordinate $r_{O}=15M_{\text{ADM}}$ and light sources distributed on a two-sphere $S_{L}^2$ at the radius coordinate $r_{L}=30M_{\text{ADM}}$ (note that for the Kerr spacetime the ADM [Arnowitt-Deser-Misner] mass corresponds to the mass parameter $m$) for three different Kerr black holes with $J_{\text{ADM}}/M_{\text{ADM}}^2\in \left\{0.85,0.894,0.999\right\}$ and thus between $a=3m/4$ and $a=m$. However, these differences are particularly interesting since they allow us to draw conclusions in how far the choice of the observer affects the observed lensing patterns. \\
We first compare our results to the result of Bohn \emph{et al.} \cite{Bohn2015}. In their Fig.~4 in the bottom right panel the authors show a lens map for an observer in the equatorial plane. When we compare their lens map with the middle right panel and the bottom left panel of Fig.~5 we immediately see that the overall structure is very similar, however, due to the different observer-lens-source geometries and the different orthonormal tetrads the details in both figures are difficult to compare. Luckily the lens maps presented in the work of Cunha et al. \cite{Cunha2015} were calculated for an observer-lens-source geometry which is more similar to ours. When we compare both lens maps with the top middle panel ($J_{\text{ADM}}/M_{\text{ADM}}^2=0.999$), the right panel in the second row ($J_{\text{ADM}}/M_{\text{ADM}}^2=0.85$), and the bottom right panel ($J_{\text{ADM}}/M_{\text{ADM}}^2=0.894$) of Fig.~5 in Cunha \emph{et al.} \cite{Cunha2015} we see that again the overall structure is very similar (note though that their field of view is much more narrow than ours). In the lens map in the first panel for corotating light rays we can easily see bands with images up to fifth order and then a coarsely resolved region with images of higher orders while for counterrotating light rays and gravitational waves their lens map only shows bands with images up to second order. Similarly for corotating light rays and gravitational waves the other two lens maps show bands with images up to fourth order while for counterrotating light rays and gravitational waves they again only show bands with images up to second order. Since the results presented in this paper were calculated for different radius coordinates for the observer $r_{O}$ and the two-sphere of sources $r_{S}$ we also calculated lens maps for the standard observer using their specifications (the lens maps are not shown).  The results are overall very similar but there also seem to be small differences. Unfortunately, without having both lens maps in the same high resolution and in the same format it is difficult to draw a definite conclusion. Here, it would be particularly interesting and also important for the correct interpretation of observations to investigate in more detail how the choice of the observer and different observer-lens-source geometries affect the details of the observed lensing patterns. Unfortunately, this is beyond the scope of this paper and may be part of future work.\\
The next question which needs to be addressed is the question where we can find the critical curves in the lens maps for the Kerr spacetime. As mentioned above in the lens map for the Schwarzschild spacetime they are located at the boundaries between the rings with images of different orders. However, we saw that in the lens maps for the Kerr spacetime images of first order generated by counterrotating light rays and gravitational waves are separated from images of second order generated by corotating light rays and gravitational waves emitted by sources on the same quadrant of the two-sphere of sources $S_{S}^2$ by sharp lines, and that these lines mark light rays and gravitational waves crossing one of the axes. However, it is clear that these lines cannot be part of the critical curves. Therefore, from the lens maps in Fig.~5 alone it is not completely straightforward to identify the positions of the critical curves. However, a comparison with the lens maps presented in Bohn \emph{et al.} \cite{Bohn2015} and Cunha \emph{et al.} \cite{Cunha2015} can shed light onto this question. In their lens maps they plot a coordinate grid and mark a reference light source spot in the direction in which the observer looks. In their obtained lens maps the images of this reference light source spot form a ring-like structure around the shadow. When we look closely we see that on the western hemisphere around $\Psi=\pi/2$ and on the eastern hemisphere around $\Psi=3\pi/2$ the centre of the ring-like structure seems to be aligned with the boundaries between the images of first and second order. When we now follow this structure from the western to the eastern hemisphere and compare the lens maps of Bohn \emph{et al.} and Cunha \emph{et al.} with our lens maps for the standard observer we can easily see that it has to cross through areas with images of first and second order. Thus for the Kerr spacetime in the lens maps the critical curves are not always aligned with the boundaries between images of different orders anymore and thus when we want to know their exact positions we have to explicitly calculate them.\\
The last question we have to address is now whether the regions with images of higher orders generated by corotating light rays and gravitational waves shown in Fig.~6 represent real image structures or not. Here we would like to note that besides the bands of images shown in Fig.~6 the code also gave results far beyond what is visible in Fig.~6. Here, from the lens map alone there are already two indicators that this is the case. First, the structures visible in Fig.~6 and also the structures in the nonvisible part of the output of the code are consistent in themselves and do not show random behaviour which should appear if the solutions to the equations of motion derived in this paper would break down due to diverging terms. Note that these diverging terms inevitably appear when we approach the photon orbit or the horizon. The second reason is that the obtained results are consistent with what one would physically expect for the Kerr black hole, namely that when we approach the shadow for a fixed celestial latitude light rays and gravitational waves travelling around the black hole on orbits with a low inclination to the equatorial plane make more turns around the black hole before crossing the equatorial plane than light rays and gravitational waves travelling around the black hole on orbits with a higher inclination.\\
In addition, before the codes were used to calculate the lens maps, the redshift maps, and the travel time maps a thorough intercomparison with numerical results was performed, and this intercomparison showed a good agreement. However, visible images beyond the 15th order were initially not expected. Thus for selected individual geodesics associated with higher order images an initial comparison between the analytical results and numerical results was performed and again both showed a good agreement. However, before the results for higher order images are used to search for observational features for these images a much more thorough and systematic investigation should be performed. Here, the main reason is simply that for $0<a<m$ the $\varphi$ coordinate diverges when we move backward in time along two different types of lightlike geodesics. In the first case we have lightlike geodesics asymptotically going to the photon orbit at the radius coordinate $r_{\text{ph}}$. In the second case we have lightlike geodesics asymptotically going to the event horizon. Now when we have $a=m$ for corotating light rays and gravitational waves for a certain range around the equatorial plane the photon orbit and the horizon are located at the same radius coordinate $r_{\text{H}}=r_{\text{ph}}$. As mentioned before although this is just a projection effect which arises from the choice of the coordinates and in the embedding diagram the photon orbit and the horizon are distinct we have to very carefully investigate whether the visibility of the higher order images is just due to the inadequate projection or whether they are still observable when we use another coordinate system in which the photon orbit and the horizon are located at distinct spacetime coordinates.\\

\begin{figure}\label{fig:LEKerr}
  \begin{tabular}{cc}
    Schwarzschild Spacetime: $\vartheta_{O}=\pi/2$ & Kerr Spacetime: $a=m/4$ and $\vartheta_{O}=\pi/2$\\
\\
    \hspace{-0.5cm}\includegraphics[width=85mm]{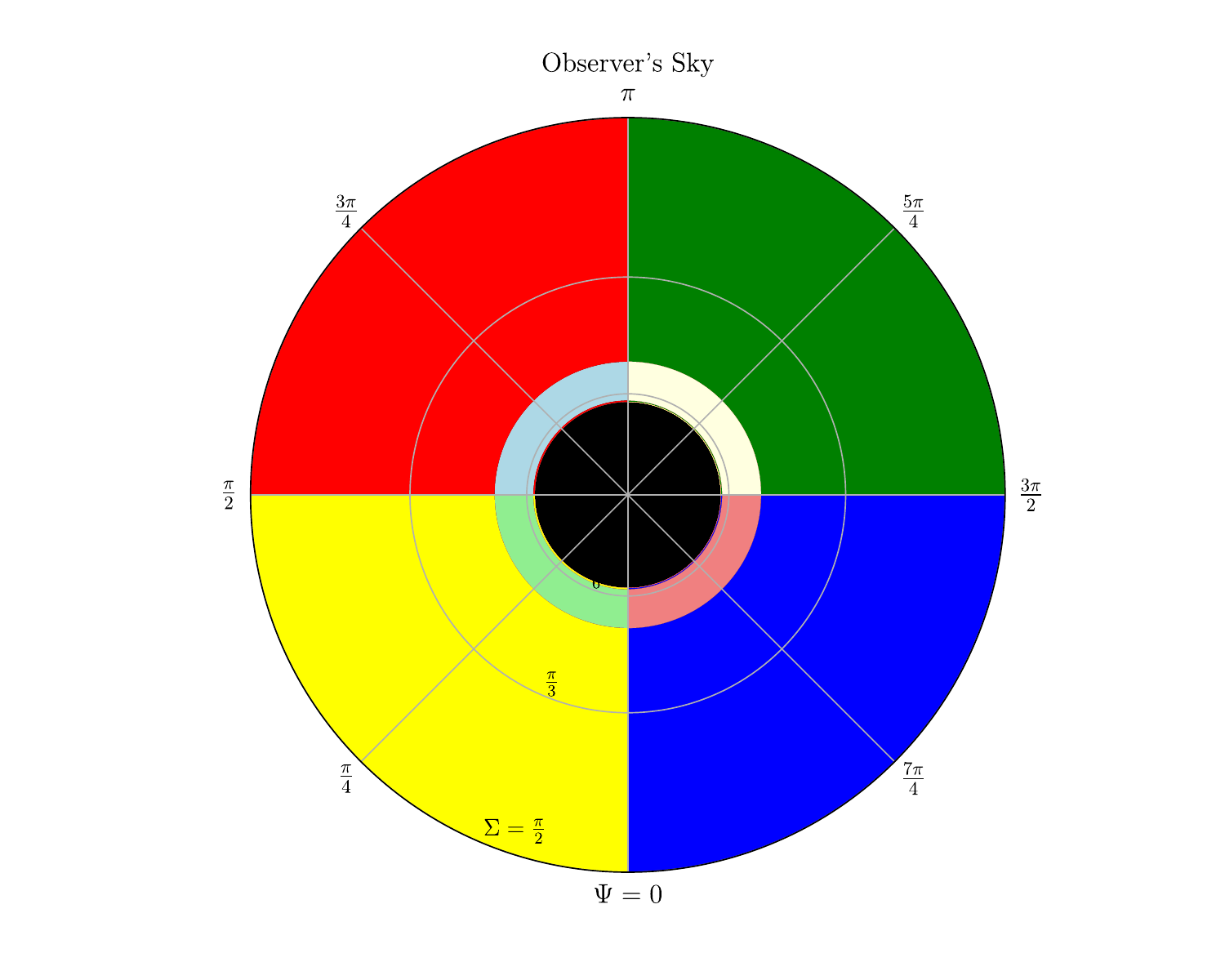} &   \hspace{-0.5cm}\includegraphics[width=85mm]{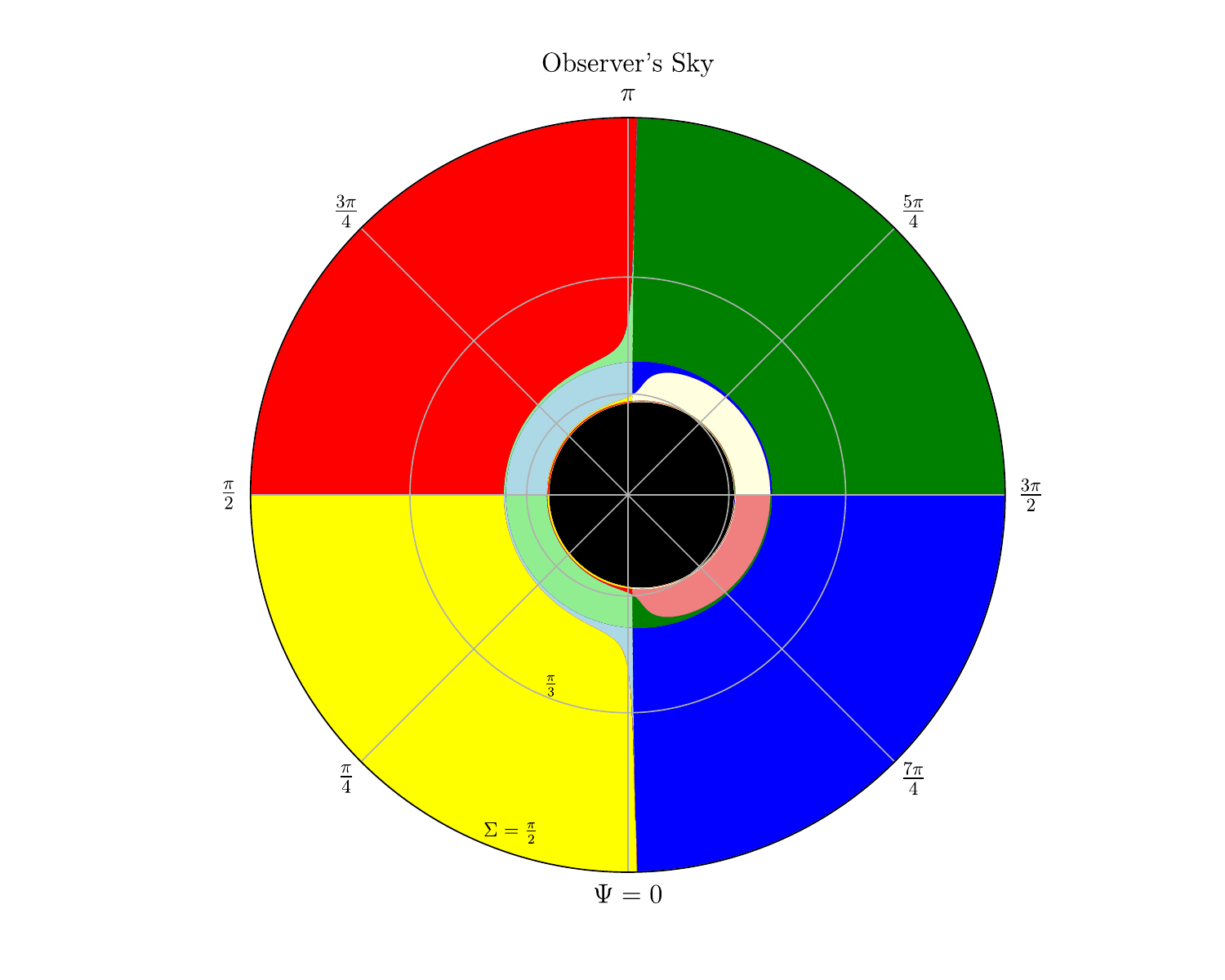} \\
\\
    Kerr Spacetime: $a=m/2$ and $\vartheta_{O}=\pi/2$ & Kerr Spacetime: $a=3m/4$ and $\vartheta_{O}=\pi/2$\\
\\
    \hspace{-0.5cm}\includegraphics[width=85mm]{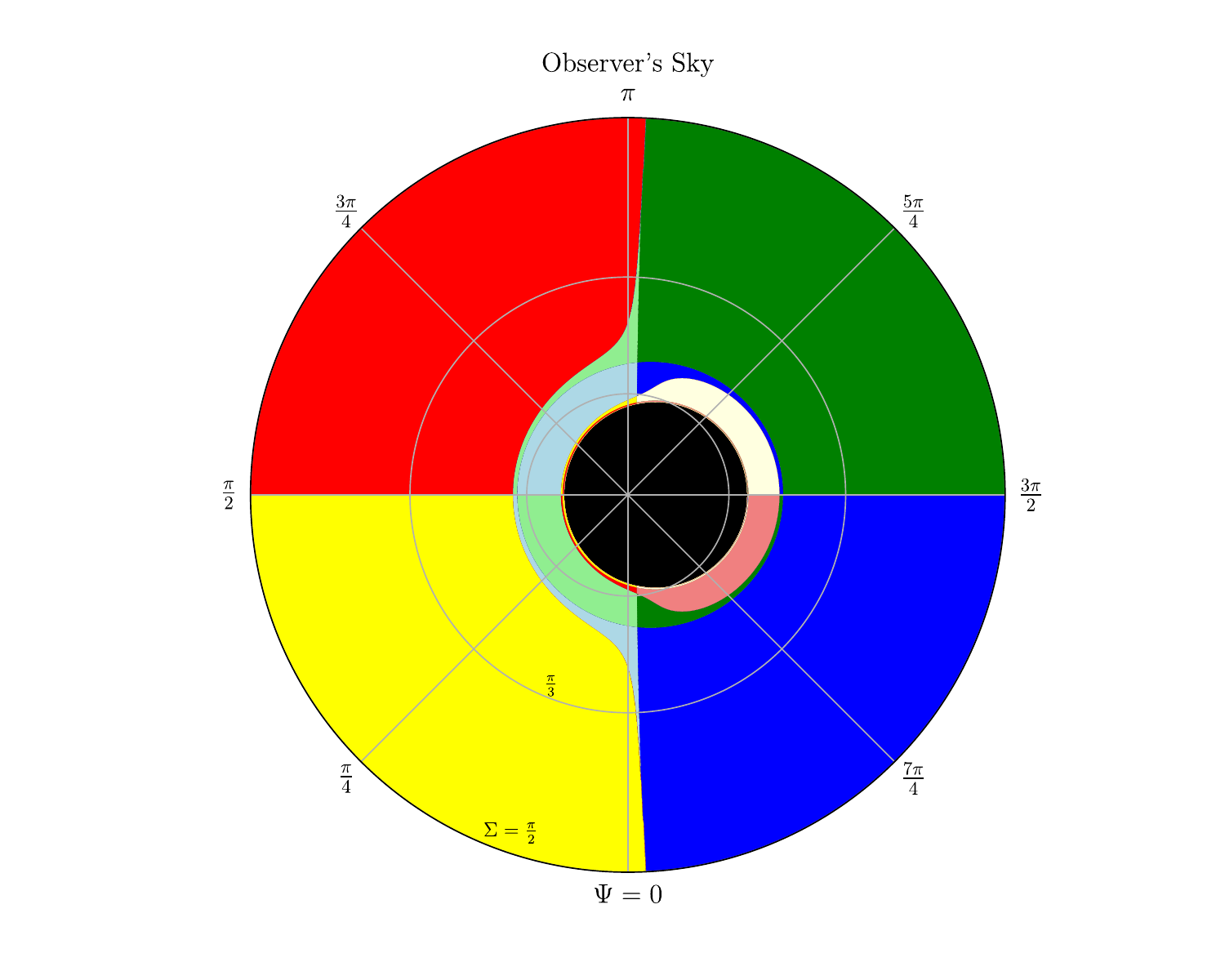} &   \hspace{-0.5cm}\includegraphics[width=85mm]{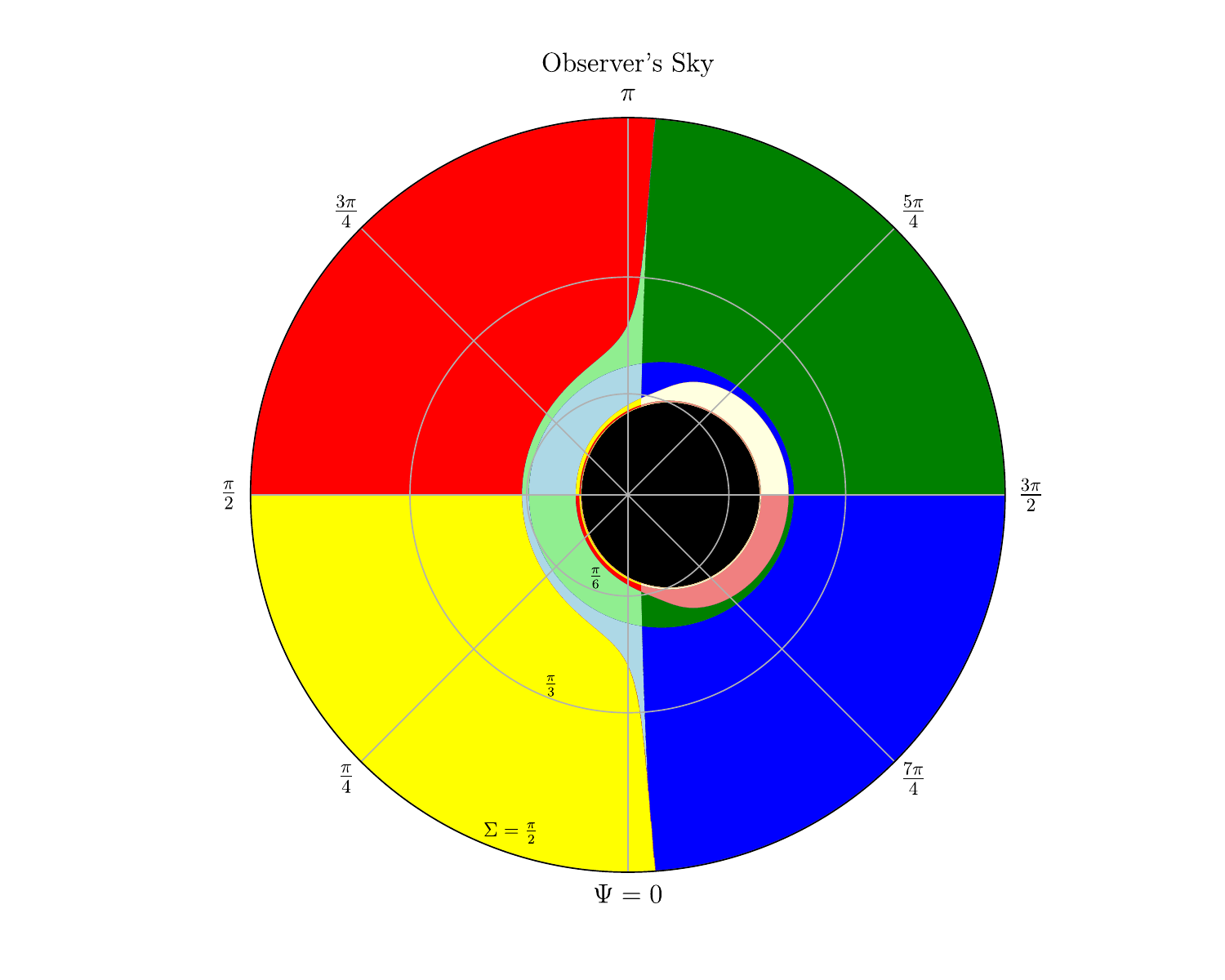} \\
    Kerr Spacetime: $a=m$ and $\vartheta_{O}=\pi/2$ & Kerr Spacetime: $a=m$ and $\vartheta_{O}=\pi/4$\\
    \\
    \hspace{-0.5cm}\includegraphics[width=85mm]{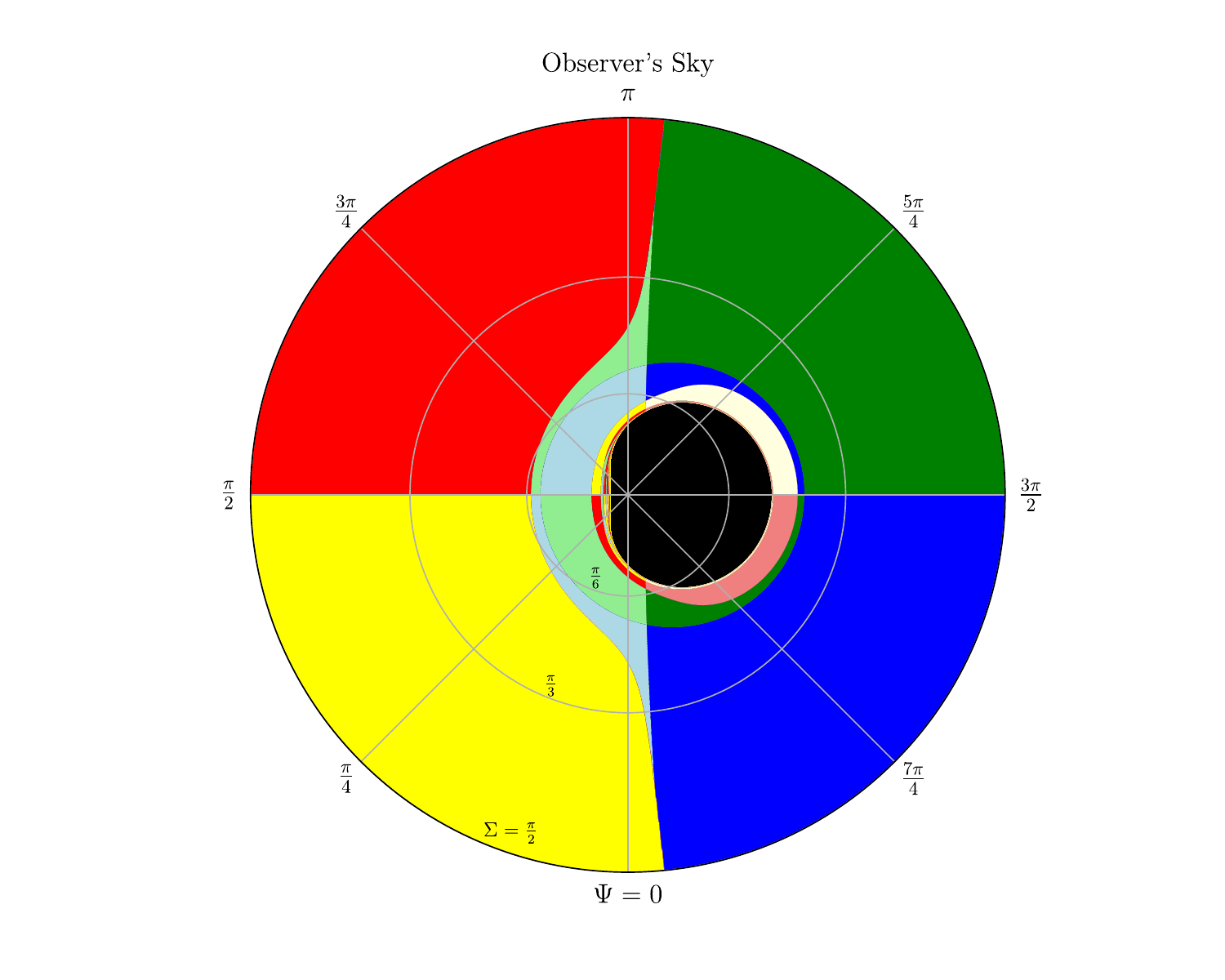} &   \hspace{-0.5cm}\includegraphics[width=85mm]{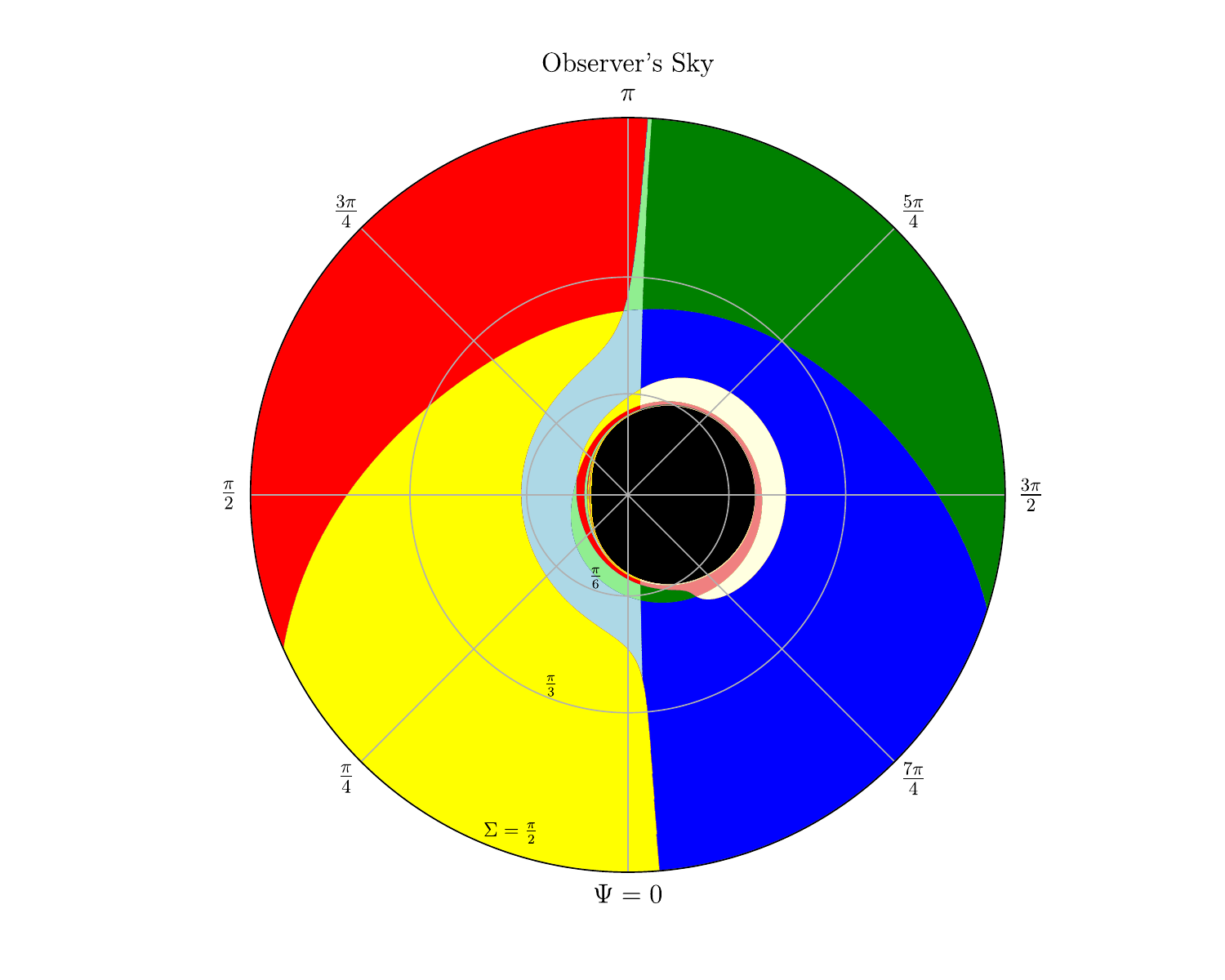} \\
  \end{tabular}
	\caption{Lens maps for light rays and gravitational waves emitted by sources on the two-sphere of sources $S_{S}^2$ at the radius coordinate $r_{S}=20m$ and detected by a standard observer at $r_{O}=10m$ in the equatorial plane ($\vartheta_{O}=\pi/2$) for the Schwarzschild spacetime (top left panel) and the Kerr spacetime with $a=m/4$ (top right panel), $a=m/2$ (middle left panel), $a=3m/4$ (middle right panel), and $a=m$ (bottom left panel). The bottom right panel shows a lens map for light rays and gravitational waves emitted by sources on the two-sphere of sources $S_{S}^2$ at the radius coordinate $r_{S}=20m$ and detected by a standard observer located at $r_{O}=10m$ and $\vartheta_{O}=\pi/4$ for the Kerr spacetime. For this map the spin parameter is $a=m$.}
\end{figure}

\begin{figure}\label{fig:LEZoomKerr}
\includegraphics[width=0.5\textwidth]{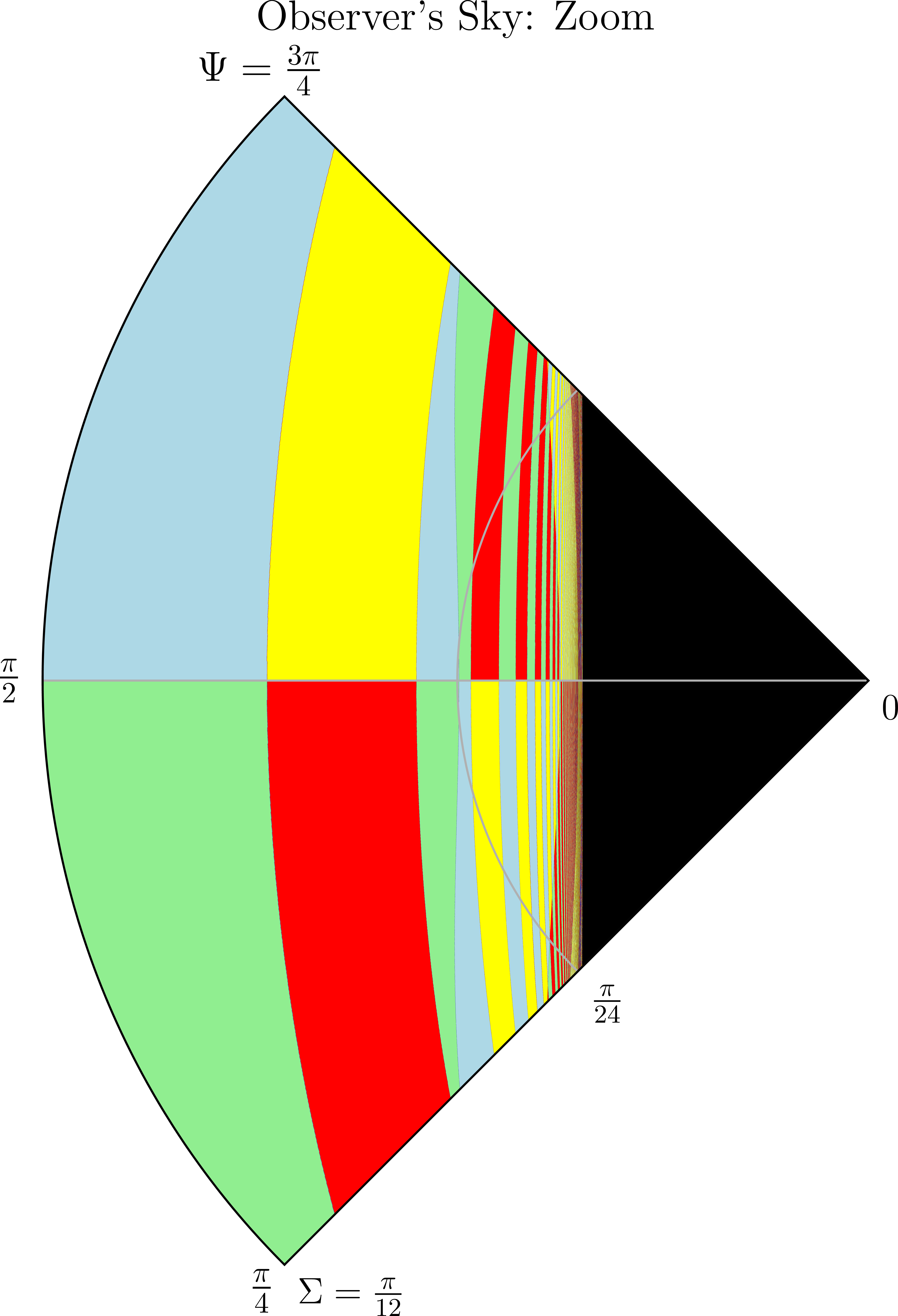} 
	\caption{Enlarged view (zoom) of the lens map for the Kerr spacetime with $a=m$ in the bottom left panel of Fig.~5 for the region $0\leq\Sigma\leq \pi/12$ between $\Psi=\pi/4$ and $\Psi=3\pi/4$. The light rays and gravitational waves are emitted by sources on the two-sphere of sources $S_{S}^2$ at the radius coordinate $r_{S}=20m$ and detected by a standard observer at $r_{O}=10m$ in the equatorial plane ($\vartheta_{O}=\pi/2$).}
\end{figure}

\subsection{The Redshift}\label{Subsec:Redshift}
The redshift is one of the most easily accessible observables in astronomy. When we observe characteristic atomic or molecular emission lines in the electromagnetic spectrum of a source we can compare it to the same atomic or molecular emission lines measured in a laboratory on Earth. In terms of the energies of a light ray or a gravitational wave at the source $E_{S}$ and at the observer $E_{O}$ the redshift reads
\begin{eqnarray}\label{eq:DefRS}
z=\frac{E_{S}}{E_{O}}-1.
\end{eqnarray}
We now want to determine the redshift a local observer measures when he detects a light ray or a gravitational wave emitted by a source in the Kerr spacetime. For this purpose we first need to determine the energies of the light ray or the gravitational wave in the local frames of the source and the observer. For this purpose, let us first write them down in terms of the four-momentum $p_{\mu}$ of the light ray or the gravitational wave and the four-velocities of the source $u_{S}$ and the observer $u_{O}$. In its most general form the local energy $E_{\text{loc}}$ of a light ray or a gravitational wave measured by a detector moving with the velocity $u_{\text{loc}}$ at an event $x$ reads, see, e.g., Wald \cite{Wald1984}, p.~69,
\begin{eqnarray}\label{eq:Eloc}
E_{\text{loc}}=-p_{\mu}u_{\text{loc}}^{\mu},
\end{eqnarray}
where in our case we have $u_{\text{loc}}=u_{S}$ at the position of the source and $u_{\text{loc}}=u_{O}$ at the position of the observer.
In our case we consider a standard observer. Its four-velocity $u_{O}$ is given by the tetrad vector $e_{0}$ in (\ref{eq:ONT1}). Thus we have as nonvanishing components of the four-velocity
\begin{eqnarray}\label{eq:uobs}
u_{O}^{t}=\frac{r_{O}^2+a^2}{\sqrt{\rho(r_{O},\vartheta_{O})P(r_{O})}}~~~\text{and}~~~u_{O}^{\varphi}=\frac{a}{\sqrt{\rho(r_{O},\vartheta_{O})P(r_{O})}}.
\end{eqnarray}
Therefore, the energy the standard observer measures in his local frame is given by
\begin{eqnarray}\label{eq:ElocO}
E_{O}=-p_{t}u_{O}^{t}-p_{\varphi}u_{O}^{\varphi}.
\end{eqnarray}
Now we insert the components of the four-velocity and rewrite the conjugated momenta in terms of the constants of motion using $p_{t}=-E$ and $p_{\varphi}=L_{z}$ and get
\begin{eqnarray}\label{eq:Eobs}
E_{O}=\frac{(r_{O}^2+a^2)E-aL_{z}}{\sqrt{\rho(r_{O},\vartheta_{O})P(r_{O})}}.
\end{eqnarray}
In the last step we use (\ref{eq:ConsE}) and (\ref{eq:ConsLz}) to rewrite $E$ and $L_{z}$ in terms of the latitude-longitude coordinates on the observer's celestial sphere and get 
\begin{eqnarray}\label{eq:EobsC}
E_{O}=1.
\end{eqnarray}
In this paper we will now consider three different types of sources. These are sources on $t$-lines (we again recall that for simplicity and to distinguish them from the other stationary sources we refer to them as static), zero angular momentum (or Bardeen) sources, and standard (or Carter) sources. In the following we will now calculate the redshift for light rays and gravitational waves emitted by the three different sources and detected by a standard observer in the domain of outer communication.
\subsubsection{Static Sources}
The first type of sources we are going to consider are sources on $t$-lines. While in spherically symmetric and static spacetimes the four-velocity of these sources is orthogonal to the spacelike hypersurfaces and thus they are static, in axisymmetric and stationary spacetimes such as the Kerr spacetime this is generally not the case and thus these sources are only stationary. However, since in this paper we consider three different stationary sources and it may be confusing to refer to sources on $t$-lines simply as stationary, we recall that for the discussion in this paper we agreed on the following convention. Whenever a source moves along a $t$-line we refer to it as \emph{static} even though in the Kerr spacetime these sources are only stationary. Note that some authors such as Straumann, see, e.g., p. 471 in Ref.~\cite{Straumann2013}, seem to generally refer to observers and sources on $t$-lines as static, however, in this paper we only use this term as a label. From the normalisation condition for timelike motion we can easily see that the four-velocity of these sources is given by
\begin{eqnarray}\label{eq:ulsstc}
u_{S_{\text{stc}}}=u_{S_{\text{stc}}}^{t}\partial_{t}=\sqrt{\frac{\rho\left(r_{S},\vartheta_{S}\right)}{P(r_{S})-a^2\sin^2\vartheta_{S}}}\partial_{t}.
\end{eqnarray}
Since $u_{S_{\text{stc}}}^{t}$ is the only nonvanishing component of the four-velocity the energy of the light ray or the gravitational wave at the source given by (\ref{eq:Eloc}) now reads 
\begin{eqnarray}\label{eq:Elocstat}
E_{S_{\text{stc}}}=-p_{t}u_{S_{\text{stc}}}^{t}.
\end{eqnarray}
Now we insert $p_{t}=-E$ and $u_{S_{\text{stc}}}^{t}$ and obtain as relation between the energy of the light ray or the gravitational wave measured at the source and the energy $E$ along the lightlike geodesic
\begin{eqnarray}\label{eq:ELstc}
E_{S_{\text{stc}}}=\sqrt{\frac{\rho\left(r_{S},\vartheta_{S}\right)}{P(r_{S})-a^2\sin^2\vartheta_{S}}}E.
\end{eqnarray}
In the second step we use (\ref{eq:ConsE}) to rewrite the energy $E$ in terms of the latitude-longitude coordinates on the observer's celestial sphere. We take the result for $E_{S_{\text{stc}}}$ and the result for the energy $E_{O}$ measured by the observer given by (\ref{eq:EobsC}) and insert them in the general redshift formula (\ref{eq:DefRS}) and obtain for the redshift of light rays and gravitational waves emitted by a static source and detected by a standard observer
\begin{eqnarray}\label{eq:zstc}
z_{\text{stc}}=\sqrt{\frac{\rho(r_{S},\vartheta_{S})}{\rho(r_{O},\vartheta_{O})\left(P(r_{S})-a^2\sin^2\vartheta_{S}\right)}}\left(\sqrt{P(r_{O})}+a\sin\vartheta_{O}\sin\Sigma\sin\Psi\right)-1.
\end{eqnarray}
We can easily see that for $a=0$ the obtained relation for the redshift reduces to the redshift for a static observer and a static source in the Schwarzschild spacetime
\begin{eqnarray}\label{eq:zstcSch}
z_{\text{Schw}}=\frac{r_{S}}{r_{O}}\sqrt{\frac{P(r_{O})}{P(r_{S})}}-1.
\end{eqnarray}
When we compare both relations for the redshift we can easily see that there are two major differences. In the Schwarzschild spacetime besides the mass parameter $m$ the redshift only depends on the radius coordinates of the source and the observer $r_{S}$ and $r_{O}$ (note though that in this paper we measure all quantities in units of $m$ and thus the mass parameter is just a scaling parameter). This is different for the Kerr spacetime. Here, the redshift does not only have an additional dependency on the spin parameter $a$, but it also depends on the spacetime latitude $\vartheta_{O}$ of the observer, and directly and indirectly, through the spacetime latitude of the source $\vartheta_{S}(\Sigma,\Psi)$, on the celestial coordinates on the observer's celestial sphere $\Sigma$ and $\Psi$. 

\subsubsection{Zero Angular Momentum Sources}
The second type of sources we are going to consider are zero angular momentum (or Bardeen) sources. As the sources on $t$-lines they are stationary. The four-velocity of these sources is given by \cite{Bardeen1972}
\begin{eqnarray}
&u_{S_{0}}=u_{S_{0}}^{t}\partial_{t}+u_{S_{0}}^{\varphi}\partial_{\varphi}=\frac{\left(\left(r_{S}^2+a^2\right)^2-a^2\sin^2\vartheta_{S}P(r_{S})\right)\partial_{t}+a\left(\left(r_{S}^2+a^2\right)-P(r_{S})\right)\partial_{\varphi}}{\sqrt{\rho(r_{S},\vartheta_{S})P(r_{S})\left(\left(r_{S}^2+a^2\right)^2-a^2\sin^2\vartheta_{S}P(r_{S})\right)}}.
\end{eqnarray}
Since this time both $u_{S_{0}}^{t}$ and $u_{S_{0}}^{\varphi}$ are nonzero the energy of the light rays and gravitational waves at the source given by (\ref{eq:Eloc}) reads
\begin{eqnarray}\label{eq:Eloc0}
E_{S_{0}}=-p_{t}u_{S_{0}}^{t}-p_{\varphi}u_{S_{0}}^{\varphi}.
\end{eqnarray}
Now we insert the components of the four-velocity, $p_{t}=-E$, and $p_{\varphi}=L_{z}$ and obtain
\begin{eqnarray}\label{eq:EL0}
E_{S_{0}}=\frac{(r_{S}^2+a^2)\left((r_{S}^2+a^2)E-aL_{z}\right)-\left(a^2\sin^2\vartheta_{S}E-aL_{z}\right)P(r_{S})}{\sqrt{\rho(r_{S},\vartheta_{S})P(r_{S})\left(\left(r_{S}^2+a^2\right)^2-a^2\sin^2\vartheta_{S}P(r_{S})\right)}}.
\end{eqnarray}
Now we use (\ref{eq:ConsE}) and (\ref{eq:ConsLz}) to rewrite the energy $E$ and the angular momentum about the $z$-axis $L_{z}$ in terms of the latitude-longitude coordinates on the observer's celestial sphere. We take the obtained result for $E_{S_{0}}$ and the result for the energy $E_{O}$ measured by the standard observer given by (\ref{eq:EobsC}) and insert them in the general redshift formula (\ref{eq:DefRS}). We obtain for the redshift of light rays and gravitational waves emitted by zero angular momentum sources and detected by a standard observer
\begin{eqnarray}\label{eq:z0}
&z_{0}=\frac{\left(\left(r_{S}^2+a^2\right)\rho(r_{S},\vartheta_{O})+\left(\rho(r_{O},\vartheta_{S})-\rho(r_{O},\vartheta_{O})\right)P(r_{S})\right)\sqrt{P(r_{O})}+\left(\left(r_{S}^2+a^2\right)\left(r_{S}^2-r_{O}^2\right)+\rho(r_{O},\vartheta_{S})P(r_{S})\right)a\sin\vartheta_{O}\sin\Sigma\sin\Psi}{\sqrt{\rho(r_{O},\vartheta_{O})\rho(r_{S},\vartheta_{S})P(r_{S})\left(\left(r_{S}^2+a^2\right)^2-a^2\sin^2\vartheta_{S}P(r_{S})\right)}}-1.
\end{eqnarray}
We can easily see that also for zero angular momentum sources for $a=0$ the redshift reduces to the redshift for light rays and gravitational waves emitted by a static source and detected by a static observer in the Schwarzschild spacetime (\ref{eq:zstcSch}). As for static sources the obtained redshift depends on the spin parameter $a$ and the spacetime latitude of the observer $\vartheta_{O}$, and directly and indirectly, through the spacetime latitude of the source $\vartheta_{S}(\Sigma,\Psi)$, on the celestial coordinates on the observer's celestial sphere $\Sigma$ and $\Psi$. However, for zero angular momentum sources this dependency is slightly more complicated than for static sources.

\subsubsection{Standard Sources}
The third type of sources are standard (or Carter) sources. As for the standard observer their four-velocity is given by $e_{0}$ (with $x_{O}$ replaced by $x_{S}$) and reads
\begin{eqnarray}\label{eq:uLC}
u_{S_{\text{C}}}=u_{S_{\text{C}}}^{t}\partial_{t}+u_{S_{\text{C}}}^{\varphi}\partial_{\varphi}=\frac{\left(r_{S}^2+a^2\right)\partial_{t}+a\partial_{\varphi}}{\sqrt{\rho(r_{S},\vartheta_{S})P(r_{S})}}.
\end{eqnarray}
The energy of the light rays and the gravitational waves measured at the position of the source is given by the same relation as for the standard observer. It reads
\begin{eqnarray}\label{eq:ELCmom}
E_{S_{\text{C}}}=-p_{t}u_{S_{\text{C}}}^{t}-p_{\varphi}u_{S_{\text{C}}}^{\varphi}.
\end{eqnarray}
Now we insert the components of the four-velocity, $p_{t}=-E$, and $p_{\varphi}=L_{z}$ and obtain 
\begin{eqnarray}\label{eq:ELC}
E_{S_{\text{C}}}=\frac{(r_{S}^2+a^2)E-aL_{z}}{\sqrt{\rho(r_{S},\vartheta_{S})P(r_{S})}}.
\end{eqnarray}
Again we use (\ref{eq:ConsE}) and (\ref{eq:ConsLz}) to rewrite the energy $E$ and the angular momentum about the $z$-axis $L_{z}$ in terms of the latitude-longitude coordinates on the observer's celestial sphere. Then we take the obtained relations for $E_{S_{\text{C}}}$ and for the energy of the light rays and gravitational waves measured at the position of the standard observer $E_{O}$ given by (\ref{eq:EobsC}) and insert them in the general redshift formula (\ref{eq:DefRS}). The result for the redshift of light rays and gravitational waves emitted by a standard source and detected by a standard observer reads
\begin{eqnarray}\label{eq:zC}
z_{\text{C}}=\frac{\rho(r_{S},\vartheta_{O})\sqrt{P(r_{O})}+a(\rho(r_{S},\vartheta_{O})-\rho(r_{O},\vartheta_{O}))\sin\vartheta_{O}\sin\Sigma\sin\Psi}{\sqrt{\rho(r_{O},\vartheta_{O})\rho(r_{S},\vartheta_{S})P(r_{S})}}-1.
\end{eqnarray}
Again in the limit $a=0$ we obtain the redshift for light rays and gravitational waves emitted by a static source and detected by a static observer in the Schwarzschild spacetime given by (\ref{eq:zstcSch}). As for the static and zero angular momentum sources the obtained redshift for light rays and gravitational waves emitted by a standard source and detected by a standard observer in the Kerr spacetime depends on the spin parameter $a$, the spacetime latitude of the observer $\vartheta_{O}$, and directly and indirectly, through the spacetime latitude of the source $\vartheta_{S}(\Sigma,\Psi)$, on the celestial coordinates on the observer's celestial sphere $\Sigma$ and $\Psi$. 

\subsubsection{Redshift Maps}
\begin{figure}\label{fig:zstcKerr}
  \begin{tabular}{cc}
    Schwarzschild Spacetime: $\vartheta_{O}=\pi/2$ & Kerr Spacetime: $a=m/4$ and $\vartheta_{O}=\pi/2$\\
\\
    \hspace{-0.7cm}\includegraphics[width=78mm]{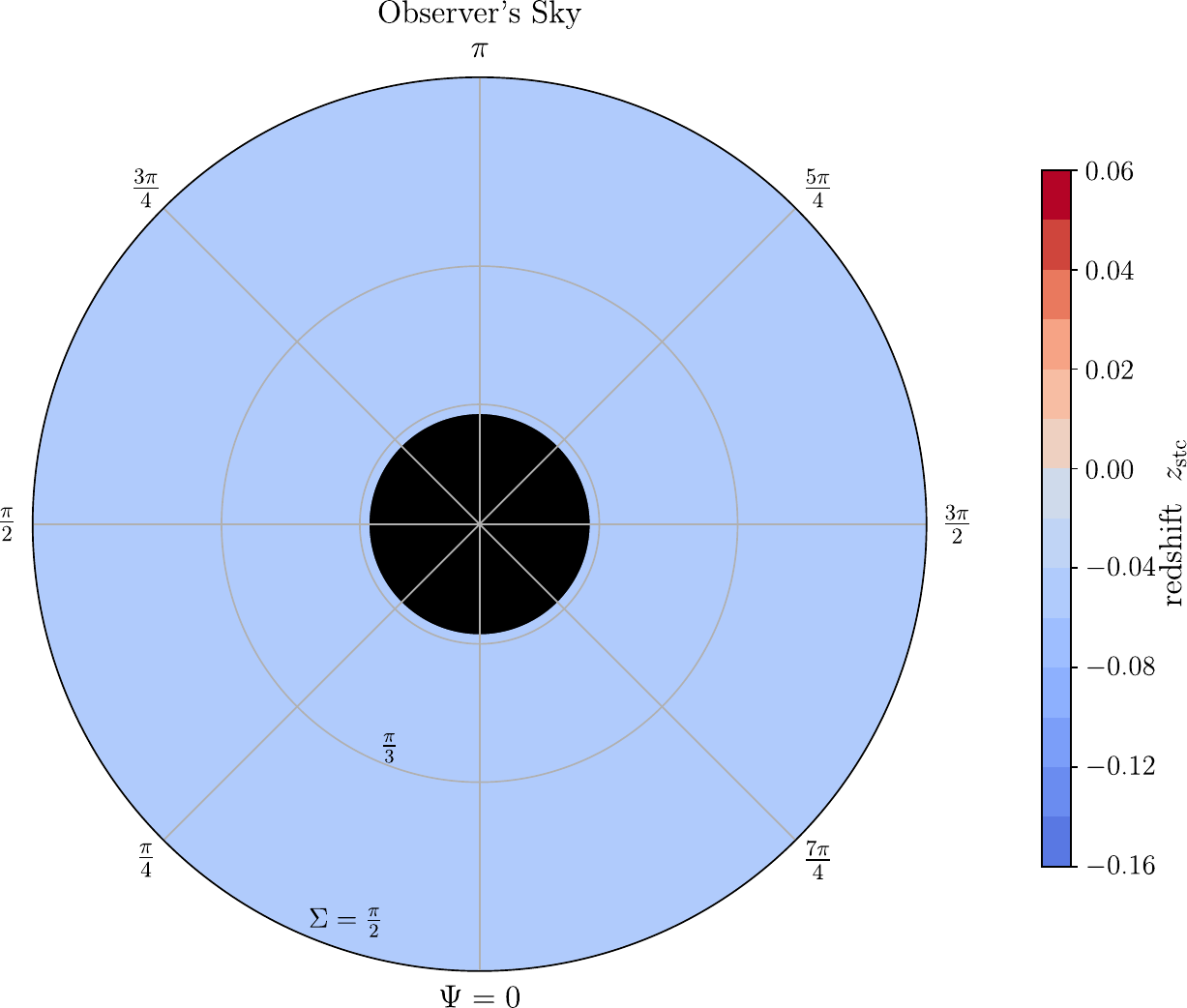} &   \hspace{0.7cm}\includegraphics[width=78mm]{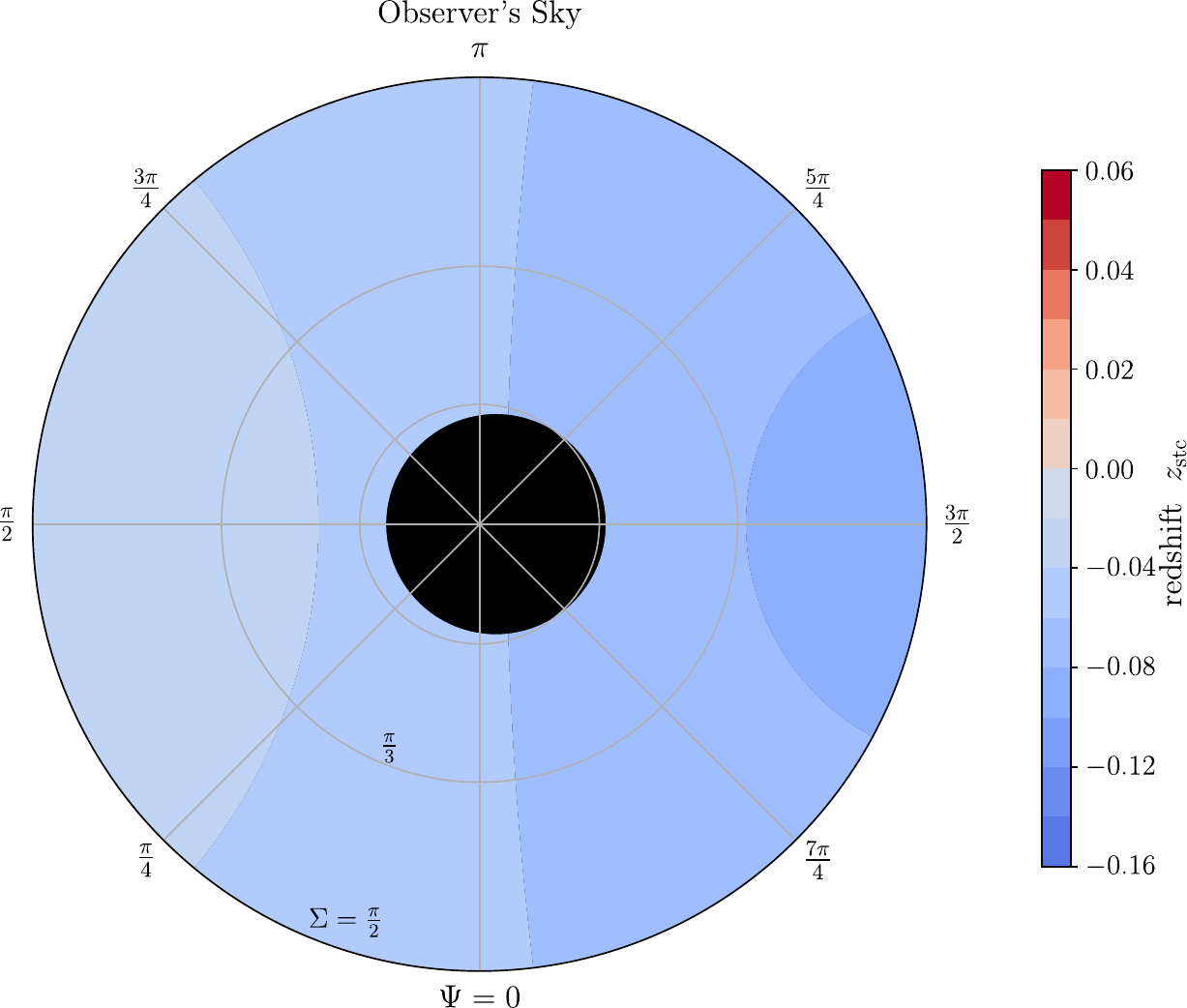} \\
\\
    Kerr Spacetime: $a=m/2$ and $\vartheta_{O}=\pi/2$ & Kerr Spacetime: $a=3m/4$ and $\vartheta_{O}=\pi/2$\\
\\
    \hspace{-0.7cm}\includegraphics[width=78mm]{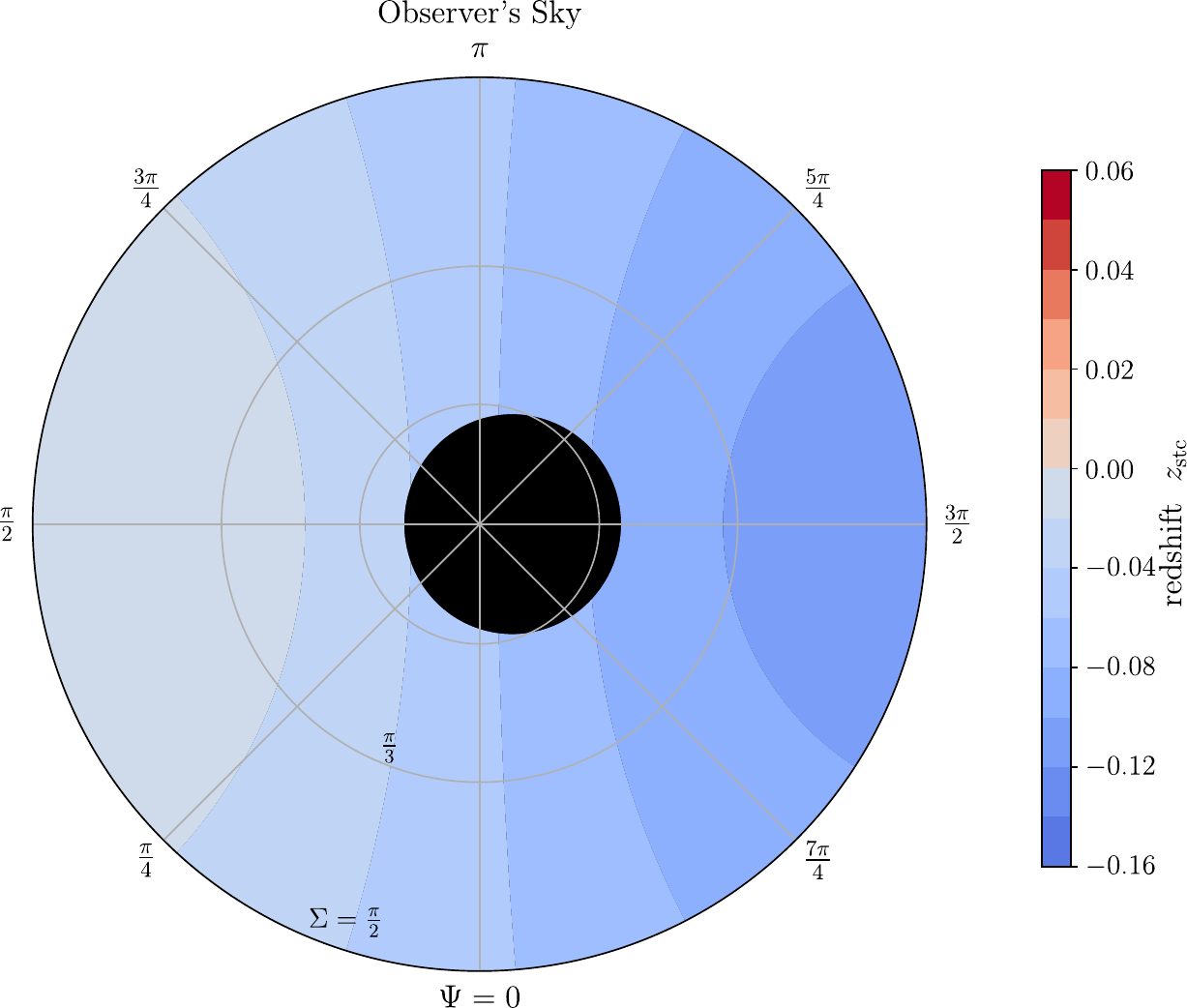} &   \hspace{0.7cm}\includegraphics[width=78mm]{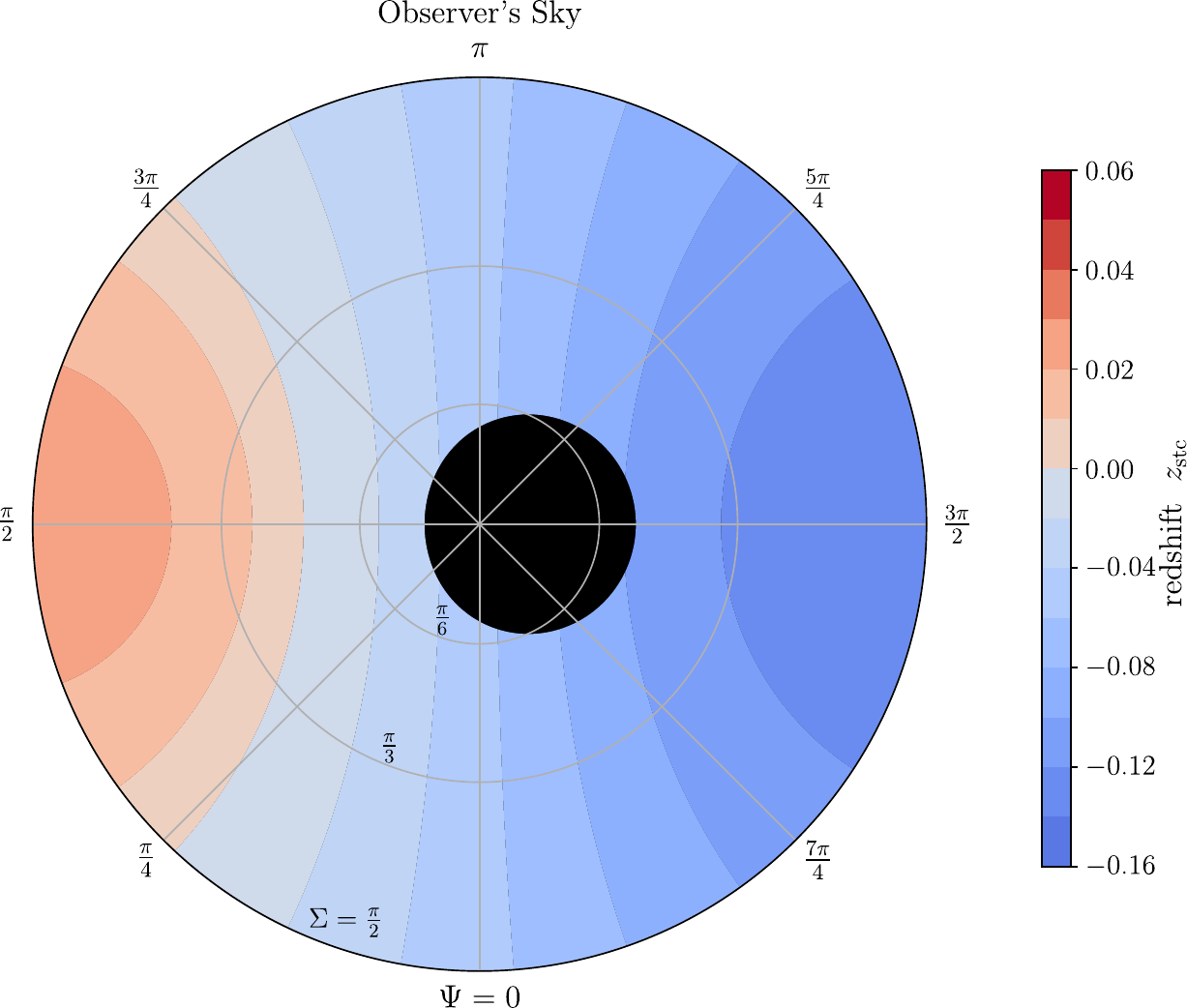} \\
    Kerr Spacetime: $a=m$ and $\vartheta_{O}=\pi/2$ & Kerr Spacetime: $a=m$ and $\vartheta_{O}=\pi/4$\\
    \\
    \hspace{-0.7cm}\includegraphics[width=78mm]{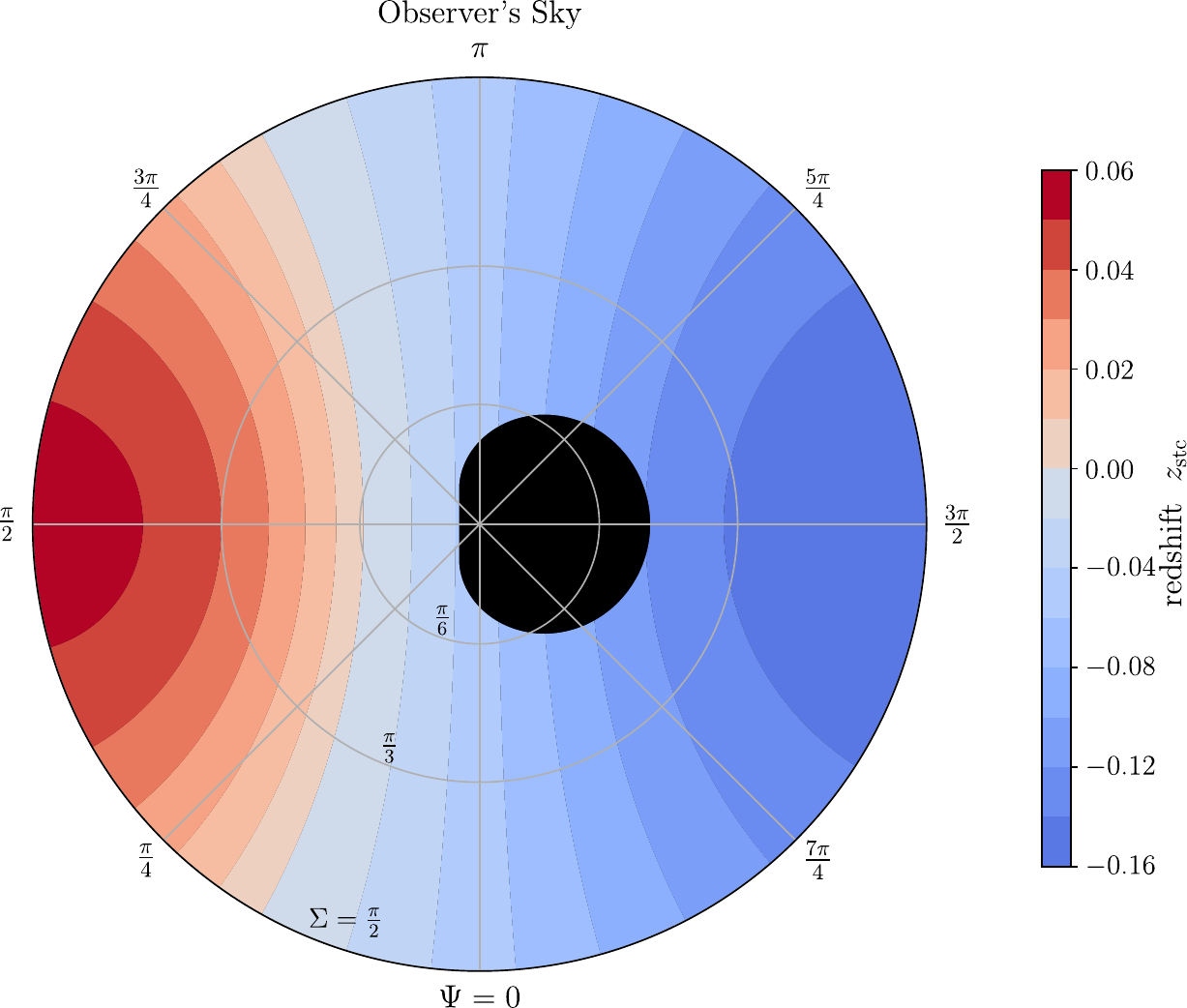} &   \hspace{0.7cm}\includegraphics[width=78mm]{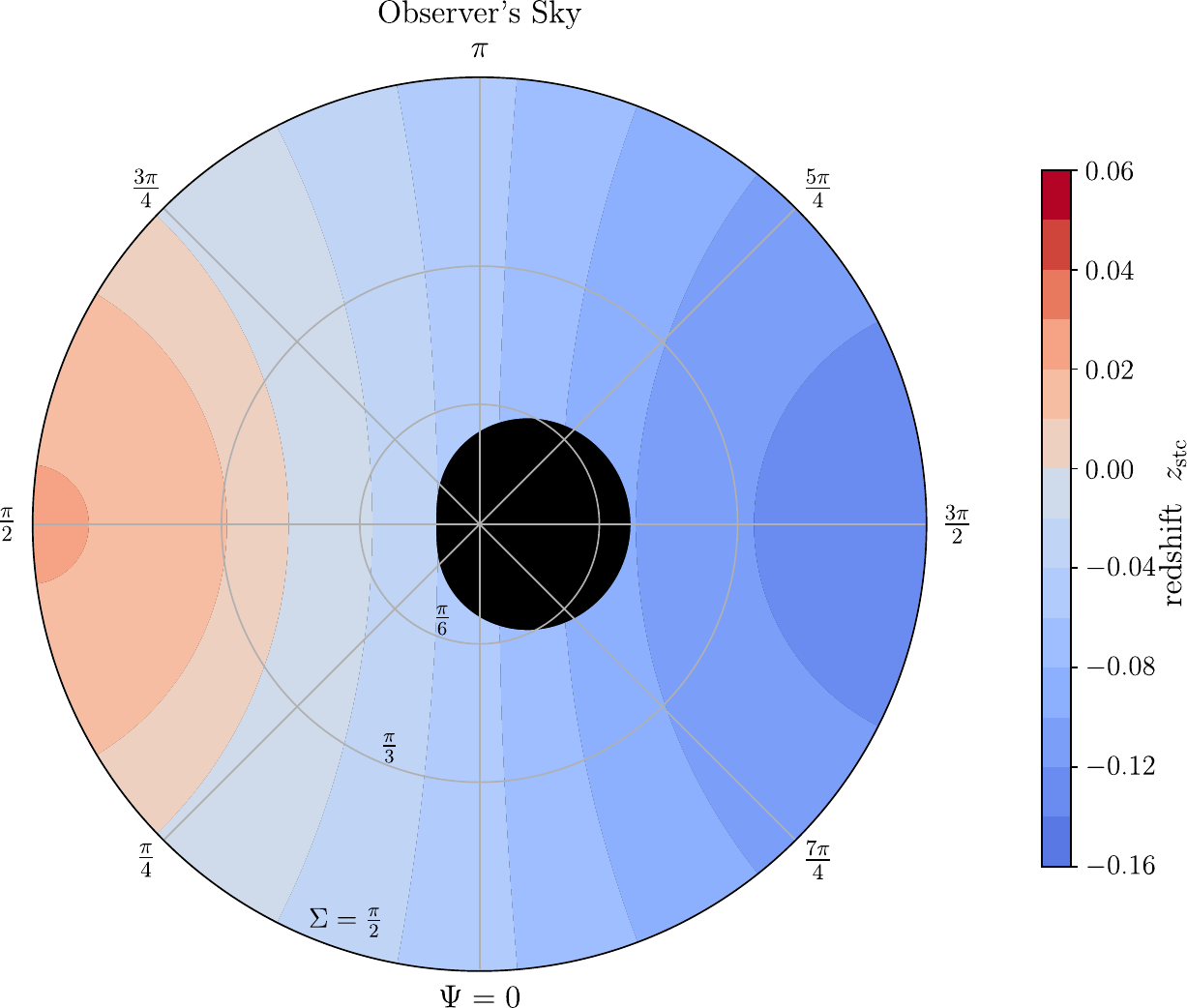} \\
  \end{tabular}
	\caption{Redshift maps for light rays and gravitational waves emitted by static sources on the two-sphere of sources $S_{S}^2$ at the radius coordinate $r_{S}=20m$ and detected by a standard observer at $r_{O}=10m$ in the equatorial plane ($\vartheta_{O}=\pi/2$) for the Schwarzschild spacetime (top left panel) and the Kerr spacetime with $a=m/4$ (top right panel), $a=m/2$ (middle left panel), $a=3m/4$ (middle right panel), and $a=m$ (bottom left panel). The bottom right panel shows a redshift map for light rays and gravitational waves emitted by static sources on the two-sphere of sources $S_{S}^2$ at the radius coordinate $r_{S}=20m$ and detected by a standard observer located at $r_{O}=10m$ and $\vartheta_{O}=\pi/4$ for the Kerr spacetime. For this map the spin parameter is $a=m$.}
\end{figure}

\begin{figure}\label{fig:redshiftKerrMetric}
  \begin{tabular}{cc}
    Static Sources\\
    \\
    \includegraphics[width=160mm]{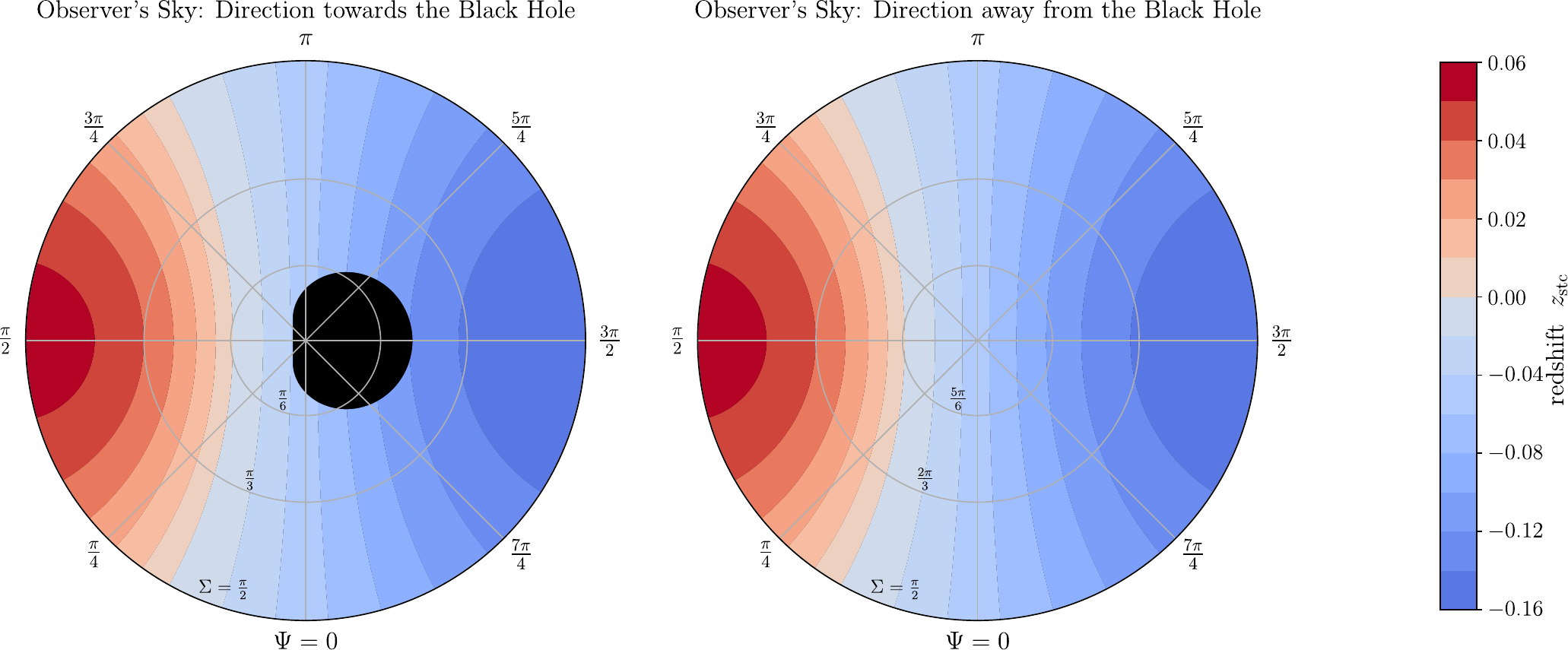} 
    \\
    \\
    Zero Angular Momentum Sources\\
    \\
    \includegraphics[width=160mm]{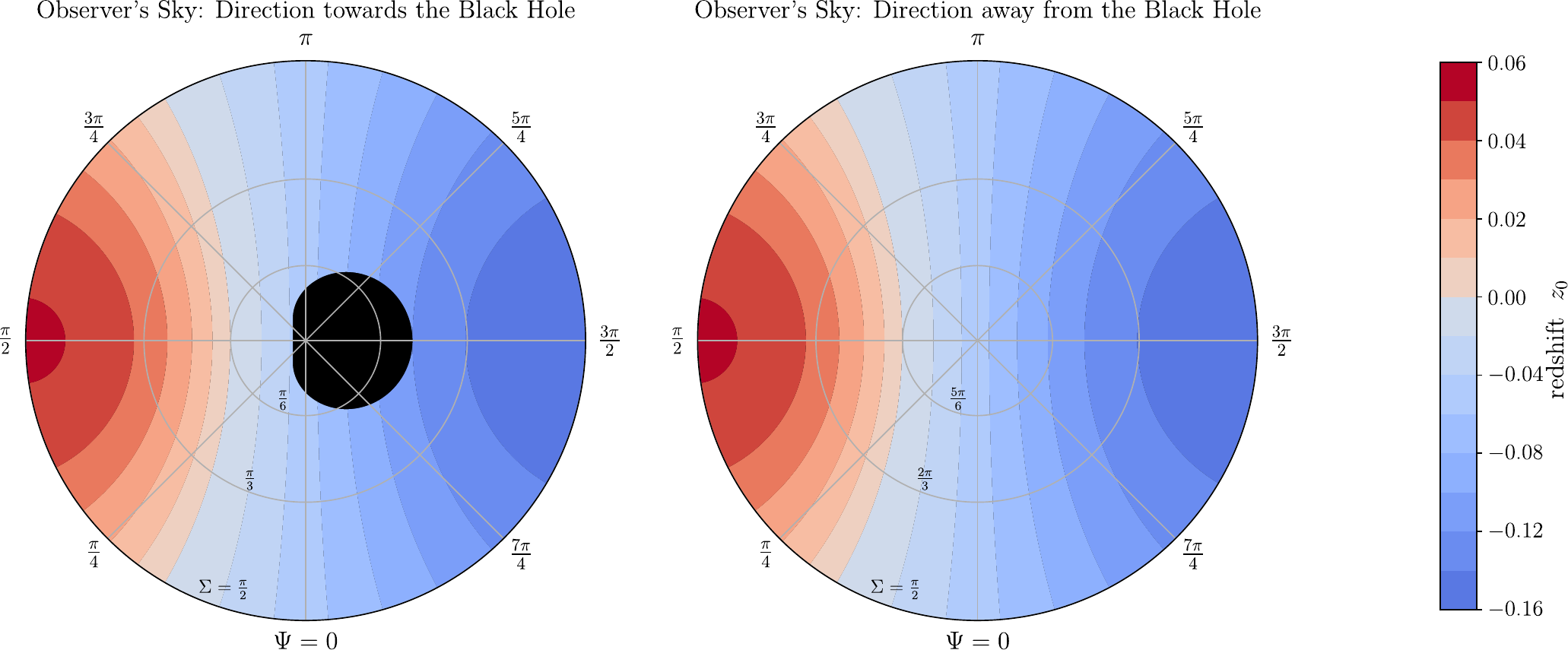}\\
    \\
    Standard Sources\\
    \\
    \includegraphics[width=160mm]{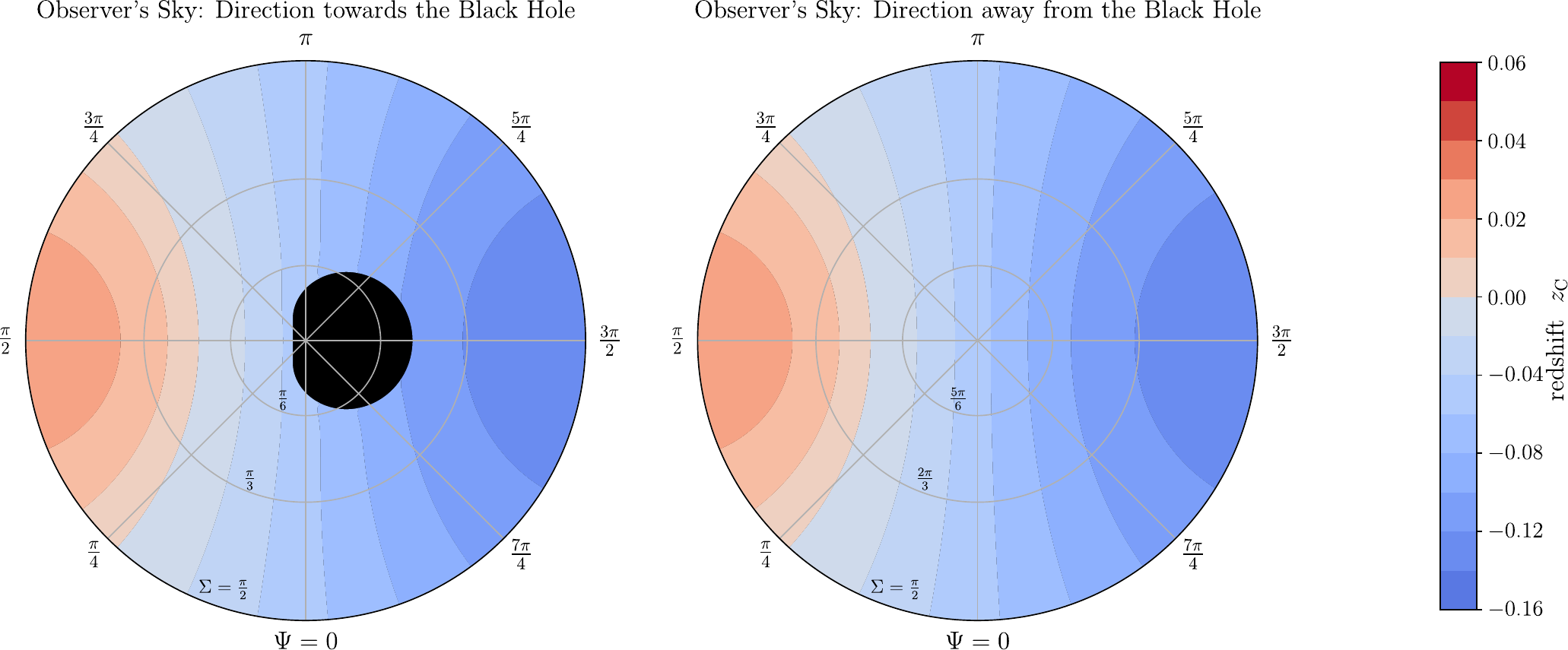}\\
  \end{tabular}
	\caption{Redshift maps for light rays and gravitational waves emitted by static (top panel), zero angular momentum (middle panel), and standard (bottom panel) sources on the two-sphere of sources $S_{S}^2$ at the radius coordinate $r_{S}=20m$ and detected by a standard observer at $r_{O}=10m$ in the equatorial plane ($\vartheta_{O}=\pi/2$) for the Kerr spacetime. The spin parameter is $a=m$.}
\end{figure}
The computational evaluation of the redshifts was carried out using the same set of Julia codes which was also used for the evaluation of the lens equation. As for the lens equation we visualise the redshifts as maps on the observer's celestial sphere. Fig.~7 shows redshift maps for light rays and gravitational waves emitted by static sources and detected by a standard observer. The first five panels show redshift maps for the Schwarzschild spacetime (top left panel) and the Kerr spacetime with $a=m/4$ (top right panel), $a=m/2$ (middle left panel), $a=3m/4$ (middle right panel), and $a=m$ (bottom left panel). The observer is located in the equatorial plane ($\vartheta_{O}=\pi/2$) at the radius coordinate $r_{O}=10m$ and the two-sphere of sources $S_{S}^2$ is located at the radius coordinate $r_{S}=20m$. The sixth panel (the bottom right panel) shows a redshift map for the Kerr spacetime with $a=m$ and an observer located at the coordinates $r_{O}=10m$ and $\vartheta_{O}=\pi/4$. Again the sources are distributed on a two-sphere $S_{S}^2$ located at the radius coordinate $r_{S}=20m$. In all maps the observer looks in the direction of the black hole and the black area at the centre of each map is the shadow of the black hole.\\
Again we start our discussion with the Schwarzschild spacetime. As we can read from (\ref{eq:zstcSch}) for the Schwarzschild spacetime the redshift is independent of the spacetime latitudes of the observer and the light source $\vartheta_{O}$ and $\vartheta_{S}$, respectively. Thus the redshift is constant. In our specific case we have $r_{O}=10m<r_{S}=20m$ and thus the redshift is $z_{\text{stc}}=-0.057$ and thus actually a blueshift. When we turn on the spin and set it to $a=m/4$ (top right panel of Fig.~7) first differences start to appear. Now the observer orbits the black hole with a constant angular velocity at the radius coordinate $r_{O}=10m$ and the spacetime latitude $\vartheta_{O}=\pi/2$ in the direction of the spin. On the eastern hemisphere in the direction of motion (centered around $\Psi=3\pi/2$) the blueshift increases. On the western hemisphere on the other hand opposite to the direction of motion (centered around $\Psi=\pi/2$) the blueshift decreases. The reason for this is that in the first case the combined gravitational blueshift and the Doppler blueshift due to the motion of the observer result in an overall increase of the total blueshift. On the other hand in the second case the Doppler redshift due to the motion of the observer reduces the gravitational blueshift and thus the overall blueshift decreases. When we increase the spin to $a=m/2$ (middle left panel of Fig.~7) this effect increases and when we increase the spin to $a=3m/4$ (middle right panel of Fig.~7) on the western hemisphere redshifts start to occur. When we now increase the spin to $a=m$ (bottom left panel of Fig.~7) the blueshifts on the eastern hemisphere and the redshifts on the western hemisphere continue to increase. In addition, on the western hemisphere the area covered with redshifts grows. As a result now more than half of it is covered by redshifts. When we now move the observer to $\vartheta_{O}=\pi/4$ the angular velocity of the observer decreases. As a consequence on the western hemisphere the area covered with redshifts shrinks and the observed redshifts decrease. Similarly on the eastern hemisphere the blueshifts decrease. However, the overall area covered with blueshifts grows.\\
In addition to these relatively obvious changes, we can also find an interesting effect close to the shadow. When we turn on the spin and zoom in close to the boundary of the shadow, we see small wave-like patterns. They occur at the positions at which the isocontours marking the boundaries between the areas with different blueshifts approach the shadow. Here, when we increase the spin from $a=m/4$ to $a=m$ the size of the wave-like patterns increases. On the other hand when we move the observer out of the equatorial plane and keep the spin parameter constant the size of the wave-like patterns remains roughly the same. Here, the dependency on the spin indicates that the wave-like patterns can be an effect caused by the rotation of the black hole. However, since in the Kerr spacetime the standard observer orbits the black hole with a constant angular velocity, and the observer's angular velocity also increases with increasing spin, it can also be an effect caused by the motion of the observer. We will discuss this in more detail below.\\
Now for the redshift three questions remain. The first two questions are how the redshift map looks like for the direction away from the black hole and how the overall pattern in the redshift map changes when the sources themselves move. To address these questions in Fig.~8 we show redshift maps on the full celestial sphere of a standard observer in the equatorial plane ($\vartheta_{O}=\pi/2$) at the radius coordinate $r_{O}=10m$ for light rays and gravitational waves emitted by static (top panel), zero angular momentum (middle panel), and standard (bottom panel) sources on two-spheres of sources $S_{S}^2$ at the radius coordinate $r_{S}=20m$ for a Kerr black hole with the spin $a=m$.\\
Let us start with the first question: How does the redshift map look like in the direction away from the black hole? When we look at the top panel of Fig.~8 we can easily see that the answer to this question is quite simple. The overall pattern is roughly the same. The main difference is that we do not have the shadow at the centre of the map. The answer to the second question is also quite easy. When we consider zero angular momentum and standard sources also the sources orbit the black hole with a constant angular velocity at a constant radius coordinate $r_{S}$ and a constant spacetime latitude $\vartheta_{S}$. When we have zero angular momentum sources (middle panel of Fig.~8) instead of static sources we see two major changes. First the area on the celestial sphere covered by redshifts shrinks slightly while the area covered by blueshifts grows. In addition overall the redshifts and the blueshifts reduce. The reason for this change is quite simple. In this case also the sources orbit the black hole in the direction of the spin and thus the Doppler red- and blueshifts decrease. When we have standard sources (bottom panel of Fig.~8) we see a similar effect. In comparison to the redshift maps for static and zero angular momentum sources the area on the observer's celestial sphere covered by redshifts shrinks and the area covered with blueshifts grows. In addition, overall the redshifts and the blueshifts decrease. As for the zero angular momentum sources the reason for this effect is simply that the sources orbit the black hole at a constant angular velocity at a fixed radius coordinate $r_{S}$ and a fixed spacetime latitude $\vartheta_{S}$ in the direction of the spin. Here, the standard sources move faster than zero angular momentum sources and thus the changes in the redshift maps are more pronounced. Furthermore, when we have standard light sources the size of the wave-like patterns, which can be observed at celestial latitudes close to the shadow, increases. \\
The third question is now which effects lead to the formation of the wave-like patterns close to the shadow of the black hole. They occured in the redshift maps for static, zero angular momentum, and standard light sources for the Kerr black hole. As already mentioned above the size of the wave-like patterns increases when we increase the spin parameter $a$ and also when we increase the angular velocity of the source. Here, the second dependency indicates that the angular velocity of the source and thus very likely also the angular velocity of the observer have an effect on the size of the wave-like patterns. Unfortunately, for the spin parameter it is much more difficult to draw a conclusion. The reason is quite simple. When we increase the spin parameter $a$ we also increase the angular velocity of the standard observer and thus the effects of the spin and the motion of the observer are not completely separable. This would require to completely reimplement the lensing code for an observer on a $t$-line. While in general the solutions to the equations of motion derived in this paper can also be used to implement a gravitational lensing code for an observer on a $t$-line, with respect to the relations between the constants of motion and the celestial angles on the observer's celestial sphere this will be a much more complicated task since for an observer on a $t$-line in general $\Sigma=0$ and $\Sigma=\pi$ will not be associated with $K=0$ anymore. While in general it will be a worthwhile task to investigate this question since it may lead to the identification of a new observable effect which may allow to determine the spin of rotating black holes, it by far exceeds the context of this study and thus will be left for future work. 

\subsection{The Travel Time}\label{Subsec:TravTime}
\begin{figure}\label{fig:TravKerr}
  \begin{tabular}{cc}
    Schwarzschild Spacetime: $\vartheta_{O}=\pi/2$ & Kerr Spacetime: $a=m/4$ and $\vartheta_{O}=\pi/2$\\
\\
    \hspace{-0.7cm}\includegraphics[width=78mm]{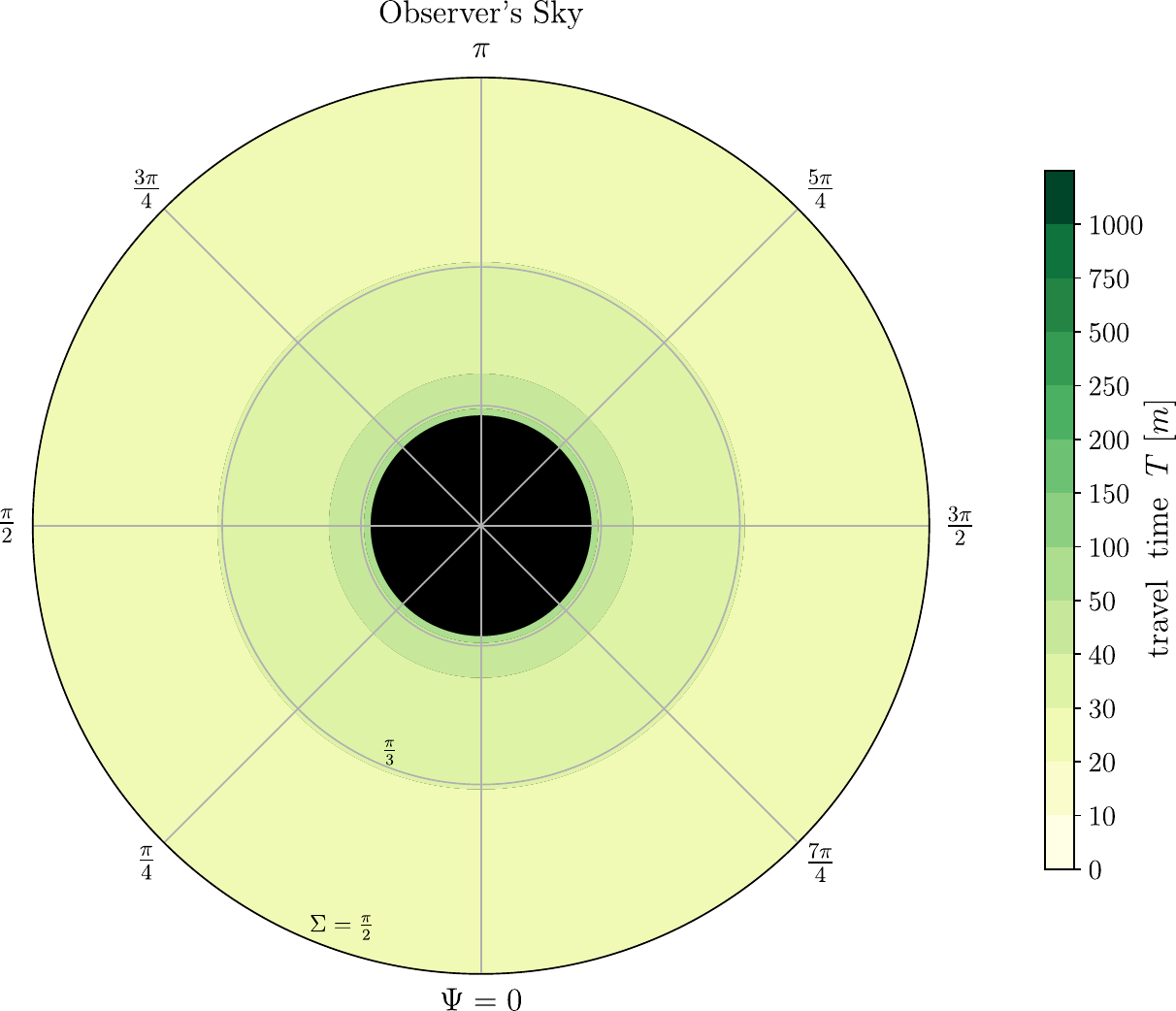} &   \hspace{0.7cm}\includegraphics[width=78mm]{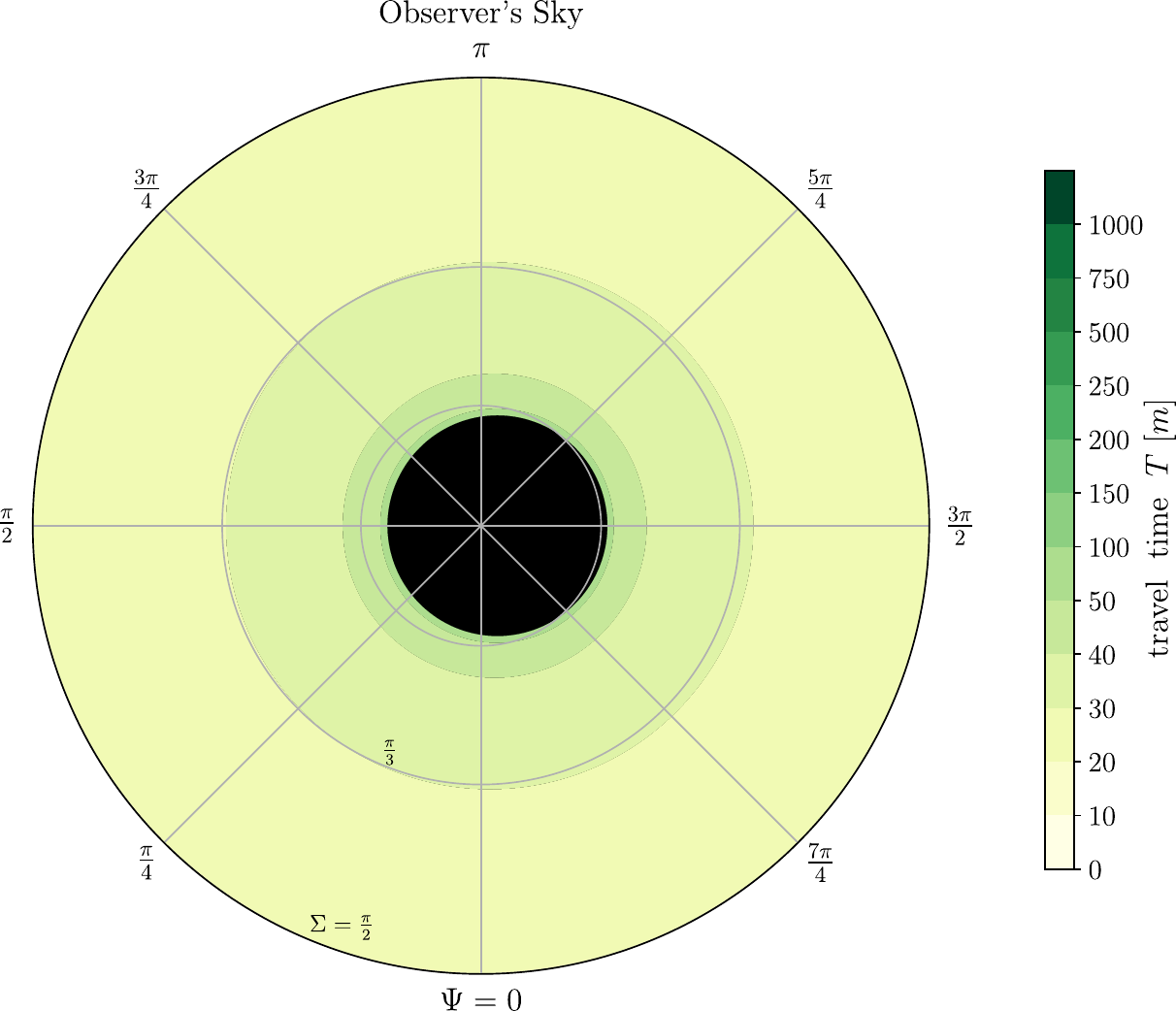} \\
\\
    Kerr Spacetime: $a=m/2$ and $\vartheta_{O}=\pi/2$ & Kerr Spacetime: $a=3m/4$ and $\vartheta_{O}=\pi/2$\\
\\
    \hspace{-0.7cm}\includegraphics[width=78mm]{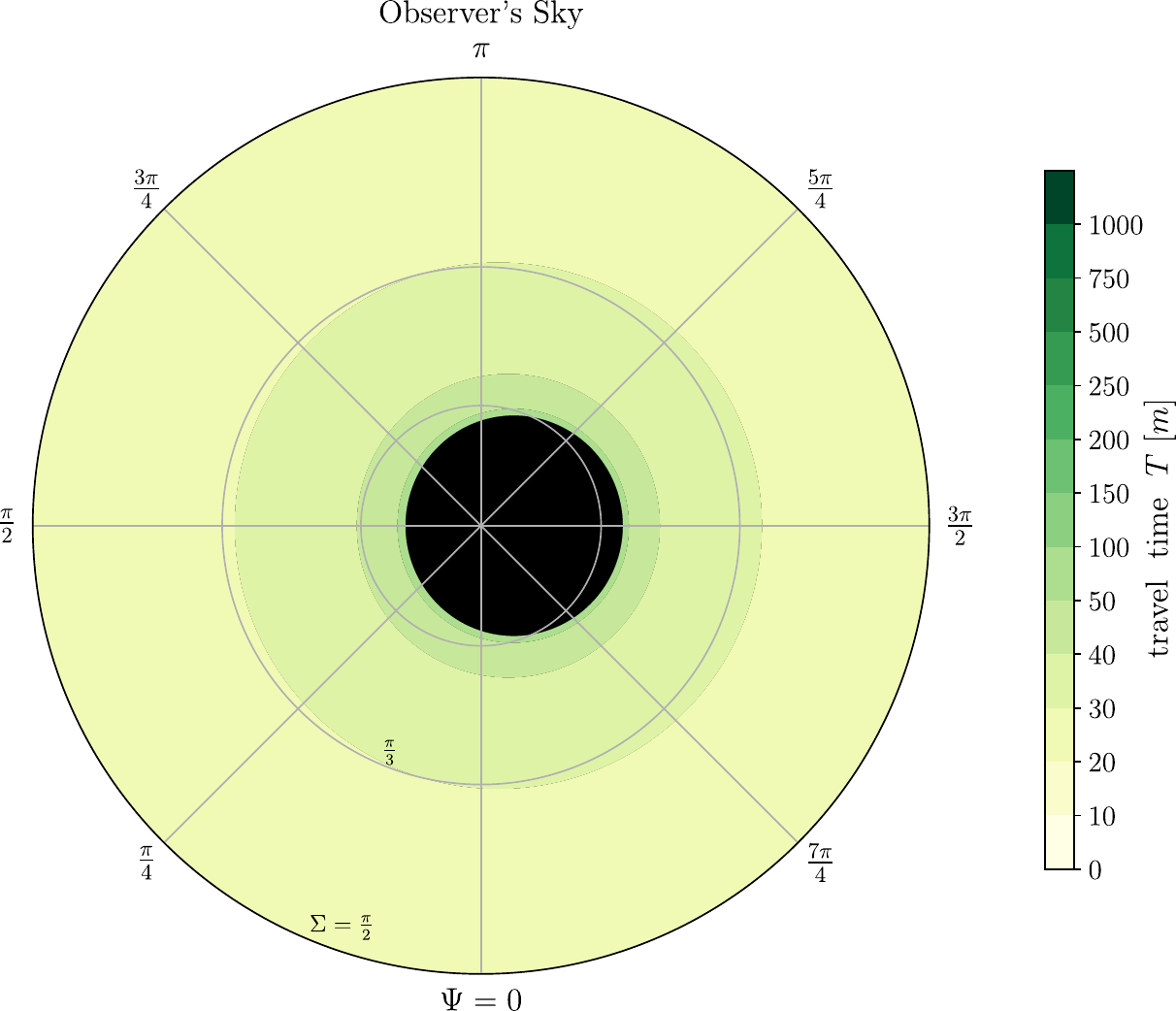} &   \hspace{0.7cm}\includegraphics[width=78mm]{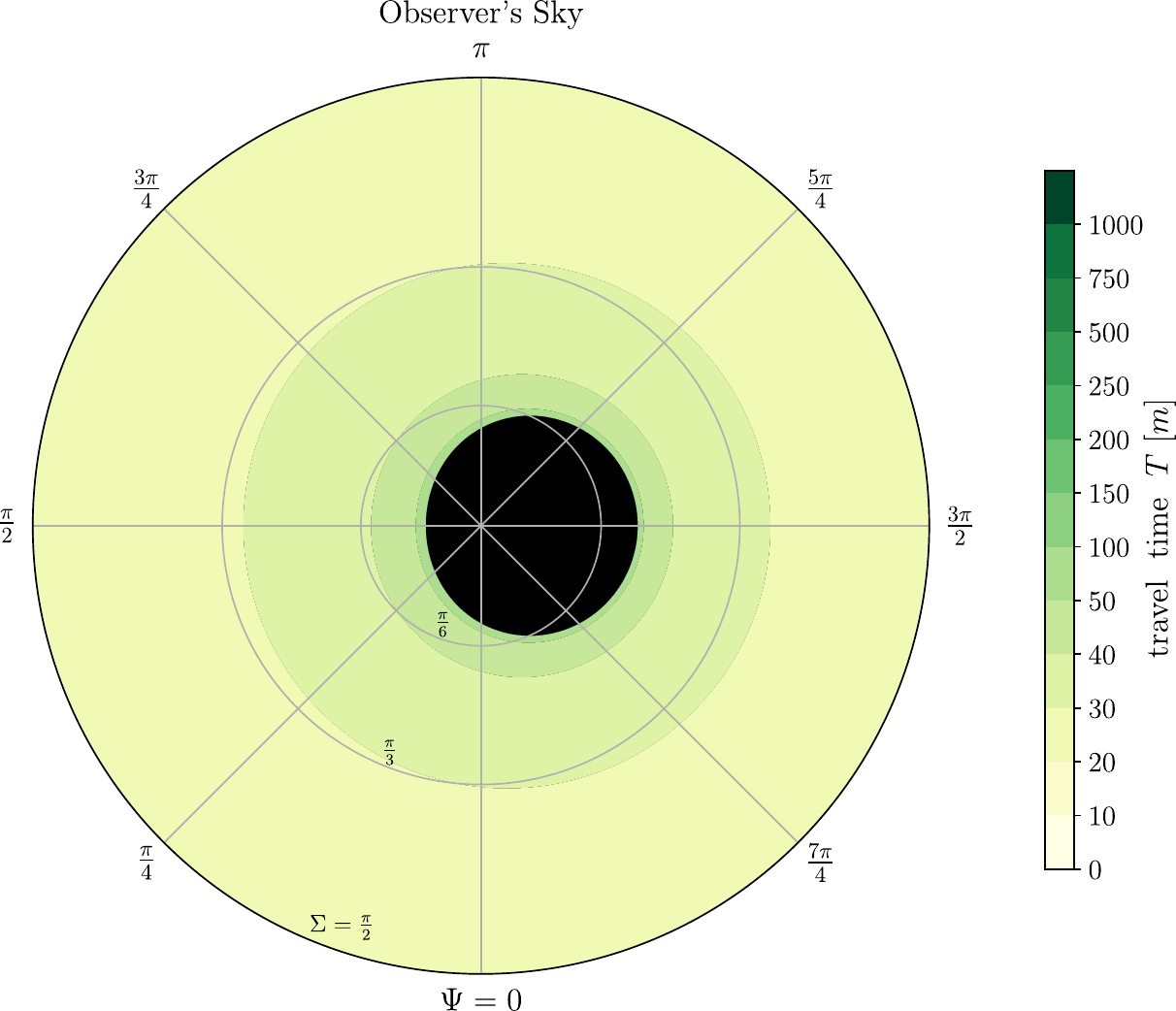} \\
    Kerr Spacetime: $a=m$ and $\vartheta_{O}=\pi/2$ & Kerr Spacetime: $a=m$ and $\vartheta_{O}=\pi/4$\\
    \\
    \hspace{-0.7cm}\includegraphics[width=78mm]{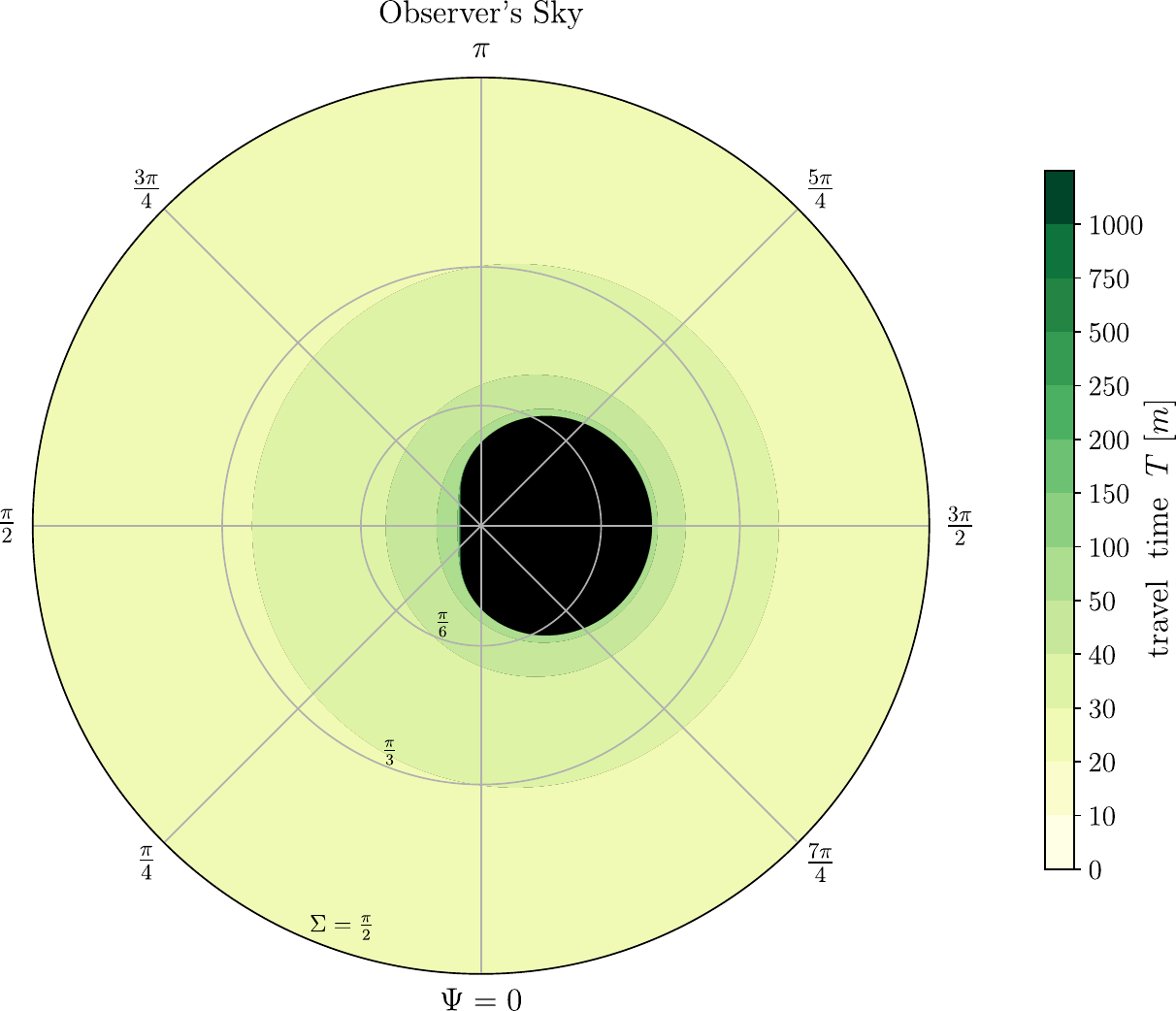} &   \hspace{0.7cm}\includegraphics[width=78mm]{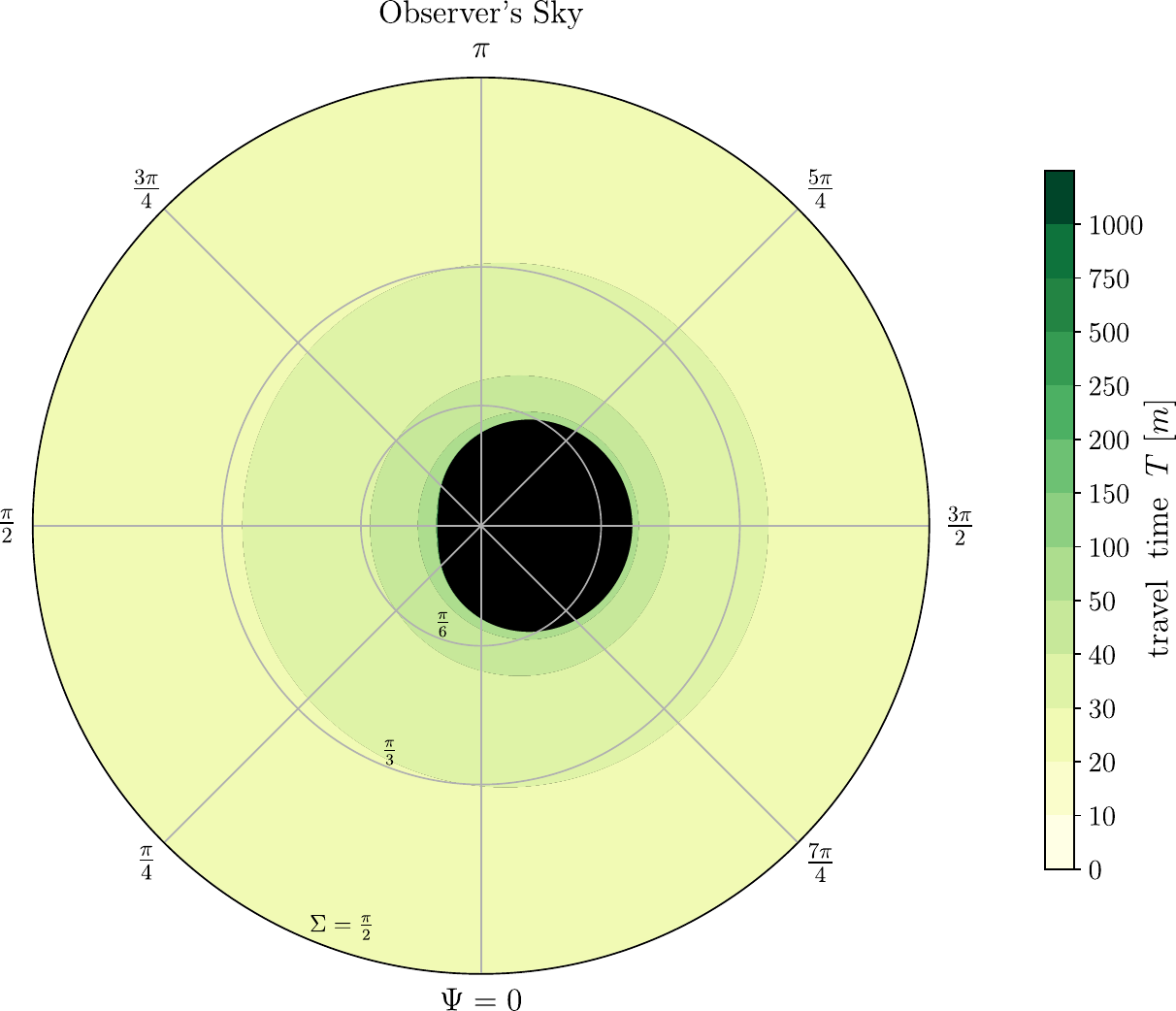} \\
  \end{tabular}
	\caption{Travel time maps for light rays and gravitational waves emitted by sources on the two-sphere of sources $S_{S}^2$ at the radius coordinate $r_{S}=20m$ and detected by a standard observer at $r_{O}=10m$ in the equatorial plane ($\vartheta_{O}=\pi/2$) for the Schwarzschild spacetime (upper left panel) and the Kerr spacetime with $a=m/4$ (top right panel), $a=m/2$ (middle left panel), $a=3m/4$ (middle right panel), and $a=m$ (bottom left panel). The bottom right panel shows a travel time map for light rays and gravitational waves emitted by sources on the two-sphere of sources $S_{S}^2$ at the radius coordinate $r_{S}=20m$ and detected by a standard observer located at $r_{O}=10m$ and $\vartheta_{O}=\pi/4$ for the Kerr spacetime. For this map the spin parameter is $a=m$.}
\end{figure}
The last quantity we want to discuss is the travel time. It measures in terms of the time coordinate $t$ the time a light ray or a gravitational wave needs to travel from its source to an observer. In terms of the time coordinate $t_{S}$ at which the light ray or the gravitational wave was emitted by the source and the time coordinate $t_{O}$ at which the light ray or the gravitational wave is detected by the observer it reads
\begin{eqnarray}
T=t_{O}-t_{S}.
\end{eqnarray}
The travel time itself is not directly measurable. However, when we observe two or more images of the same source and detect characteristic signatures in the associated light curves, we can derive travel time differences between them. Similarly we can also calculate travel time differences when we detect two or more gravitational wave signals emitted by the same source. In both cases we can then compare the calculated travel times with theoretical predictions for a black hole model, e.g., in our case the Kerr spacetime.\\
In the following we now want to calculate travel time maps on the observer's celestial sphere. For this purpose we first set $t_{O}=0$. Then we rewrite the integrals in the two components of the time coordinate given by (\ref{eq:trlam}) and (\ref{eq:EoMtinttheta}) in terms of the latitude-longitude coordinates on the observer's celestial sphere. Here, we rewrite the first integral as an integral over $r$ and keep the second integral in its general form. Then we insert them in (\ref{eq:tint}) and get
\begin{eqnarray}
&T(\Sigma,\Psi)=\int_{r_{O}...}^{...r_{S}}\frac{\left(r'^2+a^2\right)\left((r'^2+a^2\cos^2\vartheta_{O})\sqrt{P(r_{O})}+a(r'^2-r_{O}^2)\sin\vartheta_{O}\sin\Sigma\sin\Psi\right)\text{d}r'}{P(r')\sqrt{\left((r'^2+a^2\cos^2\vartheta_{O})\sqrt{P(r_{O})}+a(r'^2-r_{O}^2)\sin\vartheta_{O}\sin\Sigma\sin\Psi\right)^2-P(r')\rho\left(r_{O},\vartheta_{O}\right)^2\sin^2\Sigma}}\\
&+a\int_{0}^{\lambda_{S}}\frac{\left((r_{O}^2+a^2\cos^2\vartheta(\lambda'))\sin\vartheta_{O}\sin\Sigma\sin\Psi+a\left(\cos^2\vartheta(\lambda')-\cos^2\vartheta_{O}\right)\sqrt{P(r_{O})}\right)\text{d}\lambda'}{\sqrt{\rho(r_{O},\vartheta_{O})}},\nonumber
\end{eqnarray}
where the dots in the limits of the first integral shall again indicate that when the light ray or the gravitational wave passes through a turning point we have to split the integral into one integral from the radius coordinate of the observer $r_{O}$ to the turning point and one integral from the turning point to the radius coordinate of the source $r_{S}$. Here, the sign of the root in the denominator has to be chosen such that it agrees with the direction of the motion along the different parts of the geodesic. The second integral was kept in its general form since depending on the case we consider we either rewrite it as a linear function of the Mino parameter, or as an integral of nonlinear functions of the Mino parameter. For the explicit calculation of the travel time we now rewrite both integrals in terms of elementary functions and Legendre's elliptic integrals of the first, second, and third kind as described in Sec.~\ref{sec:timec}. The computational evaluation of the travel time was carried out using the same set of Julia codes as for the lens equation and the redshifts. Again we visualise the travel time as maps on the observer's celestial sphere.\\
Fig.~9 shows travel time maps for the Schwarzschild spacetime (top left panel) and the Kerr spacetime with $a=m/4$ (top right panel), $a=m/2$ (middle left panel), $a=3m/4$ (middle right panel), and $a=m$ (bottom left panel) for an observer in the equatorial plane ($\vartheta_{O}=\pi/2$) at the radius coordinate $r_{O}=10m$ and sources distributed on a two-sphere $S_{S}^2$ at the radius coordinate $r_{S}=20m$. The bottom right panel shows a travel time map for the Kerr spacetime with $a=m$ for an observer at $r_{O}=10m$ and $\vartheta_{O}=\pi/4$. The two-sphere of sources $S_{S}^2$ is again located at $r_{S}=20m$. As for the lens maps and the redshift maps the observer looks in the direction of the black hole and in each map the black area at the centre is the shadow of the black hole.\\
Again we start our discussion with the travel time map for the Schwarzschild spacetime in the top left panel. As we would expect for a spherically symmetric and static spacetime the travel time map is symmetric with respect to rotations about the centre of the shadow. Towards lower celestial latitudes the travel time increases and it approaches infinity for light rays and gravitational waves passing through a turning point infinitesimally close to the photon sphere. When we turn on the spin parameter $a$ and slowly increase it from $a=m/4$ to $a=m$ the most significant change we can see is that the structures in the travel time maps become asymmetric with respect to the meridian and the antimeridian. For the three spin values $a=m/4$, $a=m/2$, and $a=3m/4$ the travel time in the recognisable part of the travel time maps has roughly the same magnitude as for the Schwarzschild spacetime. However, when we increase the spin to $a=m$ (in this case the Kerr spacetime is extremal and thus for corotating light rays and gravitational waves for a certain longitude range the radius coordinate of the photon orbit $r_{\text{ph}}$ coincides with the radius coordinate of the horizon) we see on the western hemisphere close to the shadow a very narrow region with very long travel times. This is the region in which we also found very high order images in the lens map in the bottom left panel of Fig.~5. Thus towards the shadow the light rays and gravitational waves make more and more turns around the black hole which results in the observed increase of the travel time. In addition, when we slowly increase the spin for corotating light rays and gravitational waves we see a latitudinal broadening of the features close to the shadow while for counterrotating light rays and gravitational waves we see a latitudinal narrowing.\\ 
Now it is interesting to see how the features in the travel time map change when we place the observer outside the equatorial plane. This case is shown in the bottom right panel of Fig.~9. When we compare the travel time map to the travel time map in the bottom left panel we on one hand see that the asymmetry with respect to the meridian and the antimeridian is less pronounced. In addition, when we compare both travel time maps to the travel time map for the Schwarzschild spacetime we see that for the observer outside the equatorial plane the latitudinal broadening for corotating light rays and gravitational waves on the western hemisphere and the latitudinal narrowing for counterrotating light rays and gravitational waves on the eastern hemisphere is less pronounced. Furthermore, when we look closely at the meridian and the antimeridian around $\Sigma=\pi/3$ we can also see a very small asymmetry with respect to the celestial equator.\\
Before we proceed to the next subsection we want to close the discussion of the travel time with a remark. For $a=m$ the travel time maps showed very long travel times for corotating light rays and gravitational waves passing very close to the boundary of the shadow. Here, in these regions for the travel time we have the same problem as for the lens equation (here we performed the same initial comparison for the travel times associated with very high order images as for the lens equation and got the same result). When we have $0<a<m$ the travel time integral diverges for light rays and gravitational waves travelling along two different types of geodesics. Again when we move backward in time along the first type of geodesics the geodesics asymptotically approach the photon orbits. On the other hand when we move backward in time along the second type of geodesics the geodesics approach the horizon. Now, when we have $a=m$, for the region with the very long travel times close to the boundary of the shadow the radius coordinate of the photon orbit is the same as the radius coordinate of the horizon. Thus, also here we have to very carefully distinguish which part of the rapid increase of the travel time is caused because the light rays and gravitational waves pass closer and closer to the photon orbit and which part is caused because in Boyer-Lindquist coordinates the photon orbit and the horizon are projected onto the same radius coordinate and thus is a pure projection effect. As for the lens equation properly addressing this question is beyond the scope of this paper and may be part of future work.\\

\subsection{Implications for Astrophysical Observations}\label{Sec:IAO}
Now the question is how the results for the lens equation, the redshifts, and the travel time can be applied to identify the spacetime around and the spin of black holes in astrophysical environments. Let us for this purpose consider three different scenarios. In the first scenario we will consider sources only emitting electromagnetic radiation. These sources can be regular stars or neutron stars orbiting the black hole, or an accretion disk surrounding the black hole. In the second scenario we will consider pure high-frequency gravitational wave sources. Since in this paper we assumed that the gravitational waves propagate along lightlike geodesics and the geodesic approach is only valid in the high-frequency limit, our results are only valid for gravitational waves emitted by stellar mass binary black hole mergers, black hole-neutron star mergers, and binary neutron star mergers, as well as continuous gravitational waves emitted by neutron stars which are asymmetric with respect to their axis of rotation, for an overview for the latter see, e.g., the review of Riles \cite{Riles2023}, gravitationally lensed by a supermassive black hole. Therefore, we will limit our discussion to gravitational wave signals emitted by these sources. Note that there are also different scenarios which may lead to the emission of so-called ultra-high frequency gravitational waves such as the decay of cosmic strings, see, e.g., Servant and Simakachorn \cite{Servant2024}, or phase transitions, see, e.g., Athron \emph{et al.} \cite{Athron2024}, for an overview see, e.g., the review of Aggarwal \emph{et al.} \cite{Aggarwal2021}, however, since they are rather hypothetical and at the present point of time we are not able to detect them we will not discuss them in detail and only briefly note on how gravitational lensing by a supermassive black hole may help us to detect them. Finally, in the third case we will consider the combined emission of light and gravitational waves. Here, the sources can be neutron stars emitting electromagnetic radiation and continuous gravitational waves or binary neutron star mergers. 

\subsubsection{Sources only Emitting Electromagnetic Radiation}
We will start with sources only emitting electromagnetic radiation. In our case we considered a very idealised scenario in which the light sources are either fixed or orbit the black hole at a constant angular velocity. In addition, in both cases they were located on the surface of a sphere. Furthermore, in the Schwarzschild spacetime we considered a static observer while in the Kerr spacetime we considered a standard observer orbiting the black hole at a constant radius coordinate $r_{O}$ and a constant spacetime latitude $\vartheta_{O}$. While in real situations for both spacetimes we would rather expect observers and light sources on elliptic or, close to the equatorial plane, circular geodesic orbits, the presented work is a first step to a general analytic approach for observers and sources on arbitrary geodesic orbits. In addition, in our chosen scenario we fixed the spacetime coordinates at which the observer detects the light rays, and also the radius coordinate of the light sources. However, in real astrophysical scenarios we do not have these information and thus in addition to the parameters of the black hole spacetime we also have to determine them from observations.\\
Let us for now only consider a single light source, e.g., a normal star or a neutron star. We start with reviewing the observational features for the spherically symmetric and static Schwarzschild spacetime. Here, to simplify our discussion, we assume that the observer is located in the equatorial plane at $\varphi_{O}=0$. When we have a fixed source in the Schwarzschild spacetime the images of this source are located along a line on different sides of the shadow. When the source is located at spacetime longitudes smaller than $\pi$ we observe images of odd order on the eastern hemisphere while we observe images of even order on the western hemisphere. In the case that the source is located at spacetime longitudes larger than $\pi$ this situation is reversed. In the case that we have $\vartheta_{L}<\pi/2$ the images of odd order are located on the northern hemisphere while the images of even order are located on the southern hemisphere. In the case that we have $\pi/2<\vartheta_{L}$ the images of even order are located on the northern hemisphere while the images of odd order are located on the southern hemisphere. When the source is located exactly at the spacetime longitude $\varphi_{L}=\pi$ where we see odd or even order images is purely defined by the spacetime latitude. When we have $\vartheta_{L}<\pi/2$ the images of odd order are located on the northern hemisphere while the images of even order are located on the southern hemisphere. Again for $\pi/2<\vartheta_{L}$ this situation is reversed. For $\vartheta_{L}=\pi/2$ all light rays travel the same path length from the source to the observer. Here all cases have in common that the celestial latitude at which we can find the images decreases with increasing order of the image. In addition, with increasing order of the image also the travel time increases. While the travel time itself cannot be measured when the source has characteristic signatures in its light curve we can use them to calculate travel time differences between the images of different orders. Here, the travel time difference, and thus also the time delay, between the image of first order and the images of higher orders increases with increasing order of the image. In addition, all light rays emitted by the light source are redshifted. Unfortunately, in the Schwarzschild spacetime, and in general all spherically symmetric and static vacuum spacetimes, all light rays emitted by the source, independent of the fact whether they generate an image of first or an image of $n$-th order experience the same redshift. Thus although the redshift is the most easily measurable observable, observations of the redshift alone cannot provide detailed information on the nature of the underlying spacetime. In addition, in most spherically symmetric and static spacetimes images beyond second or third order can only be found very close to the boundary of the shadow and thus generally they cannot be resolved. The result is that in spherically symmetric and static spacetimes the information we can infer from the observed images is rather limited. Therefore, to obtain information about the underlying black hole spacetime we have to combine information about the positions of the images of different orders on the observer's celestial sphere, the travel time differences between these images, and the measured redshift. However, even when we have all these information they may only be enough to determine that the black hole spacetime is spherically symmetric and static, and we will very likely not be able to determine the parameters describing the black hole spacetime and the observer-lens-source geometry.\\
An interesting question is now if this is different when we have multiple sources. When we have more than one source either at fixed positions or orbiting the black hole on different (geodesic) orbits we can gain additional information. When they are all fixed on the same two-sphere of light sources $S_{L}^{2}$ we basically observe the same features as for a single light source but depending on the exact positions of the light sources on the two-sphere we observe their images at different celestial longitudes. Here, the light rays emitted by the different sources still experience the same redshift, however, when there are no other interactions of the light rays with the environment this already tells us that the black hole can be described by a spherically symmetric and static spacetime. In addition we can combine the information about the redshift with the information about the image positions and the travel time differences between the images of different orders of the single sources. However, whether this allows us to draw conclusions on the parameters describing the black hole spacetime or not strongly depends on the question if we can use the combined information to determine the distances between the observer and the black hole and the light sources and the black hole.\\
This question will become even more interesting when we consider one or more fixed light sources on two-spheres $S_{L_{i}}^2$ at different radius coordinates $r_{L_{i}}$ (here the index $i$ shall indicate that we have $i$ different light sources each on a different two-sphere) or light sources orbiting the black hole on geodesic orbits. Here, the two cases describe two different astrophysical scenarios. The first case is approximately true when we consider light sources very far away from the black hole. In this case the stars will move rather slowly and thus their positions on the observer's celestial sphere will appear as if they were nearly fixed. When the light rays emitted by these stars are gravitationally lensed by a black hole, as in the case discussed above, the observer will see the associated images of different orders at different latitude-longitude coordinates on his celestial sphere. In particular they appear at different angular distances to the boundary of the shadow. For each light source we can then determine the travel time differences between the images of different orders and the redshift of the images. Here, even when we only consider spherically symmetric and static spacetimes, unless we have exactly the same line element and exactly the same kind of geodesic motion in two different theories of gravity, the positions of the images of the different orders, the travel time differences between these images, and the redshifts for the different sources will carry unique information about the parameters describing the spacetime and the observer-lens-source geometry.\\
The second case describes light sources orbiting the black hole at relatively close distances. Let us for now only consider a single source. The positions of the images of the source on the observer's celestial sphere are now time dependent and thus already provide a much better probe of the black hole spacetime. The images of the different orders form separate trajectories on the observer's celestial sphere and these trajectories encode characteristic information about the spacetime trajectory of the star. In addition, when we can observe the trajectories of the star for an extended period of time for each trajectory we can again record a light curve. When we can at least record the light curves along the trajectories generated by images of first and second order we can use them to calculate the travel time differences between the images with the same features in the light curve. The travel time difference between the images of first and second order (and all other higher order images) is now time dependent and therefore carries much more information about the nature of the spacetime than for a fixed source. In addition, for each image of the source we can also determine the redshift. In this scenario also the redshift varies along the trajectories. Here, the rate how the redshift changes depends on one hand on the four-velocity of the light source at each point of the orbit, on the directions in which the light rays were emitted, and on the parameters describing the black hole spacetime. Due to this the redshift alone may still not be enough to determine the parameters describing the black hole spacetime, the distance between observer and black hole, and the parameters characterising the orbit of the star around the black hole, however, when we combine the positions of the image trajectories of different orders on the observer's celestial sphere with the travel time differences between and the redshift change along them it very likely provides us with enough information to determine them. In addition, adding information from additional sources will improve our ability to constrain the parameters describing the black hole spacetime. However, when the orbits of the light sources get too close to each other they start to interact. Thus in the case that we have multiple light sources we have to take potential interactions between them into account which cannot only change their orbits but also lead to microlensing effects on the emitted light rays.\\
In the scenarios we discussed above we did not yet consider that also the observer may move relative to the sources and the black hole. Here, when we know the position of the observer relative to the black hole and also the components of the observer's four-velocity we can obtain the trajectories of the light sources on the observer's celestial sphere from the trajectories on the celestial sphere of a static observer, or vice versa, by considering aberrational effects following the approach of Grenzebach \cite{Grenzebach2015b}. Here, due to the Doppler effect the light rays observed in the direction in which the observer is moving will be blueshifted while the light rays observed in the opposite direction will be redshifted. Unfortunately, usually when we observe astrophysical sources we neither know our exact distance to them nor our velocity relative to them and thus these information also have to be determined from the observations.\\
The whole situation changes for the Kerr spacetime. In this paper we considered a standard observer and thus as mentioned above the observer is orbiting the black hole at a constant angular velocity. While in particular for lower spins the observational situation is very similar to the one discussed for the Schwarzschild spacetime there are also some pronounced differences. The first is of course the well-known asymmetry of the shadow which is currently the main target of the Event Horizon Telescope observations.\\ 
When we observe other features around the shadow for a rotating black hole it is a bit more difficult to classify them with respect to the line connecting the observer and the centre of the spacetime because due to its asymmetry the centre of the shadow strongly deviates from it. Thus in a realistic observational scenario we would rather measure all angles from the centre or the boundary of the shadow. Therefore for our following discussion for the Kerr spacetime we will take the boundary of the shadow as a reference instead of the celestial coordinates. In addition, while in the Schwarzschild spacetime lightlike geodesics crossing the axes at $\vartheta=0$ or $\vartheta=\pi$ separate light rays orbiting the black hole either in clockwise (in the direction of decreasing $\varphi$) or counterclockwise (in the direction of increasing $\varphi$) direction for the Kerr spacetime these lightlike geodesics do not only separate light rays travelling in clockwise or counterclockwise direction but at the same time also between counterrotating light rays and corotating light rays.\\
Let us again start with a single light source. In principle the images of different orders show the same features as for the Schwarzschild spacetime. With increasing order the associated images can be found closer to the boundary of the shadow. However, here the spin of the black hole will lead to an asymmetry. When we discussed the lens maps in Sec.~\ref{Subsec:LensEqua} we saw that the bands with images of different orders broadened for corotating light rays while they narrowed for counterrotating light rays and that this effect becomes stronger with increasing spin of the black hole. This effect is independent of the fact whether we measure it in terms of the celestial coordinates or in relation to the boundary of the shadow. When we transfer this to the images of a single light source this means that for a certain range of orders higher order images generated by corotating light rays will be found at larger angular distances from the boundary of the shadow than comparatively lower order images generated by counterrotating light rays. In addition, we saw that close to the boundary separating corotating and counterrotating light rays, in the lens maps on the side with the counterrotating light rays the areas with images of first order stretched to celestial latitudes closer to the shadow. We also saw that the light sources which emitted the associated light rays were located on the other quadrant on the eastern hemisphere of the two-sphere of sources. Thus when we have individual light sources observing the first effect is a clear indicator that we have a rotating black hole. Similarly observing the second effect is a first indicator that we have an axisymmetric spacetime but not necessarily that the black hole is rotating. Here, in both cases also including information about the redshift and the travel time differences between images of different orders will allow to draw more precise conclusions about the nature of the black hole spacetime and the parameters describing it, in particular the spin of the black hole. Here, even when we only consider individual light sources which move close to $t$-lines (in our terminology static light sources) adding information about the redshift and the travel time differences will allow us to obtain more characteristic signatures than for the Schwarzschild spacetime. In particular for static light sources and static observers these information may already be enough to infer information about the spin of the black hole and the observation geometry.\\
In addition, for very high spins close to the boundary of the shadow the lens map showed areas with images of very higher orders generated by corotating light rays. Therefore, when we have individual light sources close to the equatorial plane in this region we will find a series of very high order images. Unfortunately, even if we have a telescope with sufficiently high angular resolution it is very doubtful that we can detect all of them. While in our case for all images the redshift was relatively similar in real astrophysical scenarios this might be different. So some of the images will very likely be red- or blueshifted out of our observational band. However, even if the frequency change is relatively small there is another factor which we have to consider. With increasing order of the image the intensity of the image decreases rapidly and thus from this aspect alone we will very likely only be able to observe the images of first to third order. However, we also saw that for some of the bands we have images of sources from the same hemisphere of the two-sphere of light sources over several subsequent orders. In particular for light sources located in or close to the equatorial plane these images can be found in very close proximity. When these images appear close enough to each other we may not be able to resolve the individual images, however, in the case that the frequency change of the photons across these images is relatively low the photon flux of all images will add up, and when the overall intensity becomes large enough we may be able to observe them as a combined image. In this case it will be very likely impossible to derive an exact redshift and travel time differences with respect to other individual images, however, the intensity of the combined image may show a characteristic signature which depends on the spin of the black hole. Thus observing this combined image may offer another way to determine the spin of a black hole and to probe gravity in the strong field regime. While the sensitivity and the angular resolution of the Event Horizon Telescope may not be high enough to resolve and distinguish this combined image from other sources it will be an interesting question to investigate if the next-generation Event Horizon Telescope \cite{Johnson2023} or the Black Hole Explorer \cite{Johnson2024} will be able to detect it.\\ 
In addition, adding observations for more light sources will allow us to gain more solid constraints on  the parameters describing the observation geometry and the black hole spacetime. However, as already mentioned for the Schwarzschild spacetime in real astrophysical settings we usually deal with moving sources and observers. Thus in a realistic scenario when we want to constrain the parameters describing the observation geometry and the black hole spacetime the motion of the light sources and the observer relative to each other and relative to the black hole have to be taken into account.\\
Unfortunately, when we consider real astrophysical scenarios we also have another problem. Most black holes will be surrounded by an accretion disk and since we are usually located at very large distances to the black hole all images of individual sources will be so close to the boundary of the shadow that the intensity of the accretion disk will be higher than the intensity of the images. Thus we will not be able to distinguish them from the bright background. However, in this situation there will be another phenomenon. Light emitted by the accretion disk will lead to the formation of what people usually refer to as \emph{photon rings}. In the case that we observe a black hole with very high spin and are not located on the spin axis we will have asymmetric photon rings. Here, the parts of the photon rings generated by corotating photons will be broader compared to the parts of the photon rings generated by counterrotating photons. In addition, depending on our bandwidth and the sensitivity of our telescopes for corotating photons we will be able to observe parts of photon rings of higher orders than for counterrotating photons. In the case that the intensity of the individual photon rings is not high enough there is still the chance that the photon fluxes of the individual higher order photon rings add up and we can observe a combined photon ring, which will then be more prominent for corotating photons than for counterrotating photons. Here, the intensity and the width of this combined photon ring will very likely also carry a very unique signature of the spin of the black hole. That the spin of the black hole has an impact on the width and the shape of the photon rings, and that in particular the asymmetry of the latter carries characteristic information about the spin is already well known, see, e.g., the work of Gralla, Lupsasca, and Marone \cite{Gralla2020c} and Paugnat \emph{et al.} \cite{Paugnat2022}. However, to the best of our knowledge that also the detectability of higher order photon rings (with orders larger than two) or the detectability and, if detected, the width of a combined photon ring may carry characteristic information about the spin of a black hole has not been discussed so far and certainly deserves more attention. While the angular resolution of the Event Horizon Telescope is still too low to observe the photon rings (note that Broderick \emph{et al.} claimed that they detected first hints for the existence of the lowest order photon ring \cite{Broderick2022}, however, this claim is strongly disputed, see, e.g., Lockhart and Gralla \cite{Lockhart2022}), detecting them will be one of the main objectives of the Black Hole Explorer \cite{Johnson2024}. Here, currently the main target is to use the width and the asymmetry of the first and second order photon rings to determine the spin of supermassive black hole candidates, see, e.g., the works of Gralla, Lupsasca, and Marone \cite{Gralla2020c} and Paugnat \emph{et al.} \cite{Paugnat2022}. However, our results now suggest that if the angular resolution of the Black Hole Explorer is high enough also the detectability of a signature from higher order photon rings or a combined photon ring, and the width of the combined photon ring, may be able to contribute to determinining the spin of a supermassive black hole candidate.\\
Here, in all the scenarios discussed above the applications for our analytical results are quite straightforward. On one hand they can be used to perform ray tracing calculations for different gravitational lensing scenarios to predict different observational effects. On the other hand we can also adapt and extend them to model specific observational scenarios. Alternatively, when we can extract the mass and the spin for a black hole candidate and also the observer-lens-source geometry from observational data, we can use them as input for our equations to model the gravitational lensing scenario and perform a consistency check. Here, in comparison to other works the analytical approach introduced in this work has the particular advantage that it allows to perform very high-resolution calculations, in particular close to the shadow.

\subsubsection{Sources only Emitting Gravitational Waves}
Let us now turn to the second scenario. In this scenario we consider gravitational wave sources in the LIGO-Virgo-KAGRA band which emit characteristic high-frequency gravitational wave signals such as binary compact object mergers containing neutron stars and black holes as well as currently predicted but not yet detected continuous gravitational waves emitted by neutron stars with asymmetries with respect to their axis of rotation. Compared to light rays they have the advantage that they can pass relatively unaffected through space, in particular through accretion disks around black holes. However, compared to light sources they are relatively rare and in particular the gravitational wave signals emitted by binary compact object systems only enter the LIGO-Virgo-KAGRA band for a very short time before the components of the system merge. In general for gravitational waves emitted by these sources and gravitationally lensed by a supermassive black hole we observe the same lensing effects as for light rays. However, in the case of gravitational waves emitted by merging binary compact objects for detecting two gravitationally lensed gravitational wave signals of different orders emitted by the same source we have to perfectly time our observing runs. This makes using gravitationally lensed gravitational waves emitted by binary compact objects for probing the spacetimes of supermassive black holes much more challenging. This will be slightly different for continuous gravitational wave sources. As the name indicates for continuous gravitational wave sources we will detect a continuous signal and thus the observational situation is similar to that for light rays (with the exception that here we have a signal instead of an image).\\ 
Here, in both cases when the gravitational wave signal is gravitationally lensed by a rotating supermassive black hole the mass and the spin of the black hole will leave a very characteristic imprint on the signal. Thus in the case that we detect two gravitational wave signals from the same source gravitationally lensed by a rotating supermassive black hole when the imprint on the signals is strong enough we can use them to determine the mass and the spin of the black hole. Here, again when we can determine the mass and the spin of the black hole and also the parameters describing the observer-lens-source geometry we can use the analytical results derived in this paper to perform a consistency check. Of course as for light rays in the most general scenario the motion of the gravitational wave source and the observer relative to each other and relative to the supermassive black hole have to be taken into account.\\
However, although compared to electromagnetic radiation gravitational wave signals have several advantages, they also have one huge disadvantage. The sky localisation of our gravitational wave detectors is only very imprecise and thus we have to be very careful in our analysis of potentially lensed signals. However, here the detection of two gravitationally lensed signals may actually also have an interesting application. When we can clearly infer from two detected gravitational wave signals that they were emitted by the same source we have for each signal a sky localisation. When we can also clearly identify that the signals were gravitationally lensed by a supermassive black hole and we can infer the mass and the spin of the black hole as well as the parameters characterising the observer-lens-source geometry we can use these signals to test the quality of the sky localisation of our gravitational wave detectors. When the uncertainty of the estimated parameters is small enough so that they do not strongly affect the outcome of our calculations, for each signal we can calculate a backward trajectory from Earth to the source. Since both signals come from the same source the spacetime coordinates as well as the time coordinate at which they were emitted have to be the same. The deviation of the maximum likelihoods of the sky localisations from the directions of the intersecting geodesics will then provide us with a good indicator for the quality of the sky localisation. When the maximum likelihoods of the sky localisations of the individual gravitational wave signals agree with the directions for which the calculated backward trajectories intersect we can say with good confidence that also for signals from other sources the sky localisation will have a high confidence level. In the case that they disagree this indicates that we should also be very careful with the sky localisations for signals from other sources. Note that here we assumed that the positions of the maximum likelihoods of the sky localisations show small differences. In real scenarios this may not be the case. However, in this situation the described approach can be adapted such that one independently varies the initial directions of the backward trajectories around the position of the maximum likelihood until the trajectories intersect at the same spacetime coordinates. However, to ensure that one has really found the right spacetime coordinates in this case the gravitational wave signals should be propagated backwards along the geodesics to check if the delensed signals are consistent.\\
The last question we now want to address is how gravitational lensing of gravitational waves may help us to detect gravitational waves emitted by more hypothetical sources like cosmic strings or during phase transitions. Generally these gravitational waves are expected to be part of a gravitational wave background and thus gravitational lensing by a supermassive black hole can have two effects. On one hand it can magnify the background as a whole and on the other hand only single signals such that they become detectable. While these events will be rather rare and still difficult to detect in the case that we detect such a signal, but also in the case that we do not detect it, it will provide us with valuable information about the associated physical processes.

\subsubsection{Sources Emitting Electromagnetic Radiation and Gravitational Waves}
The third and most interesting scenario will be the observation of electromagnetic and high-frequency gravitational radiation emitted by the same source and gravitationally lensed by a supermassive black hole. In the current accessible gravitational wave band the most likely sources for such events are merging binary neutron stars and certain isolated neutron stars which emit continuous gravitational waves and at the same time characteristic electromagnetic radiation. Unfortunately most neutron stars mainly emit in the radio band and thus the emitted radiation will be very likely weaker than the emission from the accretion disk surrounding the black hole. Thus in this case we will very likely only detect the gravitational wave signal. Therefore, merging binary neutron stars are the most promising target. However, they will be rather rare. Currently the only high-confidence gravitational wave signal emitted by a merging binary neutron star with an observed electromagnetic counterpart is the gravitational wave signal GW170817 \cite{Abbott2017a}. In addition to the gravitational wave signal follow-up observations found an electromagnetic counterpart across a very broad range of the electromagnetic spectrum including $\gamma$-rays, X-rays, the UV, and the visible range, see, e.g., Refs. \cite{Abbott2017b,Abbott2017c,SoaresSantos2017}. However, here we generally have two problems. On one hand the resolution of $\gamma$-ray and X-ray telescopes is usually rather low. On the other hand electromagnetic radiation in the UV and in the visible spectrum is strongly scattered by interstellar dust and thus cannot be used for observing images of individual sources around supermassive black holes. Therefore, even if we observe gravitational wave signals and an electromagnetic counterpart emitted by a binary neutron star merger which were gravitationally lensed by a supermassive black hole we have a similar situation as for pure gravitationally lensed gravitational wave signals discussed in the last subsection. The signals will be rather short and the angular resolution relatively coarse. The only difference will likely be that a careful analysis of the gravitational wave signals in combination with the electromagnetic counterpart may allow us to place slightly better constraints on the mass and the spin of the black hole and the observer-lens-source geometry. Also here the analytical results derived in this paper will allow us to perform a consistency check for the measured parameters describing the black hole spacetime and the observer-lens-source geometry. 

\section{Summary}\label{Sec:Sum}
Due to its astrophysical relevance the Kerr spacetime is one of the most well-investigated black holes spacetimes from general relativity. However, from time to time we can still find new aspects and observational features. In the present paper we performed a thorough gravitational lensing analysis for a standard observer in the domain of outer communication. For this purpose we first derived the equations of motion for lightlike geodesics. Then we related the three constants of motion $E$, $L_{z}$, and $K$ characterising these geodesics to latitude-longitude coordinates on the observer's celestial sphere. Here, these relations are valid for arbitrary values of the energy $E$, however, in this paper we only used them to describe lightlike geodesic motion with $0<E$. In the next step we derived a conditional equation for the radius coordinates of the photon orbits in the stationary regions of the Kerr spacetime, and their latitudinal projections onto the observer's celestial sphere as functions of the mass parameter $m$, the spin parameter $a$, the spacetime latitude of the observer $\vartheta_{O}$, and the celestial longitude $\Psi$. We obtained the conditional equation in form of a polynomial of sixth order which in general cannot be solved analytically. However, we managed to identify three special cases for which we can analytically derive the radius coordinates of the photon orbits and their latitudinal projections onto the observer's celestial sphere, and thus also the angular radius of the shadow. Here, in two of them these analytical solutions even allow to calculate the angular radius for the whole boundary of the shadow. In the first of the two cases we have $\vartheta_{O}=0$ or $\vartheta_{O}=\pi$ and thus the standard observer is located on the axis of symmetry. In the second case we have $\vartheta_{O}=\pi/2$ and thus the standard observer is located in the equatorial plane. While it is already well known that we can analytically derive the radius coordinates of the photon orbits and the angular radius of the shadow for an observer on the axis of symmetry and for an observer at infinity, for the latter see the work of Cunha and Herdeiro \cite{Cunha2018}, to our knowledge this paper is the first to show that this is also possible for a standard observer at arbitrary radius coordinates in the equatorial plane. Since in this paper we only derived this result for a standard observer it is now an interesting question if this is also possible for other observers such as an observer on a $t$-line or a zero angular momentum observer. In addition, it will also be an interesting task to show whether this result can only be derived for axisymmetric and stationary spacetimes with equatorial reflection symmetry and separable equations of motion for which the equation of motion for $r$ has a similar structure as for the Kerr spacetime or if we can also derive it for general stationary and axisymmetric spacetimes for which the equatorial reflection symmetry is broken but we still have a Carter constant.\\  
In the next step we used the latitudinal projections of the radius coordinates of the stable and unstable photon orbits in the stationary regions of the Kerr spacetime onto the observer's celestial sphere to classify the different types of $r$ motion. Here, based on the photon orbit sturctures and the latitudinal projections of the photon orbits we found a straightforward classification scheme. However, because it turned out that some types of motion have the same solutions to the equation of motion for $r$ the full complexity of the solution structure only became visible when we also considered the solutions for the $r$ dependent parts of the $\varphi$ coordinate and the time coordinate $t$. Similarly we used the celestial coordinates in combination with the constants of motion to classify the different types of $\vartheta$ motion. Here the $\vartheta$ motion can be roughly divided into five different types and even when we include the $\vartheta$ dependent components of the $\varphi$ coordinate and the time coordinate $t$ the solution structure is far less complex than for the $r$ motion.\\
Then we solved the equations of motion analytically using elementary as well as Jacobi's elliptic functions and Legendre's elliptic integrals of the first, second, and third kind. The derived solutions basically fall into two different cathegories. For the types of motion in the first cathegory the right-hand side of the equations of motion for $r$ and $\vartheta$ have at least one multiple root. In this case the equations of motion for $r$ and $\vartheta$ can always be solved in terms of elementary functions. Similarly the corresponding $r$ and $\vartheta$ dependent parts of the $\varphi$ coordinate and the time coordinate $t$ can always be written down in terms of elementary functions. For the types of motion in the second cathegory on the other hand the right-hand sides of the equations of motion for $r$ or $\vartheta$ have only distinct roots. In this case the equations of motion can only be solved in terms of Jacobi's elliptic functions. Similarly the solutions for the $r$ and $\vartheta$ dependent parts of the $\varphi$ coordinate and the time coordinate $t$ can only be written down in terms of elementary functions and Legendres elliptic integrals of the first, second, and third kind.\\
Here, the solutions for the most general cases and also a few special cases have already been discussed by Gralla and Lupsasca \cite{Gralla2020b} and by Slez\'{a}kov\'{a} \cite{Slezakova2006}. However, compared to these works the approach presented in this paper has two advantages. On one hand it offers a unified classification scheme almost entirely based on the celestial coordinates on the celestial sphere of a standard observer which makes it much easier to use the solutions to the equations of motion for astrophysical applications. On the other hand it can also be easily transferred to light rays and high-frequency gravitational waves travelling along lightlike geodesics characterised by $E<0$, and to massive particles.\\
In the second part of the paper we then used the derived solutions to the equations of motion to investigate gravitational lensing in the Kerr spacetime. For this purpose we first placed a standard observer in the domain of outer communication outside the ergoregion between the outer boundary of the photon region and a two-sphere of sources $S_{S}^2$. Then we used the exact solutions to the equations of motion to derive three different lensing quantities, namely the lens equation, the redshift, and the travel time. While this paper is not the first to calculate a lens equation for the Kerr spacetime, numerically calculated lens equations can be found, e.g., in the works of Bohn \emph{et al.} \cite{Bohn2015} and Cunha \emph{et al.} \cite{Cunha2015}, to our knowledge it is the first paper which uses the analytic solutions to the equations of motion in combination with the angular radius of the shadow as a function of the celestial longitude $\Psi$, and the latitude-longitude coordinates on the observer's celestial sphere to calculate it. In general the derived lens maps show the same features as the numerically calculated lens maps in the works of Bohn \emph{et al.} \cite{Bohn2015} and Cunha \emph{et al.} \cite{Cunha2015}, however, our approach allowed us to calculate them with a much higher resolution.\\
The most prominent and well-investigated feature in the lens maps is certainly the shadow of the black hole. In the lens map for the Schwarzschild spacetime it is circular and in the lens maps for the Kerr spactime it takes its characteristic crescent-like shape. For the Schwarzschild spacetime otherwise the lens map shows the characteristic features of a spherically symmetric and static spacetime. The bands with images of different orders form concentric rings around the shadow and with increasing order the images are located at lower celestial latitudes closer to the shadow. In addition, images of different orders of sources from the same quadrant on the two-sphere of sources are clearly separated from each other and the boundaries between the rings with images of different orders mark the positions of the critical curves.\\
The lens maps for the Kerr spacetime on the other hand show a general asymmetry between corotating and counterrotating light rays and gravitational waves. Here, when we increase the spin the features in the lens maps, in particular close to the shadow, show a characteristic latitudinal broadening for corotating light rays and gravitational waves and a latitudinal narrowing for counterrotating light rays and gravitational waves. As a result in general the areas with images of higher orders are much better visible for corotating light rays and gravitational waves than for counterrotating light rays and gravitational waves. This effect becomes particularly pronounced for the extremal Kerr spacetime with $a=m$. When we have an observer in the equatorial plane in this case for corotating light rays and gravitational waves one can easily see bands with images up to fourth order. In addition, close to the shadow one can also see a region composed of bands with images of higher orders. Here an enlarged view of this region allowed to distinguish bands with images up to 15th order. Beyond the 15th order the orders of the images became difficult to count, however, the enlarged view likely showed images roughly up to 30th order or slightly higher. This region was already present in the lens map for $a=0.999$ created by Cunha \emph{et al.} \cite{Cunha2015}, however, in their lens maps it was only very coarsely resolved and thus one could not distinguish the individual image bands. Thus this work is the first to show this area in higher resolution. In addition, the enlarged view also showed a different effect. For the image bands between the 5th and the 15th order the light rays and gravitational waves generating the parts of the image bands on the northern and the southern celestial hemisphere were all emitted by sources on the northern and the southern spacetime hemisphere, respectively. Here, this effect is particularly interesting for observations. While we will not be able to resolve the single images when the images of a single source can be observed very close to each other and the red- or blueshift of the associated photons is small enough so that they are not shifted out of the observational band of our telescopes, the photon fluxes of the individual images will add up and if its intensity becomes large enough the combined signal may be detectable. Here, the intensity of the combined image may encode characteristic information about the spin of the black hole. Thus detecting this combined image has on one hand the potential to allow us to determine the spin of a black hole candidate and on the other hand it may also allow us to probe gravity in the strong field regime. Therefore, this aspect certainly deserves a more thorough investigation. In addition, it will be an interesting task to investigate if the next-generation Event Horizon Telescope \cite{Johnson2023} or the Black Hole Explorer \cite{Johnson2024} will be sensitive enough to detect such a combined image.\\
Besides these higher order images in the lens map for the extremal Kerr spacetime one can also find a second feature which is characteristic for the spin. Close to the lines marking light rays and gravitational waves travelling along lightlike geodesics crossing one of the axes one can find areas with images of second order surrounded by an area with images of third order both generated by counterrotating light rays and gravitational waves emitted by sources on different quadrants on the two-sphere of sources. Here, for the former the spin of the black hole effectively reduces the angle $\Delta\varphi$ the light rays and gravitational waves travel before they reach the observer.\\
We also addressed the question where in the lens maps we can find the critical curves. Here, a comparison with the lens maps in Bohn \emph{et al.} \cite{Bohn2015} and Cunha \emph{et al.} \cite{Cunha2015} showed that around $\Psi=\pi/2$ and $\Psi=3\pi/2$ they are very likely aligned with the boundaries between the images of first and second order, however, the comparison also revealed that between these points they run through the areas with images of first and second order. Unfortunately, neither the lens maps presented in this work nor the lens maps in the works of Bohn \emph{et al.} and Cunha \emph{et al.} allowed to determine the exact positions. This will only be possible by explicitly calculating them.\\
In addition, this result very likely also provides us with an answer to the open question where we can find the critical curves associated with the images of first and second order, the images of third and fourth order, and the images of fifth and sixth order in the lens maps for light rays and massive particles for the charged NUT-de Sitter spacetimes in Ref.~\cite{Frost2022} and for the charged NUT spacetimes in Ref.~\cite{Frost2023}, respectively. It is likely that they are circles which touch the outermost boundaries of the areas with, e.g., the images of second order and pass through the intermittent regions with images of first order (and so on for the higher order images). Consequentially similar structures may generally occur for all spacetimes for which the lens maps contain regions with images of odd and even order from the same quadrant on the two-sphere of sources $S_{S}^2$ with a common boundary and this certainly deserves a more thorough investigation.\\
The second quantity we discussed was the redshift. Here, we derived redshift maps on the celestial sphere of the standard observer for three different types of sources. The first type were static sources moving along $t$-lines, the second type were zero angular momentum sources, and the third type were standard sources. While for the Schwarzschild spacetime the standard observer and also the light sources are truly static and thus the redshift, which in our case was actually a blueshift, has the same constant value for all light rays and gravitational waves emitted by sources on the two-sphere of sources $S_{S}^2$, this is different when we have a standard observer in the Kerr spacetime. Here, the observer orbits the black hole in the direction of the spin and thus for most counterrotating light rays and gravitational waves the blueshift increases. On the other hand for most corotating light rays and gravitational waves the blueshift decreases and in the case of large spins becomes a redshift. In the case the source is moving in the same direction as the observer this effect reduces, however, the general pattern remains the same. In addition, when we turn on the spin close to the shadow small wave-like patterns become visible. The size of these patterns increases with increasing spin and also with increasing angular velocity of the source. Here, this allowed us to draw the following conclusions. First the size of the wave-like patterns increases with the angular velocity of the source. This indicates that it depends on the angular velocity of the source and thus very likely also on the angular velocity of the observer. Second, the size of the wave-like patterns also increases with the spin. However, since also the angular velocity of the observer increases with increasing spin the redshift maps presented in this work did not allow us to draw a final conclusion whether the size of the wave-like patterns depends on the spin or not. Also this question deserves a more thorough investigation because in the case that the size of the wave-like patterns depends on the spin and we observe such a feature it has the potential to serve as a probe for the spin of a black hole. \\
The last quantity we discussed was the travel time. Again we plotted it as maps on the observer's celestial sphere. Compared to the travel time maps for the Schwarzschild spacetime the travel time maps for the Kerr spacetime did not show very significant changes. The travel time maps show the same asymmetry as the lens and redshift maps and the travel time itself has roughly the same magnitude as for the Schwarzschild spacetime. The most pronounced differences occured for the extremal Kerr spacetime. Here, in the region close to the boundary of the shadow where we found images beyond fourth order we also found particularly long travel times. However, this was not really surprising since with increasing order of the images the light rays and gravitational waves make more and more turns around the black hole and thus the time they need to travel from the source to the observer gets longer.\\
The presence of images beyond 15th order close to the boundary of the shadow in the lens map and the associated long travel times in the travel time map for the extremal Kerr spacetime with $a=m$ also lead to a different question. How accurate are actually the analytical results we presented in this work? Before the codes for the lens equation, the redshifts, and the travel time were run, for lower order images a thorough intercomparison with numerical results was performed and this intercomparison indicated a good agreement. While initially images beyond 15th order were not expected, a first comparison of the analytical results with numerical results for these higher order images indicated that the analytical solutions may also allow to calculate gravitational lensing effects for images of orders well beyond those which are visible in the lens map for the extremal Kerr spacetime. However, this result has still to be confirmed by a much more thorough investigation. In the same context it will also be important to investigate how the results for the extremal Kerr spacetime are affected by the fact that in Boyer-Lindquist coordinates for corotating light rays and gravitational waves for a certain range of the celestial longitude $\Psi$ the photon orbit at the radius coordinate $r_{\text{ph}}$ is located at the same radius coordinate as the horizon. \\
While the main focus of this paper was to present a thorough analytical investigation of gravitational lensing in the Kerr spacetime, as last part of our investigation we also discussed several implications for astrophysical observations. Here, we also discussed several applications. We will summarise the most important of them in the following.\\
The first application is quite obvious and already present in the literature. The analytical solutions to the equations of motion can be implemented in ray tracing codes for a faster and more efficient calculation of photon trajectories, see, e.g., the work of Wang and Yang \cite{Yang2013}. However, although this application is not new, our approach of parameterising the lightlike geodesics using latitude-longitude coordinates on the observer's celestial sphere has still one advantage compared to other approaches. On one hand we classify the different types of motion using the angular projections of the stable and unstable photon orbits in the stationary regions of the spacetime onto the observer's celestial sphere. Thus we have precise information on the exact solutions to the equations of motion which have to be used for a specific combination of latitude-longitude coordinates. On the othern hand using the analytical solutions in combination with the celestial angles as well as the angular radius of the shadow also allows us to perform very high-resolution calculations, in particular close to the boundary of the shadow.\\
The second application of our results is related to stars orbiting supermassive black hole candidates. As discussed in Sec.~\ref{Sec:IAO} when we have stars orbiting a supermassive black hole candidate at the centre of a galaxy like the S-stars in our Milky Way and we observe either two or more image trajectories of the same star or the image trajectories of several stars we can measure the redshift variations along each trajectory, and potentially the travel time differences between the images of different orders of the individual stars. When we can determine the mass and the spin of the supermassive black hole and the parameters describing the observer-lens-source geometry from the motion of the stars around the black hole, the redshift, and the travel time differences we can then use them in combination with our analytic solutions to perform a consistency check by modelling the gravitational lensing situation.\\
The third application of our results is again relatively straightforward. While in the present work we investigated the lens equation for sources on a two-sphere, many of the observed features, in particular the latitudinal broadening effect for images generated by corotating light rays, will also occur when we have an accretion disk. Here, one specific feature is of particular interest, the photon rings. In the case that we are not located on the axis of rotation, when we observe the photon rings around Kerr black holes, the parts of the photon rings which are generated by corotating photons will be slightly broader than the parts of the photon rings generated by counterrotating photons. In addition, depending on the spin of the black hole the photon rings will be asymmetric. That both effects alone and together carry a characteristic imprint of the spin is not particularly new, see, e.g., the works of Gralla, Lupsasca, and Marrone \cite{Gralla2020c} and Paugnat \emph{et al.} \cite{Paugnat2022}. However, from the features in the lens map for the extremal Kerr spacetime we also concluded that the detectability of higher order photon rings itself or alterantively the detectability and the width of a combined photon ring may carry information about the spin of a black hole. To our knowledge this effect has not been investigated so far and since it may strongly contribute to our efforts to determine the spin of supermassive black hole candidates when the next-generation Event Horizon Telescope and the Black Hole Explorer become operational, it deserves a much more thorough investigation.\\
The last two applications involve gravitational lensing of high-frequency gravitational waves. When a merging binary black hole orbits a supermassive black hole the emitted gravitational waves are gravitationally lensed by the supermassive black hole and depending on the strength of the imprint of the supermassive black hole the gravitational wave signals will carry characteristic information about the mass and the spin of the supermassive black hole. When we can infer these information from the signals we can now derive the mass and the spin of the supermassive black hole. Similarly, when we have gravitational waves and electromagnetic radiation emitted by, e.g., a binary neutron star merger, when this binary neutron star is orbiting a supermassive black hole both signals can be gravitationally lensed. Again depending on the strength of the imprint from the supermassive black hole the gravitationally lensed gravitational wave signals themselves will carry a characteristic signature of the mass and the spin of the supermassive black hole. Here, when we combine the information from the gravitational wave signals with the information from the electromagnetic counterpart this may allow us to place slightly tighter constraints on the mass and the spin of the black hole and on the parameters describing the observer-lens-source geometry than when we only detect two gravitationally lensed gravitational wave signals. Here, in both scenarios the analytical solutions derived in this paper can again be used to check if the results obtained from the gravitational wave signals and, in the second scenario, the electromagnetic counterpart are consistent. In addition, when we can infer very precise information about the observer-lens-source geometry and the mass and the spin of the supermassive black hole from the gravitational wave signals we can use the gravitationally lensed signals in combination with our analytical solutions to test the sky localisation of our gravitational wave detectors. \\

\section*{Acknowledgments}
I would like to thank Volker Perlick, Oleg Tsupko, Che-Yu Chen, Hsu-Wen Chiang, and Xian Chen for their valuable comments and our discussions. In addition, I would like to thank the developers of the programming language Julia and its packages for their continuous effort to provide fast and easily accessible tools for mathematical evaluations. I also acknowledge funding from the China Postdoctoral Science Foundation (Grant No. 2023M740111) and the National Natural Science Foundation of China (Grant No. 12473037). 
\appendix
\section{Elementary Integrals}\label{Sec:ELI}
While we integrated the equation of motion for $r$, and the $r$ dependent parts of the $\varphi$ coordinate and the time coordinate $t$ in Section~\ref{Sec:SolEoM} we encountered several elementary integrals. In this appendix we will briefly outline how to evaluate them and summarise the results. Note that throughout this whole appendix we will write down the antiderivatives without the integration constant.

\subsection{$r$ Motion with a Double Root at $r_{\text{ph}_{-}}$}\label{Sec:ELI1}
In this case we have lightlike geodesics characterised by $\Sigma=\Sigma_{\text{ph}_{-}}$ or $\Sigma=\pi-\Sigma_{\text{ph}_{-}}$. Light rays and gravitational waves travelling along these geodesics are characterised by the same constants of motion as light rays and gravitational waves on the unstable photon orbit at the radius coordinate $r_{\text{ph}_{-}}$. We recall that in this case the right-hand side of (\ref{eq:EoMr}) has a real double root at $r_{\text{ph}_{-}}=r_{1}=r_{2}<r_{\text{H}_{\text{i}}}\leq r_{\text{H}_{\text{o}}}$ and a pair of complex conjugate roots given by $r_{3}=\bar{r}_{4}=R_{3}+iR_{4}$, where we chose $0<R_{4}$. In this case we have four elementary integrals. They read
\begin{eqnarray}\label{eq:I1}
I_{1}=\int\frac{r\text{d}r}{\sqrt{(R_{3}-r)^2+R_{4}^2}}=\sqrt{\left(R_{3}-r\right)^2+R_{4}^2}+R_{3}\text{arsinh}\left(\frac{r-R_{3}}{R_{4}}\right),
\end{eqnarray}
\begin{eqnarray}\label{eq:I2}
I_{2}=\int\frac{\text{d}r}{\sqrt{(R_{3}-r)^2+R_{4}^2}}=\text{arsinh}\left(\frac{r-R_{3}}{R_{4}}\right),
\end{eqnarray}
\begin{eqnarray}\label{eq:I3}
I_{3}=\int\frac{\text{d}r}{(r-r_{a})\sqrt{(R_{3}-r)^2+R_{4}^2}}=-\frac{1}{\sqrt{(R_{3}-r_{a})^2+R_{4}^2}}\text{arsinh}\left(\frac{(r_{a}-R_{3})(r-r_{a})+(R_{3}-r_{a})^2+R_{4}^2}{R_{4}(r-r_{a})}\right),
\end{eqnarray}
\begin{eqnarray}\label{eq:I4}
&I_{4}=\int\frac{\text{d}r}{(r-r_{a})^2\sqrt{(R_{3}-r)^2+R_{4}^2}}=\frac{r_{a}-R_{3}}{\left((R_{3}-r_{a})^2+R_{4}^2\right)^{\frac{3}{2}}}\text{arsinh}\left(\frac{(r_{a}-R_{3})(r-r_{a})+(R_{3}-r_{a})^2+R_{4}^2}{R_{4}(r-r_{a})}\right)\\
&-\frac{\sqrt{(R_{3}-r)^2+R_{4}^2}}{\left((R_{3}-r_{a})^2+R_{4}^2\right)(r-r_{a})},\nonumber
\end{eqnarray}
where in (\ref{eq:I3}) the parameter $r_{a}$ can be $r_{\text{ph}_{-}}$, $r_{\text{H}_{\text{o}}}$, $r_{\text{H}_{\text{i}}}$, or $r_{\text{H}}$, and in (\ref{eq:I4}) the parameter $r_{a}$ can only be $r_{\text{H}}$. For the explicit evaluation of (\ref{eq:I3}) and (\ref{eq:I4}) we now first substitute $z=r-r_{a}$. Then we evaluate all four integrals and obtain the results on the right-hand sides.

\subsection{$r$ Motion with a Double Root at $r_{\text{ph}_{0}}$, and $a=m$ and $r$ Motion with a Double Root at $r_{\text{ph}_{0}}=r_{\text{ph}_{+}}$}\label{Sec:ELI2}
In this case we have two different types of lightlike geodesics. In the first case we have lightlike geodesics characterised by $\Sigma=\Sigma_{\text{ph}_{0}}$ or $\Sigma=\pi-\Sigma_{\text{ph}_{0}}$ (case 5). Light rays and gravitational waves travelling along these geodesics are characterised by the same constants of motion as light rays and gravitational waves on the stable photon orbit at the radius coordinate $r_{\text{ph}_{0}}$. We recall that in this case the right-hand side of (\ref{eq:EoMr}) has four real roots, that we labelled and sorted them such that $r_{4}<r_{\text{ph}_{0}}=r_{3}=r_{2}<r_{1}<r_{\text{H}_{\text{i}}}$, and that all of them lie inside the Cauchy horizon. In the second case we have $a=m$ and lightlike geodesics characterised by $\Sigma_{\text{ph}}<\Sigma=\Sigma_{\text{ph}_{0}}=\Sigma_{\text{ph}_{+}}$ or $\Sigma=\pi-\Sigma_{\text{ph}_{+}}=\pi-\Sigma_{\text{ph}_{0}}<\pi-\Sigma_{\text{ph}}$ (case 7c). Light rays and gravitational waves travelling along these geodesics are characterised by the same constants of motion as light rays and gravitational waves on the photon orbit at the radius coordinate $r_{\text{ph}_{0}}=r_{\text{ph}_{+}}$. We recall that in this case the right-hand side of (\ref{eq:EoMr}) has four real roots, that we labelled and sorted them such that $r_{4}<r_{\text{ph}_{0}}=r_{\text{ph}_{+}}=r_{\text{H}}=r_{3}=r_{2}<r_{1}$, and that light rays and gravitational waves travelling along these geodesics can pass through a turning point at $r_{\text{min}}=r_{1}$. In both cases integrating the equations of motion requires to calculate five elementary integrals. They read
\begin{eqnarray}\label{eq:I5}
I_{5}=\int\frac{r\text{d}r}{\sqrt{(r-r_{1})(r-r_{4})}},
\end{eqnarray}
\begin{eqnarray}\label{eq:I6}
I_{6}=\int\frac{\text{d}r}{\sqrt{(r-r_{1})(r-r_{4})}},
\end{eqnarray}
\begin{eqnarray}\label{eq:I7}
I_{7}=\int\frac{\text{d}r}{(r-r_{a})\sqrt{(r-r_{1})(r-r_{4})}},
\end{eqnarray}
\begin{eqnarray}\label{eq:I8}
I_{8}=\int\frac{\text{d}r}{(r-r_{a})^2\sqrt{(r-r_{1})(r-r_{4})}},
\end{eqnarray}
\begin{eqnarray}\label{eq:I9}
I_{9}=\int\frac{\text{d}r}{(r-r_{a})^3\sqrt{(r-r_{1})(r-r_{4})}}.
\end{eqnarray}
Here in (\ref{eq:I7}) the parameter $r_{a}$ can be $r_{\text{ph}_{0}}$, $r_{\text{H}_{\text{o}}}$, $r_{\text{H}_{\text{i}}}$, or $r_{\text{H}}$, while in (\ref{eq:I8}) and (\ref{eq:I9}) the parameter $r_{a}$ can only be $r_{\text{H}}$. In the cases of (\ref{eq:I5}) and (\ref{eq:I6}) we can integrate directly. The results read
\begin{eqnarray}\label{eq:I5S}
I_{5}=\sqrt{(r-r_{1})(r-r_{4})}+\frac{r_{1}+r_{4}}{2}\text{ln}\left(2\sqrt{(r-r_{1})(r-r_{4})}+2r-r_{1}-r_{4}\right)
\end{eqnarray}
and
\begin{eqnarray}\label{eq:I6S}
I_{6}=\text{ln}\left(2\sqrt{(r-r_{1})(r-r_{4})}+2r-r_{1}-r_{4}\right).
\end{eqnarray}
In the cases of (\ref{eq:I7}), (\ref{eq:I8}), and (\ref{eq:I9}) we first substitute $z=r-r_{a}$. Then we integrate and resubstitute. In the case of the integral $I_{7}$ we have to distinguish two different cases. In the first case we have $-(r_{1}-r_{a})(r_{a}-r_{4})<0$ and the result reads
\begin{eqnarray}\label{eq:I71S}
I_{7_{1}}=\frac{1}{\sqrt{(r_{1}-r_{a})(r_{a}-r_{4})}}\arcsin\left(\frac{(2r_{a}-r_{1}-r_{4})(r-r_{a})-2(r_{1}-r_{a})(r_{a}-r_{4})}{(r_{1}-r_{4})(r-r_{a})}\right),
\end{eqnarray}
where the parameter $r_{a}$ can be $r_{\text{ph}_{0}}$ or $r_{\text{H}}$ (for case 7c with $r_{\text{ph}_{0}}=r_{\text{ph}_{+}}=r_{\text{H}}$). In the second case we have $(r_{a}-r_{1})(r_{a}-r_{4})>0$ and the result reads
\begin{eqnarray}\label{eq:I72S}
I_{7_{2}}=-\frac{1}{\sqrt{(r_{a}-r_{1})(r_{a}-r_{4})}}\text{arcosh}\left(\frac{2(r_{a}-r_{1})(r_{a}-r_{4})+(2r_{a}-r_{1}-r_{4})(r-r_{a})}{(r_{1}-r_{4})(r-r_{a})}\right),
\end{eqnarray}
where $r_{a}$ can be $r_{\text{H}_{\text{o}}}$, $r_{\text{H}_{\text{i}}}$, or $r_{\text{H}}$. For $I_{8}$ we have the same cases and proceed analogously. When we have $-(r_{1}-r_{a})(r_{a}-r_{4})<0$ the result reads
\begin{eqnarray}\label{eq:I81S}
&I_{8_{1}}=\frac{\sqrt{(r-r_{1})(r-r_{4})}}{(r_{1}-r_{a})(r_{a}-r_{4})(r-r_{a})}+\frac{2r_{a}-r_{1}-r_{4}}{2\left((r_{1}-r_{a})(r_{a}-r_{4})\right)^{\frac{3}{2}}}\arcsin\left(\frac{(2r_{a}-r_{1}-r_{4})(r-r_{a})-2(r_{1}-r_{a})(r_{a}-r_{4})}{(r_{1}-r_{4})(r-r_{a})}\right),
\end{eqnarray}
while when we have $(r_{a}-r_{1})(r_{a}-r_{4})>0$ the result reads
\begin{eqnarray}\label{eq:I82S}
&I_{8_{2}}=\frac{2r_{a}-r_{1}-r_{4}}{2\left((r_{a}-r_{1})(r_{a}-r_{4})\right)^{\frac{3}{2}}}\text{arcosh}\left(\frac{(2r_{a}-r_{1}-r_{4})(r-r_{a})+2(r_{a}-r_{1})(r_{a}-r_{4})}{(r_{1}-r_{4})(r-r_{a})}\right)-\frac{\sqrt{(r-r_{1})(r-r_{4})}}{(r_{a}-r_{1})(r_{a}-r_{4})(r-r_{a})}.
\end{eqnarray}
In both cases the parameter $r_{a}$ can only be $r_{\text{H}}$. Note that the result given by $I_{8_{1}}$ only occurs when we have $r_{\text{ph}_{0}}=r_{\text{ph}_{+}}=r_{\text{H}}$. \\
In the case of the integral $I_{9}$ we only have the case characterised by $-(r_{1}-r_{a})(r_{a}-r_{4})<0$. For this case we proceed as outlined above and get as result
\begin{eqnarray}\label{eq:I9S}
&I_{9}=\frac{3(2r_{a}-r_{1}-r_{4})\sqrt{(r-r_{1})(r-r_{4})}}{4(r_{1}-r_{a})^2(r_{a}-r_{4})^2(r-r_{a})}+\frac{\sqrt{(r-r_{1})(r-r_{4})}}{2(r_{1}-r_{a})(r_{a}-r_{4})(r-r_{a})^2}\\
&+\frac{3(2r_{a}-r_{1}-r_{4})^2+4(r_{1}-r_{a})(r_{a}-r_{4})}{8\left((r_{1}-r_{a})(r_{a}-r_{4})\right)^{\frac{5}{2}}}\arcsin\left(\frac{(2r_{a}-r_{1}-r_{4})(r-r_{a})-2(r_{1}-r_{a})(r_{a}-r_{4})}{(r_{1}-r_{4})(r-r_{a})}\right),\nonumber
\end{eqnarray}
where as for $I_{8_{1}}$ and $I_{8_{2}}$ the parameter $r_{a}$ can only be $r_{\text{H}}$. Note that the result given by $I_{9}$ only occurs when we have $r_{\text{ph}_{0}}=r_{\text{ph}_{+}}=r_{\text{H}}$.

\subsection{$0<a<m$ and $r$ Motion with a Double Root at $r_{\text{ph}_{+}}=r_{\text{ph}_{0}}$}\label{Sec:ELI3}
In this case we have $0<a<m$ and lightlike geodesics characterised by $\Sigma=\Sigma_{\text{ph}_{0}}=\Sigma_{\text{ph}_{+}}$ or $\Sigma=\pi-\Sigma_{\text{ph}_{+}}=\pi-\Sigma_{\text{ph}_{0}}$ (case 7b). Light rays and gravitational waves travelling along these geodesics are characterised by the same constants of motion as light rays and gravitational waves on the photon orbit at the radius coordinate $r_{\text{ph}_{0}}=r_{\text{ph}_{+}}$. We recall that in this case the right-hand side of (\ref{eq:EoMr}) has four real roots, that we labelled and sorted them such that $r_{4}<r_{\text{ph}_{0}}=r_{\text{ph}_{+}}=r_{3}=r_{2}=r_{1}<r_{\text{H}_{\text{i}}}$, and that all of them lie inside the Cauchy horizon. In this case we have four different elementary integrals. They read
\begin{eqnarray}\label{eq:I10}
I_{10}=\int\frac{r\text{d}r}{\sqrt{(r-r_{\text{ph}_{+}})(r-r_{4})}},
\end{eqnarray}
\begin{eqnarray}\label{eq:I11}
I_{11}=\int\frac{\text{d}r}{\sqrt{(r-r_{\text{ph}_{+}})(r-r_{4})}},
\end{eqnarray}
\begin{eqnarray}\label{eq:I12}
I_{12}=\int\frac{\text{d}r}{\sqrt{(r-r_{\text{ph}_{+}})^3(r-r_{4})}},
\end{eqnarray}
\begin{eqnarray}\label{eq:I13}
I_{13}=\int\frac{\text{d}r}{(r-r_{a})\sqrt{(r-r_{\text{ph}_{+}})(r-r_{4})}},
\end{eqnarray}
where in $I_{13}$ $r_{a}$ can be $r_{\text{H}_{\text{o}}}$ or $r_{\text{H}_{\text{i}}}$. We can easily see that with the exception of $I_{12}$ all integrals are very similar to the integrals in App.~\ref{Sec:ELI2}. The evaluation of $I_{12}$ is straighforward and thus in this case we get as results
 \begin{eqnarray}\label{eq:I10S}
I_{10}=\sqrt{(r-r_{\text{ph}_{+}})(r-r_{4})}+\frac{r_{\text{ph}_{+}}+r_{4}}{2}\text{ln}\left(2\sqrt{(r-r_{\text{ph}_{+}})(r-r_{4})}+2r-r_{\text{ph}_{+}}-r_{4}\right),
\end{eqnarray}
\begin{eqnarray}\label{eq:I11S}
I_{11}=\text{ln}\left(2\sqrt{(r-r_{\text{ph}_{+}})(r-r_{4})}+2r-r_{\text{ph}_{+}}-r_{4}\right),
\end{eqnarray}
 \begin{eqnarray}\label{eq:I12S}
I_{12}=-\frac{2}{r_{\text{ph}_{+}}-r_{4}}\sqrt{\frac{r-r_{4}}{r-r_{\text{ph}_{+}}}},
\end{eqnarray}
\begin{eqnarray}\label{eq:I13S}
&I_{13}=-\frac{1}{\sqrt{(r_{a}-r_{\text{ph}_{+}})(r_{a}-r_{4})}}\text{arcosh}\left(\frac{2(r_{a}-r_{\text{ph}_{+}})(r_{a}-r_{4})+(2r_{a}-r_{\text{ph}_{+}}-r_{4})(r-r_{a})}{(r_{\text{ph}_{+}}-r_{4})(r-r_{a})}\right),
\end{eqnarray}
where in (\ref{eq:I13S}) $r_{a}$ can be $r_{\text{H}_{\text{o}}}$ or $r_{\text{H}_{\text{i}}}$.
\subsection{$r$ Motion with a Double Root at $r_{\text{ph}}$, $r_{\text{ph}_{+}}$, or $r_{\text{ph}_{+}}=r_{\text{ph}}$}\label{Sec:ELI4}
In this case we have three different types of lightlike geodesics. The first type of lightlike geodesics is characterised by $\Sigma=\Sigma_{\text{ph}_{+}}$ or $\Sigma=\pi-\Sigma_{\text{ph}_{+}}$ (case 7a). Light rays and gravitational waves travelling along these geodesics have the same constants of motion as light rays and gravitational waves on the unstable photon orbit at the radius coordinate $r_{\text{ph}_{+}}$. We recall that in this case the right-hand side of (\ref{eq:EoMr}) has four real roots, that we labelled and sorted them such that $r_{4}<r_{3}<r_{\text{ph}_{+}}=r_{2}=r_{1}<r_{\text{H}_{\text{i}}}$, and that all lie inside the Cauchy horizon. The second type of lightlike geodesics is characterised by $\Sigma=\Sigma_{\text{ph}}$ or $\Sigma=\pi-\Sigma_{\text{ph}}$ (case 9a). In this case we have light rays and gravitational waves asymptotically coming from or going to the unstable photon orbit at the radius coordinate $r_{\text{ph}}$. They are characterised by the same constants of motion as light rays and gravitational waves on the photon orbit. We recall that in this case the right-hand side of (\ref{eq:EoMr}) has four real roots and that we labelled and sorted them such that $r_{4}<r_{3}<r_{\text{H}_{\text{i}}}\leq r_{\text{H}_{\text{o}}}<r_{\text{ph}}=r_{2}=r_{1}$. Note that in both cases we do not have degenerate photon orbits. The third type of lightlike geodesics is characterised by $\Sigma=\Sigma_{\text{ph}_{+}}=\Sigma_{\text{ph}}$ or $\Sigma=\pi-\Sigma_{\text{ph}}=\pi-\Sigma_{\text{ph}_{+}}$ (case 9b). Light rays and gravitational waves travelling along these geodesics are either asymptotically coming from or going to the unstable photon orbit at the radius coordinate $r_{\text{ph}_{+}}=r_{\text{ph}}$. We recall that in this case the right-hand side of (\ref{eq:EoMr}) has four real roots and that we labelled and sorted them such that $r_{4}<r_{3}<r_{\text{ph}_{+}}=r_{\text{H}}=r_{\text{ph}}=r_{2}=r_{1}$. We have in total three structurally different integrals. They read
\begin{eqnarray}\label{eq:I14}
I_{14}=\int\frac{\text{d}y}{(y-y_{a})\sqrt{y-y_{1}}},
\end{eqnarray}
\begin{eqnarray}\label{eq:I15}
I_{15}=\int\frac{\text{d}y}{(y-y_{a})^2\sqrt{y-y_{1}}},
\end{eqnarray}
\begin{eqnarray}\label{eq:I16}
I_{16}=\int\frac{\text{d}y}{(y-y_{a})^3\sqrt{y-y_{1}}},
\end{eqnarray}
where in (\ref{eq:I14}) the parameter $y_{a}$ can be $a_{2r}/12$, $y_{\text{ph}_{+}}$, $y_{\text{ph}}$, $y_{\text{H}_{\text{o}}}$, $y_{\text{H}_{\text{i}}}$, or $y_{\text{H}}$, in (\ref{eq:I15}) the parameter $y_{a}$ can be $a_{2r}/12$ or $y_{\text{H}}$, and in (\ref{eq:I16}) the parameter $y_{a}$ can only be $y_{\text{ph}_{+}}=y_{\text{H}}=y_{\text{ph}}$. Here $y_{1}$, $y_{\text{ph}_{+}}$, $y_{\text{ph}}$, $y_{\text{H}_{\text{o}}}$, $y_{\text{H}_{\text{i}}}$, and $y_{\text{H}}$ are related to $r_{4}$, $r_{\text{ph}_{+}}$, $r_{\text{ph}}$, $r_{\text{H}_{\text{o}}}$, $r_{\text{H}_{\text{i}}}$, and $r_{\text{H}}$ by the coordinate transformation (\ref{eq:elm1}), respectively. When we integrate (\ref{eq:I14}) and (\ref{eq:I15}) we now have to distinguish the cases $y_{a}=a_{2r}/12$ and $y_{a}=y_{\text{ph}}$ for $r_{\text{H}_{\text{o}}}<r<r_{\text{ph}}$ [the latter only for (\ref{eq:I14})] from all other cases.\\
We start with deriving the solution to (\ref{eq:I14}). When we have $y_{a}=a_{2r}/12$ or $y_{a}=y_{\text{ph}}$ for $r_{\text{H}_{\text{o}}}<r<r_{\text{ph}}$ we have $y_{a}<y$. We first substitute $z=y-y_{a}$ and integrate. We obtain two structurally different results. For $y_{a}=a_{2r}/12$ the result reads
\begin{eqnarray}\label{eq:I141}
I_{14_{1}}=-4\sqrt{\frac{r_{3}-r_{4}}{a_{3r}}}\text{arcoth}\left(\sqrt{\frac{r-r_{4}}{r-r_{3}}}\right), 
\end{eqnarray}
where here and in all following results $a_{3r}$ is given by (\ref{eq:coeffa3r}). In the case $y_{a}=y_{\text{ph}}$ for $r_{\text{H}_{\text{o}}}<r<r_{\text{ph}}$ on the other hand the result reads
\begin{eqnarray}\label{eq:I142}
I_{14_{2}}=-4\sqrt{\frac{(r_{a}-r_{3})(r_{3}-r_{4})}{a_{3r}(r_{a}-r_{4})}}\text{arcoth}\left(\sqrt{\frac{(r_{a}-r_{3})(r-r_{4})}{(r_{a}-r_{4})(r-r_{3})}}\right),
\end{eqnarray}
where the parameter $r_{a}$ can only be $r_{\text{ph}}$.\\
In all other cases we have $r_{a}<r$ and thus $y<y_{a}$. In these cases we first substitute $z=y-y_{1}$. Then we integrate and obtain as result
\begin{eqnarray}\label{eq:I143}
I_{14_{3}}=-4\sqrt{\frac{(r_{a}-r_{3})(r_{3}-r_{4})}{a_{3r}(r_{a}-r_{4})}}\text{artanh}\left(\sqrt{\frac{(r_{a}-r_{3})(r-r_{4})}{(r_{a}-r_{4})(r-r_{3})}}\right),
\end{eqnarray}
where the parameter $r_{a}$ can be $r_{\text{ph}}$, $r_{\text{ph}_{+}}$, $r_{\text{H}_{\text{o}}}$, $r_{\text{H}_{\text{i}}}$, or $r_{\text{H}}$.\\
Now we turn to (\ref{eq:I15}). Here we only have to distinguish the cases $y_{a}=a_{2r}/12$ and $y_{a}=y_{\text{H}}$. In the first case we again substitute $z=y-y_{a}$ and integrate. The result reads
\begin{eqnarray}\label{eq:I151}
I_{15_{1}}=\frac{8\sqrt{r_{3}-r_{4}}}{a_{3r}^{\frac{3}{2}}}\left((r_{3}-r_{4})\text{arcoth}\left(\sqrt{\frac{r-r_{4}}{r-r_{3}}}\right)-\sqrt{(r-r_{3})(r-r_{4})}\right).
\end{eqnarray}
In the second case on the other hand we first substitute $z=y-y_{1}$. We integrate and obtain as result
\begin{eqnarray}\label{eq:I152}
\hspace{-0.2cm}I_{15_{2}}=\frac{8}{a_{3r}^{\frac{3}{2}}}\left(\frac{(r_{a}-r_{3})^2\sqrt{(r_{3}-r_{4})(r-r_{3})(r-r_{4})}}{(r_{a}-r_{4})(r-r_{a})}+\left(\frac{(r_{3}-r_{4})(r_{a}-r_{3})}{r_{a}-r_{4}}\right)^{\frac{3}{2}}\text{artanh}\left(\sqrt{\frac{(r_{a}-r_{3})(r-r_{4})}{(r_{a}-r_{4})(r-r_{3})}}\right)\right),
\end{eqnarray}
where the parameter $r_{a}$ can only be $r_{\text{H}}$.\\
In the case of the integral $I_{16}$ we only have one case. Again we first substitute $z=y-y_{1}$. Then we integrate and obtain as result
\begin{eqnarray}\label{eq:I16S}
&I_{16}=-8\sqrt{\frac{r_{3}-r_{4}}{a_{3r}^5}}\left(\frac{3(r_{a}-r_{3})^3(r_{3}-r_{4})\sqrt{(r-r_{3})(r-r_{4})}}{(r_{a}-r_{4})^2(r-r_{a})}+\frac{2(r_{a}-r_{3})^3\sqrt{(r-r_{3})^3(r-r_{4})}}{(r_{a}-r_{4})(r-r_{a})^2}\right.\\
&\left.+3(r_{3}-r_{4})^2\left(\frac{r_{a}-r_{3}}{r_{a}-r_{4}}\right)^{\frac{5}{2}}\text{artanh}\left(\sqrt{\frac{(r_{a}-r_{3})(r-r_{4})}{(r_{a}-r_{4})(r-r_{3})}}\right)\right),\nonumber
\end{eqnarray}
where the parameter $r_{a}$ can only be $r_{\text{ph}_{+}}=r_{\text{H}}=r_{\text{ph}}$.

\subsection{Elementary Integrals for Triple Degenerate Photon Orbits}\label{Sec:ELI5}
In this case we have $a=m$ and lightlike geodesics characterised by $\Sigma=\Sigma_{\text{ph}_{0}}=\Sigma_{\text{ph}_{+}}=\Sigma_{\text{ph}}$ or $\Sigma=\pi-\Sigma_{\text{ph}}=\pi-\Sigma_{\text{ph}_{+}}=\pi-\Sigma_{\text{ph}_{0}}$. Light rays and gravitational waves travelling along these geodesics are either asymptotically coming from or going to the photon orbit at the radius coordinate $r_{\text{ph}_{0}}=r_{\text{ph}_{+}}=r_{\text{H}}=r_{\text{ph}}$. We recall that in this case the right-hand side of (\ref{eq:EoMr}) has four real roots and that we labelled and sorted them such that $r_{4}<r_{\text{ph}_{0}}=r_{\text{ph}_{+}}=r_{\text{H}}=r_{\text{ph}}=r_{3}=r_{2}=r_{1}$. In this case we have five structurally different elementary integrals. They read 
\begin{eqnarray}\label{eq:I17}
I_{17}=\int\frac{r\text{d}r}{\sqrt{(r-r_{\text{H}})(r-r_{4})}},
\end{eqnarray}
\begin{eqnarray}\label{eq:I18}
I_{18}=\int\frac{\text{d}r}{\sqrt{(r-r_{\text{H}})(r-r_{4})}},
\end{eqnarray}
\begin{eqnarray}\label{eq:I19}
I_{19}=\int\frac{\text{d}r}{\sqrt{(r-r_{\text{H}})^3(r-r_{4})}},
\end{eqnarray}
\begin{eqnarray}\label{eq:I20}
I_{20}=\int \frac{\text{d}r}{\sqrt{(r-r_{\text{H}})^5(r-r_{4})}},
\end{eqnarray}
\begin{eqnarray}\label{eq:I21}
I_{21}=\int \frac{\text{d}r}{\sqrt{(r-r_{\text{H}})^7(r-r_{4})}}.
\end{eqnarray}
The first two integrals are structurally the same as $I_{10}$ and $I_{11}$ given by (\ref{eq:I10}) and (\ref{eq:I11}) in Appendix~\ref{Sec:ELI3}, respectively. Thus we simply have to replace $r_{\text{ph}_{+}}\rightarrow r_{\text{H}}$ in (\ref{eq:I10S}) and (\ref{eq:I11S}) and get
\begin{eqnarray}\label{eq:I17S}
I_{17}=\sqrt{(r-r_{\text{H}})(r-r_{4})}+\frac{r_{\text{H}}+r_{4}}{2}\text{ln}\left(2\sqrt{(r-r_{\text{H}})(r-r_{4})}+2r-r_{\text{H}}-r_{4}\right),
\end{eqnarray}
\begin{eqnarray}\label{eq:I18S}
I_{18}=\text{ln}\left(2\sqrt{(r-r_{\text{H}})(r-r_{4})}+2r-r_{\text{H}}-r_{4}\right).
\end{eqnarray}
For the evaluation of $I_{19}$, $I_{20}$, and $I_{21}$ we now first substitute $z=r-r_{\text{H}}$. Then we integrate and get as results
\begin{eqnarray}\label{eq:I19S}
I_{19}=-\frac{2}{r_{\text{H}}-r_{4}}\sqrt{\frac{r-r_{4}}{r-r_{\text{H}}}},
\end{eqnarray}
\begin{eqnarray}\label{eq:I20S}
I_{20}=\frac{4}{3(r_{\text{\text{H}}}-r_{4})^2}\sqrt{\frac{r-r_{4}}{r-r_{\text{\text{H}}}}}-\frac{2}{3(r_{\text{\text{H}}}-r_{4})}\sqrt{\frac{r-r_{4}}{(r-r_{\text{\text{H}}})^3}},
\end{eqnarray}
\begin{eqnarray}\label{eq:I21S}
I_{21}=-\frac{16}{15(r_{\text{\text{H}}}-r_{4})^3}\sqrt{\frac{r-r_{4}}{r-r_{\text{H}}}}+\frac{8}{15(r_{\text{H}}-r_{4})^2}\sqrt{\frac{r-r_{4}}{(r-r_{\text{H}})^3}}-\frac{2}{5(r_{\text{H}}-r_{4})}\sqrt{\frac{r-r_{4}}{(r-r_{\text{H}})^5}}.
\end{eqnarray}

\section{Elliptic Integrals}\label{Sec:ELIL}
While integrating the equations of motion in Sec.~\ref{Sec:SolEoM} we also encountered several different elliptic integrals. All of these integrals can be rewritten in terms of elementary functions and Legendre's elliptic integrals of the first, second, and third kind. Since not all readers may be familiar with Legendre's elliptic integrals, in this appendix we will briefly define them and summarise their most important basic properties. In addition, we will use them in combination with elementary functions to rewrite six different nonstandard elliptic integrals.\\
We start by defining Legendre's elliptic integrals of the first, second, and third kind. They read
\begin{eqnarray}\label{eq:LF}
F_{L}(\chi,k)=\int_{0}^{\chi}\frac{\text{d}\chi'}{\sqrt{1-k\sin^2\chi'}},
\end{eqnarray}
\begin{eqnarray}\label{eq:LE}
E_{L}(\chi,k)=\int_{0}^{\chi}\sqrt{1-k\sin^2\chi'}\text{d}\chi',
\end{eqnarray}
and 
\begin{eqnarray}\label{eq:LPi}
\Pi_{L}(\chi,k,n)=\int_{0}^{\chi}\frac{\text{d}\chi'}{(1-n\sin^2\chi')\sqrt{1-k\sin^2\chi'}}.
\end{eqnarray}
Here, $\chi$ is called the amplitude of the elliptic integrals, $k$ is the square of the elliptic modulus, and $n\in \mathbb{R}$ is a real parameter. Note that in the forms given above Legendre's elliptic integrals are usually also referred to as incomplete elliptic integrals. When the amplitude $\chi$ takes the value $\chi=\pi/2$ on the other hand they are referred to as complete elliptic integrals. In this case one commonly omits the amplitude in the argument and writes Legendre's elliptic integral of the first kind as $K_{L}(k)=F_{L}(\pi/2,k)$. Furthermore, we can easily see that when we add or subtract an integer multiple $j$ of $\pi$ in the limits and transform $\chi\Rightarrow\chi\mp j\pi$ the integrands of the elliptic integrals remain invariant. In addition, when we replace $\chi\rightarrow-\chi$ we see that the elliptic integrals have the property
\begin{eqnarray}
F_{L}(-\chi,k)=-F_{L}(\chi,k),~~~E_{L}(-\chi,k)=-E_{L}(\chi,k),~~~\Pi_{L}(-\chi,k,n)=-\Pi_{L}(\chi,k,n).
\end{eqnarray}
Furthermore, for $1\leq n$ the integrand of Legendre's elliptic integral of the third kind has a singularity. In this paper we will encounter several integrals for which we have to integrate over this singularity. In these cases we will use (17.7.7) and (17.7.8) from Ref.~\cite{MilneThomson1972} to rewrite (\ref{eq:LPi}) as
\begin{eqnarray}\label{eq:PiN}
\Pi_{L}(\chi,k,n)=F_{L}(\chi,k)-\Pi_{L}\left(\chi,k,\frac{k}{n}\right)+\frac{1}{2p}\text{ln}\left(\frac{\cos\chi\sqrt{1-k\sin^2\chi}+p\sin\chi}{\left|\cos\chi\sqrt{1-k\sin^2\chi}-p\sin\chi\right|}\right),
\end{eqnarray}
where the parameter $p$ is defined by
\begin{eqnarray}
p=\sqrt{\frac{(n-1)(n-k)}{n}}.
\end{eqnarray}
Note that here rewriting Legendre's elliptic integral of the third kind only prevents that it diverges when we integrate over a singularity and the lower and upper limits lie below and above the singularity, respectively. When the upper limit is located at the singularity, in our case, e.g., a horizon, using (\ref{eq:PiN}) will lead to the same result as evaluating Legendre's elliptic integral of the third kind directly.\\
While after a coordinate transformation some of the elliptic integrals which we encountered in the process of solving the equations of motion either directly or after very few rearrangements take one of Legendre's standard forms given by (\ref{eq:LF}), (\ref{eq:LE}), and (\ref{eq:LPi}) we also encountered six different general elliptic integrals which cannot be easily rewritten. They read 
\begin{eqnarray}\label{eq:EI1}
I_{L_{1}}(\chi_{i},\chi,k_{4})=\int_{\chi_{i}}^{\chi}\frac{\text{d}\chi'}{\left(1-k_{4}\sin^2\chi'\right)^{\frac{3}{2}}},
\end{eqnarray}
\begin{eqnarray}\label{eq:EI2}
I_{L_{2}}(\chi_{i},\chi,k_{1},n)=\int_{\chi_{i}}^{\chi}\frac{\text{d}\chi'}{(1+n\tan\chi')\sqrt{1-k_{1}\sin^2\chi'}},
\end{eqnarray}
\begin{eqnarray}\label{eq:EI3}
I_{L_{3}}(\chi_{i},\chi,k_{1},n)=\int_{\chi_{i}}^{\chi}\frac{\text{d}\chi'}{(1+n\tan\chi')^2\sqrt{1-k_{1}\sin^2\chi'}},
\end{eqnarray}
\begin{eqnarray}\label{eq:EI4}
I_{L_{4}}(\chi_{i},\chi,k_{2},n)=\int_{\chi_{i}}^{\chi}\frac{\text{d}\chi'}{(1+n\cos\chi')\sqrt{1-k_{2}\sin^2\chi'}},
\end{eqnarray}
\begin{eqnarray}\label{eq:EI5}
I_{L_{5}}(\chi_{i},\chi,k_{2},n)=\int_{\chi_{i}}^{\chi}\frac{\text{d}\chi'}{(1+n\cos\chi')^2\sqrt{1-k_{2}\sin^2\chi'}},
\end{eqnarray}
\begin{eqnarray}\label{eq:EI6}
I_{L_{6}}(\chi_{i},\chi,k,n)=\int_{\chi_{i}}^{\chi}\frac{\text{d}\chi'}{(1-n\sin^2\chi')^2\sqrt{1-k\sin^2\chi'}}.
\end{eqnarray}
In the following we will now rewrite them in terms of elementary functions and Legendre's elliptic integrals of the first, second, and third kind.\\
We start with $I_{L_{1}}(\chi_{i},\chi,k_{4})$. It occured when we integrated $t_{\vartheta}(\lambda)$ in Section~\ref{Sec:ttheta}. We can rewrite it in terms of elementary functions and Legendre's elliptic integral of the second kind. Then it reads
\begin{eqnarray}\label{eq:EI1S}
I_{L_{1}}(\chi_{i},\chi,k_{4})=\frac{1}{1-k_{4}}\left(E_{L}\left(\chi,k_{4}\right)-E_{L}\left(\chi_{i},k_{4}\right)+\frac{k_{4}\sin\chi_{i}\cos\chi_{i}}{\sqrt{1-k_{4}\sin^2\chi_{i}}}-\frac{k_{4}\sin\chi\cos\chi}{\sqrt{1-k_{4}\sin^2\chi}}\right),
\end{eqnarray}
where $\chi_{i}$ and $\chi$ are related to $\cos\vartheta_{i}$ and $\cos\vartheta(\lambda)$ by (\ref{eq:elchi5}) and the square of the elliptic modulus $k_{4}$ is given by (\ref{eq:EM4}). Note that we omitted $\lambda$ in the argument for brevity.\\
We continue with $I_{L_{2}}(\chi_{i},\chi,k_{1},n)$ and $I_{L_{3}}(\chi_{i},\chi,k_{1},n)$ given by (\ref{eq:EI2}) and (\ref{eq:EI3}), respectively. We recall that in this case the right-hand side of (\ref{eq:EoMr}) has two distinct pairs of complex conjugate roots and that we labelled and sorted them such that $r_{1}=\bar{r}_{2}=R_{1}+iR_{2}$ and $r_{3}=\bar{r}_{4}=R_{3}+iR_{4}$, where we chose $R_{1}<R_{3}$, and $0<R_{2}$ and $0<R_{4}$. When we apply the substitution (\ref{eq:elsub1}) to $\varphi_{r}(\lambda)$ and $t_{r}(\lambda)$ we first reduce the integrals to forms similar to the ones given by (267.01) and (267.02) in the book of Byrd and Friedman \cite{Byrd1954}. The resulting terms contain on one hand Legendre's elliptic integral of the first kind and on the other hand either $I_{L_{2}}(\chi_{i},\chi,k_{1},n)$ alone or together with $I_{L_{3}}(\chi_{i},\chi,k_{1},n)$. Here $\chi_{i}$ and $\chi$ are related to $r_{i}$ and $r(\lambda)$ by (\ref{eq:elchi1}) and the square of the elliptic modulus $k_{1}$ is given by (\ref{eq:EM1}). Note that again we will omit the argument $\lambda$ in the following for brevity.\\
Now we define a new constant of motion $g_{1}$. It reads
\begin{eqnarray}
g_{1}=\frac{R_{2}+g_{0}(R_{1}-r_{\hat{\text{H}}})}{R_{1}-g_{0}R_{2}-r_{\hat{\text{H}}}},
\end{eqnarray}
where $g_{0}$ is given by (\ref{eq:g0}), the coefficients $R$ and $\bar{R}$ in $g_{0}$ are given by (\ref{eq:CR1}), and the parameter $r_{\hat{\text{H}}}$ can be $r_{\text{H}_{\text{o}}}$, $r_{\text{H}_{\text{i}}}$, or $r_{\text{H}}$. In both integrals we can now have $n=g_{0}$ or $n=g_{1}$, however, in the case of (\ref{eq:EI3}) the parameter $r_{\hat{\text{H}}}$ in $g_{1}$ can only be $r_{\text{H}}$. Using elementary functions and Legendre's elliptic integrals of the first, second, and third kind we can now rewrite (\ref{eq:EI2}) and (\ref{eq:EI3}) in the following forms
\begin{eqnarray}\label{eq:EI2S}
I_{L_{2}}\left(\chi_{i},\chi,k_{1},n\right)=I_{L_{2}}\left(\chi,k_{1},n\right)-I_{L_{2}}\left(\chi_{i},k_{1},n\right)~~~\text{and}~~~I_{L_{3}}\left(\chi_{i},\chi,k_{1},n\right)=I_{L_{3}}\left(\chi,k_{1},n\right)-I_{L_{3}}\left(\chi_{i},k_{1},n\right),
\end{eqnarray}
where $I_{L_{2}}\left(\chi_{i},k_{1},n\right)$, $I_{L_{2}}\left(\chi,k_{1},n\right)$, $I_{L_{3}}\left(\chi_{i},k_{1},n\right)$, and $I_{L_{3}}\left(\chi,k_{1},n\right)$ are given by
\begin{eqnarray}\label{eq:EI2PS}
I_{L_{2}}\left(\chi',k_{1},n\right)=\frac{F_{L}(\chi',k_{1})+n^2\Pi_{L}\left(\chi',k_{1},1+n^2\right)}{1+n^2}+\frac{n\tilde{I}_{L_{1}}(\chi',k_{1},n)}{2\sqrt{(1+n^2)(1-k_{1}+n^2)}}
\end{eqnarray}
and
\begin{eqnarray}\label{eq:EI3PS}
&I_{L_{3}}\left(\chi',k_{1},n\right)=\frac{F_{L}(\chi',k_{1})}{(1+n^2)^2}+\frac{n^2}{(1+n^2)(1-k_{1}+n^2)}\left(n+\frac{\sin\chi'-n\cos\chi'}{\cos\chi'+n\sin\chi'}\sqrt{1-k_{1}\sin^2\chi'}-E_{L}(\chi',k_{1})\right)\\
&+\frac{2(1-k_{1}+n^2)-n^2k_{1}}{(1+n^2)(1-k_{1}+n^2)}\left(\frac{n^2\Pi_{L}\left(\chi',k_{1},1+n^2\right)}{1+n^2}+\frac{n\tilde{I}_{L_{1}}(\chi',k_{1},n)}{2\sqrt{(1+n^2)(1-k_{1}+n^2)}}\right),\nonumber
\end{eqnarray}
where in both terms the function $\tilde{I}_{L_{1}}(\chi',k_{1},n)$ is given by
\begin{eqnarray}
\tilde{I}_{L_{1}}(\chi',k_{1},n)=\ln\left(\left|\frac{\left(1+\sqrt{\frac{1+n^2}{1-k_{1}+n^2}}\right)\left(1-\sqrt{\frac{1+n^2}{1-k_{1}+n^2}\left(1-k_{1}\sin^2\chi'\right)}\right)}{\left(1-\sqrt{\frac{1+n^2}{1-k_{1}+n^2}}\right)\left(1+\sqrt{\frac{1+n^2}{1-k_{1}+n^2}\left(1-k_{1}\sin^2\chi'\right)}\right)}\right|\right).
\end{eqnarray}
Note that here in (\ref{eq:EI2PS}) and (\ref{eq:EI3PS}) for the parameter of Legendre's elliptic integral of the third kind we have $1+n^2>1$ and thus we now use (\ref{eq:PiN}) to rewrite it.\\
Now we turn to $I_{L_{4}}(\chi_{i},\chi,k_{2},n)$ and $I_{L_{5}}(\chi_{i},\chi,k_{2},n)$ given by (\ref{eq:EI4}) and (\ref{eq:EI5}), respectively. We recall that in this case the right-hand side of (\ref{eq:EoMr}) has two distinct real roots and a pair of complex conjugate roots and that we labelled and sorted them such that $r_{2}<r_{1}$ and $r_{3}=\bar{r}_{4}=R_{3}+iR_{4}$, where we chose $0<R_{4}$. $\chi_{i}$ and $\chi$ are related to $r_{i}$ and $r(\lambda)$ by (\ref{eq:elchi2}) (note that again we will omit the argument $\lambda$ in the following) and the square of the elliptic modulus $k_{2}$ is given by (\ref{eq:EM2}). When we apply the substitution (\ref{eq:elsub2}) to $\varphi_{r}(\lambda)$ and $t_{r}(\lambda)$ we see that they can be rewritten in terms of Legendre's elliptic integral of the first kind and $I_{L_{4}}(\chi_{i},\chi,k_{2},n)$ either alone or in combination with $I_{L_{5}}(\chi_{i},\chi,k_{2},n)$, where the parameter $n$ can be given by 
\begin{eqnarray}
n_{3}=\frac{\hat{R}+\tilde{R}}{\hat{R}-\tilde{R}}
\end{eqnarray}
or
\begin{eqnarray}
n_{4}=\frac{(r_{\hat{\text{H}}}-r_{1})\hat{R}+(r_{\hat{\text{H}}}-r_{2})\tilde{R}}{(r_{\hat{\text{H}}}-r_{1})\hat{R}-(r_{\hat{\text{H}}}-r_{2})\tilde{R}},
\end{eqnarray}
where $\hat{R}$ and $\tilde{R}$ are given by (\ref{eq:CR2}). Note that here in general in $n_{3}$ the parameter $r_{\hat{\text{H}}}$ can be $r_{\text{H}_{\text{o}}}$, $r_{\text{H}_{\text{i}}}$, or $r_{\text{H}}$, however, in the case of (\ref{eq:EI5}) we can only have $r_{\hat{\text{H}}}=r_{\text{H}}$.\\
For both integrals it is relatively straightforward to rewrite them in terms of elementary functions and Legendre's elliptic integrals of the first, second, and third kind. We start by rewriting (\ref{eq:EI4}) and (\ref{eq:EI5}) in the following forms
\begin{eqnarray}\label{eq:EI3S}
I_{L_{4}}\left(\chi_{i},\chi,k_{2},n\right)=I_{L_{4}}\left(\chi,k_{2},n\right)-I_{L_{4}}\left(\chi_{i},k_{2},n\right)~~~\text{and}~~~I_{L_{5}}\left(\chi_{i},\chi,k_{2},n\right)=I_{L_{5}}\left(\chi,k_{2},n\right)-I_{L_{5}}\left(\chi_{i},k_{2},n\right),
\end{eqnarray}
where $I_{L_{4}}\left(\chi_{i},k_{2},n\right)$, $I_{L_{4}}\left(\chi,k_{2},n\right)$, $I_{L_{5}}\left(\chi_{i},k_{2},n\right)$, and $I_{L_{5}}\left(\chi,k_{2},n\right)$ are calculated in the following. In the case of $I_{L_{4}}\left(\chi',k_{2},n\right)$ we first expand with $1-n\cos\chi''$ and obtain 
\begin{eqnarray}
&I_{L_{4}}(\chi',k_{2},n)=\int_{0}^{\chi'}\frac{\text{d}\chi''}{(1+n\cos\chi'')\sqrt{1-k_{2}\sin^2\chi''}}=\int_{0}^{\chi'}\frac{(1-n\cos\chi'')\text{d}\chi''}{(1-n^2\cos^2\chi'')\sqrt{1-k_{2}\sin^2\chi''}}\\
&=\frac{1}{n^2-1}\left(n\int_{0}^{\chi'}\frac{\cos\chi''\text{d}\chi''}{\left(1-\frac{n^2}{n^2-1}\sin^2\chi''\right)\sqrt{1-k_{2}\sin^2\chi''}}-\Pi_{L}\left(\chi',k_{2},\frac{n^2}{n^2-1}\right)\right),\nonumber
\end{eqnarray}
where we already rewrote the second term in terms of Legendre's elliptic integral of the third kind. We can easily see that the first term is an elementary integral. Finding its antiderivative requires a case by case analysis which is quite lengthy and thus it will not be reproduced here. After the integration the final result can be written in terms of elementary functions and Legendre's elliptic integral of the third kind. It reads 
\begin{eqnarray}\label{eq:EI4PS}
I_{L_{4}}(\chi',k_{2},n)=\frac{\Pi_{L}\left(\chi',k_{2},\frac{n^2}{n^2-1}\right)}{1-n^2}+\frac{n\tilde{I}_{L_{2}}(\chi',k_{2},n)}{2\sqrt{(n^2-1)(n^2(1-k_{2})+k_{2})}},
\end{eqnarray}
where the function $\tilde{I}_{L_{2}}(\chi',k_{2},n)$ is given by (\ref{eq:Itil2}) below.\\
For the evaluation of $I_{L_{5}}\left(\chi',k_{2},n\right)$ we proceed analogously. We first expand by $(1-n\cos\chi'')^2$ and obtain
\begin{eqnarray}
&I_{L_{5}}(\chi',k_{2},n)=\int_{0}^{\chi'}\frac{\text{d}\chi''}{(1+n\cos\chi'')^2\sqrt{1-k_{2}\sin^2\chi''}}=\int_{0}^{\chi'}\frac{(1-n\cos\chi'')^2\text{d}\chi''}{(1-n^2\cos^2\chi'')^2\sqrt{1-k_{2}\sin^2\chi''}}\\
&=\frac{2}{(n^2-1)^2}\left(\int_{0}^{\chi'}\frac{\text{d}\chi''}{\left(1-\frac{n^2}{n^2-1}\sin^2\chi''\right)^2\sqrt{1-k_{2}\sin^2\chi''}}-n\int_{0}^{\chi'}\frac{\cos\chi''\text{d}\chi''}{\left(1-\frac{n^2}{n^2-1}\sin^2\chi''\right)^2\sqrt{1-k_{2}\sin^2\chi''}}\right)+\frac{\Pi_{L}\left(\chi',k_{2},\frac{n^2}{n^2-1}\right)}{n^2-1}.\nonumber
\end{eqnarray}
Here we have three different terms. The first term is again a nonstandard elliptic integral and has the form of (\ref{eq:EI6}). The second term is an elementary integral. Again finding the antiderivative requires a case by case analysis, which will not be reproduced here. The third term is again an elliptic integral and we already rewrote it as Legendre's elliptic integral of the third kind. After evaluating the elementary integral we use (\ref{eq:EI6PS}) to evaluate the first term. We simplify all terms and obtain as result
\begin{eqnarray}\label{eq:EI5PS}
&I_{L_{5}}(\chi',k_{2},n)=\frac{n^3\sin\chi'\sqrt{1-k_{2}\sin^2\chi'}}{(n^2-1)(n^2(1-k_{2})+k_{2})(1+n\cos\chi')}-\frac{n(n^2(1-2k_{2})+2k_{2})\tilde{I}_{L_{2}}(\chi',k_{2},n)}{2\left((n^2-1)(n^2(1-k_{2})+k_{2})\right)^{\frac{3}{2}}}+\frac{F_{L}(\chi',k_{2})}{n^2-1}\\
&-\frac{n^2E_{L}(\chi',k_{2})}{(n^2-1)(n^2(1-k_{2})+k_{2})}+\frac{(n^2(1-2k_{2})+2k_{2})\Pi_{L}\left(\chi',k_{2},\frac{n^2}{n^2-1}\right)}{(n^2-1)^2(n^2(1-k_{2})+k_{2})},\nonumber
\end{eqnarray}
where here and in (\ref{eq:EI4PS}) $\tilde{I}_{L_{2}}(\chi',k_{2},n)$ is given by
\begin{eqnarray}\label{eq:Itil2}
\tilde{I}_{L_{2}}(\chi',k_{2},n)=\ln\left(\frac{\sin\chi'\sqrt{\frac{n^2(1-k_{2})+k_{2}}{n^2-1}}+\sqrt{1-k_{2}\sin^2\chi'}}{\left|\sin\chi'\sqrt{\frac{n^2(1-k_{2})+k_{2}}{n^2-1}}-\sqrt{1-k_{2}\sin^2\chi'}\right|}\right).
\end{eqnarray}
Before we proceed to the last integral we conclude the derivation of (\ref{eq:EI4PS}) and (\ref{eq:EI5PS}) with a remark. In both results the parameter of Legendre's elliptic integral of the third kind is given by $n^2/(n^2-1)>1$. Therefore, we use (\ref{eq:PiN}) to rewrite Legendre's elliptic integral of the third kind.\\
The last integral we have to rewrite is $I_{L_{6}}(\chi_{i},\chi,k,n)$. On one hand it occurred in a slightly different form when we rewrote (\ref{eq:EI5}) in terms of elementary functions and Legendre's elliptic integrals of the first, second, and third kind. On the other hand it also occurred when we rewrote $\varphi_{r}(\lambda)$ and $t_{r}(\lambda)$ in terms of elementary functions and Legendre's elliptic integrals of the first, second, and third kind for case 6 and case 10. We recall that in the first case $\chi_{i}$ and $\chi$ are related to $r_{i}$ and $r(\lambda)$ by (\ref{eq:elchi2}) (note that again we omit the argument $\lambda$ in the following) and that in this case we have $k=k_{2}$, where $k_{2}$ is given by (\ref{eq:EM2}). In the second case on the other hand for case 6 $\chi_{i}$ and $\chi$ are related to $r_{i}$ and $r(\lambda)$ by (\ref{eq:elchi3}) (as above we will omit the argument $\lambda$ in the following) while for case 10 $\chi_{i}$ and $\chi$ are related to $r_{i}$ and $r(\lambda)$ by (\ref{eq:elchi3}) or (\ref{eq:elchi4}). In both cases we have $k=k_{3}$, where $k_{3}$ is given by (\ref{eq:EM3}). Now we rewrite (\ref{eq:EI6}) as
\begin{eqnarray}\label{eq:EI4S}
I_{L_{6}}\left(\chi_{i},\chi,k,n\right)=I_{L_{6}}\left(\chi,k,n\right)-I_{L_{6}}\left(\chi_{i},k,n\right),
\end{eqnarray}
where $I_{L_{6}}\left(\chi_{i},k,n\right)$ and $I_{L_{6}}\left(\chi,k,n\right)$ are given in terms of elementary functions and Legendre's elliptic integrals of the first, second, and third kind by
\begin{eqnarray}\label{eq:EI6PS}
&I_{L_{6}}(\chi',k,n)=\frac{n^2\sin(2\chi')\sqrt{1-k\sin^2\chi'}}{4(n-k)(n-1)(1-n\sin^2\chi')}+\frac{F_{L}(\chi',k)}{2(n-1)}-\frac{nE_{L}(\chi',k)}{2(n-k)(n-1)}+\frac{n(n-2)-(2n-3)k}{2(n-k)(n-1)}\Pi(\chi',k,n).
\end{eqnarray}
Note that when we use (\ref{eq:EI6PS}) to rewrite (\ref{eq:EI5}) in terms of (\ref{eq:EI5PS}) we have to replace $n\rightarrow n^2/(n^2-1)$. In addition, for case 6 we have $r_{4}<r_{3}<r_{2}<r_{1}<r_{\text{H}_{\text{i}}}\leq r_{\text{H}_{\text{o}}}$ and thus for the terms associated with $r_{\text{H}_{\text{o}}}$, $r_{\text{H}_{\text{i}}}$, and $r_{\text{H}}$ we now use (\ref{eq:PiN}) to rewrite Legendre's elliptic integral of the third kind. For case 10 on the other hand we can use (\ref{eq:EI6PS}) as is. 
\section{Jacobi's Elliptic Functions and Their Application to Solving Differential Equations}\label{Sec:EFD}
In Sections~\ref{Sec:EoMr} and \ref{Sec:EoMtheta} we use Jacobi's elliptic functions to solve the equations of motion for $r$ and $\vartheta$. Since not all readers may be familiar with these functions in the following we will briefly define them and introduce their basic properties (for a more in-depth overview we recommend the interested reader to consult the introductory book of Hancock \cite{Hancock1917}). For this purpose let us first revisit Legendre's incomplete elliptic integral of the first kind in its general form
\begin{eqnarray}\label{eqn:EIJG}
\tilde{\lambda}=F_{L}\left(\chi,k\right)=\int_{0}^{\chi}\frac{\text{d}\chi'}{\sqrt{1-k\sin^2\chi'}},
\end{eqnarray}
where we recall that $\chi$ and $k$ are the amplitude and the square of the elliptic modulus. In addition, we introduced a new variable $\tilde{\lambda}$, which for now is not related to the Mino parameter $\lambda$. As we can see the variable $\tilde{\lambda}$ is linked to the amplitude $\chi$ via Legendre's incomplete elliptic integral of the first kind and to indicate this one also says $\chi$ is the amplitude of $\tilde{\lambda}$ and writes $\chi=\text{am}\tilde{\lambda}$. Using this notation we can now introduce Jacobi's elliptic functions via the sine and the cosine functions from trigonometry. We start with Jacobi's elliptic sn function. It is often also referred to as \emph{sinus amplitudinis} and is defined by 
\begin{eqnarray} 
\sin\chi=\sin\text{am}\tilde{\lambda}=\text{sn}\left(\tilde{\lambda},k\right).
\end{eqnarray}
Analogously we define Jacobi's elliptic cn function, also referred to as \emph{cosinus amplitudinis}, via
\begin{eqnarray} 
\cos\chi=\cos\text{am}\tilde{\lambda}=\text{cn}\left(\tilde{\lambda},k\right).
\end{eqnarray}
Finally, we have Jacobi's elliptic dn function, sometimes also referred to as \emph{delta amplitude}. It does not have a trigonometric analog and is defined via
\begin{eqnarray}
\sqrt{1-k\sin^2\chi}=\sqrt{1-k\sin^2\text{am}\tilde{\lambda}}=\text{dn}\left(\tilde{\lambda},k\right).
\end{eqnarray}
These three functions form a set of basic elliptic functions. In addition to Jacobi's elliptic sn, cn, and dn functions we can also define several associated elliptic functions. Let us for this purpose introduce the notation $\text{p},\text{q}\in \{\text{s},\text{c},\text{d}\}~\text{with}~\text{p}\neq \text{q}$. Then we can write the associated elliptic functions as
\begin{eqnarray}
\text{pq}\left(\tilde{\lambda},k\right)=\frac{\text{pn}\left(\tilde{\lambda},k\right)}{\text{qn}\left(\tilde{\lambda},k\right)}.
\end{eqnarray}
In this paper we only need two of them, Jacobi's elliptic sc and sd functions (note that in analogy to the tangent in the older literature the former is also often written as tn).\\
Jacobi's elliptic functions are periodic with respect to Legendre's complete elliptic integral of the first kind and fulfill the following periodicity relations \cite{Hancock1917}
\begin{eqnarray}
&\text{sn}\left(\tilde{\lambda},k\right)=-\text{sn}\left(\tilde{\lambda}\pm 2K_{L}(k),k\right)=\text{sn}\left(\tilde{\lambda}\pm 4K_{L}(k),k\right),\\
&\text{cn}\left(\tilde{\lambda},k\right)=-\text{cn}\left(\tilde{\lambda}\pm 2K_{L}(k),k\right)=\text{cn}\left(\tilde{\lambda}\pm 4K_{L}(k),k\right),\\
&\text{dn}\left(\tilde{\lambda},k\right)=\text{dn}\left(\tilde{\lambda}\pm 2K_{L}(k),k\right).
\end{eqnarray}
In addition, all of Jacobi's elliptic functions have the property to solve the differential equation
\begin{eqnarray}\label{eq:LDE}
\left(\frac{\text{d}\chi}{\text{d}\tilde{\lambda}}\right)^2=1-k\sin^2\chi.
\end{eqnarray}
In this paper we now want to use Jacobi's elliptic functions to solve differential equations of the type
\begin{eqnarray}\label{eqn:DEJG}
\left(\frac{\text{d}z}{\text{d}\lambda}\right)^2=a_{4}z^4+a_{3}z^3+a_{2}z^2+a_{1}z+a_{0},
\end{eqnarray}
where the coefficients $a_{0}$, $a_{1}$, $a_{2}$, $a_{3}$, and $a_{4}$ are all real, and $\lambda$ is a variable which parameterises the evolution of $z$. In our case it corresponds to the Mino parameter. For this purpose let us now assume that (\ref{eqn:DEJG}) does not have real or complex double or multiple roots (in these cases we can solve the differential equation using elementary functions). As first step we now use appropriate coordinate transformations $z=f(\sin\chi)$, $z=f(\cos\chi)$, or $z=f(\sin\chi,\cos\chi)$ to bring (\ref{eqn:DEJG}) in a form which is already very similar to (\ref{eq:LDE}). It reads
\begin{eqnarray}\label{eq:LFD}
\left(\frac{\text{d}\chi}{\text{d}\lambda}\right)^2=a_{4}C_{L}\left(1-k\sin^2\chi\right),
\end{eqnarray}
where $C_{L}$ is a new coefficient whose exact form depends on the chosen coordinate transformation. Now we separate variables and integrate from $\chi(\lambda_{i})=\chi_{i}$ to $\chi(\lambda)=\chi$ and get
\begin{eqnarray}\label{eq:LFI}
\lambda-\lambda_{i}=\frac{i_{\chi_{i}}}{\sqrt{a_{4}C_{L}}}\int_{\chi_{i}}^{\chi}\frac{\text{d}\chi'}{\sqrt{1-k\sin^2\chi'}},
\end{eqnarray}
where $i_{\chi_{i}}=\text{sgn}\left(\left.\text{d}\chi/\text{d}\lambda\right|_{\chi=\chi_{i}}\right)$. Now we bring all terms containing the initial values to one side and get
\begin{eqnarray}
\int_{0}^{\chi}\frac{\text{d}\chi'}{\sqrt{1-k\sin^2\chi'}}=i_{\chi_{i}}\sqrt{a_{4}C_{L}}\left(\lambda-\lambda_{i}\right)+\int_{0}^{\chi_{i}}\frac{\text{d}\chi'}{\sqrt{1-k\sin^2\chi'}}.
\end{eqnarray}
When we compare the obtained result with (\ref{eqn:EIJG}) we immediately recognise the similarity of both equations. Now we introduce a new variable $\tilde{\lambda}(\lambda)$ which depends on the original parameter $\lambda$ and get
\begin{eqnarray}
\tilde{\lambda}(\lambda)=i_{\chi_{i}}\sqrt{a_{4}C_{L}}\left(\lambda-\lambda_{i}\right)+\lambda_{\chi_{i},k}=\int_{0}^{\chi}\frac{\text{d}\chi'}{\sqrt{1-k\sin^2\chi'}},
\end{eqnarray}
where we defined 
\begin{eqnarray}
\lambda_{\chi_{i},k}=F_{L}(\chi_{i},k)=\int_{0}^{\chi_{i}}\frac{\text{d}\chi'}{\sqrt{1-k\sin^2\chi'}}.
\end{eqnarray}
Now we can easily see that depending on the coordinate transformation we used to put (\ref{eqn:DEJG}) into the Legendre form (\ref{eq:LFD}) the solutions to (\ref{eqn:DEJG}) are given in terms of different elliptic functions with $\tilde{\lambda}(\lambda)$ as first argument. In our case we will need $\text{sn}(\tilde{\lambda}(\lambda),k)$, $\text{cn}(\tilde{\lambda}(\lambda),k)$, $\text{dn}(\tilde{\lambda}(\lambda),k)$, $\text{sc}(\tilde{\lambda}(\lambda),k)$, and $\text{sd}(\tilde{\lambda}(\lambda),k)$. Using these elliptic functions we can then write the solutions to (\ref{eqn:DEJG}) as $z(\lambda)=f(\text{sn}(\tilde{\lambda}(\lambda),k))$, $z(\lambda)=f(\text{cn}(\tilde{\lambda}(\lambda),k))$, $z(\lambda)=f(\text{dn}(\tilde{\lambda}(\lambda),k))$, $z(\lambda)=f(\text{sc}(\tilde{\lambda}(\lambda),k))$, and $z(\lambda)=f(\text{sd}(\tilde{\lambda}(\lambda),k))$.
\bibliography{Kerr_Metric.bib}
\end{document}